%% file: main.tex
\documentclass[11pt]{book}
\usepackage[dutch, english]{babel}
\usepackage{a4wide}
\usepackage{graphicx}
\usepackage{amsfonts}
\usepackage{amssymb}
\usepackage{latexsym}
\usepackage{url}
\usepackage{type1cm}

\clearpage{\pagestyle{empty}\cleardoublepage}

\newread\testifexists
\def\GetIfExists #1 {\immediate\openin\testifexists=#1
    \ifeof\testifexists\immediate\closein\testifexists\else
    \immediate\closein\testifexists\input #1\fi}

\immediate\openin\testifexists=e
    \ifeof\testifexists\immediate\closein\testifexists\else
    \immediate\closein\testifexists\input e\fipsf

\def\Bbb#1{\setbox0=\hbox{$\tt #1$}  \copy0\kern-\wd0\kern .1em\copy0}

\def\bbf#1{\setbox0=\hbox{$#1$} \kern-.025em\copy0\kern-\wd0
        \kern.05em\copy0\kern-\wd0 \kern-.025em\raise.0433em\box0}

\def\a{\alpha}      \def\b{\beta}         
\def\d{\delta}        \def\e{\varepsilon}
\def\h{\eta}          \def\l{\lambda}     \def\L{\Lambda}
\def\m{\mu}             \def\F{\Phi}        
\def\n{\nu}         \def\j{\psi}    
       
\def\t{\tau}        \def\th{\theta}  
\def\x{\xi}              
\def\w{\omega}        

 \def\LL{{\cal L}} 
\def\pa{\partial} \def\ra{\rightarrow}
\def\na{\nabla}
 
\def\dd{{\rm d}}     \def\ket{\rangle}

\def\deff{\ {\buildrel{\rm def}\over{=}}\ }

\def\fract#1#2{{\textstyle{#1\over#2}}}
\def\ffract#1#2{\raise .3 em\hbox{$\scriptstyle#1$}\kern-.25em/
                \kern-.2em\lower .2 em \hbox{$\scriptstyle#2$}}

\def\half{\fract12}  

\def\part#1#2{{\partial#1\over\partial#2}}

\newcommand{\be}{\begin{eqnarray}}
\newcommand{\ee}{\end{eqnarray}}
\newcommand{\eqn}[1]{(\ref{#1})}
\newcommand{\nn}{\nonumber\\}

\newcommand{\itm}[1]{\item[#1]}

 \newcommand {\eel}[1]{\label{#1}\end{eqnarray}} \newcommand{\crl}[1]{\label{#1}\\} 

\begin{document}

\include{titlemain}
\thispagestyle{empty} \tableofcontents
\clearpage{\pagestyle{empty}\cleardoublepage}
\include{preface}
\mainmatter

\include{chapter1}
\include{chapter2}
\include{chapter3}
\include{chapter31}
\include{chapter4}
\include{chapter41}
\include{chapter5}
\include{chapter6}

\chapter{Conclusions}\label{Conclusions}

In this thesis we critically discussed different approaches to the
cosmological constant problem. The many ways in which the problem
can be phrased often blur the road to a possible solution and the
wide variety of approaches makes it difficult to distinguish real
progress.

So far we can only conclude that in fact none of the approaches
described above is a real outstanding candidate for a solution of
the `old' cosmological constant problem. The most elegant solution
would be a symmetry, that protects the cosmological constant. All
possible candidates we can think of were treated in chapter~3 and
(\ref{imaginaryspace}). However, no symmetry, consistent with
established results, was found. The symmetry analytically
continuing quantum field theory and general relativity to the full
complex space (chapter (\ref{imaginaryspace})) is interesting, but
as it stands, not yet sufficient.

Back-reaction effects, as studied in chapter four are typically
very weak, since they depend on quantum gravitational
interactions. The model proposed by Tsamis and Woodard, chapter
\ref{TsamisWoodardbr}, appeared promising at first, but a closer
study has revealed major obstacles.

To power their influence, almost all back-reaction approaches need
an inflationary background, with a very large bare cosmological
constant, which could take on its `natural' value $\sim M_{P}^2$.
Consequently, an enormously large number of e-folds is needed to
build up any significant effect. This is another reflection of the
fact that there are roughly 120~orders of magnitude between
$M_{P}^4$ and the observed value of the vacuum energy density
which have to be accounted for. Since there exists no convincing
upper bound on the number of e-folds (our universe would just be
enormously bigger than in usual inflationary scenarios) as
discussed in chapter~4, they cannot be ruled out completely.
However, we have seen that each one of them at least so far,
suffers from serious drawbacks.

Perhaps more promising are approaches that suggest a modification
of general relativity, in such a way that the graviton no longer
universally couples to all sources. The infinite volume DGP-model
of section \ref{infiniteed}, appeared as most serious candidate in
this category. However, also this model has some serious
difficulties to overcome. General relativity is a very constrained
theory, and generally even infrared modifications will also be
felt as strongly interacting degrees of freedom at much shorter
distances.

Since the cosmological constant problem lies at the heart of a
fusion between general relativity and quantum mechanics, it is
reasonable to look for modifications of either one, or even both.
Interpreting the data as pointing towards a very small effective
cosmological constant, a modification of GR seems more plausible.
As argued before in this thesis, distances of a tenth of a
millimeter are very important in understanding a small
cosmological constant and quantum mechanics has been tested
thoroughly on much smaller distance scales.

However, since even the sometimes very drastic modifications,
advocated in the proposals we discussed, do not lead to a
satisfactory answer, this could imply that the ultimate theory of
quantum gravity might very well be based on very different
foundations than imagined so far. It certainly is conceivable that
this is partly due to a misunderstanding of quantum mechanics.
Perhaps ultimately the world can perfectly be described by a
local, deterministic theory, describing the fundamental degrees of
freedom of nature, see
\cite{'tHooft:2001ar,'tHooft:2001fb,'tHooft:2001ct,'tHooft:2006sy}
for interesting steps in this direction. Quantum mechanics would
of course still be a perfect description at low energies, but it
would be a statistical theory, not describing the ontological
degrees of freedom. It is clear, that such a radical step would
completely change the nature of a cosmological constant. Whether
or not one should look in this direction for a solution to the
cosmological constant problem, is much too early to say.

Since there is no convincing argument that naturally sets the
cosmological constant to its small value, the anthropic argument
of section~(\ref{AP}) is considered more and more seriously.
However, as stated before, we continue to search for a more
satisfactory solution.

\section{Future Evolution of the Universe}

If indeed the acceleration of the universe will turn out to be the
consequence of a pure non-vanishing cosmological constant, with
equation of state $w=-1$, the universe will accelerate forever.
But, because of the existence of an event horizon in de Sitter
space, the observable universe will continue to shrink, as other
sources, not gravitationally bound to our local group, will vanish
beyond the horizon. Simulations based on the $\Lambda$CDM model
have shown that roughly 100~billion years from now the observable
universe, from our point of view, will consist of only one single
galaxy, a merger product of the Milky Way and the Andromeda
galaxies \cite{Nagamine:2002wi}.

For variable dark energy models, there are several possibilities,
even a future collapse into a Big Crunch can be part of the
scenario \cite{Alam:2003rw,Kallosh:2003mt,Kallosh:2002gf}.

\appendix
\include{appendixa}
\include{appendixb}

\backmatter

\fontsize{10}{10}\selectfont

\bibliography{main}
\addcontentsline{toc}{chapter}{Bibliography}
\bibliographystyle{utcaps}

\end{document}

%% file: titlemain.tex
\thispagestyle{empty}

\begin{flushright}
{\small
ITP-UU-06/34\\
SPIN-06/30\\
gr-qc/yyxxxxx\\
}
\end{flushright}

\begin{center}
{\huge \setlength\baselineskip{30pt} {\bf The Cosmological
Constant Problem,\\
an Inspiration for New Physics}
\par}
\vspace{3\baselineskip}
\end{center}
\begin{center}
{\huge \setlength\baselineskip{30pt} {Stefan Nobbenhuis}}\\
\vspace{1\baselineskip}
{\huge \setlength\baselineskip{30pt} {Ph.D. Thesis, defended June 15. 2006}}\\
\vspace{1\baselineskip} {\huge
\setlength\baselineskip{30pt}{Supervisor: Prof. Dr. G. 't Hooft.}}
\end{center}
\vspace{1\baselineskip}
\begin{center}
Institute for Theoretical Physics \\
Utrecht University, Leuvenlaan 4\\ 3584 CC Utrecht, the
Netherlands\medskip \\ and
\medskip \\ Spinoza Institute \\ Postbox 80.195 \\ 3508 TD
Utrecht, the Netherlands \\
S.J.B.Nobbenhuis@phys.uu.nl\\
\end{center}
\vspace{2\baselineskip}
\begin{center}
\begin{minipage}[h]{13cm}
We have critically compared different approaches to the
cosmological constant problem, which is at the edge of elementary
particle physics and cosmology. This problem is deeply connected
with the difficulties formulating a theory of quantum gravity.
After the 1998 discovery that our universe's expansion is
accelerating, the cosmological constant problem has obtained a new
dimension. We are mainly interested in the question why the
cosmological constant is so small. We have identified four
different classes of solutions: a symmetry, a back-reaction
mechanism, a violation of (some of) the building blocks of general
relativity, and statistical approaches.\\
In this thesis we carefully study all known potential candidates
for a solution, but conclude that so far none of the approaches
gives a satisfactory solution. A symmetry would be the most
elegant solution and we study a new symmetry under transformation
to imaginary spacetime.
\end{minipage}
\end{center}

\newpage
\thispagestyle{empty}
\begin{quote}
\end{quote}

%% file: chapter1.tex
\chapter{The Cosmological Constant Problem}

In this chapter we carefully discuss the different aspects of the
cosmological constant problem. How it occurs, what the assumptions
are, and where the difficulties lie in renormalizing vacuum
energy. Different interpretations of the cosmological constant
have been proposed and we briefly discuss these. We subsequently,
as a warm-up, point at some directions to look at for a solution,
before ending this chapter by giving an outline of the rest of the
thesis.

\section{The `Old' Problem, Vacuum Energy as Observable Effect}

In quantum mechanics, the energy spectrum, $E_n$, of a simple
harmonic oscillator\footnote{See Appendix A for conventions and
some definitions used throughout this thesis, e.g. $\hbar=1$.} is
given by $E_n=(n+1/2)\omega$. The energy of the ground state, the
state with lowest possible energy, is non-zero, contrary to
classical mechanics. This so-called zero-point energy is usually
interpreted by referring to the uncertainty principle: a particle
can never completely come to a halt.

Free quantum field theory is formulated as an infinite series of
simple harmonic oscillators. The energy density of the vacuum in
for example free scalar field theory therefore receives an
infinite positive contribution from the zero-point energies of the
various modes of oscillation:
\begin{equation}\label{infgen}
\langle\rho\rangle =
\int_{0}^{\infty}\frac{d^3k}{(2\pi)^{3}}\frac{1}{2}\sqrt{k^{2} +
m^{2}},
\end{equation}
and we set $\omega (k) = \sqrt{k^2+m^2}$. With a UV-cutoff
$k_{\max}$, regularizing the a priori infinite value for the
vacuum energy density, this integral diverges\footnote{This is
just the leading divergence, see section (\ref{Renormalization})
for a more precise discussion.} as $k_{max}^{4}$.

The filling of the `Dirac sea' in the quantization of the free
fermion theory leads to a downward shift in the vacuum energy with
a similar ultraviolet divergence. The Hamiltonian is:
\begin{equation}
H=\int\frac{d^3k}{(2\pi)^3}\;\omega\sum_s\left(a^{s\dag}a^s -
b^{s\dag}b^s\right),
\end{equation}
where the $b_{k}^{s\dag}$ creates negative energy. Therefore we
have to require that these operators satisfy anti-commutation
relations:
\begin{equation}
\{b_{k}^r,b_{k'}^{s\dag}\}=(2\pi)^3\delta^{(3)}(k-k')\delta^{rs},
\end{equation}
since this is symmetric between $b^r$ and $b^{r\dag}$, we can
redefine the operators as follows:
\begin{equation}
\tilde{b}^{s}\equiv b^{s\dag};\quad\quad \tilde{b}^{s\dag}\equiv
b^s,
\end{equation}
which obey the same anti-commutation relations. Now the second
term in the Hamiltonian becomes:
\begin{equation}
-\omega b^{s\dag}b^{s} = +\omega \tilde{b}^{s\dag}\tilde{b}^{s} -
(\mbox{const.});\quad\quad(\mbox{const})=\int^{k_{max}}\frac{d^3k}{(2\pi)^3}\omega
(k).
\end{equation}
Now we choose $|0\rangle$ to be the vacuum state that is
annihilated by the $a^s$ and $\tilde{b}^s$ and all excitations
have positive energy. All the negative energy states are filled;
this is the Dirac sea. A hole in the sea corresponds to an
excitation of the ground state with positive energy, compared to
the ground state. The infinite constant contribution to the vacuum
energy density has the same form as in the bosonic case, but
enters with opposite sign.

Spontaneous symmetry breaking gives a finite but still possibly
large shift in the vacuum energy density. In this case,
\begin{equation}
\mathcal{L}=\frac{1}{2}g^{\mu\nu}\partial_\mu\phi\partial_\nu\phi
- V(\phi)
\end{equation}
where the potential $V(\phi)$ is given by:
\begin{equation}
V(\phi)= -\mu^2(\phi^{\dag}\phi) + \lambda(\phi^{\dag}\phi)^2
+\epsilon_0,
\end{equation}
where $\mu$ and $\lambda$ both are positive constants and
$\epsilon_0$ is a constant with arbitrary sign. The potential is
at its minimum value for $|\phi|=\sqrt{\mu^2/2\lambda}$, leading
to a shift in the energy density of the ground state:
\begin{equation}
\langle\rho\rangle = \epsilon_0 - \frac{\mu^4}{4\lambda}.
\end{equation}
The spontaneous breaking of the weak interaction $SU(2) \times
U(1)$ symmetry and of the strong interaction chiral symmetry both
would be expected to shift the vacuum energy density in this way.
Through the additive constant $\epsilon_0$ the minimum of the
potential either after, or before symmetry breaking, can be tuned
to any value one likes, but within field theory, this value is
completely arbitrary.

More generally, in elementary particle physics experiments, the
absolute value of the vacuum energy is unobservable.
Experimentally measured particle masses, for example, are energy
differences between the vacuum and certain excited states of the
Hamiltonian, and the absolute vacuum energy cancels out in the
calculation of these differences.

In GR however, each form of energy contributes to the
energy-momentum tensor $T_{\mu\nu}$, hence gravitates and
therefore reacts back on the spacetime geometry, as can be seen
from Einstein's equations:
\begin{equation}\label{formbelow}
R_{\mu\nu} - \frac{1}{2}Rg_{\mu\nu} - \Lambda g_{\mu\nu} = -8\pi
GT_{\mu\nu}.
\end{equation}
As it stands, the cosmological constant is a free parameter that
can be interpreted as the curvature of empty spacetime. However,
Lorentz invariance tells us that the energy-momentum tensor of the
vacuum looks like:
\begin{equation}
T^{\mu\nu}_{vac} = -\langle\rho\rangle g^{\mu\nu},
\end{equation}
since the absence of a preferred frame for the vacuum means that
$T^{\mu\nu}_{vac}$ must be the same for all observers connected by
Lorentz transformations. Apart from zero, there is only one
isotropic tensor of rank two, which is the metric tensor. If we
move the CC term to the right-hand-side of the equation
(\ref{formbelow}), it looks exactly the same:
\begin{equation}
T^{\mu\nu}_{vac} = -\frac{\Lambda}{8\pi G}g^{\mu\nu}.
\end{equation}
This interpretation of a cosmological constant measuring the
energy density of the vacuum, was first explicitly given by
Zel'dovich in 1968 \cite{Zeldovich1,Zeldovich2}.

We can therefore define an effective cosmological
constant\footnote{Note that using this definition we use units in
which the cosmological constant has dimension $\mbox{GeV}^2$
throughout.}:
\begin{equation}
\Lambda_{eff} = \Lambda + 8\pi
G\langle\rho\rangle,\quad\quad\mbox{or}\quad\quad\rho_{vac}=\langle\rho\rangle
+ \Lambda/8\pi G.
\end{equation}
The vacuum energy density calculated from normal quantum field
theory thus has a potentially significant observable effect
through the coupling with gravity. It contributes to the effective
cosmological constant and an infinite value would possibly
generate an infinitely large spacetime curvature through the
semiclassical Einstein equations:
\begin{equation}
G_{00} = -8\pi G\langle T_{00}\rangle.
\end{equation}
Note at this point that only the effective cosmological constant,
$\Lambda_{eff}$, is observable, not $\Lambda$, so the latter
quantity may be referred to as a `bare', or perhaps classical
quantity that has to be `dressed' by quantum corrections,
analogously to all other physical parameters in ordinary quantum
field theory. If one believes quantum field theory to be correct
all the way up to the Planck scale, at $10^{19}$~GeV, then this
scale provides a natural ultraviolet cutoff to all field theory
processes like eqn.~(\ref{infgen}). Such a cutoff however, would
lead to a vacuum energy density of
$(10^{19}~\mbox{GeV})^4=10^{76}~\mbox{GeV}^4$, which is roughly
123~orders of magnitude larger than the currently observed value:
\begin{equation}
\rho_{vac} \simeq 10^{-47}~\mbox{GeV}^{4} \sim
10^{-29}~\mbox{g/cm}^{3}\sim (0,1~\mbox{mm})^{-4} \sim
(10^{-12}~\mbox{s})^{-4}.
\end{equation}
This means that the bare cosmological constant $\Lambda$ has to be
fine-tuned to a stunning 123 decimal places, in order to yield the
correct physical result\footnote{This often mentioned factor
$\sim$~120, depends on the dimension used for energy density. In
Planckian units, the factor~120 is the correct one, relating
dimensionless numbers.}. As is well known, even if we take a TeV
scale cut-off the difference between experimental and theoretical
results still requires a fine-tuning of about 50 orders of
magnitude. Even a cutoff at for example the QCD scale (at $\sim
200$~MeV), worrying only about zero-point energies in quantum
chromodynamics, would not help much; such a cutoff would still
lead to a discrepancy of about 40~orders of magnitude. The answer
clearly has to lie somewhere else. Even non-perturbative effects,
like ordinary QCD instantons, would give far too large a
contribution if not cancelled by some fundamental mechanism.

Physicists really started to worry about this in the
mid-seventies, after the success of spontaneous symmetry breaking.
Veltman \cite{Veltman2} in 1975 ``\ldots~concluded that this
undermines the credibility of the Higgs mechanism.''

We have no understanding of why the effective cosmological
constant, $\Lambda_{eff}$ is so much smaller than the vacuum
energy shifts generated in the known phase transitions of particle
physics, and so much smaller again than the underlying field
zero-point energies. No symmetry is known that can protect the
cosmological constant to such a small value\footnote{See however
chapter (\ref{imaginaryspace}) for an attempt.}. The magnificent
fine-tuning needed to obtain the correct physical result seems to
suggest that we miss an important point. In this thesis we give an
overview of the main ideas that have appeared in trying to figure
out what this point might be. For instance, one might suspect that
it was naive to believe that vacuum energy, like any other form of
energy contributes to the energy-momentum tensor and gravitates.
It is however an assumption that is difficult to avoid.

In conclusion, the question is why is the effective cosmological
constant so close to zero? Or, in other words, why is the vacuum
state of our universe (at present) so close to the classical
vacuum state of zero energy, or perhaps better, why is the
resulting four-dimensional curvature so small, or why does Nature
prefer a flat spacetime? Apparently spacetime is such, that it
takes a lot of energy to curve it, while stretching it is (almost)
for free, since the cosmological constant is (almost) zero. This
is quite contrary to properties of objects from every day
experience, where bending requires much less energy than
stretching.

As a nice example, Pauli in the early 1920's was way ahead of his
time when he wondered about the gravitational effect of the
zero-point energy of the radiation field. He used a cutoff for the
radiation field at the classical electron radius and realized that
the entire universe ``could not even reach to the moon"
\cite{Paulimoon}. The calculation is straightforward and also
restated in \cite{Straumann1}. The vacuum energy-density of the
radiation field is:
\begin{equation}
\langle\rho\rangle
=\frac{8\pi}{(2\pi)^3}\int^{\omega_{max}}_0\frac{1}{2}\omega^3d\omega
=\frac{1}{8\pi^2}\omega_{max}^4
\end{equation}
With the cutoff $\omega_{max} = 2\pi/\lambda_{max} = \frac{2\pi
m_e}{\alpha}$. This implies for the radius of the universe:
\begin{equation}
R \sim \frac{\alpha^2}{\pi}\frac{M_{Pl}}{m_e^2}\sim 31~ \mbox{km},
\end{equation}
indeed far less then the distance to the moon.

Nobody else seems to have been bothered by this, until the late
1960's when Zel'dovich \cite{Zeldovich1,Zeldovich2} realized that,
even if one assumes that the zero-point contributions to the
vacuum energy density are exactly cancelled by a bare term, there
still remain very problematic higher order effects. On dimensional
grounds the gravitational interaction between particles in vacuum
fluctuations would be of order $G\mu^6$, with $\mu$ some cutoff
scale. This corresponds to two-loop Feynman diagrams. Even for
$\mu$ as low as 1 GeV, this is about 9~orders of magnitude larger
than the observational bound.

This illustrates that all 'naive' predictions for the vacuum
energy density of our universe are greatly in conflict with
experimental facts, see table (\ref{energiestoCC}) for a list of
order-of-magnitude contributions from different sources. All we
know for certain is that the unification of quantum field theory
and gravity cannot be straightforward, that there is some
important concept still missing from our understanding. Note that
the divergences in the cosmological constant problem are even more
severe than in the case of the Higgs mass: The main divergences
here are quartic, instead of quadratic. It is clear that the
cosmological constant problem is one of the major obstacles for
quantum gravity and cosmology to further progress.

\begin{table}[h!]\label{energiestoCC}
\caption{Energy density of the Quantum Vacuum}
\begin{tabular}{|l|l|c|}
\hline &&\\
  Observed & $\rho\simeq 10^{-30}$ & $\mbox{gcm}^{-3}$ \\
 \hline &&\\
  Quantum Field Theory &  $\rho=\infty$ & $\mbox{gcm}^{-3}$ \\
  \hline &&\\
  Quantum Gravity &  $\rho\simeq 10^{+90}$ & $\mbox{gcm}^{-3}$ \\
   \hline &&\\
  Supersymmetry &  $\rho\approx 10^{+30}$&$\mbox{gcm}^{-3}$ \\
   \hline &&\\
  Higgs Potential &  $\rho\approx -10^{+25}$ & $\mbox{gcm}^{-3}$ \\
   \hline &&\\
  Other Sources &  $\rho\approx \pm 10^{+20} $&$ \mbox{gcm}^{-3}$ \\ \hline
\end{tabular}
\end{table}

\subsection{Reality of Zero-Point Energies}

The reality of zero-point energies, i.e. their observable effects,
has been a source of discussion for a long time and so far without
a definite conclusion about it. Besides the gravitational effect
in terms of a cosmological constant, there are two other
observable effects often ascribed to the existence of zero-point
energies: the Lamb shift and the Casimir effect.

A recent investigation by Jaffe \cite{Jaffe2005} however,
concluded that neither the experimental confirmation of the
Casimir effect, nor of the Lamb shift, established the reality of
zero-point fluctuations. However, no completely satisfactory
description of, for example QED, is known in a formulation without
zero-point fluctuations.

Note that in the light-cone formulation, a consistent description
of e.g. QED can be given, with the vacuum energy density
automatically set to zero. This is accomplished by selecting two
preferred directions, and thus breaks Lorentz invariance. However,
there is no physical reason to select these coordinates, and
moreover, a discussion of tadpole diagrams and cosmological terms
in de Sitter spacetime becomes problematic in these coordinates.

\subsection{Repulsive Gravity}

A positive cosmological constant gives a repulsive gravitational
effect. This can easily be seen by considering the static
gravitational field created by a source mass $M$ at the origin,
with density $\rho_{M} = M\delta^{3}(r)$. For weak fields
$g_{\mu\nu}\simeq\eta_{\mu\nu}$ is the usual Lorentz metric. If we
further assume a non-relativistic regime with
$T_{00}\simeq\rho_{M}$ the $(0,0)$-component of the Einstein
equation reads:
\begin{equation}
G_{00} + \Lambda = -8\pi G_{N}\rho_{M},
\quad\quad\mbox{with}\quad\quad G_{00} = R_{00} + \frac{1}{2}R.
\end{equation}
In the non-relativistic case we also expect that $T_{ij}\ll
T_{00}$ which is equivalent to saying that we neglect pressure and
stress compared to matter density. Therefore we can set $G_{ij} -
\Lambda g_{ij}\simeq 0$, implying $R_{ij}\simeq(1/2R +
\Lambda)g_{ij}$ and thus the curvature scalar becomes: $R =
g^{\mu\nu}R_{\mu\nu}\simeq -R_{00} + 3(1/2R + \Lambda)$ or
$R\simeq 2R_{00} - 6\Lambda$. Substituting this back in the
equation for the $(0,0)$ component we find:
\begin{equation}
R_{00} - \Lambda = -4\pi G_{N}\rho_{M}.
\end{equation}
Furthermore we can derive that within our approximation
$R_{00}\simeq(-1/2)\nabla^{2}g_{00}$. Finally recalling that
Newton's potential $\phi$ is related to the deviation of the
$(0,0)$-component of the metric tensor from $-\eta_{00} = 1$
(through $-g_{00}\simeq 1 + 2\phi$) we are led to the equation:
\begin{equation}\label{nonrellambda}
\nabla^{2}\phi = 4\pi G_{N}\left(\rho_{M} - \frac{\Lambda}{4\pi
G_{N}}\right).
\end{equation}
This is of course nothing but Poisson's equation for the Newton
potential with an additional term from the CC whose sign depends
on that of $\Lambda$. For $\Lambda > 0$ the original gravitational
field is diminished as though there were an additional repulsive
interaction. These features are confirmed by explicitly solving
the above equation:
\begin{equation}
\phi = -\frac{G_{N}M}{r} - \frac{1}{6}\Lambda r^{2}.
\end{equation}
We are thus led to the expected gravitational potential plus a new
contribution which is like a ``harmonic oscillator'' potential,
repulsive for $\Lambda > 0$.

\section{Two Additional Problems}

Actually, after the remarkable discoveries and subsequent
confirmations starting in 1997
(SN)\cite{Riess1998,Garnavich1998,Filippenko1998,Perlmutter1998,Perlmutter1998-2,Riess2000,Riess2001,Tonry2003,Knop2003,Barris2003,Riess2004}
(WMAP) \cite{Bennett2003,Spergel2003} (Boomerang)
\cite{Netterfield2001,deBernardis2000} (SdSS)
\cite{Tegmark2003,Afshordi2003} (Hubble) \cite{Freedman2000} that
the universe really is accelerating its expansion (more about this
in the next chapter), there appear to be at present at least three
cosmological constant problems.

The first, or sometimes called ``old" cosmological constant
problem is why is the effective cosmological constant so
incredibly small, as described in the previous section.

The second problem is, if it is so small, then why is it not
exactly equal to zero? Often in physics it is a lot easier to
understand why a parameter is identically zero, than why it is a
very small number.

And a third question may be posed, based on the measured value of
the effective cosmological constant. The energy density of the
vacuum that it represents, appears to be of the same order of
magnitude as the present matter energy density in the universe.
This is quite peculiar, since, as we we will see in chapter
(\ref{Cosmology}), vacuum energy density, denoted
$\Omega_{\Lambda}$, remains constant during the evolution of the
universe, whereas the matter energy density, $\Omega_{m}$
obviously decreases as the universe grows larger and larger. If
the two energy densities are of the same order of magnitude
nowadays, this means that their ratio,
$\Omega_{\Lambda}/\Omega_{m}$ had to be infinitesimal in the early
universe, but fine-tuned to become equal now. Therefore, one
obviously starts to wonder whether we are living in some special
epoch, that causes these two forms of energy density to be roughly
equal in magnitude. This has become known as the ``cosmic
coincidence problem" and is also sometimes phrased as the ``Why
now?" problem. In this thesis we mainly concentrate on the first
problem.

\section{Renormalization}\label{Renormalization}

In quantum field theory the value of the vacuum energy density has
no observational consequences. We can simply rescale the zero
point of energy, which amounts to adding or subtracting a constant
to the action, without changing the equations of motion. It can
also be done more elegantly, by a book-keeping method called
``normal ordering'', denoted by placing a quantity between
semicolons, i.e. $:T_{\mu\nu}:$. In this prescription, one demands
that wherever a product of creation and annihilation operators
occurs, it is understood that all creation operators stand to the
left of all annihilation operators. This differs from the original
notation by commutator terms which renormalize vacuum energy and
mass, since:
\begin{equation}
\langle\psi |:T_{\mu\nu}:|\psi\rangle = \langle\psi
|T_{\mu\nu}|\psi\rangle - \langle 0|T_{\mu\nu}|0\rangle.
\end{equation}
However, it is not a very satisfactory way of dealing with the
divergences, considering that we have seen that vacuum energy is
observable, when we include gravity. By way of comparison, we note
that in solid state physics, through X-ray diffraction, zero-point
energy is measurable. Here, a natural UV-cutoff to infinite
integrals like (\ref{infgen}) exists, because the system is
defined on a lattice. And indeed, this zero-point energy turns out
to be $\omega/2$ per mode.

So let us return to eqn. (\ref{infgen}). In fact, there are not
just quartic divergences in the expression for the vacuum energy,
but also quadratic and logarithmic ones. More precisely, for a
field with spin $j>0$:
\begin{eqnarray}\label{infgen2}
\langle\rho\rangle &=&
\frac{1}{2}(-1)^{2j}(2j+1)\int_{0}^{\Lambda_{UV}}\frac{d^3k}{(2\pi)^{3}}\sqrt{\mathbf{k}^{2}
+ m^{2}} \nonumber\\
&=& \frac{(-1)^{2j}(2j+1)}{16\pi^2}\Big(\Lambda_{UV}^4 +
m^2\Lambda_{UV}^2 -
\frac{1}{4}m^4\left[\log\left(\Lambda_{UV}^2/m^2\right)
+\frac{1}{8}-\frac{1}{2}\log 2\right]\nonumber\\
&+& \mathcal{O}\left(\Lambda_{UV}^{-1}\right)\Big).
\end{eqnarray}
where we have imposed an ultraviolet cutoff $\Lambda_{UV}$ to the
divergent integral. This shows that quartic divergences to the
vacuum energy come from any field, massive or massless\footnote{It
should be noted however, that since this cutoff is only on spatial
momenta, it violates Lorentz invariance. It was shown in
\cite{DeWitt:1975ys, Akhmedov:2002ts} that the quartic divergence
for the energy density and pressure of a free scalar field in fact
do not describe vacuum energy, but a homogeneous sea of background
radiation. However, when one does the calculation correctly, the
quartic divergences are again recovered \cite{Ossola:2003ku}. When
evaluating the integral using dimensional regularization, one
finds: $\rho=-m^4/(64\pi^2)\left[\log\Lambda_{UV}^2/m^2 +
3/2\right]$; no quartic or quadratic term.}.

In combination with gravity, these divergences must be subtracted,
as usual, by a counterterm, a `bare' cosmological constant, as we
will see in the following. Then there are quadratic and
logarithmic divergences, but these only arise for massive fields.
The quadratic divergences have to be treated in the same way as
the quartic ones. The logarithmic divergences are more
problematic, because, even after having been cancelled by
counterterms, their effect is still spread through the
renormalization group. Since the theory is not renormalizable, the
infinities at each loop order have to be cancelled separately with
new counterterms, introducing new free parameters in the theory.
Because of dimensional reasons, the effects of these new terms are
small, and are only felt in the UV. However, this is an
unsatisfactory situation, since much predictability is lost.
Moreover, knowledge from the past has taught us that this signals
that we do not understand the short distance behavior of the
theory, so clearly something better is needed.

The subtraction procedure between bare terms and quantum
corrections depends on the energy scale $\mu$ in the divergent
integral. The renormalization of the cosmological constant
therefore must be performed such that at energy scales where it is
measured today, we get the observed value. The renormalization
condition, in other words, has to be chosen at some fixed energy
scale $\mu_c$. For the cosmological constant the renormalization
condition becomes:
\begin{equation}
\Lambda_{eff}(\mu_c)= \Lambda_{vac}(\mu_c) + \Lambda_{ind}(\mu_c),
\end{equation}
where $\Lambda_{eff}$ is the observable, physical cosmological
constant, $\Lambda_{vac}$ is the bare cosmological constant in
Einstein's equations, and $\Lambda_{ind}$ are the quantum
corrections. The coupling constants of the theory become a
function of $\mu$, where $\mu$ is called the renormalization
scale. Physical observables must be independent of $\mu$ and this
is expressed by the renormalization group equations, or more
precisely, the one-dimensional subgroup of scale transformations,
sending $\mu\rightarrow \alpha\mu$, with $\alpha$ some constant.

Despite the fact that there exists no renormalizable theory
combining gravity and particle physics, the hope is that under
certain restrictions, these considerations still make sense. It
has been tried, in vain, to argue that this type of running with
energy scale $\mu$ under the renormalization group, could explain
the very small value $\Lambda_{eff}$ nowadays. We will discuss
this scenario in section (\ref{RenormalizationGroup}).

The restrictions mentioned above are quite severe. General
relativity is a non-linear theory, which implies that gravitons
not only transmit the force of gravity, but also set up a
gravitational field themselves, similar, mutatis mutandis, to
gluons in QCD. Gravitons `feel' a gravitational field just as much
as for example photons do. The approach usually taken, also in
most parts of this thesis, is to consider these graviton
contributions as a metric perturbation on a background spacetime,
$g_{\mu\nu} = g_{\mu\nu}^{bg} + h_{\mu\nu}$, where $h_{\mu\nu}$
represents the gravitons, or gravitational waves. The
gravitational waves are then treated as a null fluid, and
considered to contribute to the energy momentum tensor, i.e. to
the right-hand-side of Einstein's equation, instead of the
left-hand-side. The combined action for gravity and matter fields
can be expanded in both $g^{bg}$ and $h$, and Feynman rules can be
derived, see e.g. the lectures by Veltman \cite{Veltman:1975vx}.

As discussed above, at higher orders the gravitational corrections
are out of control. However, if one truncates the expansion at a
finite number of loops, then the finite number of divergent
quantities that appear can be removed by renormalizing a finite
number of physical quantities, see e.g.
\cite{Wald:1992fr,BirrelDavies,Wald:1995yp} for a more detailed
description of this setup.

This procedure provides a way to use the semi-classical Einstein
equations:
\begin{equation}
G_{\mu\nu} - \Lambda g_{\mu\nu} = -8\pi G\langle
T_{\mu\nu}\rangle,
\end{equation}
where $\langle T_{\mu\nu}\rangle$ is the quantum expectation value
of the energy momentum tensor. For this to make any sense at all,
usually the expansion is truncated already at one-loop level. In
this semi-classical treatment, the gravitational field is treated
classically, while the matter fields, including the graviton to
one-loop order, are treated quantum mechanically.

To say something meaningful about $\langle T_{\mu\nu}\rangle$ one
now has to renormalize this parameter, since the expectation value
in principle diverges terribly. To do this in a curved background
is not trivial. First one has to introduce a formal regularization
scheme, which renders the expectation value finite, but dependent
on an arbitrary regulator parameter. Several choices for
regularization are available, see e.g. \cite{BirrelDavies}.

One option is to separate the spacetime points at which the fields
in $T_{\mu\nu}$ are evaluated and then average over the direction
of separation. This is a covariant regularization scheme which
leaves $\langle T_{\mu\nu}\rangle$ dependent on an invariant
measure of the distance between the two points. The price to pay
for any regularization scheme is the breaking of conformal
invariance, so massless fields no longer have traceless
stress-tensors. Often this distance is chosen to be one-half times
the square of the geodesic distance between them, denoted by
$\sigma$. Asymptotically, the regularized expression then becomes:
\begin{equation}
\langle T_{\mu\nu}\rangle\sim A\frac{g_{\mu\nu}}{\sigma^2} +
B\frac{G_{\mu\nu}}{\sigma} + \left(C_1H_{\mu\nu}^{(1)} +
C_2H_{\mu\nu}^{(2)}\right)\ln\sigma,
\end{equation}
where $A,B,C_1$ and $C_2$ are constants, $G_{\mu\nu}$ is the
Einstein tensor and the $H_{\mu\nu}^{(1),(2)}$ are covariantly
conserved tensors, quadratic in the Riemann tensor. So this gives
a correction to the bare cosmological constant present in the
Einstein-Hilbert action (term linear in $g_{\mu\nu}$), a
correction to Newton's constant (term $\propto G_{\mu\nu}$) plus
higher order corrections to $T_{\mu\nu}$. The tensors
$H_{\mu\nu}^{(1),(2)}$ are the functional derivatives with respect
to the metric tensor of the square of the scalar curvature and of
the Ricci tensor, respectively. With $\delta/\delta g_{\mu\nu}$
the Euler-derivative, one arrives at (see e.g.
\cite{BirrelDavies}):
\begin{eqnarray}
H_{\mu\nu}^{(1)}&\equiv& \frac{1}{\sqrt{-g}}\frac{\delta}{\delta
g^{\mu\nu}}\left[\sqrt{-g}R^2\right]\nonumber\\
&=& 2\nabla_{\nu}\nabla_{\mu}R
-2g_{\mu\nu}\nabla_{\rho}\nabla^{\rho}R -
\frac{1}{2}g_{\mu\nu}R^2+2RR_{\mu\nu},
\end{eqnarray}
and
\begin{eqnarray}
H_{\mu\nu}^{(2)}&\equiv& \frac{1}{\sqrt{-g}}\frac{\delta}{\delta
g^{\mu\nu}}\left[\sqrt{-g}R_{\alpha\beta}R^{\alpha\beta}\right]=2\nabla_{\alpha}\nabla_{\nu}R_{\mu}^{\alpha} -\nabla_{\rho}\nabla^{\rho}R_{\mu\nu}\nonumber\\
&& -\frac{1}{2}g_{\mu\nu}\nabla_{\rho}\nabla^{\rho}R
-\frac{1}{2}g_{\mu\nu}R_{\alpha\beta}R^{\alpha\beta} +
2R_{\mu}^{\rho}R_{\rho\nu}.
\end{eqnarray}
The divergent parts of $\langle T_{\mu\nu}\rangle$ can then be
taken into account by adding counterterms to the standard
Einstein-Hilbert action:
\begin{equation}
S_{G}=\frac{1}{16\pi G_0}\int
d^4x\sqrt{-g}\left(R-2\Lambda_0+\alpha_0R^2+\beta_0R_{\alpha\beta}R^{\alpha\beta}\right).
\end{equation}
We did not include a term
$R_{\alpha\beta\gamma\delta}R^{\alpha\beta\gamma\delta}$, since it
can be absorbed in the other counterterms, using that the
combination:
\begin{equation}\label{topinv}
\int d^4x\sqrt{-g}\left(R^2 - 4R_{\mu\nu}^2 +
R_{\alpha\beta\mu\nu}^2\right),
\end{equation}
is a topological invariant and does not affect the field
equations. Furthermore, in pure gravity, with no matter fields,
the Einstein equations tell us that $R=0$ and $R_{\mu\nu}=0$. In
this case the counterterms are unphysical, which means, together
with the observation that (\ref{topinv}) is a topological
invariant, that pure gravity has no infinities at one-loop level
and thus is one-loop renormalizable.

If we now again include the matter action and vary both with
respect to the metric in order to obtain the field equations, and
replace the classical $T_{\mu\nu}$ with the quantum mechanical
expectation value $\langle T_{\mu\nu}\rangle$ we obtain:
\begin{equation}
G_{\mu\nu} - \Lambda_0g_{\mu\nu} = -8\pi G_0\langle
T_{\mu\nu}\rangle - \alpha_0H_{\mu\nu}^{(1)}
-\beta_0H_{\mu\nu}^{(2)}.
\end{equation}
The divergent parts in $\langle T_{\mu\nu}\rangle$ may be removed
by renormalizing the bare coupling constants,
$G_0,\Lambda_0,\alpha_0$ and $\beta_0$ after which they become the
physical parameters of the theory. Of course, in order for this
semiclassical treatment to be a good approximation, the terms
$H_{\mu\nu}^{(1),(2)}$ are assumed to be very small.

This explicitly clarifies some of the statements we made in this
section. At the one-loop level, renormalization of $G$, $\Lambda$,
and the coupling constants of two new geometrical tensors,
suffices to render the theory finite. At one-loop order, we had to
introduce already two new free parameters to absorb the infinities
and this only becomes worse at higher orders.

\section{Interpretations of a Cosmological Constant}

Since the 1970's the standard interpretation of a CC is as vacuum
energy density. It is an interesting philosophical question,
closely connected with the reality of spacetime, whether this
interpretation really differs from the original version where the
CC was simply a constant, one of the free parameters of the
universe. Perhaps the CC is determined purely by the fabric
structure of spacetime. However, we won't go into those
discussions in this thesis. In this section we consider different
interpretations that have sometimes been put forward.

\subsection{Cosmological Constant as Lagrange Multiplier}

The action principle for gravity in the presence of a CC can be
written:
\begin{eqnarray}
A &=& \frac{1}{16\pi G}\int d^{4}x\sqrt{-g}(R+2\Lambda)\nonumber\\
&=& \frac{1}{16\pi G}\int d^4x \sqrt{-g}R + \frac{1}{8\pi G}\int
d^4x\sqrt{-g}\Lambda
\end{eqnarray}
This can be viewed as a variational principle where the integral
over $R$ is extremized, subject to the condition that the 4-volume
of the universe remains constant. The second term has the right
structure, to mathematically think of the CC as a Lagrange
multiplier ensuring the constancy of the 4-volume of the universe
when the metric is varied.

It does not help at all in solving the cosmological constant
problem, but it might be useful to have a different perspective.

\subsection{Cosmological Constant as Constant of Integration}

If one assumes that the determinant $g$ of $g_{\mu\nu}$ is not
dynamical we could admit only those variations which obey the
condition $g^{\mu\nu}\delta g_{\mu\nu} = 0$ in the action
principle. The trace part of Einstein's equation then is
eliminated. Instead of the standard result, after varying the
standard action, without a cosmological constant, we now obtain:
\begin{equation}
R^{\mu\nu} - \frac{1}{4}g^{\mu\nu}R = -8\pi G\left(T^{\mu\nu} -
\frac{1}{4}g^{\mu\nu}T^{\lambda}_{\lambda}\right)
\end{equation}
just the traceless part of Einstein's equation. The general
covariance of the action still implies that $T^{\mu\nu}_{;\mu} =0$
and the Bianchi identities $(R^{\mu\nu} -
\frac{1}{2}g^{\mu\nu}R)_{;\mu} = 0$ continue to hold. These two
conditions imply:
\begin{equation}
\partial_{\mu}R = 8\pi G\partial_{\mu}T\quad\quad \Rightarrow\quad\quad R - 8\pi
GT = \mbox{constant} \equiv -4\Lambda
\end{equation}
Defining the constant term this way, we arrive at:
\begin{equation}
R^{\mu\nu} - \frac{1}{2}g^{\mu\nu}R - \Lambda g^{\mu\nu} = -8\pi
GT^{\mu\nu}
\end{equation}
which is precisely Einstein's equation in the presence of a CC,
only in this approach is has not so much to do with any term in
the action or vacuum fluctuations. It is merely a constant of
integration to be fixed by invoking ``suitable'' boundary
conditions for the solutions.

There are two main difficulties in this approach, which has become
known as the `unimodular' approach. The first is how to interpret
the assumption that $g$ must remain constant when the variation is
performed. See \cite{Sorkin} for some attempts in this direction
and section (\ref{unimodular}) for more details on this. A priori,
the constraint keeping the volume-element constant, is just a
gauge restriction in the coordinate frame chosen. The second
problem is that it still does not give any control over the value
of the CC which is even more worrying in case of a non-zero value
for the CC.

\section{Where to Look for a Solution?}\label{wheresol}

The solution to the cosmological constant problem may come from
one of many directions. As we will argue in chapter
(\ref{Symmetry}) a symmetry would be the most natural candidate.
Let us therefore first investigate what the symmetries of the
gravity sector are, with and without a cosmological constant,
since these are quite different.

The unique vacuum solution to Einstein's equations in the presence
of a (negative) cosmological constant is (anti-) de Sitter
spacetime, (A)dS for short. Many lectures on physics of de Sitter
space are available, for example
\cite{HawkingEllis,WittenQGinDS,Goheer:2002vf,Klemm:2004mb,Strominger:2001gp,Padmanabhan2002}.
The cosmological constant is a function of $\alpha$, the radius of
curvature of de Sitter space, and in D-dimensions given by:
\begin{equation}
\Lambda =
\frac{(D-1)(D-2)}{2\alpha^2}\stackrel{(D=4)}{=}\frac{3}{\alpha^2}.
\end{equation}
The explicit form of the metric is most easily obtained by
thinking of de Sitter space as a hypersurface embedded in
$D+1$-dimensional Minkowski spacetime. The embedding equation is:
\begin{equation}
-z_0^2 + z_1^2+\ldots + z_D^2 = \alpha^2.
\end{equation}
This makes it manifest that the symmetry group of dS-space is the
ten parameter group $SO(1,D)$ of homogeneous `Lorentz
transformations' in the D-dimensional embedding space, and the
metric is the induced metric from the flat Minkowski metric on the
embedding space. In the literature, one encounters several
different coordinate systems, which, after quantization, all lead
to what appear to be different natural choices for a vacuum state.
The metric often used in cosmology is described by coordinates
$(t,\vec{x})$ defined as:
\begin{eqnarray}
z_0&=& \alpha\sinh(t/\alpha) +
\frac{1}{2}\alpha^{-1}e^{t/\alpha}|\vec{x}|^2\nonumber\\
z_4&=& \alpha\cosh(t/\alpha) -
\frac{1}{2}\alpha^{-1}e^{t/\alpha}|\vec{x}|^2\nonumber\\
z_i&=& e^{t/\alpha}x_i,\quad i=1,2,3,\quad\quad
-\infty<t,x_i<\infty
\end{eqnarray}
in which the metric becomes:
\begin{equation}
ds^2 = -dt^2 + e^{\pm 2t/\alpha}d\vec{x}^2.
\end{equation}
covering only half the space, with $z_0+z_4>0$, describing either
a universe originating with a big bang, or one ending with a big
crunch, depending on the signs. This portion of dS-space is
conformally flat. The apparent time-dependence of the metric is
just a coordinate artifact. In the absence of any source other
than a cosmological constant, there is no preferred notion of
time. The translation along the time direction merely slides the
point on the surface of the hyperboloid. This time-independence
can be made explicit, by choosing so-called static coordinates,
defined by:
\begin{eqnarray}
z_0&=&(\alpha^2-r^2)^{\frac{1}{2}}\sinh(t/\alpha)\nonumber\\
z_1&=&(\alpha^2-r^2)^{\frac{1}{2}}\cosh(t/\alpha)\nonumber\\
z_2&=&r\sin\theta\cos\phi\nonumber\\
z_3&=&r\sin\theta\sin\phi\nonumber\\
z_4&=&r\cos\theta,\quad\quad 0\leq r<\infty.
\end{eqnarray}
Then the metric takes the form:
\begin{equation}\label{staticdS}
ds^2 = -\left[1-(r^2/\alpha^2)\right]dt^2 +
\frac{dr^2}{\left[1-(r^2/\alpha^2)\right]} + r^2(d\theta^2 +
\sin^2\theta d\phi^2).
\end{equation}
These coordinates also cover only half of the de Sitter manifold,
with $z_0+z_1>0$. The key feature of the manifold, is that it
possesses a coordinate singularity at
$r=\alpha=\sqrt{3/\Lambda}=H^{-1}$, with $H$ the Hubble parameter.
This represents the event horizon for an observer situated at
$r=0$, following the trajectory of the Killing vector $\partial_t$
(obviously not a global Killing vector, since $t$ is a timelike
coordinate only in the region $r<H^{-1}$). In these coordinates,
the metric looks very similar to the Schwarzschild metric:
\begin{equation}
ds^2=-\left(1-\frac{2M}{r}\right)dt^2 +
\frac{dr^2}{\left(1-\frac{2M}{r}\right)} + r^2(d\theta^2 +
\sin^2\theta d\phi^2),
\end{equation}
describing empty spacetime around a static spherically symmetric
source, with a singularity at $r=2M$, which turns out to be a
coordinate artefact, and a `real' singularity at $r=0$.

In the coordinate system (\ref{staticdS}), the notion of a vacuum
state is especially troublesome. The energy-momentum tensor
diverges at the event horizon $r=\alpha$, but this horizon is an
observer dependent quantity, i.e. it depends on the origin of the
radial coordinates. The vacuum state is not even translationally
invariant, each observer must be associated with a different
choice of vacuum state and comoving observers appear to perceive a
bath of thermal radiation \cite{BirrelDavies}.

Moreover, like the Bekenstein-Hawking entropy of a black hole, one
can assign an entropy $S$ to de Sitter spacetime:
\begin{equation}
S = \frac{A}{4G}
\end{equation}
with $A$ the area of the horizon. Its size, generalizing to
$D$-dimensions again, is given by
\cite{WittenQGinDS,Goheer:2002vf,Klemm:2004mb,Strominger:2001gp,Padmanabhan2002}:
\begin{equation}
A=\frac{2\pi^{n/2}}{\Gamma\left(\frac{n}{2}\right)}\left(\frac{(D-1)(D-2)}{2\Lambda}\right)^{\frac{D-2}{2}}
\stackrel{(D=4)}{=}\frac{12\pi}{\Lambda}.
\end{equation}
As noted, however, the horizon in this case is observer-dependent
and it is not immediately clear which concepts about black holes
carry over to de Sitter space.

One can also derive the Hawking temperature of de Sitter space, by
demanding that the metric be regular across the horizon. This
yields:
\begin{equation}
T_H=\frac{1}{4\pi}\sqrt{\frac{8\Lambda}{(D-1)(D-2)}}\stackrel{(D=4)}{=}\frac{1}{\pi}\sqrt{\frac{\Lambda}{12}}.
\end{equation}

There are a number of difficulties when one tries to formulate
quantum theories in de Sitter space. One is that since there is no
globally timelike Killing vector in dS-space, one cannot define
the Hamiltonian in the usual way. There is no positive conserved
energy in de Sitter space. Consequently, there cannot be unbroken
supersymmetry, since, if there were, there would be a non-zero
supercharge $Q$ that we can take to be Hermitian\footnote{Possibly
after replacing $Q$ by $Q+Q^{\dag}$ or $i(Q-Q^{\dag})$.} and $Q^2$
would have to be non-zero, a non-negative bosonic conserved
quantity and we arrive at a contradiction. For a long time, this
was a serious difficulty for string theory.

Based on the finite entropy, it has been argued that the Hilbert
space in de Sitter has finite dimension \cite{Banks2000-1}. This
could imply that the standard Einstein-Hilbert action with
cosmological constant cannot be quantized for general values of
$G$ and $\Lambda$, but that it must be derived from a more
fundamental theory, which determines these values
\cite{WittenQGinDS}. Another issue, especially encountered in
string theory, is that since the de Sitter symmetry group is
non-compact, it cannot act on a finite dimensional Hilbert space
\cite{WittenQGinDS}.

We will return to these interesting issues in much more detail in
section (\ref{Holography}), where we will also discuss arguments
from holography. In chapter (\ref{Back-reaction}) we will
encounter arguments that intend to show that de Sitter space is
unstable, leading to a decaying effective cosmological constant.

\subsection{Weinberg's No-Go Theorem}\label{nogowein}

Another route, often tried to argue that the effective
cosmological constant would gradually decay over time, involves a
screening effect using the potential of a scalar field $\phi$.

Note that in order to make the effective cosmological constant
time dependent, this necessarily involves either introducing extra
degrees of freedom, or an energy-momentum tensor that is not
covariantly constant. This follows directly by taking the
covariant derivative on both sides of the Einstein equation. One
arrives at:
\begin{equation}
\partial_{\mu}\Lambda - 8\pi GT_{\mu ;\nu}^{\nu}=0
\end{equation}
since the covariant derivative on the Einstein tensor vanishes
because of the contracted Bianchi-identity. The energy-momentum
tensor is obtained from varying the matter action with respect to
$g_{\mu\nu}$, and if general covariance is unbroken,
$T_{\mu;\nu}^{\nu}$ is zero, as a result of the equations of
motion. Hence, $\Lambda$ must be a constant. Therefore, to make
the cosmological constant time-dependent, the best one can do is
introduce a new dynamical field.

Consider the source of this field to be proportional to the trace
of the energy-momentum tensor, or the curvature scalar. Suppose
furthermore, that $T_{\mu}^{\mu}$ depends on $\phi$, and vanishes
at some field value $\phi_0$. Then $\phi$ will evolve until it
reaches its equilibrium value $\phi_0$, where $T_{\mu}^{\mu}$ is
zero, and the Einstein equations have a flat space solution. We
will consider these approaches in chapter~5. However, on rather
general terms Weinberg has derived a no-go theorem, first given in
his 1989 review \cite{Weinbergreview}, stating that many of these
approaches are fatally flawed. We follow the `derivation' of this
no-go theorem as given by Weinberg in \cite{Weinberg:1996xe}, see
also \cite{Weinberg2000}.

He assumes that there will be an equilibrium solution to the field
equations in which $g_{\mu\nu}$ and all matter fields $\phi_n$ are
\textit{constant} in spacetime. The field equations are:
\begin{equation}
\partial\mathcal{L}/\partial g_{\mu\nu} =
0\quad\quad\mbox{and}\quad\quad
\partial\mathcal{L}/\partial\phi_n =0
\end{equation}
With $N$ $\phi$'s, there are $N+6$ equations for $N+6$ unknowns,
since the Bianchi identities remove four of the ten metric
coefficients $g_{\mu\nu}$. So one might expect to find a solution
without fine-tuning.

The problem Weinberg argues, is in satisfying the trace of the
gravitational field equation, which receives a contribution from
$\rho_{vac}$ which for $\rho_{vac}\neq 0$ prevents a solution. The
trace of the left-hand-side of the Einstein equation is $-R +
4\Lambda$. This contribution of the cosmological constant to the
trace of the gravitational field equations, should be cancelled in
the screening. Therefore, one tries to make the trace a linear
combination of the $\phi_n$ field equations, as follows:
\begin{equation}\label{Weinberg4}
g_{\lambda\nu}\frac{\partial\mathcal{L}(g,\phi)}{\partial
g_{\lambda\nu}}=\sum_n\frac{\partial\mathcal{L}(g,\phi)}{\partial\phi_n}f_n(\phi)
\end{equation}
for all constant $g_{\mu\nu}$ and $\phi_n$, and $f(\phi)$
arbitrary, except for being finite.

Now, if there is a solution of the the field equation
$\partial\mathcal{L}/\partial\phi = 0$ for constant $\phi$, then
the trace $g_{\mu\nu}\partial\mathcal{L}/\partial g_{\mu\nu} = 0$
of the Einstein field equation for a spacetime-independent metric
is also satisfied, despite the fact there is a bare cosmological
constant term in the Einstein equation.

However, Weinberg points out that under these assumptions, the
Lagrangian has such a simple dependence on $\phi$ that it is not
possible to find a solution of the field equation for $\phi$. With
the action stationary with respect to variations of all other
fields, general covariance and (\ref{Weinberg4}) imply that the
following transformations are a symmetry:
\begin{equation}\label{symnogo}
\delta g_{\lambda\nu} = 2\epsilon g_{\lambda\nu}, \quad\quad
\delta\phi = -\epsilon f(\phi).
\end{equation}
The Lagrangian density for spacetime-independent fields
$g_{\mu\nu}$ and $\phi$ can therefore be written as\footnote{Note
the different pre-factor in the exponential.}:
\begin{equation}
\mathcal{L} =
c\sqrt{-g}\exp\left(2\int^{\phi}\frac{d\phi'}{f(\phi')}\right),
\end{equation}
where $c$ is a constant whose value depends on the lower limit
chosen for the integral. Only when $c=0$ is this stationary with
respect to $\phi$.

Another way to see how this scenario fails, originally due to
Polchinski \cite{Weinbergreview}, is that the above symmetry
$\delta g_{\lambda\nu} = 2\epsilon g_{\lambda\nu}$ ensures that
for constant fields, the Lagrangian can depend on $g_{\lambda\nu}$
and $\phi$ only in the combination $e^{2\phi}g_{\lambda\nu}$,
which can be considered as just a coordinate rescaling of the
metric $g_{\lambda\nu}$ and therefore cannot have any physical
effects. In terms of the new metric $\hat{g}_{\lambda\nu} =
e^{2\phi}g_{\lambda\nu}$, $\phi$ is just a scalar field with only
derivative couplings.

Many proposals have been put forward based on such spontaneous
adjustment mechanism using one or more scalar fields. However, on
closer inspection, they either do not satisfy (\ref{Weinberg4}),
in which case a solution for $\phi$ does not imply a vanishing
vacuum energy, or they do satisfy (\ref{Weinberg4}), but no
solution for $\phi$ exists. We will return to these types of
`solutions' in chapter~5 and see that in many proposals not only
the cosmological constant is screened to zero value, but also
Newton's constant. A flat space solution of course always can be
obtained if there is no gravity.

This argument can be cast in the form of a no-go theorem. This
theorem states that the vacuum energy density cannot be cancelled
without fine-tuning in any effective four-dimensional theory that
satisfies the following conditions
\cite{DvaliGabadadzeShifman20022}:
\begin{enumerate}
\item General Covariance;
\item Conventional four-dimensional gravity is mediated by a
\textit{massless} graviton;
\item Theory contains a finite number of fields below the cutoff
scale;
\item Theory contains no negative norm states,
\item The fields are assumed to be spacetime independent at late
times.
\end {enumerate}

In section (\ref{traceanomaly}) quantum anomalies to the energy
momentum tensor are discussed. These cannot circumvent this no-go
theorem.

\subsection{Some Optimistic Numerology}

One often encounters some optimistic numerology trying to relate
the value of the effective cosmological constant to other
`fundamental' mass scales. For example \cite{Hsu:2004jt}:
\begin{equation}
\rho_{vac} \sim \left(\frac{M_P}{L}\right)^2\quad\quad\rightarrow
M_\Lambda\sim\left(M_PM_U\right)^{\frac{1}{2}},
\end{equation}
with $L=M_{U}^{-1}=10^{-33}$~eV the size of the universe, and its
`Compton mass'. The mass scale of the cosmological constant is
given by the geometric mean of the UV cutoff $M_P$ and an IR
cutoff $M_U$.

Another relation arises if one includes the scale of supersymmetry
breaking $M_{SUSY}$, and the Planck mass $M_P$, to the value of
the effective cosmological constant \cite{Banks2000little}.
Experiment indicates:
\begin{equation}
M_{susy}\sim
M_P\left(\frac{\Lambda}{M_P^2}\right)^{\alpha},\quad\quad\mbox{with}\quad\alpha=\frac{1}{8}
\end{equation}
The standard theoretical result however indicates
$M_{susy}\sim\Lambda ^{1/2}$.

Another relation often mentioned, see for example
\cite{Carroll2003why} is:
\begin{equation}
\Lambda^{1/2}\sim\left(\frac{M_{susy}}{M_P}\right)M_{susy},
\end{equation}
where it is guessed that the supersymmetry breaking scale is the
geometric mean of the vacuum energy and the Planck energy. In both
relations, the experimental input is used that $\Lambda^{1/2}\sim
10^{-15}M_{susy}$.

Another non-SUSY numerological match can be given
\cite{Starobinsky:1998mj}, related to the fine-structure constant
$\alpha =1/137$:
\begin{equation}
\rho_\Lambda = \frac{M_{P}^4}{(2\pi^2)^3} e^{-2/\alpha} \sim
10^{-123}M_{P}^4.
\end{equation}
Note that this looks very similar to 't Hooft's educated guess
originating from gravitational instantons, for the electron mass
\cite{'tHooft:1988qq}:
\begin{equation}
Gm_{e}^2\cong\left(\alpha\sqrt{2}\right)^{-1}e^{-\pi/4\alpha}
\end{equation}
Whether any of these speculative relations holds, is by no means
certain but they may be helpful guides in looking for a solution.

\section{Outline of this thesis}

In this thesis we will consider different scenario's that have
been put forward as possible solutions of the cosmological
constant problem. The main objective is to critically compare them
and see what their perspectives are, in the hope to get some
better idea where to look for a solution. We have identified as
many different, credible mechanisms as possible and provided them
with a code for future reference. They can roughly be divided in
five categories: Fine-tuning, symmetry, back-reaction, violating
the equivalence principle and statistical approaches.

In the next chapter we describe the cosmological consequences of
the latest experimental results, since they have dramatically
changed our view of the universe. This is standard cosmological
theory and is intended to demonstrate the large scale implications
of a non-zero cosmological constant.

The remaining chapters are subsequently devoted to the above
mentioned five different categories of proposals to solve the
cosmological constant problem and is largely based on my paper
\cite{Nobbenhuis:2004wn}. We will discuss each of these proposals
separately, indicating where the difficulties lie and what the
various prospects are. See table (\ref{idcode}) for a list.

Three approaches are studied in great detail. The first can be
found in chapter \ref{imaginaryspace} and is based on my paper
written together with Gerard 't Hooft \cite{'tHooft:2006rs}, in
which we explore a new symmetry based on a transformation to
imaginary space. The idea is that the laws of nature have a much
wider symmetry than previously expected, and that quantum field
theory can be analytically continued to the full complex plane.
This generally leads to negative energy states. Positivity of
energy arises only after imposing hermiticity and boundary
conditions, which opens the way for a vacuum state invariant under
these transformations to have zero energy, leading to zero
cosmological constant.

The second one is the proposal by Tsamis and Woodard, which we
carefully analyze in chapter \ref{TsamisWoodardbr}. This scenario
is based on a purely quantum gravitational screening of the
cosmological constant. However, in our opinion there are several
fatal flaws in their arguments.

Thirdly, starting in section \ref{infiniteed}, we carefully study
the so-called DGP-gravity model. This string-theory inspired model
requires at least three extra infinite volume spatial dimensions
in order to solve the cosmological constant problem. As a result,
general relativity is modified at both very short and very long
distances. At first sight this is a very interesting prospect,
however, modifying GR without destroying its benefits at distance
scales where the theory is tested, is a very difficult task and we
will see that also the DGP-model faces serious obstacles. It shows
just how difficult it is to modify general relativity.

The conclusions of this work, as well as an outlook on further
research that still needs to be done in this field, are given in
chapter (\ref{Conclusions}). Finally, there are two appendices,
Appendix A gives the definitions and conventions, used throughout
this thesis. Appendix B provides a full list of best fit values
for the different cosmological parameters.

\begin{table}[h]
\caption{Classification of different approaches. Each of them can
also be thought of as occurring 1) Beyond 4D, or 2) Beyond Quantum
Mechanics, or both.}\label{idcode}
\begin{center}
\begin{tabular}{|l|l|} \hline
Type 0: Just Finetuning &  \\
\hline\hline
Type I: Symmetry; A: Continuous & a) Supersymmetry \\
\hline & b) Scale invariance \\
\hline & c) Conformal Symmetry \\
\hline  $\quad\qquad\qquad\qquad\quad$B: Discrete & d) Imaginary Space \\
\hline & e) Energy $\rightarrow$ -Energy \\
\hline & f) Holography\\
\hline & g) Sub-super-Planckian \\
\hline & h) Antipodal Symmetry \\
\hline & i) Duality Transformations \\
\hline\hline
Type II: Back-reaction Mechanism & a) Scalar \\
\hline
& b) Gravitons\\
\hline
& c) Running CC from Renormalization Group  \\
\hline
& d) Screening Caused by Trace Anomaly  \\
\hline\hline Type III: Violating Equiv. Principle & a) Non-local Gravity, Massive Gravitons\\
\hline
& b) Ghost Condensation \\
\hline
& c) Fat Gravitons \\
\hline & d) Composite graviton as Goldst. boson \\
\hline\hline Type IV: Statistical Approaches& a) Hawking Statistics \\
\hline
& b) Wormholes \\
\hline
& c) Anthropic Principle, Cont.\\
\hline & d) Anthropic Principle, Discrete \\ \hline
\end{tabular}
\end{center}
\end{table}

For reviews on the history of the cosmological constant (problem)
and many phenomenological considerations, see
\cite{Dolgov:1989vb,Weinbergreview,Sahni1999,Carroll2000,Weinberg2000,Padmanabhan2002,Peebles2002,Straumann2002,Ellwanger2002,Yokoyama2003}.

%% file: chapter2.tex
\chapter{Cosmological Consequences of Non-Zero $\Lambda$}\label{Cosmology}

Recent experiments may have shown that the cosmological constant
is non-zero. In this chapter we will study the cosmological
effects of a non-zero cosmological constant, from the introduction
of this term by Einstein, to the evolution of the universe as
unravelled by these recent results. They indicate that the
universe started a phase of accelerated expansion, about 5~billion
years ago. This expansion could very well be driven by a non-zero
cosmological constant. We give a brief review of supernovae
measurements since these are most important in tracking down the
expansion history of the universe.

In this chapter we are concerned with the cosmological effects of
a non-zero cosmological constant and assume that general
relativity is the correct theory of gravity, also at the largest
distance scales. Experimental results have lead to the
introduction, not only of dark energy, which may be due to a
non-zero cosmological constant, but also of dark matter. Together
these two spurious forms of energy seem to make up roughly 96\% of
the total energy density of the universe. In chapter
(\ref{equivalenceviolatie}) we discuss modifications of GR,
intended to explain the observed phenomena without introducing any
new form of matter or energy density.

\section{The Expanding Universe}

The cosmological constant was introduced by Einstein in 1917 when
he first applied his equations of GR to cosmology, assuming that
the universe even at the largest scales can be described by these
equations. Without a cosmological constant, they read:
\begin{equation}\label{Einsteinseqs}
R_{\mu\nu} - \frac{1}{2}g_{\mu\nu}R = -8\pi GT_{\mu\nu}
\end{equation}
Furthermore, at large scales, larger than a few hundred
Megaparsecs, the universe appears to be very homogeneous and
isotropic; our position in the universe seems in no way
exceptional. This observation is known as the `cosmological
principle' and can be formulated more precisely as follows
\cite{Weinbergbook}:
\begin{enumerate}
\item[1.] The hypersurfaces with constant comoving time coordinate
are maximally symmetric subspaces of the whole of spacetime.
\item[2.] Not only the metric $g_{\mu\nu}$, but all cosmic
tensors, such as the energy-momentum tensor $T_{\mu\nu}$, are
form-invariant with respect to the isometries of these subspaces.
\end{enumerate}
To see that this formal definition says the same, consider a
general coordinate transformation:
\begin{equation}
x^{\mu}\rightarrow \tilde{x}^{\mu}=x^{\mu} + \xi^{\mu}(x),
\end{equation}
under which the metric transforms as:
\begin{equation}
\delta g_{\mu\nu} = -\left(D_{\mu}\xi_{\nu} +
D_{\nu}\xi_{\mu}\right),
\end{equation}
where $D_{\mu}$ is the covariant derivative. Therefore, if the
vector $\xi^{\mu}$ satisfies:
\begin{equation}
D_{\mu}\xi_{\nu} + D_{\nu}\xi_{\mu} = 0,
\end{equation}
the metric is unchanged by the coordinate transformation, which is
then called an isometry and the associated vector $\xi^{\mu}$ is
called a Killing vector. A space that admits the maximum number of
Killing vectors, given by $d(d+1)/2$ in $d$ dimensions, is called
a maximally symmetric space.

A tensor is called form-invariant if the transformed tensor is the
same function of $\tilde{x}^{\mu}$ as the original tensor was of
$x^{\mu}$. Specifically, the metric tensor is form-invariant under
an isometry and a tensor is called maximally form-invariant if it
is form-invariant under all isometries of a maximally symmetric
space.

The above mathematical definition of the cosmological principle
therefore first states that the universe is spatially homogeneous
and isotropic and secondly that our spacetime position is in no
way special since cosmic observables are invariant under
isometries like translations.

Note that both homogeneity and isotropy are symmetries only of
space, not of spacetime. Homogeneous, isotropic spacetimes have a
family of preferred three-dimensional spatial slices on which the
three dimensional geometry is homogeneous and isotropic. In
particular, these solutions are not Lorentz invariant.

Assuming the cosmological principle to be correct, and ignoring
local fluctuations, the metric takes the Robertson-Walker form:
\begin{equation}
ds^2 = -dt^2 + a^2(t)\left[\frac{dr^2}{1-kr^2} +
r^2\left(d\theta^2 + \sin^2\theta d\phi^2\right)\right],
\end{equation}
where $a(t)$, called the scale-factor, characterizes the relative
size of the spatial sections, and $k$ is the curvature parameter.
Coordinates can always be chosen such, that $k$ takes on the value
-1, 0 or +1, indicating respectively negatively curved, flat or
positively curved spatial sections. Robertson-Walker metrics are
the most general homogeneous, isotropic metrics one can write
down. They are called Friedman-Robertson-Walker (FRW) metrics if
the scale factor obeys the Einstein equation. If the scale factor
increases in time this line element describes an expanding
universe.

Matter and energy in FRW-model is modelled as a perfect
cosmological fluid, with an energy-density $\rho$ and a pressure
$p$. An individual galaxy behaves as a particle in this fluid with
zero velocity, since otherwise it would establish a preferred
direction, in contradiction with the assumption of isotropy. The
coordinates are comoving, an individual galaxy has the same
coordinates at all times.

The energy-momentum tensor for a perfect fluid is:
\begin{equation}
T_{\mu\nu} = (\rho + p)U_{\mu}U_{\nu} + pg_{\mu\nu},
\end{equation}
where $U_{\mu}$ is the fluid four-velocity. The rest-frame of the
fluid must coincide with a comoving observer in the FRW-metric and
in that case, the Einstein equations (\ref{Einsteinseqs}) reduce
to the two Friedmann equations. The $(00)$-component gives:
\begin{equation}\label{Friedmann1}
H^2=\left(\frac{\dot{a}}{a}\right)^2 = \frac{8\pi G}{3}\rho -
\frac{k}{a^2},
\end{equation}
where $H$ is called the Hubble parameter. Using conservation of
energy-momentum, we derive:
\begin{equation}\label{consenergy}
T^{\mu\nu}_{;\nu}=0 \quad\Rightarrow\quad \frac{d}{dt}\left(\rho
a^3\right) + p\frac{d}{dt}\left(a^3\right) = 0\quad
\Rightarrow\quad \dot{\rho} + 3\frac{\dot{a}}{a}(\rho+p)=0,
\end{equation}
where the first equation is direct consequence of the
Bianchi-identity, we find:
\begin{equation}\label{Friedmann2}
\frac{\ddot{a}}{a}=-\frac{4\pi G}{3}(\rho + p)
\end{equation}
Einstein was interested in finding static solutions, with
$\dot{a}=0$, since the universe was assumed to be static at the
time. One reason for this belief was that the relative velocities
of the stars were known to be very small. Furthermore, he assumed
that space is both finite and globally closed, as he believed this
was the only way to incorporate Mach's principle stating that the
metric field should be completely determined by the
energy-momentum tensor. A static universe with $\dot{a}=0$ and
positive energy density is compatible with (\ref{Friedmann1}) if
the spatial curvature is negative and the density is appropriately
tuned. However, (\ref{Friedmann2}) implies that $\ddot{a}$ will
never vanish in such a spacetime if the pressure $p$ is also
non-negative, which is indeed true for most forms of matter such
as stars and gas. However, Einstein realized that mathematically
his equations allowed an extra term, that could become important
at very large distances. It is invisible locally, and therefore
also not noticed earlier, by for example Newton. So he proposed a
modification to:
\begin{equation}\label{Einsteinseqslambda}
R_{\mu\nu} - \frac{1}{2}g_{\mu\nu}R - \Lambda g_{\mu\nu} = -8\pi
GT_{\mu\nu}
\end{equation}
where $\Lambda$ is a new free parameter, the cosmological
constant, and is interpreted as the curvature of empty spacetime.
The left-hand-side of (\ref{Einsteinseqslambda}) now indeed is the
most general local, coordinate-invariant, divergenceless,
symmetric two-index tensor one can construct solely from the
metric and its first and second derivatives. In other words,
mathematically there was no reason not to put it there right from
the start. With this modification, the two Friedmann equations
become:
\begin{eqnarray}\label{Friedmannmodi}
H^2=\left(\frac{\dot{a}}{a}\right)^2 &=& \frac{8\pi G}{3}\rho +
\frac{\Lambda}{3}- \frac{k}{a^2}\nonumber\\
\frac{\ddot{a}}{a}&=&-\frac{4\pi G}{3}(\rho + p) +
\frac{\Lambda}{3}.
\end{eqnarray}
These equations now do admit a static solution with both $\dot{a}$
and $\ddot{a}$ equal to zero, and all parameters $\rho$, $p$ and
$\Lambda$ non-negative. This solution is called the 'Einstein
static universe'.

However, it is not a stable solution; any slight departure of any
of the terms from their balanced equilibrium value, leads to a
rapid runaway solution. Therefore, even with a cosmological
constant a genuinely stationary universe cannot be a solution of
the Einstein equations. From the first moment on, the acceptance
of the cosmological constant and its physical implications have
been the topic of many discussions, which continue till this day.

\subsection{Some Historic Objections}

Already in the same year 1917, De Sitter found that with a
cosmological constant, an expanding cosmological model as a
solution of Einstein's equations could be obtained, which is
`anti-Machian'. This model universe contains no matter at all.

At about the same time Slipher had observed that most galaxies
show redshifts of up to $6\%$, whereas only a few show blueshifts
\cite{Slipher}. However, the idea of an expanding universe was
accepted much later, only after about 1930, despite the
breakthrough papers of Friedmann in 1922 and 1924
\cite{Friedmann1922,Friedmann1924} and Lemaitre in 1927. Einstein
also found this hard to swallow, according to Lemaitre, Einstein
was telling him at the Solvay conference in 1927: ``Vos calculs
sont correct, mais votre physique est abominable"
\cite{Straumann1}. Moreover, in order to model an expanding
universe, one does not need a cosmological constant.

Even Hubble's stunning discovery in 1924 at first did not really
change this picture, for he also did not interpret his data as
evidence for an expanding universe. It was Lemaitre's
interpretation of Hubble's results that finally changed the
paradigm and at this point Einstein rejected the cosmological
constant as superfluous and no longer justified \cite{Einstein1}.
Rumors go that Einstein rejected the CC term in his equations
calling it the biggest blunder in his life\footnote{Actually there
is only one reference about this, by Gamow \cite{Gamow1970}
referring to a private conversation.} where he might have been
referring to the missed opportunity of predicting the expansion of
the universe. There are however also indications that Einstein
already had doubts at an earlier stage. A postcard has been found
from 1923 where Einstein writes to Weyl: ``If there is no
quasi-static world, then away with the cosmological term"
\cite{Straumann1}.

It should be noted, that there was a problem interpreting Hubble's
data as evidence for an expanding universe, the so-called age
problem. The age of the universe derived from Hubble's
distance-redshift relation\footnote{See section (\ref{redshift})
for more details on this.} was a mere two billion years, which
clearly cannot be correct, since already the Earth itself is
older. For some, for example Eddington, this was reason to keep
the cosmological constant alive. For a detailed history of the
cosmological constant problem, see \cite{Straumann1}.

This history of acceptance and refusal, of struggling to
understand the constituents and the evolution of the universe,
goes on to this very day and especially concerns the role and
interpretation of the cosmological constant. There were some major
turning points on the road, some of them we will encounter in this
thesis.

\section{Some Characteristics of FRW Models}

For many details on cosmology, one can check for example
\cite{Peacock99,KolbTurner,Garcia-Bellido99, Garcia-Bellido2005};
we especially use the lectures by Garcia-Bellido. In these
sections we will review those issues that are most important to
us.

To find explicit solutions to the Friedmann equations
(\ref{Friedmannmodi}), we need to know the matter and energy
content of the universe and how they evolve with time. Recall eqn.
(\ref{consenergy}):
\begin{equation}\label{energycons}
\dot{\rho} + 3\frac{\dot{a}}{a}(\rho + p) =0,
\end{equation}
an expression for the total energy density and pressure. For the
individual components $i$, one has:
\begin{equation}\label{evolprhoind}
\dot{\rho}_i + 3\frac{\dot{a}}{a}(\rho_i + p_i)
=X_i,\quad\quad\sum_iX_i=0,
\end{equation}
with $X_i$ a measure of the interactions between the different
components. For most purposes in cosmology, the interaction can be
set to zero and we can explicitly solve (\ref{evolprhoind}). To do
so, we need a relation between $\rho_i$ and $p_i$, known as the
equation of state. The most relevant fluids in cosmology are
barotropic, i.e. fluids whose pressure is linearly proportional to
the density: $p_i=w_i\rho_i$, with $w_i$ a constant, the equation
of state parameter. In these fluids the speed of sound
$c_{s}^{2}=dp/d\rho$ is constant. The solution to
(\ref{evolprhoind}) now becomes:
\begin{equation}
\rho_i =\rho_0a^{-3(1+w_i)}
\end{equation}
where $\rho_0$ is an integration constant, set equal to $\rho_i$,
when $a^{-3(1+w_i)}=1$. Furthermore, it should be noted that we
assume here that there is no interaction between the different
components $\rho_i$.

In cosmology the number of barotropic fluids is often restricted
to only three:
\begin{itemize}

\item{Radiation:} $w=1/3$, associated with relativistic degrees of
freedom, kinetic energy much greater than the mass energy.
Radiation energy density decays as $\rho_R\sim a^{-4}$ with the
expansion of the universe.

\item{Matter:} $w=0$, associated with non-relativistic degrees of
freedom, energy density is the matter energy density. It decays as
$\rho_M\sim a^{-3}$. Also called 'dust'.

\item{Vacuum energy:} $w=-1$, associated with energy density
represented by a cosmological constant. Due to this peculiar
equation of state, vacuum energy remains constant throughout the
expansion of the universe.

\end{itemize}

From the Friedmann equations it can be seen that an accelerating
universe $\ddot{a}>0$ is possible, not only for non-zero
cosmological constant $w=-1$, but more generally for `fluids' with
$w<-1/3$. Fluids satisfying $\rho + 3p\geq 0$, or $w\geq -1/3$ are
said to satisfy the `strong energy condition' (SEC). Dark energy
thus violates this SEC. The `weak energy condition' (WEC), is
satisfied if $\rho + p \geq 0$, or $w\geq -1$. This condition is
usually assumed to hold at all times, but recently been called
into question in so-called `phantom dark energy' models, see
\cite{Caldwell99,Gibbons:2003yj,Singh:2003vx,Nojiri:2003vn}. The
effective speed of sound in such a medium $v=\sqrt{|dp/d\rho|}$
can become larger than the speed of light. A universe dominated by
phantom energy has some bizarre properties. For example, it
culminates in a future curvature singularity (`Big Rip'). Models
constructed simply with a wrong sign kinetic term, are plagued
with instabilities at the quantum level, but it has been argued
that braneworld models can be constructed devoid of these troubles
\cite{Alam:2002dv,Sahni:2002dx}.

From the Friedmann equations (\ref{Friedmannmodi}), we can define
a critical energy density $\rho_c$ that corresponds to a flat
universe:
\begin{equation}
\rho_c \equiv \frac{3H_{0}^2}{8\pi G},
\end{equation}
where the subscript~$0$ denotes parameters measured at the present
time. In terms of this critical density, we can rewrite the first
Friedmann equation of (\ref{Friedmannmodi}) in terms of density
parameters $\Omega_i \equiv \rho_i/\rho_c$ where the subscript $i$
runs over all possible energy sources. For matter, radiation,
cosmological constant and curvature, these are:
\begin{eqnarray}
\Omega_M &=& \frac{8\pi G}{3H_{0}^2}\rho_M\quad\quad \Omega_R =
\frac{8\pi G}{3H_{0}^2}\rho_R \nonumber\\
\Omega_{\Lambda} &=& \frac{\Lambda}{3H_{0}^2}\quad\quad\quad
\Omega_k = -\frac{k}{a_{0}^2H_{0}^2}.
\end{eqnarray}

With these definitions the Friedmann equation can be written as:
\begin{equation}\label{Friedmannina}
H^2(a)= H_{0}^2\left(\Omega_R\frac{a_{0}^4}{a^4} +
\Omega_M\frac{a_{0}^3}{a^3} + \Omega_{\Lambda} +
\Omega_k\frac{a_{0}^2}{a^2}\right),
\end{equation}
and therefore, the Friedmann equation today ($a=a_0$) becomes:
\begin{equation}\label{CosmicSumRule}
1=\Omega_M + \Omega_R + \Omega_{\Lambda} + \Omega_k
\end{equation}
known as the ``cosmic sum rule''. Sometimes, a dimensionless
scalefactor $R(t)$ is defined: $R(t)\equiv a(t)/a_0$, such that
$R(t)=1$ at present to make manipulations with the above
expression a little easier.

The energy density in radiation nowadays is mainly contained in
the density of photons from the cosmic microwave background
radiation (CMBR):
\begin{eqnarray}
\rho_{CMBR}&=&\pi^2k^4T^{4}_{CMBR}/(15\hbar^3c^3)=4.5\times
10^{-34}~\mbox{g/cm}^3,\nonumber\\
\Omega_{R,CMBR}&=&2.4\times 10^{-5}h^{-2}
\end{eqnarray}
where $h\sim 0.72$ is the dimensionless Hubble parameter, defined
in Appendix~\ref{AppA}. Three approximately massless neutrinos
would contribute a similar amount, whereas the contribution of
possible gravitational radiation would be much less. Therefore, we
can safely neglect the contribution of $\Omega_R$ to the total
energy density of the universe today. Moreover, CMBR measurements
indicate that the universe is spatially flat to a high degree of
precision, which means that $\Omega_k$ is also negligibly small.

From the second Friedmann equation, we can define another
important quantity, the deceleration parameter $q$, defined as
follows:
\begin{equation}
q\equiv -\frac{a\ddot{a}}{\dot{a}^2}=\frac{4\pi G}{3H^2}(\rho +
3p) - \frac{\Lambda}{3H^2}.
\end{equation}
This shows that when vacuum energy is the dominant energy
contribution in the universe, the deceleration parameter is
negative, indicating an accelerated expansion, whereas it is
positive for matter dominance. Uniform expansion corresponds to
the case $q=0$. In terms of the $\Omega_i$, the deceleration
parameter today can be written as:
\begin{equation}
q_0 = \Omega_R + \frac{1}{2}\Omega_M - \Omega_{\Lambda} +
\frac{1}{2}\sum_x(1+3w_x)\Omega_x,
\end{equation}
where we have included the option of possible other fluids, with
deviating equations of state parameters $w_x$. Recent measurements
indicate that $q_0$ is about $-0.6$, indicating that the expansion
of the universe is accelerating. Astronomers actually  measure
this quantity by making use of a different relation, where as a
time variable the redshift, denoted by $z$, is used.

\subsection{Redshift}\label{redshift}

Since FRW models are time dependent, the energy of a particle will
change as it moves through this geometry, similarly to moving in a
time dependent potential. The trajectory of a particle moving in a
gravitational field obeys the following geodesic equation:
\begin{equation}
\frac{du^{\mu}}{d\lambda} +
\Gamma^{\mu}_{\nu\alpha}u^{\nu}u^{\alpha} = 0,
\end{equation}
where $u^{\mu}\equiv dx^{\mu}/ds$ and $\lambda$ is some affine
parameter, that we can choose to be the proper length
$g_{\mu\nu}dx^{\mu}dx^{\nu}$. The $\mu=0$ component of the
geodesic equation then is very simple, since the only
non-vanishing $\mu=0$ Christoffel is
$\Gamma^{0}_{ij}=(\dot{a}/a)g_{ij}$. Also using that $g_{ij}u^iu^j
= |\vec{u}|^2$ we have:
\begin{equation}\label{redhiftder}
\frac{du^0}{ds} + \frac{\dot{a}}{a}|\vec{u}|^2 =
0\quad\quad\Rightarrow\quad\quad
\frac{1}{u^0}\frac{d|\vec{u}|}{ds} + \frac{\dot{a}}{a}|\vec{u}|
=0,
\end{equation}
and since $u^0=dt/ds$ this reduces to:
\begin{equation}
\frac{|\vec{\dot{u}}|}{|\vec{u}|}=-\frac{\dot{a}}{a}\quad\quad\quad\Rightarrow\quad\quad\quad
|\vec{u}|\propto\frac{1}{a}.
\end{equation}
Moreover, $p^{\mu}=mu^{\mu}$, therefore the 3-momentum of a freely
propagating particle redshifts as~$1/a$. The factors $ds$ in
(\ref{redhiftder}) cancel, so this also applies to massless
particles for which $ds$ is zero. This can be derived in a similar
way, choosing a different affine parameter.

This momentum-redshift can also be derived by specifying a
particular metric and writing down the Hamilton-Jacobi equations,
see \cite{KolbTurner}.

In quantum mechanics, the momentum of a photon is inversely
proportional to the wavelength of the radiation, thus a similar
shift occurs. For a photon this results in a redshift of its
wavelength, hence the name. Photons travel on null geodesics of
zero proper time and thus travel between emission time $t_e$ and
observation time $t_o$ a distance $R$, given by:
\begin{equation}
ds^2 = 0 = -dt^2 + a^2(t)dr^2\quad\quad\Rightarrow\quad\quad R=
\int_{t_e}^{t_{o}}\frac{dt}{a(t)}.
\end{equation}
Furthermore, since $R$ is a comoving quantity, changing the upper
and lower limits to account for photons emitted, and observed, at
later times, does not affect the result. In other words,
$dt_e/dt_o = a(t_e)/a(t_o)$. The redshift is then simply defined
as:
\begin{equation}
1+z \equiv \frac{\lambda_{obs}}{\lambda_{emit}} =
\frac{\nu_{emit}}{\nu_{obs}}.
\end{equation}
In practice, astronomers observe light emitted from objects at
large distances and compare their spectra with similar ones known
in their restframe. This can be done since our galaxy is a
gravitationally bound object that has decoupled from the expansion
of the universe. The distance between galaxies changes with time,
not the sizes of galaxies, or measuring rods within them. Stars
move with respect to their local environments at typical
velocities of about $10^{-3}$~times the speed of light
\cite{Prokopec:2002jn}, leading to a special relativistic Doppler
redshift of about $\Delta z\sim 10^{-3}$. If spacetime were not
expanding, this would be the only source of non-zero redshift, and
averaging over many stars at the same luminosity distance would
give zero redshift. This is precisely what happens for stars
within our galaxy, but changes for stars further away.

In a similar manner we can calculate the distance to some far away
object. The assumptions of homogeneity and isotropy give us the
freedom to choose our position at the origin of our spatial
section and to ignore the angular coordinates. In a general
FRW-geometry we thus have:
\begin{equation}\label{distance}
\int_{t_1}^{t_0}{dt\over a(t)} =
\int_0^{r_1}{dr\over\sqrt{1-kr^2}}
\equiv f(r_1) = \left\{\begin{array}{ll}\arcsin r_1 \hspace{1cm} & k=1\\
r_1 & k=0\\{\rm arcsinh}\, r_1 & k=-1\end{array}\right.
\end{equation}
If we now Taylor expand the scale factor to third order
\cite{KolbTurner,Peacock99,Garcia-Bellido99,Garcia-Bellido2005},
\begin{equation}
{a(t)\over a_0} = 1 + H_0(t-t_0) - {q_0\over2!}H_0^2(t-t_0)^2 +
{j_0\over3!}H_0^3(t-t_0)^3 + {\cal O}(t-t_0)^4\,,
\end{equation}
where, using the second Friedmann equation:
\begin{eqnarray}
&&q_0=-\,{\ddot a\over a H^2}(t_0) =
\frac{1}{2}\sum_i(1+3w_i)\Omega_i=
\frac{1}{2}\Omega_M-\Omega_\Lambda\nonumber\\
&&j_0=+\,{\stackrel{\dots}{a}\over a H^3}(t_0) =
\frac{1}{2}\sum_i(1+3w_i)
(2+3w_i)\Omega_i=\Omega_M+\Omega_\Lambda\,,\label{JerkParameter}
\end{eqnarray}
where we have set $\Omega_R$ and $\Omega_k$ to zero. We find, to
first approximation,
\begin{equation}
r_1 \approx f(r_1) = {1\over a_0}(t_0 - t_1) + \dots = {z\over
a_0H_0} + \dots
\end{equation}
This yields the famous Hubble law:
\begin{equation}\label{HubbleLaw}
H_0\,d = a_0H_0r_1 = z \simeq vc\,,
\end{equation}
Astronomers track the expansion history of the universe by
plotting redshift $z$ versus a quantity $d_L$, the luminosity
distance. This luminosity distance is defined as the distance at
which a source of absolute luminosity $\mathcal{L}$ gives a flux
$\mathcal{F}=\mathcal{L}/4\pi d_{L}^2$. The expression for $d_L$
as a function of $z$ for a FRW-metric, returning to general
$\Omega_{k}$, but keeping $\Omega_R=0$, is
\cite{KolbTurner,Peacock99,Garcia-Bellido99,Garcia-Bellido2005}:
\begin{equation}\label{LuminosityDistance}
H_0d_L(z) = \frac{1+z}{|\Omega_k|^{1/2}}{\rm sinn}\left[\int_0^z
\frac{|\Omega_k|^{1/2}\ dz'}{\sqrt{(1+z')^2(1+z'\Omega_M) -
z'(2+z')\Omega_\Lambda}}\right],
\end{equation}
where ${\rm sinn}(x)= x\ {\rm if}\ k=0;\ \sin(x)\ {\rm if}\ k=+1\
{\rm and}\ \sinh(x)\ {\rm if}\ k=-1$, and we have used the cosmic
sum rule (\ref{CosmicSumRule}). Substituting eqn's
(\ref{JerkParameter}) into Eqn.~(\ref{LuminosityDistance}) we
find:
\begin{equation}\label{DLZ}
H_0\,d_L(z) = z + {1\over2}(1-q_0)\,z^2 -
{1\over6}(1-q_0-3q_0^2+j_0)\,z^3 + {\cal O}(z^4)\,.
\end{equation}
The leading term yields Hubble's law, which is just a kinematical
law, whereas the higher order terms are sensitive to the
cosmological parameters $\Omega_{\Lambda}$ and $\Omega_M$.

An interesting point in the evolution of the universe was the
transition between deceleration and acceleration. This transition
must have occurred since the early universe had to be matter
dominated in order to form structure. Only recently clear evidence
of the existence of this transition, or {\em coasting point} has
been obtained from supernovae data \cite{Riess2004}. The coasting
point is defined as the time, or redshift, at which the
deceleration parameter vanishes,
\begin{equation}
q(z) = -1 + (1+z)\,{d\over dz}\ln H(z) = 0,
\end{equation}
using $q=-1-d/dt H(a)$, where
\begin{equation}
H(z) = H_0\Big[\Omega_M (1+z)^3 +
\Omega_x\,e^{3\int_0^z(1+w_x(z')) {dz'\over1+z'}}\ + \Omega_k
(1+z)^2\Big]^{1/2}\,,
\end{equation}
from eqn. (\ref{Friedmannina}) and using that $a_0/a=1+z$, and we
have assumed that the dark energy is parameterized by a density
$\Omega_x$ today, with a redshift-dependent equation of state,
$w_x(z)$, not necessarily equal to $-1$.

\begin{figure}[htb]
\label{fig:RiessOMOL}
\begin{center}
\includegraphics[width=8cm]{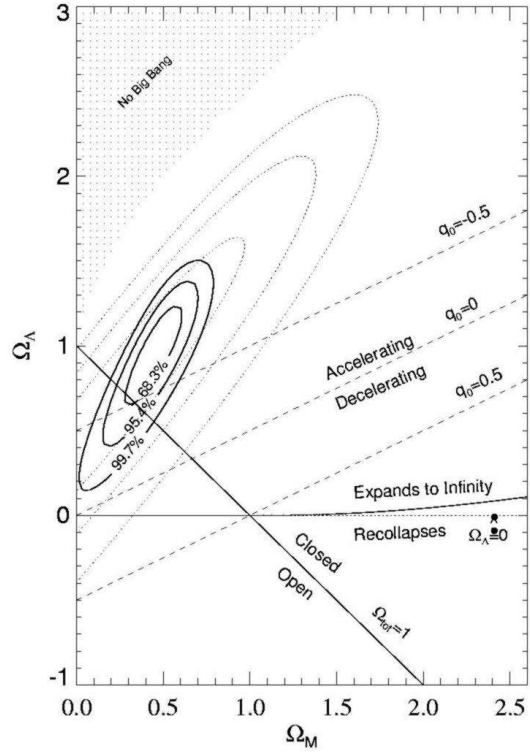}
\caption{The recent supernovae data on the
$(\Omega_M,\,\Omega_\Lambda)$ plane. Shown are the 1-, 2- and
3-$\sigma$ contours, as well as the data from 1998, for
comparison. It is clear that the old EdS cosmological model at
$(\Omega_M=1,\,\Omega_\Lambda=0)$ is many standard deviations away
from the data. From Ref.~[12].}
\end{center}
\end{figure}

Assuming that $w$ is constant, the coasting redshift can be
determined from
\begin{eqnarray}
&&q(z) = \frac{1}{2}\Big[{\Omega_M +
(1+3w)\,\Omega_x\,(1+z)^{3w}\over
\Omega_M + \Omega_x\,(1+z)^{3w} + \Omega_k (1+z)^{-1}}\Big] = 0\,,\\
&&\hspace{1cm}\Rightarrow z_c =
\left({(3|w|-1)\Omega_x\over\Omega_M} \right)^{1\over3|w|}-1\,,
\label{DecelerationParamz}
\end{eqnarray}
which, in the case of a true cosmological constant, reduces to
\begin{equation}\label{CoastingPoint}
z_c = \Big({2\Omega_\Lambda\over\Omega_M}\Big)^{1/3}-1\,.
\end{equation}
When substituting $\Omega_\Lambda\simeq0.7$ and
$\Omega_M\simeq0.3$, one obtains $z_c \simeq 0.6$, in excellent
agreement with recent observations~\cite{Riess2004}. The plane
$(\Omega_M,\,\Omega_\Lambda)$ can be seen in Fig.
(\ref{fig:RiessOMOL}), which shows a significant improvement with
respect to previous data. The best determination of the Hubble
parameter $H_0$ was made by the Hubble Space Telescope Key
Project, \cite{Freedman2000} to be $H_0 = 72 \pm 8$~km/s/Mpc,
based on objects at distances up to 500~Mpc, corresponding to
redshifts $z\leq 0.1$. In Appendix B, we provide a full list of
the best fit values for the different cosmological parameters.

As a nice example, we can calculate the effect of a cosmological
constant term, on our motion in the Milky Way \cite{Peebles2002}.
Using the non-relativistic limit, we have:
\begin{equation}
\frac{d^2\vec{r}}{dt^2}=\vec{g} + \Omega_\Lambda H_{0}^{2}\vec{r},
\end{equation}
where $\vec{g}$ is the relative gravitational acceleration
produced by the distribution of ordinary matter. Our solar system
is moving in the Milky Way galaxy at speed roughly $v_c=220$~km/s
at radius $r=8$~kpc. The ratio of the acceleration $g_\Lambda$
produced by the cosmological constant, to the total gravitational
acceleration $g=v_{c}^{2}/r$ is:
\begin{equation}
g_\Lambda/g = \Omega_\Lambda H_{0}^{2}r^2/v_{c}^{2}\sim 10^{-5},
\end{equation}
a small number. The precision of celestial dynamics in the solar
system is much better, but of course, the effect of a cosmological
constant is much smaller, $g_\Lambda/g\sim 10^{-22}$.

\section{The Early Universe}

We have already used the value of several cosmological parameters
in the above calculations. An important source of information for
these values comes from the Cosmic Microwave Background Radiation,
or CMBR.

In table (\ref{historyuniverse}) we list the major events that
took place in the history of the universe, based on the
inflationary big bang model. Important for us are the end of
inflation at $\sim10^{12}$~GeV, after which the effective
cosmological constant was very small, and the origin of the CMBR,
since many cosmological observations that are the backbone of our
theoretical models, originate from it.

\begin{table}[h]
\caption{The ``history'' of the universe from the Planck scale.
Some of the major events are also shown, together with the
dominant type of physics.}\label{historyuniverse}
\begin{center}
\begin{tabular}{|c|c|c|c|}
\hline
&&&\\
Energy of back- & Description  & Time & Redshift \\
ground radiation &   & &  \\
\hline \hline
&&&\\
$\sim 10^{19}$ GeV & Planck Scale  & $10^{-44}$ sec &$z>10^{46}$\\
\hline &&&\\
$\sim 10^{16}$ GeV& Breaking of GUT &  $10^{-37}$ sec&\\
\hline &&&\\
$\sim 10^{12}$ GeV& End of Inflation/Reheating  & $10^{-30}$ sec& $z\sim 10^{20}$\\
\hline &&&\\
$\sim 10^3$ GeV & EW symm. breaking  & $10^{-10}$ sec &\\
\hline &&&\\
1 GeV & quark-gluon plasma condens.  & $10^{-4}$ sec &\\
\hline &&&\\
$\sim$ 100 MeV & pion decay/annihilation & $10^{-4}$ sec &\\
& neutrino decoupling &&\\
\hline &&&\\
$\sim$ 0.1 MeV & nucleosynthesis,  & 100 sec &\\
& start creation light elements  & & $z\sim 10^4$\\
\hline &&&\\
$\sim$30 keV & end nucelosynthesis  & 15 min &\\
\hline &&&\\
$\sim$ 2 eV & matter-radiation equality  & 10.000 yr & 3500\\
\hline &&&\\
$\sim$0.35 eV & recombination, hydrogen-photon & 380.000 yr. & 1100\\
& decoupling, origin CMBR  & &\\
\hline &&&\\
&  Galaxy formation&$10^9$ years&\\
\hline &&&\\
$10^{-4}$ eV & now & $1.4\cdot 10^{10}$ yr. & z=0\\
\hline
\end{tabular}
\end{center}
\end{table}

When the CMBR was discovered in 1965, its temperature was found to
be 2.725 Kelvins, no matter which direction of the sky one looks
at; it is the best blackbody spectrum ever measured. The CMBR
appeared isotropic, indicating that the universe was very uniform
at those early times.

It can be considered as the ``afterglow'' of the Big Bang. After
the recombination of protons and electrons into neutral hydrogen,
about 380,000 years after the Big Bang, the mean free path of the
photons became larger than the horizon size, and the universe
became transparent for the photons produced in the earlier phases
of the evolution of the universe. This radiation therefore
provides a snapshot of the universe at that time. The collection
of points where the photons of the CMBR that are now arriving on
earth had their last scattering  before the universe became
transparent, is called the last-scattering surface.

The universe just before recombination is a tightly coupled fluid,
where photons scatter off charged particles and since they carry
energy, they feel perturbations imprinted in the metric during
inflation. These small perturbations propagate very similar to
sound waves, a train of slight compressions and rarefactions. The
compressions heated the gas, while the rarefactions cooled it,
leading to a shifting pattern of hot and cold spots, the
temperature anisotropies. A distinction is made between primary
and secondary anisotropies, the first arise due to effects at the
time of recombination, whereas the latter are generated by
scattering along the line of sight. There are three basic primary
perturbations, important on respectively large, intermediate, and
small angular scales (see
\cite{Peacock99,HuWhite96,Tegmark:1995kq} for many details on the
CMBR and its anisotropies):
\begin{enumerate}
\item[1.] Gravitational Sachs-Wolfe. Photons from high density
regions at last scattering have to climb out of potential wells,
and are redshifted: $\delta T/T=\delta\Phi/c^2$, with $\delta\Phi$
the (perturbations in the) gravitational potential. These
perturbations also cause a time dilation at the surface of last
scattering, so these photons appear to come from a younger, hotter
universe: $\delta T/T=-2\delta\Phi/3c^2$, so the combined effect
is $\delta T/T=\delta\Phi/3c^2$.

\item[2.] Intrinsic, adiabatic. Recombination occurs later in
regions of higher density, causing photons coming from overly
dense regions to have smaller redshift from the universal
expansion, and so appear hotter: $\delta T/T=-\delta
z/(1+z)=\delta\rho/\rho$.

\item[3.] Velocity, Doppler. The plasma has a certain velocity at
recombination, leading to Doppler shifts in frequency and hence
temperature: $\delta T/T=\delta\vec{v}\cdot\hat{r}/c$, with
$\hat{r}$, the direction along the line of sight, and $\vec{v}$
the characteristic velocity of the photons in the plasma.
\end{enumerate}

Through the Cosmic Background Explorer (COBE) and, more recently,
the Wilkinson Microwave Anisotropy Probe (WMAP)
\cite{Bennett2003,Spergel2003} satellites, the small variations in
the radiation's temperature were detected. They are perturbations
of about one part in 100,000. These tiny anisotropies are images
of temperature fluctuations on the last-scattering surface and
contain a wealth of cosmological information. Their angular sizes
depend on their physical size at this time of last scattering, but
they also depend on the geometry of the universe, through which
the light has been travelling before reaching us. Maps of the
temperature fluctuations are a picture of this last-scattering
surface processed through the geometry and evolution of a FRW
model.

During their journey through the universe, a small fraction of the
CMBR photons is scattered by hot electrons in gas in clusters of
galaxies. These CMBR photons gain energy as a result of this
Compton scattering, which is known as the Sunyaev-Zeldovich
effect, see for example \cite{Zeldovich:1969ff,Sunyaev:1980vz}. It
is observed as a deficit of about 0.05~\% of CMBR photons, as they
have shifted to higher energy, with about 2~\% increase. This
effect can be seen as a verification of the cosmological origin of
the CMBR.

A widely used code to calculate the anisotropies using linear
perturbation theory is CMBFAST \cite{Seljak:1996is}.

\subsection{Deriving Geometric Information from CMBR Anisotropies}

When looking at the CMBR, we are observing a projection of
soundwaves onto the sky. A particular mode with wavelength
$\lambda$, subtends an angle $\theta$ on the sky. The observed
spectrum of CMB anisotropies is mapped as the magnitude of the
temperature variations, versus the sizes of the hot and cold
spots, and this is usually plotted through a multipole expansion
in Legendre polynomials $P_l(\cos\theta)$, of a correlation
function $C(\theta)$. The order $l$ of the polynomial, related to
the multipole moments, plays a similar role in the angular
decomposition as the wavenumber $k\propto 1/\lambda$ does for a
Fourier decomposition. Thus the value of $l$ is inversely
proportional to the characteristic angular size of the wavemode it
describes.

There are many good lectures on CMB physics, e.g.
\cite{Peacock99,Garcia-Bellido99,Garcia-Bellido2005,HuWhite96,Reidetal2002,Tegmark:1995kq}
and especially the website by Hu \cite{WHu}.

\begin{figure}[htb]
\caption{The Angular Power Spectrum of CMB temperature
anisotropies, from the WMAP satellite. Also shown is the
correlation between the temperature anisotropies and the (E-mode)
polarization. From Ref.~\cite{WMAP}.} \label{PowerSpectrumCMB}
\begin{center}
\includegraphics[width=8cm]{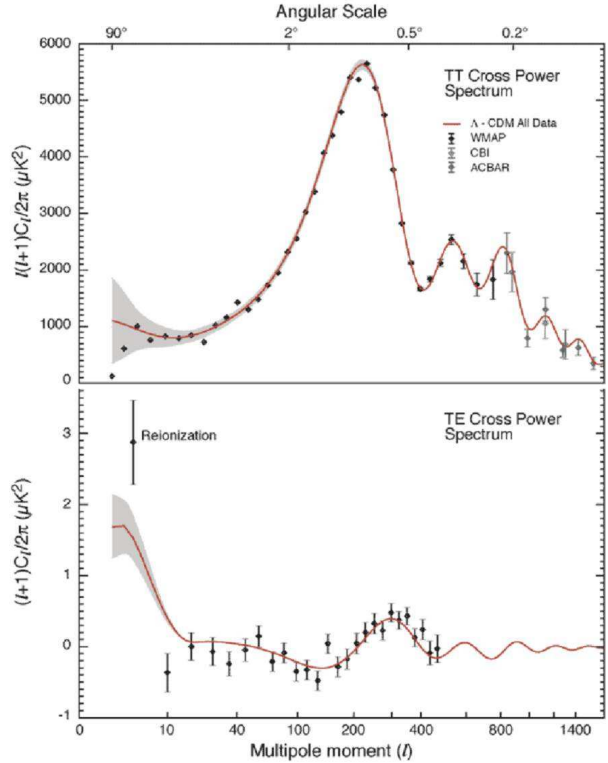}
\end{center}
\end{figure}

The correlation function $C(\theta)$, is defined as follows: Let
$\Delta T(\vec{n})/T$ be the fractional deviation of the CMBR
temperature from its mean value in the direction of a unit vector
$\vec{n}$. Take two vectors $\vec{n}$ and $\vec{n}'$ that make a
fixed angle $\theta$ with each other: $\vec{n}\cdot\vec{n'} =
\cos\theta$. The correlation function $C(\theta)$ is then defined
by averaging the product of the two $\Delta T/T$'s over the sky:
\begin{equation}
C(\theta)\equiv\left\langle\frac{\Delta T(\vec{n})}{T}\frac{\Delta
T(\vec{n}')}{T}\right\rangle,
\end{equation}
where the angle brackets denote the all-sky average over $\vec{n}$
and $\vec{n}'$ and it is assumed that the fluctuations are
Gaussian\footnote{If the $n$-point distribution is Gaussian, it is
defined by its mean vector $\langle\delta(\vec{x})\rangle$, which
is identically zero, and its covariance matrix
$C_{mn}\equiv\langle\delta(\vec{x}_m)\delta(\vec{x}_n)\rangle=\xi(|\vec{x}_m-\vec{x}_n|)$.
$\xi(x)$ is a correlation
function:$\langle\delta(\vec{x}_2)\delta(\vec{x}_1)\rangle$, that
because of homogeneity and isotropy only depends on
$x\equiv|\vec{x}_2-\vec{x}_1|$ \cite{Tegmark:1995kq}.}. In terms
of Legendre polynomials:
\begin{equation}\label{legendreCMB}
C(\theta)=\sum_{l=0}^{\infty}\frac{2l+1}{4\pi}C_lP_l(\cos\theta).
\end{equation}
Modes caught at extrema of their oscillations become the peaks in
the CMB power spectrum. They form a harmonic series based on the
distance sound can travel by recombination, called the sound
horizon.  The first peak represents the fundamental wave of the
universe, and represents the mode that compressed once inside
potential wells before recombination, the second the mode that
compressed and then rarefied, the third the mode that compressed
then rarefied then compressed, etc. These subsequent peaks in the
power spectrum represent the temperature variations caused by the
overtones. All peaks have nearly the same amplitude, as predicted
by inflation, except for a sharp drop-off after the third peak.
The physical scale of these fluctuations is so small that they are
comparable to the distance photons travel during recombination.
Recombination does not occur instantaneously, the surface of last
scattering has a certain thickness. In that short period during
which the universe recombines, the photons bounce around the
baryons and execute a random walk. If the random walk takes the
photons across a wavelength of the perturbation, then the hot and
cold photons mix and average out. The acoustic oscillations are
exponentially damped on scales smaller than the distance photons
randomly walk during recombination. See figure
(\ref{PowerSpectrumCMB}).

\begin{figure}[h]
\caption{Surface of last scattering (SLS), fundamental acoustic
mode, and the sound horizon. From \cite{Reidetal2002}.}
\label{cmbsoundhor}
\begin{center}
\includegraphics[width=8cm]{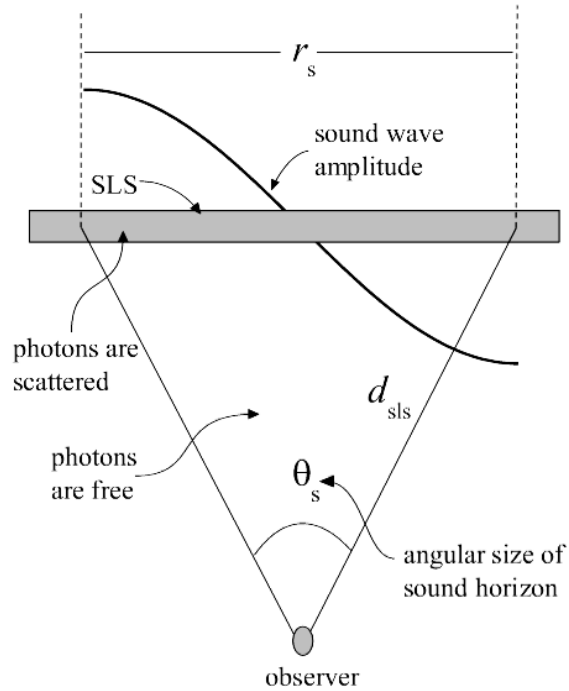}
\end{center}
\end{figure}

From the position of the peaks we can infer information about the
geometry of the universe. In the same way as the angle subtended
by say the planet Jupiter depends on both its size and its
distance from us, so depend the angular sizes of the anisotropies
on our distance to the surface of last scattering $d_{sls}$, from
which the CMBR originated, and on what is called the ``sound
horizon'' $r_s$. The sound horizon is given by the distance sound
waves could have travelled in the time before recombination, it is
a fixed scale at the surface of last scattering. The angular size
$\theta_s$ of the sound horizon thus becomes:
\begin{equation}
\theta_s \approx \frac{r_s}{d_{sls}}.
\end{equation}
The sound horizon $r_s$ and the distance to the surface of last
scattering, both depend on the cosmological parameters,
$\Omega_i$. The distance sound can travel, from the big bang to
the time of recombination is:
\begin{equation}
r_s (z_{\ast}, \Omega_i) \approx\int_0^{t_{\ast}}c_s dt,
\end{equation}
where $z_{\ast}, t_{\ast}$ are the redshift and time at
recombination. See figure (\ref{cmbsoundhor}).

The speed of sound in the baryon-photon plasma is given by
\cite{HuWhite96}:
\begin{equation}\label{speedsound}
c_s \approx
c\left[3\left(1+3\Omega_b/4\Omega_R\right)\right]^{-1/2},
\end{equation}
where $\Omega_b$ is the density of baryons and $c$ the speed of
light.

The distance to the surface of last scattering, corresponding to
its angular size, is given by what is called the angular diameter
distance. It is proportional to the luminosity distance $d$, and
explicitly given by:
\begin{equation}
d_{sls}=\frac{d(z_{\ast};\Omega_M;\Omega_{\Lambda})}{(1+z_{\ast})^2}.
\end{equation}
The location of the first peak is given by $l\approx d_{sls}/r_s$
and is most sensitive to the curvature of the universe.

In \cite{Reidetal2002}, a very instructive calculation is
presented, considering just the first peak of the spectrum, that
gives a clear idea how all of this works in practise. We will
review that calculation here. Taking the speed of sound
(\ref{speedsound}) to leading order equal to $c/\sqrt{3}$ and
assuming that the early universe was matter dominated, we obtain
for the sound horizon $r_s$:
\begin{equation}\label{soundhorizon}
r_s =
\frac{c/\sqrt{3}}{H_0\sqrt{\Omega_M}}\int_{z_{\ast}}^{\infty}(1+z)^{-5/2}dz
= \frac{2c/\sqrt{3}}{3H_0\sqrt{\Omega_M}}(1+z_{\ast})^{-3/2}.
\end{equation}
The distance to the surface of last scattering $d_{sls}$, assuming
a flat universe i.e. $\Omega_k=0$, depends only on $\Omega_M$ and
$\Omega_{\Lambda}$. It can be derived, using that
$d_{sls}=r_{sls}/(1+z_{\ast})$, where $r_{sls}$ is the radial
coordinate of the surface of last scattering. It is given by
\cite{Reidetal2002}:
\begin{equation}
r_{sls} = \frac{c}{H_0}\int_0^{\infty}\left[\Omega_M(1+z)^3 +
\Omega_{\Lambda}\right]^{-1/2}dz.
\end{equation}
This integral can be solved making use of a binomial expansion
approximating the integrand to:
\begin{equation}
r_{sls} =
\frac{c}{H_0}\int_0^{\infty}\left(\Omega_{M}^{-1/2}(1+z)^{-3/2} -
\frac{\Omega_{\Lambda}}{2\Omega_M^{3/2}}(1+z)^{-9/2}\right)dz.
\end{equation}
We obtain:
\begin{equation}
d_{sls} = \frac{2c}{7H_0(1+z_{\ast})}\left(7\Omega_M^{-1/2} -
2\Omega_{\Lambda}\Omega_M^{-3/2} +
\mathcal{O}\left[(1+z_{\ast})^{-1/2}\right]\right).
\end{equation}
With our assumption of a flat universe the cosmic sum rule simply
becomes $\Omega_{\Lambda}=1-\Omega_M$. Using this, and ignoring
higher order terms, we arrive at:
\begin{equation}
d_{sls} \approx
\frac{2c\Omega_M^{-1/2}}{7H_0(1+z_{\ast})}\left(9-2\Omega_M^3\right).
\end{equation}
Together with eqn. (\ref{soundhorizon}), this gives the prediction
for the first acoustic peak in a flat universe, ignoring density
in radiation:
\begin{equation}
l\approx \frac{d_{sls}}{r_s}\approx
0.74\sqrt{(1+z_{\ast})}\left(9-2\Omega_M^3\right)\approx 221.
\end{equation}
Consistent with the more accurate result, for example
\cite{Hananyetal2000}, from the MAXIMA-1 collaboration:
\begin{equation}
l\approx 200/\sqrt{1-\Omega_k}.
\end{equation}

To summarize, the expansion in multipoles
(\ref{PowerSpectrumCMB}), shows that for some feature at angular
size $\Delta\theta$ in radians, the $C_l$'s will be enhanced for a
value $l$ inversely related to $\Delta\theta$. For a flat
universe, it shows up at lower $l$, than it would in an open
universe, see figure. The dependence of the position of the first
peak on the spatial curvature can approximately be given by
$l_{peak}\simeq 220\;\Omega^{-1/2}$, with $\Omega=\Omega_M +
\Omega_{\Lambda}=1-\Omega_k$. With the high precision WMAP data,
this lead to $\Omega = 1.02\pm 0.02$ at 95\% confidence level.

Much more information can be gained from the power-spectrum. A
very nice place to get a feeling for the dynamics of the power
spectrum is Wayne Hu's website \cite{WHu}, whose discussion we
also closely follow in the remainder of this section.

The effect of baryons on the CMB power spectrum is threefold:
\begin{enumerate}
\item[1:] The more baryons there are, the more the second peak
will be suppressed compared to the first. This results from the
fact that the odd numbered peaks are related to how much the
plasma is compressed in gravitational potential wells, whereas the
even numbered peaks originate from the subsequent rarefaction of
the plasma.
\item[2:] With more baryons, the peaks are pushed to slightly
higher multipoles $l$, since the oscillations in the plasma will
decrease.
\item[3] Also at higher multipole moments, smaller angular scales, an
effect can be seen, due to their effect on how sound waves are
damped.
\end{enumerate}
The effects of dark matter are best identified by the higher
acoustic peaks, since they are sensitive to the energy density
ratio of dark matter to radiation in the universe and the energy
density in radiation is fairly well known. Especially the third
peak is interesting. If it is higher than the second peak, this
indicates dark matter dominance in the plasma, before
recombination. The third peak gives the best picture for this,
since in the first two peaks the self-gravity of photons and
baryons is still important.

Dark matter also changes the location of the peaks, especially the
first one. This is because the ratio of matter to radiation
determines the age of the universe at recombination,. This in turn
limits how far sound could have travelled before recombination
relative to how far light travels after recombination. The spatial
curvature has a similar effect, so to disentangle the two, at
least three peaks have to measured.

The effect of a positive cosmological constant is a small change
in how far light can travel since recombination, and hence
produces a slight shift to lower multipoles.

\subsection{Energy Density in the Universe}

\begin{figure}[htb]
\caption{The observed rotation curve, of the dwarf spiral galaxy
M33, extends considerably beyond its optical image (shown
superimposed); from \cite{Roy2000,Sahni2004}.}
\label{darkmattergalaxy}
\begin{center}
\includegraphics[width=8cm]{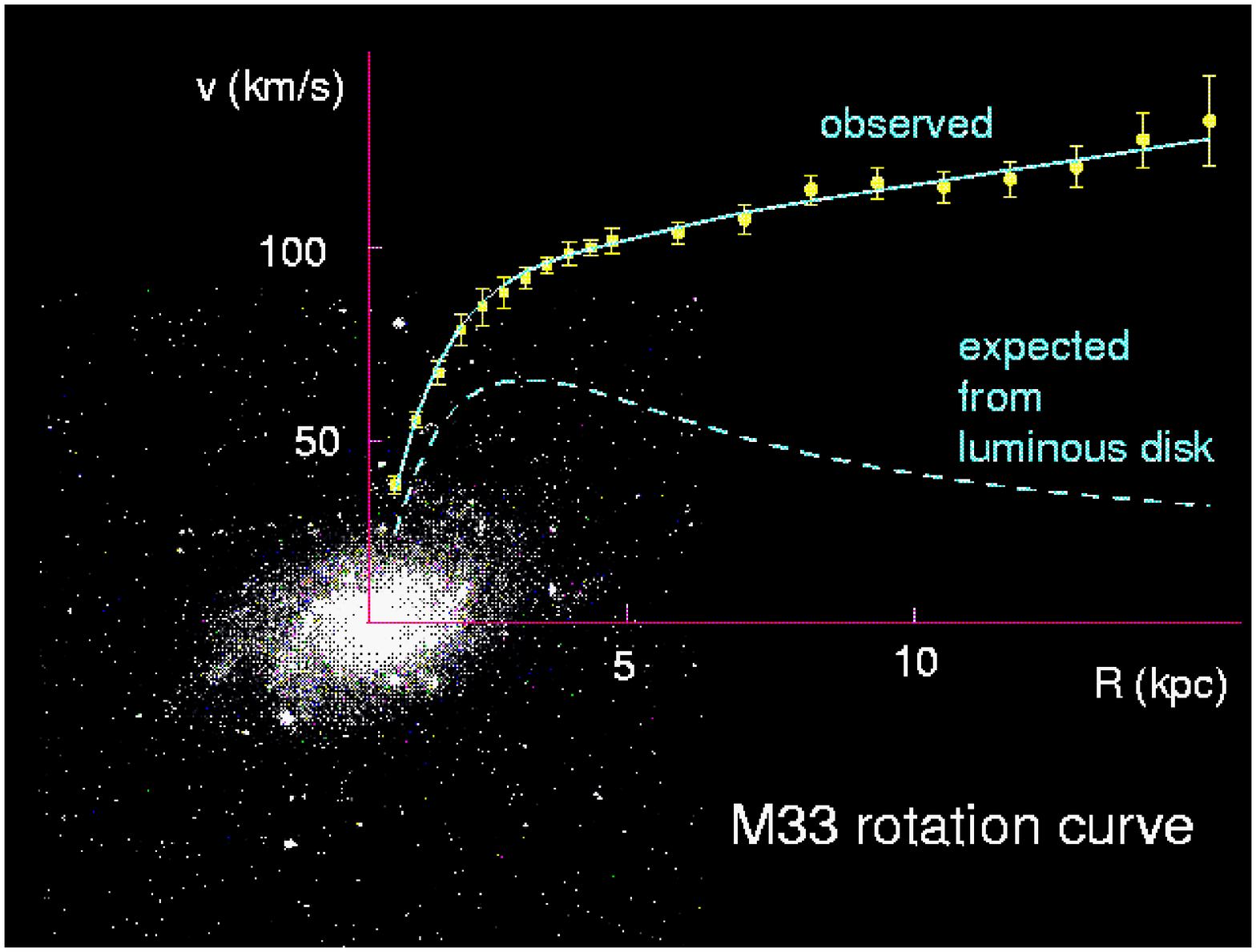}
\end{center}
\end{figure}

There is considerable evidence that most of the mass density of
the universe is neither in luminous matter, nor in radiation.
Since every mass gravitates, it can be detected by its
gravitational influence. Most direct evidence for unseen `dark
matter' comes from weighing spiral galaxies. By measuring the
Doppler shifts in the 21-cm line of neutral hydrogen the velocity
of clouds of this gas in the disk can be mapped as a function of
the distance $r$ from the center of the galaxy. One should expect
this velocity to fall off as $r^{-1/2}$, but rather it remains
constant in almost all cases, see figure (\ref{darkmattergalaxy}).

This seems to imply that, even in the outskirts of the galaxy, the
amount of mass is still growing with distance. Almost every galaxy
seems to contain a `halo' of dark, unseen matter, the amount of
which depends on the kind of galaxy. Ranging from roughly ten
times the mass seen in visible matter in spiral galaxies, to about
100 times as much in low surface brightness galaxies and dwarfs,
like the Draco dwarf spheroidal galaxy, which is nearby at only 79
kpc from the Milky Way \cite{Roy2000,Sahni2004}.

This ``missing mass" is a central problem in cosmology and
speculations about its origin range from black holes to new
species of particles, see for example \cite{Freesedarkside2005}
for a recent review. One of the most important dark matter
candidates is the lightest supersymmetric particle. From the WMAP
data and under certain model restrictions, an upper limit of the
mass of the LSP can be derived: $m_{LSP}\leq 500$~GeV
\cite{Ellisetal2003}.

However, there could also be a different explanation as propagated
by proponents of the Modified Newtonian Dynamics (MOND) approach
\cite{Milgrom:1983ca,Milgrom:1983pn,Bekenstein:1984tv,Bekenstein:2004ne}.
MOND can perfectly describe the rotation curves of individual
galaxies, but it is not without problems. Its predictions do not
agree with observations of clusters of galaxies
\cite{Pointecouteau:2005mr}.

Using measurements of CMB anisotropies, statistical analysis of
galaxy redshift surveys and measurements of the number density of
massive galaxy clusters, it has been shown that the universe is
flat to great precision and that matter, including dark matter,
makes up about a quarter of the critical energy density. Combining
these results with those of distant supernovae, it is inferred
that about 75\% of the universe is made up of dark energy. These
combined experiments moreover show that the equation of state of
this dark energy component is very close to -1. For more details
on exactly how the different parameters are measured, see any
cosmology book, like \cite{Peacock99,KolbTurner}, or more recent
reviews like \cite{Garcia-Bellido99, Garcia-Bellido2005}.

In the past two decades or so, the picture of the contents of the
universe we now have is often abbreviated as $\Lambda$~CDM. A
universe, with dominant cosmological constant, or dark energy and
Cold Dark Matter.

\section{Evolution of the Universe}

We will now consider universes dominated by matter and vacuum
energy, $\Omega = \Omega_{M} + \Omega_{\Lambda} = 1-\Omega_k$ as
given in the previous section. As discussed, a positive
cosmological constant tends to accelerate the universal expansion,
while ordinary matter and a negative cosmological constant tend to
decelerate it. The relative contributions to the energy density of
the universe scale like:
\begin{equation}
\Omega_{\Lambda}\propto a^2\Omega_k\propto a^3\Omega_M.
\end{equation}
From the Friedmann equations:
\begin{eqnarray}\label{Friedmannmodi2}
H^2=\left(\frac{\dot{a}}{a}\right)^2 &=& \frac{8\pi G}{3}\rho +
\frac{\Lambda}{3}- \frac{k}{a^2}\nonumber\\
\frac{\ddot{a}}{a}&=&-\frac{4\pi G}{3}(\rho + p) +
\frac{\Lambda}{3}.
\end{eqnarray}
we can immediately infer that, if $\Omega_{\Lambda}<0$, the
universe will always recollapse to a big crunch, either because
the matter density is sufficiently high or because eventually the
negative cosmological constant will dominate. For
$\Omega_{\Lambda}>0$ the universe will expand forever, unless
there is sufficient matter to cause a recollapse before
$\Omega_{\Lambda}$ becomes dynamically important. For
$\Omega_{\Lambda}=0$ on the hand the universe will expand forever
if $\Omega_M\leq 1$ or will recollapse if $\Omega_M
>1$. To identify these boundaries properly, let us have look again at
the equation for the time-dependent Hubble parameter:
\begin{equation}\label{Friedmannina2}
H^2(a)= H_{0}^2\left(\Omega_R\frac{a_{0}^4}{a^4} +
\Omega_M\frac{a_{0}^3}{a^3} + \Omega_{\Lambda} +
\Omega_k\frac{a_{0}^2}{a^2}\right),
\end{equation}
with dimensionless scalefactor $R(t)\equiv a(t)/a_0$ and writing
$\Omega_k = -\Omega_M - \Omega_{\Lambda} + 1$ from the cosmic sum
rule, we have:
\begin{equation}
\frac{H^2(a)}{H_{0}^2}= \Omega_{\Lambda}(1-R^{-2}) +
\Omega_M(R^{-3} - R^{-2})+ R^{-2}.
\end{equation}
Now we have to find solutions for which the left-hand-side
vanishes, since these define the turning points in the expansion.
There are four conditions, see for example
\cite{FeltenIsaacman86,Peacock99,Garcia-Bellido99,Garcia-Bellido2005}:
\begin{itemize}
\item[1:] As already stated before, negative $\Lambda$ always
implies recollapse.

\item[2:] If $\Lambda$ is positive and $\Omega_M < 1$, the model
always expands to infinity.

\item[3:] If $\Omega_M > 1$, recollapse is only avoided if
$\Omega_{\Lambda}$ exceeds a critical value:
\begin{equation}
\Omega_{\Lambda} >
4\Omega_{M}\cos^3\left[\frac{1}{3}\cos^{-1}\left(\Omega_{M}^{-1}
-1\right) + \frac{4\pi}{3}\right].
\end{equation}

\item[4:]

Conversely, if the cosmological constant is sufficiently large
compared to the matter density, there was, no initial singularity,
no Big Bang. Its early history consisted of a period of gradually
slowing contraction to a minimum radius before beginning its
current expansion; it underwent a bounce. The criterion for there
to have been no singularity in the past is
\begin{equation}
\Omega_\Lambda > 4\Omega_{M}{\rm coss}^3\left[{1\over 3}{\rm
coss}^{-1}\left(\Omega_{M}^{-1}-1\right)\right],
\end{equation}
where ``coss'' represents $\cosh$ when $\Omega_{M} < 1/2$, and
$\cos$ when $\Omega_{M} \geq 1/2$.

If the universe lies exactly on the critical line, the bounce is
at infinitely early times. Models that almost satisfy the critical
relation $\Omega_{\Lambda}(\Omega_M)$ are called loitering models,
since they spent a long time close to constant scale factor.
However, these bounce models can be ruled out quite strongly,
since the same cubic equations also give a relation for the
maximum possible redshift \cite{Peacock99}:
\begin{equation}
1+ z_{bounce} \leq 2{\rm coss}\left(\frac{1}{3}{\rm
coss}^{-1}\left(\Omega_{M}^{-1}-1\right)\right).
\end{equation}
With $\Omega_M$ as low as 0.1, a bounce is ruled out once objects
are seen at redshift $z>2$. Since galaxies have even been observed
at $z=10$, these universes are indeed ruled out.
\end{itemize}
This defines the boundaries between the different regions,
summarized in the $(\Omega_M,\Omega_\Lambda)$-plane
(\ref{fig:RiessOMOL}).

The behavior of the universe can also be seen by referring to our
previous expression for the deceleration parameter $q$:
\begin{equation}
q_0 = \Omega_R + \frac{1}{2}\Omega_M - \Omega_{\Lambda} +
\frac{1}{2}\sum_x(1+3w_x)\Omega_x \simeq \frac{1}{2}\Omega_M -
\Omega_{\Lambda}.
\end{equation}
Uniform expansion ($q_0=0$) corresponds to the line
$\Omega_\Lambda=\Omega_M/2$. Points above this line correspond to
universes that are accelerating today, while those below
correspond to decelerating universes, in particular the old
cosmological model of Einstein-de Sitter (EdS), with
$\Omega_\Lambda=0,\ \Omega_M=1$.  Since 1998, all the data from
Supernovae of type Ia appear above this line, many standard
deviations away from EdS universes. Nevertheless, sometimes it is
argued, that the EdS universe is a sound alternative to the
concordance model, and that it is in better agreement with the low
quadrupole in the CMBR \cite{Blanchardetal2003}. In such a
universe, the Hubble constant would be significantly lower,
approximately 46 km/s/Mpc, the SNIa results would have to be
discarded, for some systematical error for example, and one would
have to introduce another source of dark matter, to suppress power
on small scales. This seems a very contrived conspiracy of
effects, and has not many proponents.

The line $\Omega_\Lambda=1-\Omega_M$ corresponds with a flat
universe, $\Omega_k=0$. Points to the right of this line
correspond to closed universes, while those to the left correspond
to open ones. In the last few years we have mounting evidence that
the universe is spatially flat (in fact Euclidean).

\section{Precision cosmology in recent and upcoming
experiments}

In the past two decades or so cosmology has really grown as a very
interesting field of study. Especially observational cosmology has
made astonishing progress producing the first maps of the cosmos
and of course showing the acceleration of the expansion of the
universe. In this section we will review the supernovae
experiments, since these have been so important for the discovery
of an accelerating universe, and briefly discuss the Pioneer
anomaly, as it has been used to argue for deviations of general
relativity.

\subsection{Supernovae Type Ia}

From the end of the 1990's, two independent teams, the ``Supernova
Cosmology project", led by Perlmutter of Lawrence Berkeley
Laboratory and ``The High-Z Supernova Search Team", led by Schmidt
of Mt. Stromlo and Siding Observatories, have used the apparent
brightness of supernovae in order to study the speed of the
expansion of the universe. Both teams found that these supernovae
look fainter than expected. With the assumption, based on CMBR
experiments that the universe is flat, this could be the result of
a cosmological constant. This can be easily seen from the
expressions for the luminosity distance $d_L$, which in a matter
dominated universe $(\Omega_M=1, \Omega_\Lambda=0)$ take the form:
\begin{equation}
d_L=2H_{0}^{-1}\left[(1+z)-(1+z)^{1/2}\right],
\end{equation}
whereas in a cosmological constant dominated universe
$(\Omega_M=0, \Omega_\Lambda=1)$:
\begin{equation}
d_L=H_{0}^{-1}z(1+z).
\end{equation}
This shows that at given redshift $z$, the luminosity distance
$d_L$ is larger for a universe where the cosmological constant
dominates, and hence, a given object, at a fixed redshift, will
appear fainter.

For objects close by there is a linear relation between these two
quantities and the fixed ratio of the two is known as Hubble's
constant. For objects farther away though, deviations from this
linear dependence can be expected, either because the speed of the
expansion has changed over time, or as a result of spacetime
curvature. Therefore, to track down the history of the expansion
one has to find the distance-redshift relation of objects located
very far away. In Hubble's day, distances were determined by
assuming that all galaxies have the same intrinsic brightness. So
the fainter a galaxy really appeared, the further away from us it
is located and vice versa. This is a very crude way of determining
distances since different galaxies can have very different
properties, and therefore different intrinsic brightness.
Moreover, when looking at galaxies located far away, evolutionary
effects play an important role since light takes so long to travel
to us. We then view these galaxies as they were billions of years
ago, in their youth, and their intrinsic brightness could be very
different from the more mature ones, seen closer at home.

Distances are measured in terms of the ``distance modulus'' $m-M$,
where $m$ is the apparent magnitude of the source, and $M$ its
absolute magnitude. This distance modulus is related to the
luminosity distance:
\begin{equation}
m-M=5\log_{10}\left[d_L(Mpc)\right] + 25.
\end{equation}
A very useful relation, if you know the absolute magnitude, which
is notoriously hard to infer for a distant object. It is very
difficult to disentangle evolutionary changes from the effects of
the expansion, so astronomers have been looking for objects called
``standard candles'', whose intrinsic brightness, and therefore
distance, can be determined more unambiguously. The best candidate
for this task is a particular type of supernovae (SNe).
Supernovae, extremely bright explosions of a dead star, about
10~billion times as luminous as our sun, are among the most
violent phenomena in our universe. They come in two main classes
\cite{Turatto2003} called Type II for supernovae whose optical
spectra exhibit hydrogen lines, while hydrogen deficient SNe are
designated Type I. These SNe Type I are further subdivided
according to the detailed appearance of their early-time spectrum.
SNe Ia are characterized by strong absorption near 6150 $\AA$,
corresponding to SiII. SNe Ib lack this feature but instead show
prominent HeI lines. SNe Ic finally have neither SiII nor HeI
lines. In practise there is often little distinction between the
latter two types and they are most commonly designated as Type
Ib/c.

For the Type II SNe, four subclasses exist based on the shape of
optical light curves, see \cite{Barbonetal} for the details. For
our purposes the Type Ia are the only relevant ones.

There are two physical mechanisms that produce supernovae:
thermonuclear (the SNe Ia are the only ones of this type) and the
more common core collapse of a massive star. This last type occurs
when a massive red supergiant becomes old and produces more and
more metals in its core. Stars never fuse elements heavier than
iron, since this would cost energy, rather than produce it. Thus
iron piles up until this iron core reaches the Chandrasekhar mass,
which is about 1.4 solar masses. This is the critical mass below
which electron degeneracy pressure can stand up to the
gravitational pressure, but above this mass, no such equilibrium
is possible and the core collapses. At the extreme high pressure
in the core protons and electrons are smashed together to form
neutrons and neutrinos, while at the same time the outer layers
crash into the core and rebound, sending shockwaves outward. These
two effects together cause the entire star outside the core to be
blown apart in a huge explosion, a Type II supernova! What is left
behind of the core will be either a neutron star or, if it is
heavy enough, a black hole.

SNe Ia on the other hand originate from a white dwarf star that
accretes matter from another star orbiting nearby, until the
Chandrasekhar limit is reached. This happens for example in a
binary system, when a white dwarf can start slurping up matter
from a main sequence star that expands into a giant or supergiant.
When the critical mass is reached, the star is no longer stable
against gravitational collapse: the radius decreases, while the
density and temperature increase. The fusion of carbon and oxygen
into iron now occurs very rapidly, converting the star to a fusion
bomb until a thermonuclear firestorm ignites. The dwarf star gets
completely blown apart, spewing out material at about 10,000
kilometers per second. The amount of energy released in this
cosmic showpiece is about $10^{44}$ joules, as much as the Sun has
radiated away during its entire lifetime. The glow of this
fireball takes about three weeks to reach its maximum brightness
and then declines over a period of months. The spectrum contains
no hydrogen or helium lines, since the dwarf that is blown apart
consists of carbon and oxygen. Yet it does show silicon lines,
since that is one of the products of fusing carbon and oxygen.

In a typical galaxy such a SNe 1a lights up every 300 years or so.
In our own Milky Way therefore, it is a rare celestial event. The
last supernova in our galaxy was seen in 1604 by Kepler. However,
if you monitor a few thousand galaxies, you can expect to witness
about one type 1a supernova every month. There are actually so
many galaxies in the universe that a supernova bright enough to
study erupts every few seconds! You only have to find them.
Catching a supernova at its peak brightness however is not so easy
as their occurrence cannot be predicted and observing time at the
worlds largest telescopes therefore cannot be scheduled in
advance. Moreover, these type 1a also vary in their brilliance,
but brighter explosions last somewhat longer than fainter ones.
Therefore, they must be very carefully observed multiple times
during the first weeks, to monitor how long they last. This way
their intrinsic brightness can be deduced to within 12 percent
(absolute magnitude $M\sim -19.5$). In terms of redshift,
supernovae with redshifts as high as $z\sim 1.5$ have been
observed. This corresponds to a time when the universe was about
one-third its current age\footnote{That is, for a Hubble constant
$H_0=72$ km/s/Mpc leading to a current age of the universe of
about 14 billion years.}.

The supernova results show that although the universe is
undergoing accelerated expansion now, this has not always been the
case. Up until about 5 billion years ago, the universe was matter
dominated and decelerating. The change between these two phases
occurred at about $z\sim 0.5$. For a recent review of type Ia
supernovae in cosmology, see \cite{Filippenko:2004ub}.

\subsection{Pioneer Anomaly}

Another phenomenon that perhaps should be explained as a deviation
of GR originates from within our solar system and has become known
as the Pioneer anomaly. The two space probes Pioneer 10 and 11
were launched in the beginning of the seventies to do measurements
on Jupiter and Saturn along the outskirts of our solar system. At
distances $r$ from the sun between 20 and 70 astronomical units
(AU), the Doppler data have shown a deviation from calculations
based on GR. Analysis indicated a small, constant Doppler blue
shift drift of order $6\times 10^{-9}$~Hz/s. After accounting for
systematics, this may be interpreted as an unexplained, constant
acceleration directed towards the sun with a roughly constant
amplitude: $a_P\simeq 8\times 10^{-10}\mbox{ms}^{-2}$, or perhaps
a time acceleration of $a_t=(2.92\pm 0.44)\times
10^{-18}\mbox{s/s}^2$, see e.g. \cite{NietoAnderson2005} for a
recent review and many references.

Various mechanisms have been considered to explain this
acceleration, but so far no satisfactory scenario has been put
forward see
\cite{Anderson2002,Nieto2003,Turyshev2004,Dittus:2005re} and
references therein. The inability of conventional physics to
explain the anomaly, has triggered a growing number of new
theoretical explanations. One popular explanation was based on a
Yukawa modification of Newton's law, but it quickly became clear
that explaining the Pioneer anomaly in terms of a long-range
Yukawa correction, would also lead to very large deviations from
Kepler's law or of the precession of perihelions which obviously
have not been observed \cite{Jaekel2004,Reynaud2005}. Other
proposals suggest that we see here effects due to dark matter. If
dark matter should be held accountable for the anomalous
acceleration, it would have to be distributed in the form of a
disk in the outer solar system, with a constant density of
$\sim4\times 10^{-16}\;\mbox{kg/m}^3$. The acceleration, due to
any drag force from an interplanetary medium ia:
\begin{equation}
a_d(r)=-K_d\rho(r)v_{s}^2(r)A/m,
\end{equation}
with $K_d$ the effective reflection/absorption/transmission
coefficient of the spacecraft surface, $\rho(r)$ the density of
the medium, $v_{s}^2$ the effective relative velocity of the craft
to the medium, $A$ the spacecraft's cross-section, and $m$ its
mass. The mass of the spacecraft is 241 kg, its cross-section
about $5~\mbox{m}^2$ and its speed nowadays roughly 12 km/s. A
constant density of $\sim4\times 10^{-16}\;\mbox{kg}/\mbox{m}^3$
would therefore explain the Pioneer Anomaly. This would be in
correspondence with bounds on dark matter within the outer regions
of our solar system of about $10^{-6}~M_{\bigodot}$,
\cite{Munyaneza:1999cr}. A spherical halo of a degenerate gas of
massive neutrinos with $m_{\nu}\leq 16$~keV, around the sun
\cite{Munyaneza:1999cr} and mirror matter \cite{Foot2001,
Foot2003} have been suggested.

Another proposal suggests that the modified inertia interpretation
of gravity called `Modified Newtonian Dynamics' or MOND for short,
gives a correct explanation \cite{Milgrom2002,Moffat:2005si}. Also
certain braneworld scenarios \cite{Bertolami:2003ui} presumably
can explain the observed effect, with a right adjustment of
parameters for the potential of a scalar field.

%% file: chapter3.tex
\chapter{Type I: Symmetry Mechanism}\label{Symmetry}

One can set the cosmological constant to any value one likes, by
simply adjusting the value of the bare cosmological constant. No
further explanation then is needed. This fine-tuning has to be
precise to better than at least 55~decimal places (assuming some
TeV scale cut-off), but that is of course not a practical problem.
Since we feel some important aspects of gravity are still lacking
in our understanding and nothing can be learned from this
`mechanism', we do not consider this to be a physical solution.
However, it is a possibility that we can not totally ignore and it
is mentioned here just for sake of completeness.

A natural way to understand the smallness of a physical parameter
is in terms of a symmetry that altogether forbids any such term to
appear. This is also often referred to as `naturalness': a theory
obeys naturalness only if all of its small parameters would lead
to an enhancement of its exact symmetry group when replaced by
zero. Nature has provided us with several examples of this. Often
mentioned in this respect is the example of the mass of the
photon. The upper bound on the mass (squared) of the photon from
terrestrial measurements of the magnetic field yields:
\begin{equation}
m_{\gamma}^2\lesssim\mathcal{O}(10^{-50})\mbox{GeV}^2.
\end{equation}
The most stringent estimates on $\Lambda_{eff}$ nowadays give:
\begin{equation}
\Lambda_{eff}\lesssim\mathcal{O}(10^{-84})\mbox{GeV}^2
\end{equation}
We `know' the mass of the photon to be in principle exactly equal
to $0$, because due to the $U(1)$ gauge symmetry of QED, the
photon has only two physical degrees of freedom (helicities). In
combination with Lorentz invariance this sets the mass equal to
zero. A photon with only two transverse degrees of freedom can
only get a mass if Lorentz invariance is broken. This suggests
that there might also be a symmetry acting to keep the effective
cosmological constant an extra~34 orders of magnitude smaller.

A perhaps better example to understand the smallness of a mass is
chiral symmetry. If chiral symmetry were an exact invariance of
Nature, quark masses and in particular masses for the pseudoscalar
mesons $(\pi,K,\eta)$ would be zero. The spontaneous breakdown of
chiral symmetry would imply pseudoscalar Goldstone bosons, which
would be massless in the limit of zero quark mass. The octet
$(\pi,K,\eta)$ would be the obvious candidate and indeed the pion
is by far the lightest of the mesons. Making this identification
of the pion being the pseudo-Goldstone boson associated with
spontaneous breaking of chiral symmetry, we can understand why the
pion-mass is so much smaller than for example the proton mass.

\section{Supersymmetry}

One symmetry with this desirable feature for the energy density of
the ground state, is supersymmetry, or SUSY for short.
Contributions to the energy density of the vacuum in field theory
coming from fields with spin~$j$ are (eqn. (\ref{infgen2})):
\begin{eqnarray}\label{infgen3}
\langle\rho\rangle &=&
\frac{1}{2}(-1)^{2j}(2j+1)\int_{0}^{\Lambda_{UV}}\frac{d^3k}{(2\pi)^{3}}\sqrt{\mathbf{k}^{2}
+ m^{2}} \nonumber\\
&=& \frac{(-1)^{2j}(2j+1)}{16\pi^2}\Big(\Lambda_{UV}^4 +
m^2\Lambda_{UV}^2 -
\frac{1}{4}m^4\left[\log\left(\Lambda_{UV}^2/m^2\right)
+\frac{1}{8}-\frac{1}{2}\log 2\right]\nonumber\\
&+& \mathcal{O}\left(\Lambda_{UV}^{-1}\right)\Big).
\end{eqnarray}
so if for each mass $m$ there are an equal number of fermionic and
bosonic degrees of freedom, the net contribution to
$\langle\rho\rangle$ would be zero. Supersymmetry posits exactly
such a symmetry, and hence realizes elegantly that the energy
density of the ground state is zero.

The spin-$1/2$ generators of the supersymmetry transformations,
called supercharges are denoted by $Q$ and they satisfy
anticommutation relations:
\begin{equation}
\{Q_{\alpha},\bar{Q}_{\beta}\} =
2\sigma^{\mu}_{\alpha\beta}P_{\mu},
\end{equation}
where $\alpha$ and $\beta$ are two-component spin indices,
$\sigma^{\mu}$ are the Pauli-matrices, and $P_{\mu}$ is the
energy-momentum operator. Summing over $\alpha$ and $\beta$, we
find the Hamiltonian:
\begin{equation}
H = \frac{1}{4}\sum_{\alpha}\{Q_{\alpha}, \bar{Q}_{\alpha}\} =
P^0,
\end{equation}
where $P^0$ is the energy operator. Matrix elements of $P^0$ can
be written as:
\begin{equation}
\langle\psi|P^0|\psi\rangle =
\frac{1}{4}\sum_{\alpha}\langle\psi_{\alpha}|\psi_{\alpha}\rangle,
\end{equation}
with
\begin{equation}
\psi_{\alpha}= \left( Q_{\alpha} +
\bar{Q}_{\alpha}\right)|\psi\rangle.
\end{equation}
Thus in a supersymmetric theory, the energy of any non-vacuum
state in a positive definite Hilbert space, is positive definite.
Moreover, if supersymmetry is unbroken, the vacuum state satisfies
$Q_{\alpha}|0\rangle = \bar{Q}_{\alpha}|0\rangle =0$ for all
$\alpha$) and we see that this state has vanishing energy:
\begin{equation}
\langle 0|P^0|0\rangle =0.
\end{equation}
This includes all quantum corrections and would nicely explain a
vanishing vacuum energy.

This same result can also be obtained by looking at the potential.
The scalar field potential in supersymmetric theories takes on a
special form. Scalar fields $\phi^{i}$ must be complex, to match
the degrees of freedom coming from the fermions, and the potential
is derived from a function, called the superpotential
$W(\phi^{i})$ which is necessarily holomorphic (written in terms
of $\phi^{i}$ and not its complex conjugate $\bar{\phi}^{i}$). In
the simple Wess-Zumino models of just spin-$0$ and spin-$1/2$
fields, for example, the scalar potential is given by:
\begin{equation}
V(\phi^{i}, \bar{\phi}^{i}) = \sum_{i}|\partial_{i}W|^{2},
\end{equation}
where $\partial_{i}W = \partial W/\partial\phi^{i}$. Supersymmetry
will be unbroken only for values of $\phi^{i}$ such that
$\partial_{i}W = 0$, implying that $V$ takes it minimum value
$V(\phi^{i}, \bar{\phi}^{i}) = 0$. Quantum effects do not change
this conclusion, since the boson-fermion symmetry ensures that
boson-loops are cancelled against fermion-loops.

So if supersymmetry were an exact symmetry of nature, there would
be bosons and superpartner-fermions with the same mass and the
vacuum state of this theory would have zero energy. However, these
supersymmetric partners of the Standard Model particles have not
been found, so SUSY has to be broken at least at the TeV scale,
which again induces a large vacuum energy.

Besides, we still have the freedom to add any constant to the
Lagrangian, thereby changing the value of the effective
cosmological constant. In order to discuss the cosmological
constant problem properly, we need to bring gravity into the
picture. This implies making the supersymmetry transformations
local, leading to the theory of supergravity or SUGRA for short.
In such a theory the value of the effective cosmological constant
is given by the expectation value of the potential.

\subsection{No-Scale SUGRA}

In exact SUGRA the lowest energy state of the theory, generically
has negative energy density\footnote{This negative energy density
can also be forbidden by postulating an unbroken R-symmetry.}. Or,
in other words, the vacuum of supergravity is AdS. This has
inspired many to consider so-called no-scale supergravity models
in which supersymmetry breaking contributes precisely the amount
of positive vacuum energy to make the net result equal to zero.
See \cite{Weinbergreview} or supersymmetry textbooks such as
\cite{BailinLove} for excellent reviews. In this section we
closely follow \cite{BailinLove}.

The pure supergravity (SUGRA) Lagrangian, that is without
interactions, is:
\begin{equation}\label{pureSUGRA}
\mathcal{L}_{grav} = \mathcal{L}^{(2)} + \mathcal{L}^{(3/2)} -
\frac{e}{3}\left(S^{2} + P^{2} - A_{m}^{2}\right),
\end{equation}
where $\mathcal{L}^{(2)}$ is the Einstein-Hilbert term,
$\mathcal{L}^{(3/2)}$ is the Rarita-Schwinger Lagrangian
describing the free gravitino and $P$, a pseudoscalar, $S$ a
scalar and $A_{\mu}$ an axial vector, are auxiliary fields to
match the on- and off-shell fermionic and bosonic degrees of
freedom. The full Lagrangian is characterized by two arbitrary
functions of the scalar fields (in the presence of gravity,
renormalizability is lost anyway, so there is no reason to require
only low-order polynomials in $\phi^{i}$): a real function
$G(\phi^{i},\phi^{\ast}_{j})$, called the K\"{a}hler potential and
an analytic function $f_{ab}(\phi^{i})$, where Latin indices are
gauge indices. These functions determine the general forms allowed
for the kinetic energy terms of the scalar fields $\phi^{i}$ and
of the gauge fields $A_{\mu}^{a}$, respectively. Some of the terms
from the full Lagrangian that are important for now are:
\begin{eqnarray}\label{SUGRA}
\frac{1}{e}\mathcal{L}_{tot} = &-&\frac{1}{2}R +
G^{j}_{i}\partial_{\mu}\phi^{i}\partial^{\mu}\phi^{\ast}_{j} -
e^{G}\left[G_{i}(G^{-1})^{i}_{j}G^{j} - 3\right]\nonumber\\
&-&\frac{1}{4}Re(f_{ab})F^{a}_{\mu\nu}F^{\mu\nu b} +
e^{G/2}\bar{\psi}_{\mu}\sigma^{\mu\nu}\psi_{\nu} + \ldots
\end{eqnarray}
in which the gravitational coupling $\kappa$ has been set equal to
one and:
\begin{equation}
G_{i}\equiv\frac{\partial G}{\partial\phi^{i}},\quad\quad
G^{j}\equiv\frac{\partial G}{\partial\phi^{\ast}_{j}},\quad\quad
G^{j}_{i}\equiv\frac{\partial^{2}G}{\partial\phi^{i}\partial\phi^{\ast}_{j}}.
\end{equation}
The scalar kinetic energy term shows that $G^{j}_{i}$ plays the
role of the metric in the space spanned by the scalar fields. A
metric $G^{j}_{i}$ of this form is referred to as a K\"{a}hler
metric and $G$ is called the K\"{a}hler potential. In the absence
of gravity $G^{j}_{i}\rightarrow\delta^{j}_{i}$ and
$f_{ab}\rightarrow\delta_{ab}$. The function $G$ is invariant
under transformations of the gauge group, whereas $f_{ab}$
transforms as a symmetric product of two adjoint representations
of that group.

The Lagrangian contains a scalar potential which generally
consists of so-called D-terms and F-terms. The D-terms arise from
removing the auxiliary scalar fields contributing to the gauge
multiplets, and the F-terms from removing the auxiliary scalar
fields of the chiral matter multiplets. With $W$ denoting the
superpotential, one finds:
\begin{equation}
|F|^{2}\rightarrow\frac{\partial
W}{\partial\phi^{i}}\frac{\partial
W^{\ast}}{\partial\phi^{\ast}_{j}}(G^{-1})^{i}_{j},
\end{equation}
where we now must include the K\"{a}hler metric. This gives the
first term of the scalar potential:
\begin{equation}\label{scalarpotsugra}
V(\phi,\phi^{\ast}) = e^{G}\left[G_{i}(G^{-1})^{i}_{j}G^{j} -
3\right],
\end{equation}
In case of supergravity, there is an additional contribution form
eliminating the auxiliary scalar field terms $-|S + iP|^{2}$ in
(\ref{pureSUGRA}) and this yields the second term of
(\ref{scalarpotsugra}). The negative sign has important
consequences. The $e^{G}$ factor arises from the Weyl rescaling of
the $e_{m\mu}$ fields required to bring the first term in
(\ref{SUGRA}) into the canonical Einstein form. This rescaling
also implies a redefinition of the fermion fields and hence the
factor $e^{G/2}$ in the last term as well. Owing to this term,
when local SUSY is spontaneously broken the gravitino acquires a
mass:
\begin{equation}
m_{3/2} = e^{G/2},
\end{equation}
where $G$ then has to evaluated ar the minimum of the potential.

In general the K\"{a}hler potential has to satisfy certain
conditions for the theory to be well defined, for example
$G^{j}_{i}>0$, in order for the kinetic terms of the scalar fields
to have the correct sign. A special choice is:
\begin{equation}
G(\phi,\phi^{\ast}) = \phi^{i}\phi^{\ast}_i +
\log\left(W(\phi^{i})^{2}\right)^2,
\end{equation}
with $W$ the superpotential. This choice gives $G^{j}_{i} =
\delta^{j}_{i}$ and hence the minimal kinetic terms as in global
SUSY. The scalar potential becomes:
\begin{equation}
V =
\exp\left(\phi^{i}\phi^{\ast}_i\right)\left[\left|\frac{\partial
W}{\partial\phi^{i}} + \phi^{\ast}_{i}W\right|^{2} -
3|W|^{2}\right].
\end{equation}
Now we come to the important point. There is an elegant way of
guaranteeing a flat potential, with $V=0$ after supersymmetry
breaking, by using a nontrivial form of the K\"{a}hler potential
$G$. So far we have used the minimal form of $G$ which lead to the
above scalar potential. For a single scalar field $z$:
\begin{eqnarray}
V &=&
e^{G}\left[\frac{\partial_{z}G\partial_{z^{\ast}}G}{\partial_{z}\partial_{z^{\ast}}G}
- 3\right]\nonumber\\
&=&
\frac{9e^{4G/3}}{\partial_{z}\partial_{z^{\ast}}G}(\partial_{z}\partial_{z^{\ast}}e^{-G/3}).
\end{eqnarray}
A flat potential with $V=0$ is obtained if the expression in
brackets vanishes for all $z$, which happens if:
\begin{equation}
G = -3\log(z + z^{\ast}),
\end{equation}
and one obtains a gravitino mass:
\begin{equation}
m_{3/2} = \langle e^{G/2}\rangle = \langle (z +
z^{\ast})^{-3/2}\rangle,
\end{equation}
which, as required, is not fixed by the minimization of $V$. Thus
\textit{provided one chooses a suitable, nontrivial form for the
K\"{a}hler potential $G$, it is possible to obtain a zero CC and
to leave the gravitino mass undetermined}, just fixed dynamically
through non-gravitational radiative corrections. The minimum of
the effective potential occurs at:
\begin{equation}
V_{eff}\approx -(m_{3/2})^{4},
\end{equation}
where in this case, after including the observable sector and soft
symmetry-breaking terms, $m_{3/2}\approx M_{W}$. Such a mass is
ruled out cosmologically, see e.g. \cite{Grifols}, and so other
models with the same ideas have been constructed that allow a very
small mass for the gravitino, also by choosing a specific
K\"{a}hler potential, see \cite{Lahanas:1986uc}.

That these constructions are possible is quite interesting and in
the past there has been some excitement when superstring theory
seemed to implicate precisely the kinds of K\"{a}hler potential as
needed here, see for example \cite{Witten1985}. However, that is
not enough, these simple structures are not expected to hold
beyond zeroth order in perturbation theory.

$D=11$ SUGRA seems to be a special case; its symmetries implicitly
forbid a CC term, see \cite{DeserBautier1997}. However, also here,
the vanishing of the vacuum energy is a purely classical
phenomenon, which is spoiled by quantum corrections after
supersymmetry breaking.

\subsection{Unbroken SUSY}

To paraphrase Witten \cite{EWUnbrokensusy}: ``Within the known
structure of physics, supergravity in four dimensions leads to a
dichotomy: either the symmetry is unbroken and bosons and fermions
are degenerate, or the symmetry is broken and the vanishing of the
CC is difficult to understand''. However, as he also argues in the
same article, in $2+1$ dimensions, this unsatisfactory dichotomy
does not arise: SUSY can explain the vanishing of the CC without
leading to equality of boson and fermion masses, see also
\cite{EWCCfromString}.

The argument here is that in order to have equal masses for the
bosons and fermions in the same supermultiplet one has to have
unbroken global supercharges. These are determined by spinor
fields which are covariantly constant at infinity. The
supercurrents $J^{\mu}$ from which the supercharges are derived
are generically not conserved in the usual sense, but covariantly
conserved: $D_{\mu}J^{\mu} = 0$. However, in the presence of a
covariantly constant spinor ($D_{\mu}\epsilon =0$), the conserved
current $\bar{\epsilon}J^{\mu}$ can be constructed and therefore,
a globally conserved supercharge:
\begin{equation}
Q = \int d^3 x\bar{\epsilon}J^0.
\end{equation}
But in a $2+1$ dimensional spacetime any state of non-zero energy
produces a geometry that is asymptotically conical at infinity
(see also \cite{DeserJackiwtHooft}). The spinor fields are then no
longer covariantly constant at infinity \cite{Henneaux1984} and so
even when supersymmetry applies to the vacuum and ensures the
vanishing of the vacuum energy, it does not apply to the excited
states. Explicit examples have been constructed in
\cite{BeckerBeckerStrominger,Dvalifbdegen,Edelstein95,Edelstein96}.
Two further ideas in this direction, one in $D<4$ and one in $D>4$
are \cite{GregoryRubakovSibiryakov,CsakiErlichHollowood}, however
the latter later turned out to be internally inconsistent
\cite{DvaliGababadzePorrati}.

In any case, what is very important is to make the statement of
``breaking of supersymmetry'' more precise. As is clear, we do not
observe mass degeneracies between fermions and bosons, therefore
supersymmetry, even if it were a good symmetry at high energies
between excited states, is broken at lower energies. However, and
this is the point, as the example of Witten shows, the issue of
whether we do or do not live in a supersymmetric vacuum state is
another question. In some scenarios it is possible to have a
supersymmetric vacuum state, without supersymmetric excited
states. This really seems to be what we are looking for. The
observations of a small or even zero CC could point in the
direction of a (nearly) supersymmetric vacuum state.

Obviously the question remains how this scenario and the absence
nevertheless of a supersymmetric spectrum can be incorporated in 4
dimensions, where generically spacetime is asymptotically flat
around matter sources.

\subsection{Non-SUSY String Models}

It has been suggested that the cosmological constant might vanish
in certain string models \cite{Kachru:1998ys,Kachru:1998hd} due to
an equality between the number of boson and fermion mass states,
despite the fact that supersymmetry is broken. The one-loop
contribution to the cosmological constant therefore vanishes
trivially, but it is claimed that even higher order corrections
would not spoil this cancellation.

These ideas have been challenged in
\cite{Harvey:1998rc,Iengo:1999sm}, where it was argued that higher
loop corrections would indeed spoil the tree level results. The
most important drawbacks however for this proposed scenario, is
that the non-Abelian gauge sector, was always supersymmetric
\cite{Blumenhagen:1998uf,Angelantonj:1999gm,Angelantonj:2003hr}.
These approaches thus do not lead to viable models.

\section{Scale Invariance, e.g. Conformal Symmetry}

Another interesting symmetry with respect to the cosmological
constant problem is conformal symmetry, $g_{\mu\nu}\rightarrow
f(x^{\mu})g_{\mu\nu}$. Massless particles are symmetric under a
bigger group than just the Lorentz group, namely, the conformal
group. This group does not act as symmetries of Minkowski
spacetime, but under a (mathematically useful) completion, the
``conformal compactification of Minkowski space". This group is
15-dimensional and corresponds to $SO(2,4)$, or if fermions are
present, the covering group $SU(2,2)$. Conformal symmetry forbids
any term that sets a length scale, so a cosmological constant is
not allowed, and indeed also particle masses necessarily have to
vanish.

General coordinate transformations and scale invariance, i.e.
$g_{\mu\nu}\rightarrow fg_{\mu\nu}$, are incompatible in general
relativity. The $R\sqrt{-g}$ term in the Einstein-Hilbert action
is the only quantity that can be constructed from the metric
tensor and its first and second derivatives only, that is
invariant under general coordinate transformations. But this term
is not even invariant under a global scale transformation
$g_{\mu\nu}\rightarrow fg_{\mu\nu}$ for which $f$ is constant. $R$
transforms with Weyl weight $-1$ and $\sqrt{-g}$ with weight $+2$.
There are two ways to proceed to construct a scale invariant
action: introducing a new scalar field
\cite{Deser1970,Dirac:1973gk}, that transforms with weight $-1$,
giving rise to so-called scalar-tensor theories, or consider
Lagrangians that are quadratic in the curvature scalar. We
consider the second. See for example \cite{Booth,Barbour} for some
resent studies and many references.

Gravity can be formulated under this bigger group, leading to
``Conformal gravity", defined in terms of the Weyl tensor, which
corresponds to the traceless part of the Riemann tensor:
\begin{eqnarray}\label{weylaction}
S_G &=& -\alpha\int d^4
x\sqrt{-g}C_{\lambda\mu\nu\kappa}C^{\lambda\mu\nu\kappa}\nonumber\\
&=& -2\alpha \int d^4 x\sqrt{-g}\left(R_{\mu\nu}R^{\mu\nu}
-\frac{1}{3}R^2\right) +(\mbox{boundary}\;\mbox{terms}),
\end{eqnarray}
where $C^{\mu\nu\lambda\kappa}$ is the conformal Weyl tensor, and
$\alpha$ is a dimensionless gravitational coupling constant. Thus
the Lagrangian is quadratic in the curvature scalar and generates
field equations that are fourth-order differential equations.
Based on the successes of gauge theories with spontaneously broken
symmetries and the generation of the Fermi-constant, one may
suggest to also dynamically induce the Einstein action with its
Newtonian constant as a macroscopic limit of a microscopical
conformal theory. This approach has been studied especially by
Mannheim and Kazanas, see
\cite{Mannheim991,Mannheim992,Mannheim993,Mannheim96,Mannheim941,Mannheim942}
to solve the CC problem.

These fourth-order equations reduce to a fourth-order Poisson
equation:
\begin{equation}\label{4orderconf}
\nabla^4B(r) = f(r),
\end{equation}
where $B(r) = -g_{00}(r)$ and the source is given by:
\begin{equation}
f(r) = 3(T^{0}_{\ 0}-T^{r}_{\ r})/4\alpha B(r),
\end{equation}
For a static, spherically symmetric source, conformal symmetry
allows one to put $g_{rr} = -1/g_{00}$ and the exterior solution
to (\ref{4orderconf}) can be written \cite{Mannheim942}:
\begin{equation}
-g_{00} = 1/g_{rr}=1-\beta(2-3\beta\gamma)/r -3\beta\gamma +\gamma
r -kr^2.
\end{equation}
Assuming that the quadratic term is negligible at solar system
distance scales, the non-relativistic potential can be written:
\begin{equation}\label{nonrelpotconf}
V(r)=-\beta/r + \gamma r/2
\end{equation}
However, for a spherical source (\ref{4orderconf}) can be
completely integrated to yield:
\begin{equation}
B(r>R)=-\frac{r}{2}\int^R_0dr'f(r')r'^2
-\frac{1}{6r}\int^R_0dr'f(r')r'^4.
\end{equation}
Compared to the standard second-order equations:
\begin{equation}
\nabla^2\phi(r)=g(r)\quad\quad\rightarrow\quad\quad\phi(r>R)=-\frac{1}{r}\int^R_0dr'g(r')r'^2
\end{equation}
we see that the fourth-order equations contain the Newtonian
potential in its solution, but in addition also a linear potential
term that one would like to see dominate over Newtonian gravity
only at large distances. The factors $\beta$ and $\gamma$ in for
example (\ref{nonrelpotconf}) are given by:
\begin{equation}
\beta(2-3\beta\gamma) =
\frac{1}{6}\int_0^Rdr'f(r')r'^4\quad\quad,\quad\quad\gamma=-\frac{1}{2}\int_0^Rdr'f(r')r'^2
\end{equation}

Even if this would be correct, modifying gravity only at large
distances cannot solve the cosmological constant problem. The
(nearly) vanishing of the vacuum energy and consequently flat and
relatively slowly expanding spacetime is a puzzle already at
distance scales of say a meter. We could expect deviations of GR
at galactic scales, avoiding the need for dark matter, but at
solar system scales GR in principle works perfectly fine. It seems
hard to improve on this with conformal symmetry, since the world
simply is not scale invariant.

We also identified a more serious problem with the scenario of
Mannheim and Kazanas described above. In order for the linear term
not to dominate already at say solar system distances, the
coefficient $\gamma$ has to be chosen very small. Not only would
this introduce a new kind of fine-tuning, it is also simply not
allowed to choose these coefficients at will. The linear term will
always dominate over the Newtonian $1/r$-term, in contradiction
with the perfect agreement of GR at these scales. See also
\cite{PerlickXu} who raised the same objection.

This scenario therefore does not work.

\subsection{$\Lambda$ as Integration Constant}\label{unimodular}

Another option is to reformulate the action principle in such a
way that a scale dependent quantity like the scalar curvature,
remains undetermined by the field equations themselves. These are
the so-called `unimodular' theories of gravity, see e.g.
\cite{vanderBijvanDamNg,Unruh}. Note that although the action is
not globally scale invariant, Einstein's equations in the absence
of matter and with vanishing cosmological constant is. The
dynamical equations of pure gravity in other words, are invariant
with respect to global scale transformations, and since we have
that $R=0$, they are scale-free, i.e. they contain no intrinsic
length scale.

There is a way to keep the scale dependence undetermined also
after including matter which also generates a cosmological
constant term. This well-known procedure
\cite{AndersonFinkelstein,vanderBij:1981uw} starts by imposing a
constraint on the fluctuations in $g_{\mu\nu}$, such that the
determinant of the metric is fixed:
\begin{equation}\label{restmod}
\sqrt{-g} =
\sigma(x)\quad\quad\rightarrow\quad\quad\delta\sqrt{-g}=0,
\end{equation}
where $\sigma(x)$ is a scalar density of weight $+1$. The
resulting field equations are:
\begin{equation}
R_{\mu\nu}-\frac{1}{4}g_{\mu\nu}R = -\kappa\left(T_{\mu\nu} -
\frac{1}{4}g_{\mu\nu}T\right).
\end{equation}
The covariant derivative $D_{\mu}G_{\mu\nu}=D_{\mu}T_{\mu\nu}=0$
still vanishes and from this one obtains:
\begin{equation}
R - \kappa T = -4\Lambda,
\end{equation}
where $\Lambda$ now appears as an integration constant and the
factor of $4$ has been chosen for convenience since substituting
this back we recover the normal Einstein equations with
cosmological constant.

Recently, some arguments have been put forward in which a
unimodular theory is supposed to originate more naturally as a
result of `the quantum microstructure of spacetime being capable
of readjusting itself, soaking up any vacuum energy', see
\cite{Padmanabhan2004, Padmanabhan2002, Padmanabhan2004b}.

Obviously this does not solve much, nor does it provide a better
understanding of the cosmological constant. The restriction
(\ref{restmod}) on the variation of the metric has no deeper
motivation and the value of the integration constant $\Lambda$ has
to be inserted by hand in order to arrive at the correct value.

Besides, sometimes it is concluded that there are two inequivalent
Einstein equations for gravity, describing two theories that are
only equivalent classically, but not quantum mechanically. The
group of canonical transformations is much larger than that of
unitary transformations in Hilbert space, forcing one to quantize
in ``preferred" coordinates. We do not agree with this point of
view. The constraint $g^{\mu\nu}\delta g_{\mu\nu}=0$ just reflects
a choice of coordinates, a certain gauge.

This issue is closely related to the question of the measure of
the quantum gravity functional integral (see discussions by B.S.
DeWitt \cite{DeWitt1967,DeWitt1967b}, 't Hooft \cite{tHooft1978}
and \cite{MazurMottola1989}): Is the integration variable
$g_{\mu\nu}$, $g^{\mu\nu}$ or some other function of the metric?
The differences in the amplitudes for these theories all appear in
the one-loop diagrams, in the form of quartically divergent
momentum-independent ghost loops. These all disappear after
renormalization and therefore the theories are indistinguishable
physically.

\section{Holography}\label{Holography}

Gravitational holography \cite{tHooft1,Susskind:1994vu} limits the
number of states accessible to a system. The entropy of a region
generally grows with its covering area (in Planck units) rather
than with its volume, implying that the dimension of the Hilbert
space, i.e. the number of degrees of freedom describing a region,
is finite and much smaller than expected from quantum field
theory. Considering an infinite contribution to the vacuum energy
is not correct because states are counted that do not exist in a
holographic theory of gravity.

It is a symmetry principle since there is a projection from states
in the bulk-volume, to states on the covering surface.

In \cite{Thomas,CohenKaplanNelson} it is noted that in effective
field theory in a box of size $L$ with UV cutoff $M$ the entropy
$S$ scales extensively, as $S\sim L^3M^3$. A free Weyl fermion on
a lattice of size $L$ and spacing $1/M$ has $4^{(LM)^3}$ states
and entropy\footnote{For bosons the number of states is not
limited by a lattice cutoff alone, so in this argument one has to
limit oneself to fermions. For bosons there are an infinite number
of states, in contradiction to the conjecture of the Holographic
Principle.} $S\sim(LM)^3$. The corresponding entropy density
$s=S/V$ then is $s=M^3$. In $d=4$ dimensions quantum corrections
to the vacuum energy are therefore of order:
\begin{equation}
\rho_{vac} = \frac{\Lambda}{8\pi G} + \langle\rho\rangle =
\frac{\Lambda}{8\pi G} + \mathcal{O}(s^{4/3}),
\end{equation}
since both $\langle\rho\rangle$ and $s$ are dominated by
ultraviolet modes, (see also \cite{Hsu}). Thus finite $s$ implies
finite corrections to $\langle\rho\rangle$.

Using a cutoff $M$, $E\sim M^4L^3$ is the maximum energy for a
system of size $L$. States with $L<R_s\sim E$, with $R_s$ the
Schwarzschild radius, or $L>M^{-2}$ (in Planckian units) have
collapsed into a black-hole. If one simply requires that no state
in the Hilbert space exists with $R_s\sim E> L$, then a relation
between the size $L$ of the region, providing an IR cutoff, and
the UV cutoff $M$ can be derived, in natural units:
\begin{equation}
L^3M^4\lesssim LM_{P}^2
\end{equation}
This corresponds to the assumption that the effective theory
describes all states of the system, except those that have already
collapsed to a black hole.

Under these conditions entropy grows no faster than $A^{3/4}\sim
L^{3/2}$, with $A$ the area. If these black hole states give no
contribution to $\langle\rho\rangle$, we obtain:
\begin{equation}
\langle\rho\rangle\sim s^{4/3}\sim
\left(\frac{L^{3/2}}{L^3}\right)^{4/3}\sim L^{-2}.
\end{equation}
In \cite{Thomas} this same scaling was obtained by assuming that
$S<A$ as usual, but that the delocalized states have typical
Heisenberg energy $~1/L$:
\begin{equation}
\langle\rho\rangle\sim\frac{s}{L}\sim\frac{L^2}{L^3L}\sim L^{-2}.
\end{equation}
Plugging in for $L$ the observed size of the universe today the
quantum corrections are only of order $10^{-10}~\mbox{eV}^4$.

However, this does not yield the correct equation of state,
\cite{Hsu}. During matter dominated epochs, to which WMAP and
supernova measurements are sensitive, the horizon size grows as
the RW-scale factor, $a(t)^{3/2}$, so the above arguments imply:
\begin{equation}
\Lambda_{eff}(L)\sim a(t)^{-3},
\end{equation}
or, $w\equiv p/\rho =0$ at largest scales, since $\rho(t)\sim
a(t)^{-3(1+w)}$. The data on the other hand give $w< -0.78$~(95\%
CL) \cite{Spergel2003}. In \cite{Thomas,CohenKaplanNelson}
$\Lambda(L)$ is at all times comparable to the radiation + matter
energy density, which is also argued to give problems for
structure formation \cite{Turner2002}.

Holography-based scenarios thus naively lead to a cosmological
constant that is far less constant than what the data require.
This makes a connection between holography and dark energy a lot
harder to understand\footnote{In \cite{Kelleher2} in a different
context a similar relation between the CC and the volume of the
universe is derived, thus suffering from the same drawbacks.}.

More recently however, another proposal was made \cite{Li2004}
where instead $L$ is taken to be proportional to the size of the
future event horizon:
\begin{equation}
L(t)\sim a(t)\int_t^{\infty}\frac{dt'}{a(t')}
\end{equation}
This $L$ describes the size of the largest portion of the universe
that any observer will see. This could be a reasonable IR cutoff.
It is argued that in this case the equation of state parameter $w$
can be close enough to $-1$ to agree with the data. This relation
is rather ad hoc chosen, and its deeper meaning, if any, still has
to be discovered.

A very recent paper studying the implications of the holographic
principle for the cosmological constant problem is
\cite{Balazs:2006kc}, in which it is argued that holography makes
a tiny value for the cosmological constant `natural' in a large
universe. This is however a rather empty statement, since a large
(observable) universe necessarily must have a tiny value for the
cosmological constant. In other words, if the universe is large
the effective cosmological constant must be small, and no
holographic arguments are needed for that.

\subsection{Conceptual Issues in de Sitter Space}

Another reason to discuss holography in the context of the
cosmological constant problem lies in trying to reconcile string
theory with the apparent observation of living in a de Sitter
spacetime, see \cite{Kachru:2003aw} for a construction of de
Sitter vacua in string theory. The discussion centers around the
semi-classical result that de Sitter space has a finite entropy,
inversely related to the cosmological constant. As discussed in
section (\ref{wheresol}), the entropy is given by:
\begin{equation}
S_{dS} = \frac{A}{4G},\quad\quad\mbox{with}\quad\quad A=
\frac{12\pi}{\Lambda}
\end{equation}
with $A$ the area of the horizon. It seems natural to interpret
this entropy as the number of quantum states necessary to describe
a de Sitter universe \cite{Banks2000-1,Fischler2000}. This
interpretation is common in thermodynamics, and is also given for
the entropy of a black hole. Another motivation for this
interpretation arises from the fact that the classical phase space
of general relativity with asymptotic de Sitter boundary
conditions, both in the past and in the future, is compact. A
compact phase space yields a finite dimensional Hilbert space when
quantized. Thus one may reason that de Sitter space, or better,
asymptotic de Sitter space, since the metric fluctuates, should be
described by a theory with a finite number of independent quantum
states and that a theory of quantum gravity should be constructed
with a Hilbert space of finite dimension $\mathcal{N}$ in terms of
which the entropy is given by:
\begin{equation}\label{entropydS}
S_{dS} = \ln\mathcal{N}.
\end{equation}
In this reasoning a cosmological constant should be understood as
a direct consequence of the finite number of states in the Hilbert
space describing the world. Ergo, the larger the cosmological
constant, the smaller the Hilbert space:
\begin{equation}
\Lambda = \frac{3\pi}{\ln\mathcal{N}}.
\end{equation}
Banks \cite{Banks2000-1} therefore argues that the cosmological
constant should not be viewed as a derived, calculable quantity,
but instead as a fundamental input parameter. In
\cite{BoussoMeyersDewolfe} this reasoning is criticized. In any
case, finiteness of entropy and Hilbert space leads to several
conceptual difficulties.

One is that only compact groups can have non-trivial
finite-dimensional unitary representations, but the de Sitter
group is non-compact. Therefore, it has been claimed that either
de Sitter space has no holographic dual \cite{Goheer:2002vf},
which would make it impossible to have an analog of the successful
AdS/CFT dual for de Sitter space, or that in fact the correct
symmetry group is not the standard de Sitter group
\cite{WittenQGinDS}. However, whether these claims hold in the
future is unclear, and ways out of this conundrum have been
proposed, for example \cite{Parikh:2004wh}. See also
\cite{Strominger:2001pn,Strominger:2001gp} in which a holographic
dS/CFT correspondence is formulated, in apparent contradiction to
the above claim, since the CFT Hilbert space is in principle
infinite dimensional. It appears to be an unsettled question
whether this leads to consistent theories, especially since no
explicit example is known, in which a dS/CFT emerges directly from
string theory, see also the discussion in \cite{Klemm:2004mb}.

Another issue, as hinted on in the first chapter of this thesis,
arguments have been made that the standard Einstein-Hilbert action
with cosmological constant cannot be quantized for general values
of $G$ and $\Lambda$, but that it must be derived from a more
fundamental theory, which determines these values
\cite{WittenQGinDS}. The reasoning behind this statement is that
although the Hilbert space of quantum gravity in de Sitter space
has finite dimension, it is infinite dimensional perturbatively.
Perturbation theory is an expansion in powers $G\Lambda$, in four
dimensions. In the limit $G\Lambda\rightarrow 0$, $\mathcal{N}$
diverges exponentially, if (\ref{entropydS}) holds. In fact,
$\mathcal{N}$ will be non-trivial function of $G\Lambda$, yet it
has to take integer values, whereas the latter can vary
continuously.

However, it not immediately clear to which states the number
$\mathcal{N}$ refers to. Because of the appearance of negative
absolute temperatures, it has been argued in
\cite{Spradlin:2001pw} that instead of the usual thermodynamic
relation:
\begin{equation}
\frac{1}{T}=\frac{\partial S}{\partial
E},\quad\quad\quad\frac{1}{T}=\frac{\partial S}{\partial (-E)}
\end{equation}
the equation on the right should be used to compute the de Sitter
temperature. The idea behind this is that the de Sitter entropy
should perhaps be ascribed to states behind the horizon, which
cannot be observed.

Finally, another difficulty in de Sitter space has to to with
formulating ordinary quantum field theory. In quantum field
theory, in order to set up an S-matrix, one has to define incoming
and outgoing states, and these are only properly defined at
spatial infinity. De Sitter space however, is compact, there is no
notion of spatial infinity, only of temporal past and future
asymptotic regions. Matrix elements constructed this way are
therefore no measurable quantities. Note that also the
conventional formulation of string theory is based on the
existence of an S-matrix.

Another quantum field theory aspect is that there appear to be
instabilities in de Sitter space, for example in case of a scalar
field, but perhaps also in pure gravity. These issues are
discussed in chapter \ref{Back-reaction} and
\ref{TsamisWoodardbr}.

Much more needs to be understood about these issues to judge their
validity and possible impact on the cosmological constant problem.
It is clear however, that the measurements indicating an
accelerating universe and a dark energy equation of state
$w\approx -1$, have much more far-reaching consequences than
`just' the ordinary cosmological constant problem. In other words,
a solution to the cosmological constant problem, especially if not
by some symmetry, seems to have very deep implications for a broad
range of theoretical physics.

\section{``Symmetry" between Sub- and Super-Planckian Degrees of
Freedom}

This rather speculative reasoning originates from a comparison
with condensed matter physics and is due to Volovik, see for
example
\cite{Volovik2000,Volovik2001a,Volovik2001b,Volovik2002,Volovik2003a,Volovik2003b,Volovik2004,Volovik2005}.
The vacuum energy of superfluid $^4$Helium, calculated from an
effective theory containing phonons as elementary bosonic
particles and no fermions is:
\begin{equation}
\rho_{\Lambda} = \sqrt{-g}E^{4}_{Debye}
\end{equation}
with $g$ the determinant of the acoustic metric, since $c$ is now
the speed of sound, and $E_{Debye}=\hbar c/a$, with $a$ the
interatomic distance, which plays the role of the Planck length.
However, in the condensed matter case, the full theory exists: a
second quantized Hamiltonian describing a collection of a
macroscopic number of structureless $^4$Helium bosons or
$^3$Helium fermions, in which the chemical potential $\mu$ acts as
a Lagrange multiplier to ensure conservation of the number of
atoms:
\begin{eqnarray}
H-\mu N &=&\int
d\vec{x}\psi^{\dag}(\vec{x})\left[-\frac{\nabla^2}{2m}-\mu\right]\psi(\vec{x})\nonumber\\
&+& \int d\vec{x}d\vec{y}
V(\vec{x}-\vec{y})\psi^{\dag}(\vec{x})\psi^{\dag}(\vec{y})\psi(\vec{y})\psi(\vec{x}).
\end{eqnarray}

Using this Hamiltonian $H$ to calculate the energy density of the
ground state we get:
\begin{equation}
E_{vac} = E - \mu N = \langle\mbox{vac}|H-\mu N|\mbox{vac}\rangle
\end{equation}
An overall shift of the energy in $H$ is cancelled in a shift of
the chemical potential. Exact calculation shows that not only the
low energy degrees of freedom from the effective theory, the
phonons, but also the higher energy, ``trans-Planckian" degrees of
freedom have to be taken into account.

Besides, for a liquid of $N$ identical particles at temperature
$T$ in a volume $V$ in equilibrium, the relation between the
energy $E$ and pressure $P$ is given by the Gibbs-Duhem equation:
\begin{equation}
E = TS +\mu N - PV.
\end{equation}
Therefore at $T=0$ the energy density of the ground state becomes:
\begin{equation}
\rho_{vac}\equiv \frac{E_{vac}}{V} = -P_{vac},
\end{equation}
the same equation of state as for the vacuum state in GR. Using
just thermodynamic arguments, it is argued that in the infinite
volume, zero temperature limit, this gives exactly zero vacuum
energy density as long as there are no external forces, i.e. no
pressure acting on the quantum liquid. And assuming there is no
matter, no curvature and no boundaries which could give rise to a
Casimir effect \cite{Volovik2001a}.

The conclusion therefore is that, if these thermodynamic arguments
are also valid in a gravitational background for the universe as a
whole and up to extremely high energies, one would expect a
perfect cancellation between sub- and super-Planckian degrees of
freedom contributing to the vacuum energy, resulting in zero
cosmological constant.

Moreover, it is also argued that a non-zero cosmological constant
arises from perturbations of the vacuum at non-zero temperature.
The vacuum energy density would be proportional to the matter
energy density, solving the coincidence problem as well.

A similar result is obtained by \cite{KleinertZaanen}. In their
formulation the world is like a crystal. The atoms of the crystal
are in thermal equilibrium and exhibit therefore zero pressure,
making the cosmological constant equal to zero.

Both approaches strongly depend on the quantum systems reaching
their equilibrium state. However, in the presence of a
cosmological constant, the matter in the universe never reaches
its equilibrium state \cite{ShapiroSola2000}.

\section{Interacting Universes, Antipodal Symmetry}\label{Linde}

This is an approach developed by Linde \cite{Linde,Linde6} arguing
that the vacuum energy in our universe is so small because there
is a global interaction with another universe where energy
densities are negative. Consider the following action of two
universes with coordinates $x_{\mu}$ and $y_{\alpha}$
respectively, ($x_{\mu}, y_{\alpha} = 0,1,\dots,3$) and metrics
$g_{\mu\nu}(x)$ and $\bar{g}_{\alpha\beta}(y)$, containing fields
$\phi(x)$ and $\bar{\phi}(y)$:
\begin{equation}
S= N\int
d^{4}xd^{4}y\sqrt{g(x)}\sqrt{\bar{g}(y)}\left[\frac{M_{P}^{2}}{16\pi}R(x)
+ L(\phi(x)) - \frac{M_{P}^{2}}{16\pi}R(y) -
L(\bar{\phi}(y))\right],
\end{equation}
and where $N$ is some normalization constant. This action is
invariant under general coordinate transformations in each of the
universes separately. The important symmetry of the action is
$\phi(x)\rightarrow\bar{\phi}(x)$,
$g_{\mu\nu}(x)\rightarrow\bar{g}_{\alpha\beta}(x)$ and under the
subsequent change of the overall sign: $S\rightarrow -S$. He calls
this an antipodal symmetry, since it relates states with positive
and negative energies. As a consequence we have invariance under
the change of values of the effective potentials
$V(\phi)\rightarrow V(\phi) + c$ and $V(\bar{\phi})\rightarrow
V(\bar{\phi}) + c$ where $c$ is some constant. Therefore nothing
in this theory depends on the value of the effective potentials in
their absolute minima $\phi_{0}$ and $\bar{\phi}_{0}$. Note that
because of the antipodal symmetry $\phi_{0} = \bar{\phi}_{0}$ and
$V(\phi_{0}) = V(\bar{\phi}_{0})$.

In order to avoid the troublesome issues of theories with negative
energy states, there can be no interactions between the fields
$\phi(x)$ and $\bar{\phi}(y)$. Therefore also the equations of
motion for both fields are the same and similarly, also gravitons
from both universes do not interact.

However some interaction does occur. The Einstein equations are:
\begin{eqnarray}
R_{\mu\nu}(x) - \frac{1}{2}g_{\mu\nu}R(x) &=& -8\pi GT_{\mu\nu}(x)
- g_{\mu\nu}\langle\frac{1}{2}R(y) + 8\pi
GL(\bar{\phi}(y))\rangle\\
R_{\alpha\beta}(y) - \frac{1}{2}\bar{g}_{\alpha\beta}R(y) &=&
-8\pi GT_{\alpha\beta}(y) -
\bar{g}_{\alpha\beta}\langle\frac{1}{2}R(x) + 8\pi
GL(\phi(x))\rangle.
\end{eqnarray}
Here $T_{\mu\nu}$ is the energy-momentum tensor of the fields
$\phi(x)$ and $T_{\alpha\beta}$ the energy-momentum tensor for the
fields $\bar{\phi}(y)$ and the averaging means:
\begin{eqnarray}
\langle R(x)\rangle &=& \frac{\int d^{4}x\sqrt{g(x)}R(x)}{\int
d^{4}x\sqrt{g(x)}}\\
\langle R(y)\rangle &=& \frac{\int
d^{4}y\sqrt{\bar{g}(y)}R(y)}{\int d^{4}y\sqrt{\bar{g}(y)}}
\end{eqnarray}
and similarly for $\langle L(x)\rangle$ and $\langle L(y)\rangle$.

Thus there is a global interaction between the universes $X$ and
$Y$: The integral over the whole history of the $Y$-universe
changes the vacuum energy density of the $X$-universe. These
averages could be hard to calculate. Therefore, it is assumed that
both universes undergo a long enough period of inflation, such
that they become almost flat, and that at late times the fields
will settle near the absolute minimum of their potential. As a
result, the average of $-L(\phi(x))$ will almost coincide with the
value of $V(\phi_0)$, and the average of $R(x)$ coincides with its
value at late stages and similarly for $-L(\phi(y))$ and $R(y)$.
We arrive at:
\begin{eqnarray}
R_{\mu\nu}(x) - \frac{1}{2}g_{\mu\nu}R(x) &=& -8\pi
Gg_{\mu\nu}\left[V(\bar{\phi_0}) - V(\phi_0)\right] -\frac{1}{2}
g_{\mu\nu}R(y)\\
R_{\alpha\beta}(y) - \frac{1}{2}\bar{g}_{\alpha\beta}R(y) &=&
-8\pi Gg_{\alpha\beta}\left[V(\phi_0) - V(\bar{\phi_0})\right] -
\frac{1}{2}g_{\alpha\beta}R(x).
\end{eqnarray}
Thus at late stages the effective cosmological constant vanishes:
\begin{equation}
R(x) = -R(y) = \frac{32}{3}\pi G\left[V(\phi_0) -
V(\bar{\phi_0})\right] = 0,
\end{equation}
since because of the antipodal symmetry $\phi_0 = \bar{\phi_0}$
and $V(\phi_0) = V(\bar{\phi_0})$.

This could also be seen as a back-reaction mechanism, from one
universe at the other.

Difficulties with this approach are:
\begin{enumerate}
\item The form of the theory is completely ad hoc, devised just to
make the CC vanish and for that purpose we need to add a new
universe with a negative energy density.
\item Theory is completely classical; not obvious how to quantize
it, nor whether the cancellation of the CC survives quantum
corrections.
\item The cancellation depends on $\phi$ eventually settling down, in order to calculate the averages.
It is not clear how to generalize this.
\end{enumerate}

\section{Duality Transformations}

\subsection{S-Duality}

A different proposal was considered in \cite{Ellwanger2004}, where
S-duality acting on the gravitational field is assumed to mix
gravitational and matter degrees of freedom. The purpose is to
show that whereas the original metric may be (A)dS, de dual will
be flat. Only metrics are considered for which:
\begin{equation}\label{asssd1}
R^{a}_{\; b}\equiv R^{ca}_{\;\; bc}=\Lambda\delta^{a}_{\; b},
\end{equation}
with $\Lambda$ the cosmological constant. The mixing between
gravitational and matter degrees of freedom is obtained through a
new definition of the gravitational dual of the Riemann tensor,
including the field strength $F_{abcd}$ of a 3-form field
$A_{abc}$, which equation of motion is simply
$F_{abcd}=\omega\epsilon_{abcd}$, with $\omega$ some constant, see
also section (\ref{Hawking}):
\begin{eqnarray}
\tilde{R}_{abcd}&=&\frac{1}{2}\epsilon_{abef}\left(R^{ef}_{\;\;
cd}
+ F^{ef}_{\;\; cd}\right) + \frac{1}{12}\epsilon_{abcd}R, \nonumber\\
\tilde{F}_{abcd}&=& -\frac{1}{2}\epsilon_{abcd}R
\end{eqnarray}
such that:
\begin{eqnarray}
\tilde{\tilde{R}}_{abcd}&=& -R_{abcd}\nonumber\\
\tilde{\tilde{F}}_{abcd}&=& -F_{abcd}.
\end{eqnarray}
The equations of motion for the dual tensors become:
\begin{eqnarray}
\tilde{R}^{a}_{\; b}&=& 3\omega\delta^{a}_{\; b}\nonumber\\
\tilde{F}_{abcd}&=&
-\frac{1}{3}\Lambda\epsilon_{abcd}\equiv\tilde{\omega}\epsilon_{abcd}.
\end{eqnarray}
Therefore, if the vev $\omega$ would vanish, the dual Ricci
tensor, in casu the dual cosmological constant would also vanish.
Hence the conclusion is that if we would moreover `see' the dual
metric, determined by the dual Riemann tensor, we would observe a
flat spacetime.

But in order for this to work, one has to limit oneself to
spacetimes satisfying (\ref{asssd1}). Besides, the duality
relations have only proven to be consistent at linearized level.
There is also no particular reason for the vev of $\omega$ to
vanish.

Note that if one would constrain oneself to metrics which satisfy
$R=-4\Lambda$, the trace of the left-hand-side of Einstein's
equation vanishes by definition. In that case also the trace of
the energy-momentum tensor should vanish, which in general is not
the case. In other words, the vev of the 3-form field would then
have to be either zero or already incorporated in $\Lambda$, which
would render the duality transformations empty. Such a constraint
would be too strict, yet is very similar to (\ref{asssd1}).

Note that S-duality is an important concept in string theory. If
theories A and B are S-dual then $f_A(\alpha) = f_B(1/\alpha)$. It
relates type I string theory to the $SO(32)$ heterotic theory, and
type IIB theory to itself.

\subsection{Hodge Duality}

This duality between a $r$-form and a $(D-r)$-form in $D$
dimensions is studied in \cite{NishinoRajpoot2004}, where the
cosmological constant is taken to be represented by a 0-form field
strength, which is just a constant. This is somewhat related to
the unimodular approach of section (\ref{unimodular}) in the sense
that the intention is to introduce the cosmological constant in a
different way in the Einstein-Hilbert action. However, it does not
help in solving the cosmological constant problem.

\section{Summary}

A symmetry principle as explanation for the smallness of the
cosmological constant in itself is very attractive. A viable
mechanism that sets the cosmological constant to zero would be
great progress, even if $\Lambda$ would turn out to be nonzero.
Since supersymmetry does not really seem to help, especially some
form of scale invariance stands out as a serious option. Needless
to say, it is hard to imagine how scale invariance could be used,
knowing that the world around us is not scale invariant. Particle
masses are small, but many orders of magnitude larger than the
observed cosmological constant.

Another option might be that a symmetry condition enforcing
$\rho_{vac}$ equal to zero, could be reflected in a certain choice
of boundary conditions. In such a scenario, the vacuum state would
satisfy different boundary conditions then excited states. The
$x\rightarrow ix$ transformation of section (\ref{imaginaryspace})
could be an example of this.

%% file: chapter31.tex
\chapter{Invariance Under Complex Transformations}\label{imaginaryspace}

In this chapter\footnote{Based on our paper
\cite{'tHooft:2006rs}.} we study a new symmetry argument that
results in a vacuum state with strictly vanishing vacuum energy.
This argument exploits the well-known feature that de Sitter and
Anti- de Sitter space are related by analytic continuation. When
we drop boundary and hermiticity conditions on quantum fields, we
get as many negative as positive energy states, which are related
by transformations to complex space. The proposal does not
directly solve the cosmological constant problem, but explores a
new direction that appears worthwhile.

\section{Introduction}\label{intro.sec}

The scenario of this chapter has been introduced in
\cite{Nobbenhuis:2004wn}, and is based on a symmetry with respect
to a transformation towards imaginary values of the space-time
coordinates: \(x^\m\ra i\,x^\m\). This symmetry entails a new
definition of the vacuum state, as the unique state that is
invariant under this transformation. Since curvature switches
sign, this vacuum state must be associated with zero curvature,
hence zero cosmological constant. The most striking and unusual
feature of the symmetry is the fact that the \emph{boundary
conditions} of physical states are not invariant. Physical states
obey boundary conditions when the real parts of the coordinates
tend to infinity, not the imaginary parts. This is why all
physical states, except the vacuum, must break the symmetry. We
will argue that a vanishing cosmological constant could be a
consequence of the specific boundary conditions of the vacuum,
upon postulating this complex symmetry.

We do not address the issue of non-zero cosmological constant, nor the so-called cosmic coincidence problem. We believe that a symmetry which would
set the cosmological constant to exactly zero would be great progress.

The fact that we are transforming real coordinates into imaginary coordinates implies, \textit{inter alia}, that hermitean operators are transformed
into operators whose hermiticity properties are modified. Taking the hermitean conjugate of an operator requires knowledge of the boundary conditions
of a state. The transition from \(x\) to \(ix\) requires that the boundary conditions of the states are modified. For instance, wave functions \(\F\)
that are periodic in real space, are now replaced by waves that are exponential expressions of \(x\), thus periodic in \(ix\). But we are forced to
do more than that. Also the creation and annihilation operators will transform, and their commutator algebra in complex space is not a priori clear;
it requires careful study.

Thus, the symmetry that we are trying to identify is a symmetry of laws of nature \emph{prior} to imposing any boundary conditions. Demanding
invariance under \(x_\m\ra x_\m + a_\m\) where \(a_\m\) may be real or imaginary, violates boundary conditions at \(\F\ra\infty\), leaving only one
state invariant: the \emph{physical} vacuum.

\section{Classical Scalar Field}\label{classfield.sec}

To set our notation, consider a real, classical, scalar field
\(\F(x)\) in \(D\) space-time dimensions, with Lagrangian \be
\LL=-\half(\pa_\m\F)^2-V(\F(x))\ ,\qquad V(\F)=\half
m^2\F^2+\l\F^4\ . \eel{Lclass} Adopting the metric convention
\((-+++)\), we write the energy-momentum tensor as \be
T_{\m\n}(x)=\pa_\m\F(x)\pa_\n\F(x)+g_{\m\n}\LL(\F(x))\ .
\eel{Tmunuclass} The Hamiltonian \(H\) is \be \quad
H=\int\dd^{D-1}\vec x\,T_{00}(x)\ ;\qquad T_{00}=
\half\Pi^2+\half(\vec\pa\F)^2+V(\F)\ ;\qquad\Pi(x)=\pa_0\F(x)\ .
\eel{Hamclass}

Write our transformation as \(x^\m=iy^\m\), after which all coordinates are rotated in their complex planes such that \(y^\m\) will become real. For
redefined notions in \(y\) space, we use subscripts or superscripts \(y\), e.g., \(\pa_\m^y=i\pa_\m\). The field in \(y\) space obeys the Lagrange
equations with \be
\LL_y&=&-\LL\ =\ -\half(\pa^y_\m\F)^2+V(\F)\ ;\\
T_{\m\n}^y&=&-T_{\m\n}\ =\ \pa_\m^y\F(iy)\pa^y_\n\F(iy)+g_{\m\n}\LL_y(\F(iy))\ .
\eel{LTy} The Hamiltonian in \(y\)-space is \be H=-(i^{D-1})H_y&,&\quad
H_y=\int\dd^{D-1}y\,T_{00}^y\ ;\\T_{00}^y=\half\Pi_y^2+\half(\vec\pa_y\F)^2-V(\F)
&,&\quad\Pi_y(y)=i\Pi(iy)\ .\eel{Tyclass}

If we keep only the mass term in the potential, \(V(\F)=\half m^2\F^2\), the field obeys the Klein-Gordon equation. In the real \(x\)-space, its
solutions can be written as \be &&\F(x,\,t)\ =\int \dd^{D-1} p \left( a(p)e^{i(px)}+a^\ast(p) e^{-i(px)}\right)\ ,\crl{fieldfourrier} &&\Pi(x,\,t)\
=\int \dd^{D-1} p\ p^0\left(-ia(p)e^{i(px)}+ia^\ast(p) e^{-i(px)}\right)\ ;\crl{Pifourrier} &&p^0=\sqrt{\vec p^{\;2}+m^2}\ ,\qquad(px)\deff \vec
p\cdot\vec x-p^0t\ , \eel{pnulclass} where \(a(p)\) is just a c-number.

Analytically continuing these solutions to complex space, yields:
\be \quad\F(iy,\,i\t)&=&\int \dd^{D-1}q \left(
a_y(q)e^{i(qy)}+\hat{a_y}(q) e^{-i(qy)}\right)\
,\crl{fieldyfourrier} \quad \Pi_y(y,\,\t)\ =\
i\Pi(iy,\,i\t)&=&\int
\dd^{D-1}q\,q^0\left(-ia_y(q)e^{i(qy)}+i\hat{a_y}(q)
e^{-i(qy)}\right)\ ;\qquad{}\crl{Piyfourrier}\quad
&&q^0=\sqrt{\vec q^{\;2}-m^2}\ ,\qquad(qy)\deff \vec q\cdot\vec
y-q^0\t\ . \eel{qnul} The new coefficients could be analytic
continuations of the old ones, \be a_y(q)= (-i)^{D-1}a(p)\ ,\qquad
\hat{a_y}(q)= (-i)^{D-1}a^\ast(q)\ ,\qquad p^\m=-iq^\m\ ,\eel{aaq}
but this makes sense only if the \(a(p)\) would not have
singularities that we cross when shifting the integration contour.
Note, that, since \(D=4\) is even, the hermiticity relation
between \(a_y(q)\) and \({\hat a_y}(q)\) is lost. We can now
consider solutions where we restore them: \be {\hat
a_y}(q)=a_y^\ast(q)\ , \eel{hatstar} while also demanding
convergence of the \(q\) integration. Such solutions would not
obey acceptable boundary conditions in \(x\)-space, and the fields
would be imaginary rather than real, so these are unphysical
solutions. The important property that we concentrate on now,
however, is that, according to Eq.~\eqn{LTy}, these solutions
would have the opposite sign for \(T_{\m\n}\).

Of course, the field in \(y\)-space appears to be tachyonic, since
\(m^2\) is negative. In most of our discussions we should put
\(m=0\). A related transformation with the objective of
\(T_{\m\n}\ra -T_{\m\n}\) was made by Kaplan and Sundrum in
\cite{KaplanSundrum2005}. Non-Hermitian Hamiltonians were also
studied by Bender et al. in for example
\cite{Bender:2005pf,Bender:2005hf,Bender:1998ke,Bender2005}.
Another approach based on similar ideas which tries to forbid a
cosmological constant can be found in \cite{Bonelli:2000tz}.

\section{Gravity}

Consider Einstein's equations: \be R_{\m\n} - \half g_{\m\n}R - \L g_{\m\n} = -8\pi GT_{\m\n}. \ee Writing \be x^\m=i\,y^\m=i(\vec y,\,\t)\ , \qquad
g_{\m\n}^y(y)\ra g_{\m\n}(x=iy), \eel{compltrf} and defining the Riemann tensor in \(y\) space using the derivatives \(\pa^y_\m\), we see that \be
R_{\m\n}^y=-R_{\m\n}(iy)\ . \eel{Ricci} Clearly, in \(y\)-space, we have the equation \be R_{\m\n}^y - \half g_{\m\n}^yR^y + \L g_{\m\n}^y = +8\pi
GT_{\m\n}(iy)= -8\pi GT_{\m\n}^y. \ee Thus, Einstein's equations are invariant except for the cosmological constant term.

A related suggestion was made in \cite{Erdem2004}. In fact, we could consider formulating the equations of nature in the full complex space
\(z=x+iy\), but then everything becomes complex. The above transformation is a one-to-one map from real space \(\Re^3\) to the purely imaginary space
\(\Im^3\), where again real equations emerge.

The transformation from real to imaginary coordinates naturally relates deSitter space with anti-deSitter space, or, a vacuum solution with positive
cosmological constant to a vacuum solution with negative cosmological constant. Only if the cosmological constant is zero, a solution can map into
itself by such a transformation. None of the excited states can have this invariance, because they have to obey boundary conditions, either in real
space, or in imaginary space.

\section{Non-relativistic Particle}

The question is now, how much of this survives in a quantum
theory. The simplest example to be discussed is the
non-relativistic particle in one space dimension. Consider the
Hamiltonian \be H={p^2\over 2m}+V(x)\,,\eel{NRHamilton} where
\(p=-i\pa/\pa x\). Suppose that the function \(V(x)\) obeys \be
V(x)=-V(ix)\ ,\qquad V(x)=x^2V_0(x^4)\ , \eel{Potentl} with, for
instance, \(V_0(x^4)=e^{-\l x^4}\). Consider a wave function
\(|\j(x)\ket\) obeying the wave equation \(H|\j\ket=E|\j\ket\).
Then the substitution \be x=iy\ ,\qquad p=-ip_y\ ,\qquad
p_y=-i{\pa\over\pa y}\ ,\eel{ComplTransf} gives us a new function
\(|\j(y)\ket\) obeying \be H_y|\j(y)\ket=-E|\j(y)\ket\ ,\qquad
H_y={p_y^2\over2m}+V(y)\ .\eel{NewEq} Thus, we have here a
symmetry transformation sending the hamiltonian \(H\) into \(-H\).
Clearly, \(|\j(y)\ket\) cannot in general be an acceptable
solution to the usual Hamilton eigenvalue equation, since
\(|\j(y)\ket\) will not obey the boundary condition \(|\j(y)|^2\ra
0\) if \(y\ra \pm\infty\). Indeed, hermiticity, normalization, and
boundary conditions will not transform as in usual symmetry
transformations.

Yet, this symmetry is not totally void. If \(V=0\), a state
\(|\j_0\ket\) can be found that obeys both the boundary conditions
at \(x\ra\pm\infty\) and \(y\ra\pm\infty\). It is the ground
state, \(\j(x)=\mathrm{constant}\). To be precise, this state is
only normalizable if the boundary condition at \(x\ra\pm\infty\)
is replaced by a periodic boundary condition \(\j(x)=\j(x+L)\),
which also removes the other solution satisfying \(E=0\), namely
\(\j(x)=\alpha x\). (The important thing is that the bound \(E=0\)
on the physical states follows by comparing the solutions on the
real axis with the solutions on the imaginary axis.)

The state \(\j(x)=\mathrm{constant}\) obeys both boundary
conditions because of its invariance under transformations \(x\ra
x+a\), where \(a\) can be any complex number. Because of our
symmetry property, it obeys \(E=-E\), so the energy of this state
has to vanish. Since it is the only state with this property, it
must be the ground state. Thus, we see that our complex symmetry
may provide for a mechanism that generates a zero-energy ground
state, of the kind that we are looking for in connection with the
cosmological constant problem.

In general, if \(V(x)\ne0\), this argument fails. The reason is
that the invariance under complex translations breaks down, so
that no state can be constructed obeying all boundary conditions
in the complex plane. In our treatment of the cosmological
constant problem, we wish to understand the physical vacuum. It is
invariant under complex translations, so there is a possibility
that a procedure of this nature might apply.

As noted by Jackiw \cite{Jackiw}, there is a remarkable example in
which the potential does not have to vanish. We can allow for any
well-behaved function that depends only on \(x^4=y^4\). For
example, setting \(m=1\), \be V(x)=2x^6-3x^2=x^2(2x^4-3),\ee with
ground state wavefunction \(\exp(-x^4/2)\), indeed satisfies
condition \eqn{Potentl}, which guarantees zero energy eigenvalue.
Note that this restricts the transformation to be discrete, since
otherwise it crosses the point \(x=\sqrt{i}y\) where the potential
badly diverges. Boundary conditions are still obeyed on the real
and imaginary axis, but not for general complex values, see figure
\ref{xixfig3.fig}.

\begin{figure}[ht] \setcounter{figure}{0} \begin{quotation}
 \epsfxsize=62 mm\epsfbox{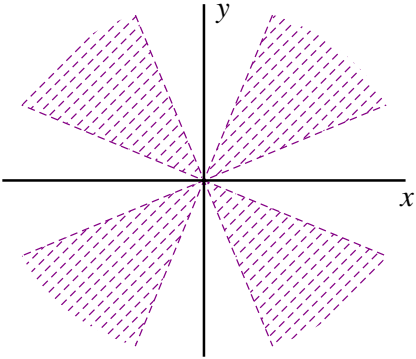}
  \caption{\footnotesize{Region in complex space where the potential is
  well-defined; the shaded region indicates where boundary conditions are not obeyed.}}
  \label{xixfig3.fig} \end{quotation}
\end{figure}

Moreover, as Jackiw also pointed out (see also
\cite{Bender:1992bk}), this example is intriguing since it reminds
us of supersymmetry. Setting again \(m=1\) for clarity of
notation, the Hamiltonian \be H=\frac{1}{2}(p+iW')(p-iW')\ee with
\(W\) the superpotential and a prime denoting a derivative with
respect to the fields, has a scalar potential \be
V=\frac{1}{2}(W'W'-W'').\ee If \(W\) satisfies condition
\eqn{Potentl}, the Hamiltonian possesses a zero energy
eigenfunction \(e^{-W}\), which obeys the correct boundary
conditions in \(x\) and \(y\). The Hamiltonian in this example is
the bosonic portion of a supersymmetric Hamiltonian, so our
proposal might be somehow related to supersymmetry.

We need to know what happens with hermiticity and normalizations. Assume the usual hermiticity properties of the bras, kets and the various operators
in \(x\) space. How do these properties read in \(y\) space? We have \be x=x^\dagger&\quad&p=p^\dagger\ ,\nn y=-y^\dagger&\quad&p_y=-p_y^\dagger\
,\eel{hermt} but the commutator algebra is covariant under the transformation: \be [p,\,x] =-i&\quad&p=-i\pa/\pa x\ ,\nn \ [p_y,\,y]
=-i&\quad&p_y=-i\pa/\pa y\ . \eel{commt} Therefore, the wave equation remains the same\ locally in \(y\) as it is in \(x\), but the boundary
condition in \(y\) is different from the one in \(x\). If we would replace the hermiticity properties of \(y\) and \(p_y\) in Eq.~\eqn{hermt} by
those of \(x\) and \(p\), then we would get only states with \(E\le0\).

\section{Harmonic Oscillator}\label{harmonic.sec}

An instructive example is the \(x\ra y\) transformation, with
\(x=iy\), in the harmonic oscillator. The Hamiltonian is \be
H={p^2\over 2m}+\half m\w^2x^2\ ,\eel{harmham}for which one
introduces the conventional annihilation and creation operators
\(a\) and \(a^\dagger\): \be &&a=\sqrt{m\w\over2}\left(x+{ip\over m\w}\right)\ ,\qquad a^\dagger=\sqrt{m\w\over2}\left(x-{ip\over m\w}\right)\ ;\\
&&H= \w(a^\dagger a+\half)\ . \eel{Hadaga} In terms of the operators in \(y\)-space, we can write \be a_y=\sqrt{m\w\over2}\left(y+{ip_y\over
m\w}\right)\ ,&\quad&
\hat{a_y}=\sqrt{m\w\over2}\left(y-{ip_y\over m\w}\right)\ =\ -a_y^\dagger\ ;\\
a_y=-ia^\dagger\ ,\quad \hat{a_y}=-i a\ ,&\quad& H= -\w(\hat{a_y} a_y+\half)\ . \eel{Hahatay}
If one were to replace the correct hermitian conjugate of \(a_y\) by \(\hat{a_y}\) instead of
\(-\hat{a_y}\), then the Hamiltonian \eqn{Hahatay} would take only the eigenvalues
\(H=-H_y=\w(-n-\half)\). Note that these form a natural continuation of the eigenstates
\(\w(n+\half)\), as if \(n\) were now allowed only to be a negative integer.

The ground state, \(|0\ket\) is not invariant. In \(x\)-space, the \(y\) ground state would be the non-normalizable state \(\exp(+\half m\w x^2)\),
which of course would obey `good' boundary conditions in \(y\)-space.

\section{Second Quantization}

\begin{figure}[ht] \begin{quotation}
 \epsfxsize=125 mm\epsfbox{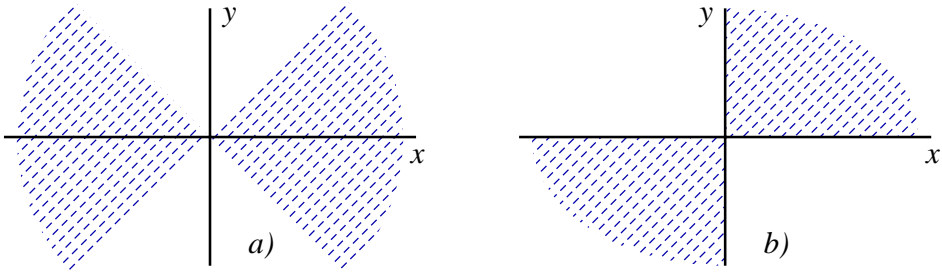}
  \caption{\footnotesize{Region in complex space where the Dirac delta function is
  well-defined, \((a)\) if \(\l\) is real, \((b)\) if \(\m\) is real and positive.}}
  \label{xixfig1.fig} \end{quotation}
\end{figure}

The examples of the previous two sections, however, are not the
transformations that are most relevant for the cosmological
constant. We wish to turn to imaginary coordinates, but not to
imaginary oscillators. We now turn our attention to
second-quantized particle theories, and we know that the vacuum
state will be invariant, at least under all complex translations.
Not only the hermiticity properties of field operators are
modified in the transformation, but now also the commutation rules
are affected. A scalar field \(\F(x)\) and its conjugate,
\(\Pi(x)\), often equal to \(\dot\F(x)\), normally obey the
commutation rules \be [\Pi(\vec x,t),\,\F({\vec
x\,}',\,t)]=-i\d^3(\vec x-{\vec x\,}')\ , \eel{fieldcomm} where
the Dirac deltafunction \(\d(x)\) may be regarded as \be
\d(x)=\textstyle{\sqrt{\l\over\pi}}\, e^{-\l x^2}\ ,
\eel{Diracdel} in the limit \(\l\uparrow\infty\). If \(\vec x\) is
replaced by \(i\vec y\), with \(\vec y\) real, then the
commutation rules are \be [\Pi(i\vec y,t),\,\F(i{\vec
y\,}',\,t)]=-i\d^3(i(\vec y-{\vec y\,}'))\ , \eel{complfieldcomm}
but, in Eq.~\eqn{Diracdel} we see two things happen:
\begin{itemize}\itm{\((i)\)} This delta function does not go to zero unless its
argument \(x\) lies in the right or left quadrant of
Fig.~\ref{xixfig1.fig}. Now, this can be cured if we add an
imaginary part to \(\l\), namely \(\l\ra -i\m\), with \(\m\) real.
Then the function \eqn{Diracdel} exists if \(x=r\,e^{i\th}\), with
\(0<\th<\half\pi\). But then, \itm{\((ii)\)} If \(x=iy\), the sign
of \(\m\) is important. If \(\m>0\), replacing \(x=iy\), the delta
function becomes \be
\d(iy)=\textstyle{\sqrt{-i\m\over\pi}}\,e^{-i\m y^2}\ra-i\d(y)\
,\eel{idel} which would be \(+i\d(y)\) had we chosen the other
sign for \(\m\).\end{itemize} We conclude that the sign of the
square root in Eq.~\eqn{Diracdel} is ambiguous.

There is another way to phrase this difficulty. The commutation rule \eqn{fieldcomm} suggests that either the field \(\F(\vec x,\,t)\) or \(\Pi(\vec
x,\,t)\) must be regarded as a distribution. Take the field \(\Pi\). Consider test functions \(f(\vec x)\), and write \be\Pi(f,\,t)\deff\int f(\vec
x)\,\Pi(\vec x,t)\,\dd^3\vec x\ ;\qquad [\Pi(f,\,t),\,\F(\vec x,\,t)]=-if(\vec x)\ . \eel{commdistr} As long as \(\vec x\) is real, the integration
contour in Eq.~\eqn{commdistr} is well-defined. If, however, we choose \(x=iy\), the contour must be taken to be in the complex plane, and if we only
wish to consider real \(y\), then the contour must be along the imaginary axis. This would be allowed if \(\Pi(\vec x,\,y)\) is holomorphic for
complex \(\vec x\), and the end points of the integration contour should not be modified.

 \begin{figure}[ht]
\begin{quotation}
 \epsfxsize=125 mm\epsfbox{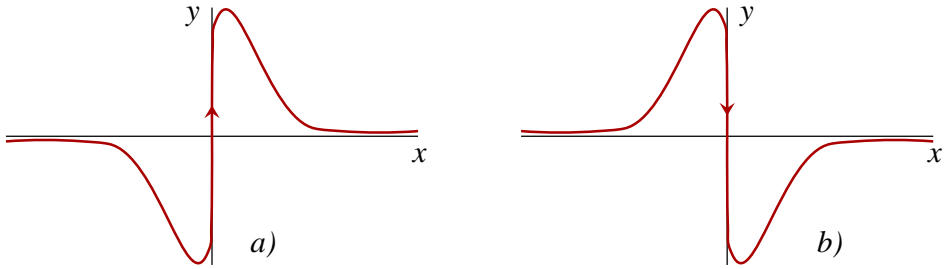}
  \caption{\footnotesize{Integration contour for the commutator algebra \eqn{commdistr},
\((a)\) and \((b)\) being two distinct choices.}}
  \label{xixfig2.fig}\end{quotation}
\end{figure}

For simplicity, let us take space to be one-dimensional. Assume that the contour becomes as in Fig.~\ref{xixfig2.fig}\textit{a}. In the \(y\) space,
we have \be \Pi(f,\,t) \deff \int_{-\infty}^\infty f(iy)\Pi(iy)\dd(iy)\ ;\qquad [\Pi(f,\,t),\,\F(i y,\,t)]=-if(i y)\ . \eel{compldistr} so that \be
[\Pi(iy,\,t),\,\F(iy',\,t)]=-\d(y-y')\ . \eel{complcomm} Note now that we could have chosen the contour of Fig.~\ref{xixfig2.fig}\textit{b} instead.
In that case, the integration goes in the opposite direction, and the commutator algebra in Eq.~\eqn{complcomm} receives the opposite sign. Note also
that, if we would be tempted to stick to one rule only, the commutator algebra would receive an overall minus sign if we apply the transformation
\(x\ra iy\) twice.

The general philosophy is now that, with these new commutation
relations in \(y\)-space, we could impose conventional hermiticity
properties in \(y\)-space, and then consider states as
representations of these operators. How do individual states then
transform from \(x\)-space to \(y\)-space or \textit{vice versa}?
We expect to obtain non-normalizable states, but the situation is
worse than that. Let us again consider one space-dimension, and
begin with defining the annihilation and creation operators \(a(
p)\) and \(a^\dagger(p)\) in \(x\)-space:

\be \F(x,\,t)=\quad\int{\dd p\over\sqrt{2\pi\cdot
2p^0}}\left(a(p)e^{i(px)}+a^\dagger(p)e^{-i(px)}\right)\ ,
\crl{fieldcreate} \Pi(x,\,t)=\ \ \int{\dd
p\sqrt{p^0}\over\sqrt{2\cdot2\pi}}\left(-ia(p)e^{i(px)}+ia^\dagger(p)e^{-i(px)}\right)\\\
\vphantom{\Big[}\qquad\qquad\qquad\qquad p^0=\sqrt{\vec
p^{\;2}+m^2}\ ,\qquad(px)\deff \vec p\cdot\vec x-p^0t\\
a(p)=\int{\dd
x\over\sqrt{2\pi\cdot2p^0}}\left(p^0\F(x,\,t)+i\Pi(x,\,t)\right)e^{-i(px)}\
,&&\crl{annihilfield} a^\dagger{(p)=\ \int{\dd
x\over\sqrt{2\pi\cdot2p^0}}\left(p^0\F(x,\,t)-i\Pi(x,\,t)\right)e^{i(px)}}\
.&&\ \eel{createfield}\\ Insisting that the commutation rules
\([a(p),\,a^\dagger(p')]=\d(p-p')\) should also be seen in
\(y\)-space operators: \be[a_y(q),\,\hat{a_y}(q')]=\d(q-q')\
,\eel{commycreate} we write, assuming \(p^0=-iq^0\) and
\(\Pi=-i\pa\F/\pa\t\) for free fields,

\be \F(iy,\,i\t)=\ \ \int{\dd q\over\sqrt{2\pi\cdot
2q^0}}\left(a_y(q)e^{i(qy)}+\hat{a_y}(q)e^{-i(qy)}\right)\
\crl{fieldycreate} \Pi(iy,\,i\t)=\qquad\int{\dd
q\sqrt{q^0}\over\sqrt{2\cdot2\pi}}\left(
-a_y(q)e^{i(qy)}+\hat{a_y}(q)e^{-i(qy)}\right)\ ,\crl{conjycreate}
\vphantom{\Big[}\qquad\qquad\qquad\qquad q^0=\sqrt{\vec
q^{\;2}-m^2}\ ,\qquad(qy)\deff \vec q\cdot\vec y-q^0\t\\
a_y(q)=\int{\dd y\over\sqrt{2\pi\cdot2q^0}}\left( q^0\F(iy,\,i\t)-
\Pi(iy,\,i\t)\right)e^{i(qy)}\ ,&&\crl{annihilyfield}
\hat{a_y}{(q)=\ \int{\dd y\over\sqrt{2\pi\cdot2q^0}}\left(
q^0\F(iy,\,i\t)+ \Pi(iy,\,i\t)\right)e^{-i(qy)}}\ ,&&\
\eel{createyfield}\\ so that the commutator \eqn{commycreate}
agrees with the field commutators \eqn{complcomm}. In most of our
considerations, we will have to take \(m=0\); we leave \(m\) in
our expressions just to show its sign switch.

In \(x\)-space, the fields \(\F\) and \(\pi\) are real, and the exponents in
Eqs~\eqn{fieldycreate}---\eqn{createyfield} are all real, so the hermiticity relations
are \(a_y^\dagger=a_y\) and \(\hat{a_y}^\dagger=\hat{a_y}\). As in the previous sections,
we replace this by \be \hat{a_y}=a_y^\dagger\ . \eel{aydagger}

The Hamiltonian for a free field reads \be
H&=&i\int_{-\infty}^\infty\dd
y\left(\half\Pi(iy)^2-\half(\pa_y\F(iy))^2+\half
m^2\F(iy)^2\right)\ =\nn &&-i\int\dd
q\,q^0\left(\hat{a_y}(q)a_y(q)+\half\right)\ =\ -i\int\dd
q\,q^0(n+\half)\ .\qquad \eel{Hamy} Clearly, with the hermiticity
condition \eqn{aydagger}, the Hamiltonian became purely imaginary,
as in Section \ref{classfield.sec}. Also, the zero point
fluctuations still seem to be there. However, we have not yet
addressed the operator ordering. Let us take a closer look at the
way individual creation and annihilation operators transform. We
now need to set \(m=0,\ p^0=|p|,\ q^0=|q|\). In order to compare
the creation and annihilation operators in real space-time with
those in imaginary space-time, substitute Eqs.~\eqn{fieldycreate}
and \eqn{conjycreate} into \eqn{annihilfield}, and the converse,
to obtain \be \quad a(p)&=&\int\int{\dd x\dd q\over
2\pi\sqrt{4p^0q^0}}\bigg\{(p^0-iq^0)\,a_y(q)\,e^{(q-ip)x}+(p^0+iq^0)\,\hat{a_y}(q)\,e^{(-q-ip)x}\bigg\}\
,\qquad \crl{aay}\quad a_y(q)&=&\int\int{\dd y\dd
p\over2\pi\sqrt{4p^0q^0}}\bigg\{(q^0+ip^0)\,a(p)\,e^{(-iq-p)y}+(q^0-ip^0)\,a^\dagger(p)\,e^{(-iq+p)y}\bigg\}
\ . \qquad\eel{aya}

The difficulty with these expressions is the fact that the \(x\)- and the \(y\)-integrals diverge. We now propose the following procedure. Let us
limit ourselves to the case that, in Eqs.~\eqn{annihilyfield} and \eqn{createyfield}, the \(y\)-integration is over a finite box only: \(|y|<L\), in
which case \(a_y(q)\sqrt{2q^0}\) will be an entire analytic function of \(q\). Then, in Eq.~\eqn{aay}, we can first shift the integration contour in
complex \(q\)-space by an amount \(ip\) up or down, and subsequently rotate the \(x\)-integration contour to obtain convergence. Now the square roots
occurring explicitly in Eqs.~\eqn{aay} and \eqn{aya} are merely the consequence of a choice of normalization, and could be avoided, but the root in
the definition of \(p^0\) and \(q^0\) are much more problematic. In principle we could take any of the two branches of the roots. However, in our
transformation procedure we actually \emph{choose} \(q^0=-ip^0\) and the second parts of Eqs.~\eqn{aay} and \eqn{aya} simply cancel out. Note that,
had we taken the other sign, i.e. \(q^0=ip^0\), this would have affected the expression for \(\F(iy,\,i\t)\) such, that we would still end up with
the same final result. In general, the \(x\)-integration yields a delta function constraining \(q\) to be \(\pm ip\), but \(q^0\) is chosen to be on
the branch \(-ip^0\), in both terms of this equation (\(q^0\) normally does not change sign if \(q\) does). Thus, we get, from Eqs.~\eqn{aay}
and \eqn{aya}, respectively, \be a(p)&=&i^{1/2}\,a_y(q)\ ,\qquad q=ip\,,\quad q^0=ip^0\ ,\\
a_y(q)&=&i^{-1/2}\,a(p)\ ,\qquad p=-iq\,,\quad p^0=-iq^0\
,\eel{apayq} so that \(a(p)\) and \(a_y(q)\) are analytic
continuations of one another. Similarly, \be\quad
a^\dagger(p)=i^{1/2}\,\hat{a_y}(q)\ ,\qquad \hat{a_y}(q)=
i^{-1/2}\,a^\dagger(p)\ ,\qquad p=-iq\,,\quad p^0=-iq^0\
.\eel{ahatdagger} There is no Bogolyubov mixing between \(a\) and
\(a^\dagger\). Note that these expressions agree with the
transformation law of the Hamiltonian \eqn{Hamy}.

Now that we have a precisely defined transformation law for the creation and annihilation operators, we can find out how the states transform. The
vacuum state \(|0\ket\) is defined to be the state upon which all annihilation operators vanish. We now see that this state is invariant under all
our transformations. Indeed, because there is no Bogolyubov mixing, all \(N\) particle states transform into \(N\) particle states, with \(N\) being
invariant. The vacuum is invariant because 1) unlike the case of the harmonic oscillator, Section~\ref{harmonic.sec}, creation operators transform
into creation operators, and annihilation operators into annihilation operators, and because 2) the vacuum is translation invariant.

The Hamiltonian is transformed into \(-i\) times the Hamiltonian (in the case \(D=2\)); the energy density \(T_{00}\) into \(-T_{00}\), and since the
vacuum is the only state that is invariant, it must have \(T_{00}=0\) and it must be the only state with this property.

\section{Pure Maxwell Fields}

This can now easily be extended to include the Maxwell action as well. In flat spacetime: \be S = -\int
d^3x\,\frac{1}{4}F_{\m\n}(x)F^{\m\n}(x),\quad\quad F_{\m\n}=\pa_\m A_\n - \pa_\n A_\m. \eel{actMaxx} The action is invariant under gauge
transformations of the form \be A_\m(x)\ra A_\m(x) + \pa_\m\x(x).\eel{gaugetrafo} Making use of this freedom, we impose the Lorentz condition
\(\pa_\m A^\m=0\), such that the equation of motion \(\pa_\m F^{\m\n}=0\) becomes \(\Box A^\m =0\). As is well known, this does not completely fix
the gauge, since transformations like \eqn{gaugetrafo} are still possible, provided \(\Box\x=0\). This remaining gauge freedom can be used to set
\(\na\cdot \vec{A}=0\), denoted Coulomb gauge, which sacrifices manifest Lorentz invariance. The commutation relations are \be [E^i(x,t),\,A_j(x',t)]
= i\d_{ij}^{tr}(\vec{x}-\vec{x}'),\eel{commxmax} where \be E^k=\frac{\pa\LL}{\pa\dot{A}_k} = -\dot{A}^k -\frac{\pa A_0}{\pa x^k},\ee is the momentum
conjugate to \(A^k\), which we previously called \(\Pi\), but it is here just a component of the electric field. The transverse delta function is
defined as \be \d_{ij}^{tr}(\vec{x}-\vec{x}')\equiv\int\frac{d^3p}{(2\pi)^3}e^{i\vec{p}(\vec{x}-\vec{x}')} \left(\d_{ij} -
\frac{p_ip_j}{\vec{p}^{\;2}}\right),\eel{deltatrx} such that its divergence vanishes. In Coulomb gauge, \(\vec{A}\) satisfies the wave equation
\(\Box\vec{A}=0\), and we write \be \vec{A}(x,t)=\int{\dd^3 p\over(2\pi)^3\sqrt{2p^0}}\sum_{\l=1}^{2}\vec{\e}(p,\,\l)\Big(
a(p,\,\l)e^{i(px)}+a^\dag(p,\,\l) e^{-i(px)}\Big)\ ,\ee where \(\vec{\e}(p,\,\l)\) is the polarization vector of the gauge field, which satisfies
\(\vec{\e}\cdot\vec{p}=0\) from the Coulomb condition \(\na\cdot\vec{A}=0\). Moreover, the polarization vectors can be chosen to be orthogonal
\(\vec{\e}(p,\,\l)\cdot\vec{\e}(p,\,\l')=\d_{\l\l'}\) and satisfy a completeness relation
\be\sum_{\l}\e_m(p,\,\l)\e_n(p\,\l)=\left(\delta_{mn}-\frac{p_mp_n}{\vec{p}^{\;2}}\right).\eel{complpol} The commutator between the creation and
annihilation operators becomes \be [a(p,\,\l),\,a^\dag(p',\,\l)]=\d(\vec{p}-\vec{p}')\d_{\l\l'},\ee in which the polarization vectors cancel out due
to their completeness relation.

In complex space, the field \(A_\m\) thus transforms analogously to the scalar field, with the only addition that the polarization vectors
\(\vec{\e}_\m(p)\) will now become function of complex momentum \(\vec{q}\). However, since they do not satisfy a particular algebra, like the
creation and annihilation operators, they do not cause any additional difficulties. The commutation relations between the creation and annihilation
operators behave similarly as in the scalar field case, since the second term in the transverse delta function \eqn{deltatrx}, and the polarization
vector completeness relation \eqn{complpol}, is invariant when transforming to complex momentum.

Thus we find \be F_{\m\n}(x,t)F^{\m\n}(x,t)\ra -F_{\m\n}(iy,i\t)F^{\m\n}(iy,i\t),\ee and again $T_{00}$ flips sign, as the energy-momentum tensor
reads: \be T_{\m\n} = -F_{\m\a}F_{\n}^{\a} + \frac{1}{4}F_{\a\b}F^{\a\b}\h_{\m\n}.\ee In term of the \(E\) and \(B\) fields, which are given by
derivatives of \(A_\m\), \(E_i=F_{0i}\), \(B_k=\half\e_{ijk}F_{jk}\), we have: \be T_{00}=\half\left(E^2 + B^2\right)\ra - T_{00},\ee which indicates
that the electric and magnetic fields become imaginary. A source term \(J^\m A_\m\) can also be added to the action \eqn{actMaxx}, if one imposes
that \(J^\m\ra -J^\m\), in which case the Maxwell equations \(\pa_\m F^{\m\n}=J^\n\) are covariant.

Implementing gauge invariance in imaginary space is also straightforward. The Max\-well action and Max\-well equations are invariant under
\(A_\m(x,t)\ra A_\m(x,t) +\pa_\m\x(x,t))\). In complex spacetime this becomes \be A_\m(iy,i\t)\ra A_\m(iy,i\t) -i\pa_\m(y,\t)\x(iy,i\t)\ee and the
Lorentz condition \be\pa_\m(x,t) A^\m(x,t) =0\quad \ra\quad -i\pa(y,\t)A^\m(iy,i\t).\ee In Coulomb gauge the polarization vectors satisfy
\be\vec{\e}(q)\cdot\vec{q}=0\ ,\ee with imaginary momentum \(q\).

Unfortunately, the Maxwell field handled this way will not be easy to generalize to Yang-Mills fields. The Yang-Mills field has cubic and quartic
couplings, and these will transform with the wrong sign. One might consider forcing vector potentials to transform just like space-time derivatives,
but then the kinetic term emerges with the wrong sign. Alternatively, one could suspect that the gauge group, which is compact in real space, would
become non-compact in imaginary space, but this also leads to undesirable features.

\section{Relation with Boundary Conditions}

All particle states depend on boundary conditions, usually imposed on the real axis. One could therefore try to simply view the \(x\ra ix\) symmetry
as a one-to-one mapping of states with boundary conditions imposed on \(\pm x\ra\infty\) to states with boundary conditions imposed on imaginary axis
\(\pm ix\ra\infty\). At first sight, this mapping transforms positive energy particle states into negative energy particle states. The vacuum, not
having to obey boundary conditions would necessarily have zero energy. However, this turns out not to be sufficient.

Solutions to the Klein-Gordon equation, with boundary conditions
imposed on imaginary coordinates are of the form: \be
\quad\F_\mathrm{im}(x,\,t)=\quad\int{\dd p\over\sqrt{2\pi\cdot
2p^0}}\left(a(p)e^{(px)}+\hat{a}(p)e^{-(px)}\right)\ ,&\quad&
p^0=\sqrt{p^2+m^2},\eel{bccompl} written with the subscript `im'
to remind us that this is the solution with boundary conditions on
the imaginary axis. With these boundary conditions, the field
explodes for real valued \(x\ra\pm\infty\), whereas for the usual
boundary conditions, imposed on the real axis, the field explodes
for \(ix\ra\pm\infty\). Note that for non-trivial \(a\) and
\(\hat{a}\), this field now has a non-zero complex part on the
real axis, if one insists that the second term is the Hermitian
conjugate of the first, as is usually the case. This is a
necessary consequence of this setup. However, we insist on writing
\(\hat{a}=a^\dag\) and, returning to three spatial dimensions, we
write for \(\F_\mathrm{im}(x,\,t)\) and
\(\Pi_\mathrm{im}(x,\,t)\): \be &\F_\mathrm{im}(\vec{x},t)& =
\int\frac{d^3p}{(2\pi)^3}\frac{1}{\sqrt{2p^0}}\left(a_pe^{(px)}
+ a^\dag_{p}e^{-(px)}\right),\nonumber\\
\dot{\F}_\mathrm{im}(\vec{x},t) = &\Pi_\mathrm{im}(\vec{x},t)& =
\int\frac{d^3p}{(2\pi)^3}(-){\sqrt{\frac{p^0}{2}}}\left(a_pe^{(
px)} - a^{\dag}_pe^{-(px)}\right),\nonumber\\
&&p^0 = \vphantom{\Big[}\sqrt{\vec p^{\;2}+m^2}\ ,\qquad(px)\deff
\vec p\cdot\vec x-p^0t,\ee and impose the normal commutation
relations between \(a\) and \(a^\dag\): \be
[a_p,a_{p'}^{\dag}]=(2\pi)^3\delta^{(3)}(\vec{p}-\vec{p}').\eel{commui}
Using Eqn. \eqn{commui}, the commutator between \(\F_\mathrm{im}\)
and \(\Pi_\mathrm{im}\) at equal times, becomes: \be
[\F_\mathrm{im}(\vec{x}),\,\Pi_\mathrm{im}(\vec{x})] =
\delta^{(3)}(\vec{x}-\vec{x}'),\ee which differs by a factor of
\(i\) from the usual relation, and by a minus sign, compared to
Eqn. \eqn{complcomm}. The energy-momentum tensor is given by \be
T_{\m\n\,\mathrm{im}}=\pa_\m\F_\mathrm{im}\pa_\n\F_\mathrm{im} -
\half \h_{\m\n}\pa^k\F_\mathrm{im}\pa_k\F_\mathrm{im}, \ee and
thus indeed changes sign, as long as one considers only those
contributions to a Hamiltonian that contain products of \(a\) and
\(a^\dagger\): \be H_\mathrm{im}^\mathrm{diag}=
\int\frac{d^3p}{(2\pi)^3}p^0\left(-a_{p}^{\dag}a_p
-\frac{1}{2}[a_p,\,a_{p}^{\dag}]\right)=-H.\ee However, the
remaining parts give a contribution that is rapidly diverging on
the imaginary axis \be T_{\m\n\,\mathrm{im}}^\mathrm{non-diag}=
a^2e^{2(px)} + (a^{\dag})^2e^{-2(px)},\ee but which blows up for
\(\pm x\ra\infty\). Note that when calculating vacuum expectation
values, these terms give no contribution.

To summarize, one can only construct such a symmetry, changing boundary conditions from real to imaginary coordinates, in a very small box. This was
to be expected, since we are comparing hyperbolic functions with their ordinary counterparts, \(\sinh(x)\) vs. \(\sin(x)\), and they are only
identical functions in a small neighborhood around the origin.

\section{Related Symmetries}

So far, we have discussed the implications of the transformation
$x_{\mu}\rightarrow ix_{\mu}$ and $g_{\mu\nu}\rightarrow
g_{\mu\nu}$. Alternatively, one might want to consider the related
discrete transformation $g_{\mu\nu}\rightarrow -g_{\mu\nu}$. The
combined transformation $x_{\mu}\rightarrow ix_{\mu}$ and
$g_{\mu\nu}\rightarrow -g_{\mu\nu}$ is a coordinate transformation
to imaginary spacetime. Under the transformation
$g_{\mu\nu}\rightarrow -g_{\mu\nu}$, the connection and Ricci
tensor are invariant, the Ricci scalar is not:
\begin{eqnarray}
\Gamma_{\mu\nu}^{\lambda} &\rightarrow& \Gamma_{\mu\nu}^{\lambda}\nonumber\\
R_{\mu\nu} &\rightarrow& R_{\mu\nu}\nonumber\\
R &\rightarrow&  -R
\end{eqnarray}
Therefore, promoting this transformation to a symmetry, a
cosmological constant term in the Einstein-Hilbert action is no
longer allowed. Only terms that transform with a minus sign are
allowed in the action. Einstein's equation transforms as:
\begin{eqnarray}
R_{\mu\nu} &-& \frac{1}{2}Rg_{\mu\nu} - \Lambda g_{\mu\nu} = -8\pi
G T_{\mu\nu} \rightarrow\nonumber\\
R_{\mu\nu} &-& \frac{1}{2}Rg_{\mu\nu} + \Lambda g_{\mu\nu} = -8\pi
G T_{\mu\nu}
\end{eqnarray}
and is invariant up to the cosmological constant term, assuming
invariance of $T_{\mu\nu}$ under this transformation. Note the
sign change compared to the action. In the field equations we
should only allow terms that are invariant under the proposed
transformation, whereas in the action only those terms are allowed
that transform with a minus sign.

Adding matter however, again causes problems. Consider for example
the Maxwell action:
\begin{eqnarray}\label{maxaction}
S_{Max} &=& \int
d^4x\sqrt{-g}\left(-\frac{1}{4}g^{\mu\alpha}g^{\nu\beta}F_{\mu\nu}F_{\alpha\beta}
+ g^{\mu\nu}J_{\mu}A_{\nu}\right)\\
F_{\mu\nu} &=& \partial_{\mu}A_{\nu} - \partial_{\nu}A_{\mu}.
\end{eqnarray}
This shows that the kinetic term for the EM-field is invariant,
but that the coupling of the source term $J_{\mu}$ is not. A
kinetic term for gauge bosons therefore would not be allowed,
unless one also demands
\begin{equation}\label{trafojena}
A_{\mu}\rightarrow \pm iA_{\mu},\quad\quad J_{\mu}\rightarrow \mp
J_{\mu}.
\end{equation}
With these additional transformations, the kinetic term is allowed
as well as the source term.

The standard energy-momentum tensor for this action is:
\begin{equation}
T_{\mu\nu} = -F_{\mu\alpha}F_{\nu}^{\alpha} +
\left(\frac{1}{4}F_{\alpha\beta}F^{\alpha\beta} -
J^{\alpha}A_{\alpha}\right)g_{\mu\nu},
\end{equation}
where the terms originating from the variation of the kinetic term
are invariant. One minus sign because of the contraction with
$g_{\mu\nu}$ and one minus sign from the $A_{\mu}\rightarrow\pm
iA_{\mu}$ transformation. This was to be expected, since the
energy momentum tensor is derived from the variation of the matter
action, with respect to $g_{\mu\nu}$:
$T_{\mu\nu}\propto\delta\mathcal{L}/\delta g_{\mu\nu}$,
neutralizing the minus sign in the action.

For a scalar field:
\begin{equation}
S = \int
d^4x\sqrt{-g}\left(g^{\mu\nu}\partial_{\mu}\phi\partial_{\nu}\phi
- m^2\phi^2\right)
\end{equation}
we have to face the same situation as in the previous section; the
kinetic part is allowed, but the potential term is not. That mass
squared terms have to transform to minus mass squared can of
course easily be seen just from special relativity:
\begin{equation}
p^2 = g^{\mu\nu}p_{\mu}p_{\nu} = m^2 \rightarrow -m^2.
\end{equation}

For fermions the situation is a bit more complicated. We normally
require:
\begin{equation}
\{\gamma^{\mu},\gamma^{\nu}\}\equiv \gamma^{\mu}\gamma^{\nu} +
\gamma^{\nu}\gamma^{\mu} = 2g^{\mu\nu},
\end{equation}
so under the transformation $g^{\mu\nu}\rightarrow -g^{\mu\nu}$
the gamma matrices would instead have to be defined as
$\gamma^{\mu}\rightarrow \pm i\gamma^{\mu}$.

The free Dirac Lagrangean in curved spacetime reads:
\begin{equation}
\mathcal{L}=\sqrt{-g}\left(i\bar{\psi}\gamma^{\mu}D_{\mu}\psi -
m\bar{\psi}\psi\right)
\end{equation}
with $D_{\mu}=\partial_{\mu}
-\frac{i}{4}\omega_{\mu}^{mn}\sigma_{mn}$ and $\sigma_{mn}$ is
defined in terms of flat space gamma-matrices:
$\sigma_{mn}=i(\gamma_n\gamma_m-\gamma_m\gamma_n)/2$. Moreover,
$\omega_{\mu}^{mn}$ is the spin-connection.

This appears to show the same behavior as the free scalar field.
The kinetic term transforms with a minus sign (because of the two
gamma-matrices in it), whereas the mass term again forces us to
consider imaginary masses.

Furthermore, the fermionic current:
\begin{equation}
J^{\mu}\equiv\bar{\psi}\gamma^{\mu}\psi\rightarrow -J^{\mu},
\end{equation}
becomes negative definite as one might have expected.

The coupling between spin-1 and spin-$\frac{1}{2}$ particles is
normally obtained by introducing the covariant derivative:
\begin{equation}
p_{\mu}\rightarrow p_{\mu} -eA_{\mu}\quad\quad\mbox{in
QM}\quad\quad i\partial_{\mu}\rightarrow i\partial_{\mu} -eA_{\mu}
\end{equation}
and this is independent of the sign of the metric tensor. However,
the requirement $A_{\mu}\rightarrow\pm iA_{\mu}$, appears to force
us to consider imaginary charges, which was to be expected,
because of (\ref{trafojena}) and consistency of Maxwell's
equations.

The main difficulty in this approach therefore is that masses as
well as general potential terms like $\lambda\phi^4$ and other
interaction, are not allowed. This transformation therefore is
similar to that discussed in the previous sections.

Yet another alternative is to consider performing both
transformations, i.e. $g_{\mu\nu}\rightarrow -g_{\mu\nu}$ and
$x_{\mu}\rightarrow ix_{\mu}$. Under this transformation the
different components of Einstein's equation transform as:
\begin{eqnarray}
\Gamma &\rightarrow& -i\Gamma\nonumber\\
R_{\mu\nu} &\rightarrow& -R_{\mu\nu}\nonumber\\
R &\rightarrow& R\nonumber\\
\Lambda g_{\mu\nu} &\rightarrow& -\Lambda g_{\mu\nu}
\end{eqnarray}
Therefore, this symmetry does not forbid a cosmological constant
term in the Einstein-Hilbert action. Einstein's equation
transforms as:
\begin{eqnarray}
R_{\mu\nu} &-& \frac{1}{2}Rg_{\mu\nu} - \Lambda g_{\mu\nu} = -8\pi
G T_{\mu\nu} \rightarrow\nonumber\\
-R_{\mu\nu} &+& \frac{1}{2}Rg_{\mu\nu} + \Lambda g_{\mu\nu} =
+8\pi G T_{\mu\nu},
\end{eqnarray}
and is invariant, assuming that also $T_{\mu\nu}$ transforms with
a minus sign. This does not seem to be helpful to control the
cosmological constant term. Moreover, the scalar field action is
invariant, but the Maxwell action (\ref{maxaction}) is not.

This is a peculiar transformation, since one would perhaps naively
expect that ordinary flat space quantum field theory would be
invariant under the combined transformations
$\eta_{\mu\nu}\rightarrow -\eta_{\mu\nu}$ and $x_{\mu}\rightarrow
ix_{\mu}$.

\subsection{Energy $\rightarrow -$ Energy}\label{Energytominus}

Another approach in which negative energy states are considered
has been recently proposed in \cite{KaplanSundrum2005}. Here the
discrete symmetry $E\rightarrow -E$ is imposed explicitly on the
matter fields by adding to the Lagrangian an identical copy of the
normal matter fields, but with an overall minus sign:
\begin{equation}
\mathcal{L} =\sqrt{-g}\left( M_{Pl}^2R - \Lambda_0 +
\mathcal{L}_{matt}(\psi,D_{\mu}) -
\mathcal{L}_{matt}(\hat{\psi},D_{\mu}) +\dots\right),
\end{equation}
where $\Lambda_0$ is the bare cosmological constant. The
Lagrangian with fields $\hat{\psi}$ occurring with the wrong sign
is referred to as the ghost sector. The two matter sectors have
equal but opposite vacuum energies, and therefore cancelling
contributions to the cosmological constant.

Crucial in this reasoning is that there is no coupling other than
gravitational between the normal matter fields and their ghost
counterparts, otherwise the Minkowski vacuum would not be stable.
Note that this approach is quite similar to Linde's antipodal
symmetry, we discussed in section (\ref{Linde}), although there
the symmetry does not involve gravity and the coupling of the two
sectors is suppressed, simply because they live in different
universes. Here any particle to ghost-particle coupling constant
$g$ in e.g a term $g\phi^2\hat{\phi}^2$ has to be postulated to be
exactly zero.

Moreover, in order to ensure stability of the vacuum, and prevent
a rapid decay of the vacuum due to negative energy particles, also
new Lorentz symmetry violating physics is required to suppress
processes where normal matter particles and ghosts emerge from the
vacuum; a process mediated by an off-shell graviton. There is no
other way to suppress the phase space integral over ghost momenta
\cite{Cline:2003gs}. In addition, one also has to assume that the
ghost sector is rather empty, compared to the normal matter
sector, in order not to spoil standard cosmology with such an
exotic type dark matter.

The gravitational coupling moreover has to be sufficiently small
in order to suppress the gravitationally induced interactions
between the two sectors and to make sure that the quantum
gravitational corrections to the bare cosmological constant are
kept very small. Processes for example in which out of nothing two
gravitons and two ghosts appear. It is therefore necessary to
impose a UV cutoff on these contributions of order $10^{-3}$~eV,
corresponding to a length scale of about 100~microns\footnote{In
section (\ref{fatgravitons}) a proposal by one of the authors of
\cite{KaplanSundrum2005} is discussed in which such a cutoff is
argued to arise from the graviton not being a point-like particle
but having this finite size.}.

\section{Summary}

It is natural to ascribe the extremely tiny value of the cosmological constant to some symmetry. Until now, the only symmetry that showed promises in
this respect has been supersymmetry. It is difficult, however, to understand how it can be that supersymmetry is obviously strongly broken by all
matter representations, whereas nevertheless the vacuum state should respect it completely. This symmetry requires the vacuum fluctuations of bosonic
fields to cancel very precisely against those of the fermionic field, and it is hard to see how this can happen when fermionic and bosonic fields
have such dissimilar spectra.

The symmetry proposed in this chapter is different. It is
suspected that the field equations themselves have a larger
symmetry than the boundary conditions for the solutions. It is the
boundary conditions, and the hermiticity conditions on the fields,
that force all physical states to have positive energies. If we
drop these conditions, we get as many negative energy as positive
energy states, and indeed, there may be a symmetry relating
positive energy with negative energy. This is the most promising
beginning of an argument why the vacuum state must have strictly
vanishing gravitational energy.

The fact that the symmetry must relate real to imaginary coordinates is suggested by the fact that De Sitter and Anti-De Sitter space are related by
analytic continuation, and that their cosmological constants have opposite sign.

Unfortunately, it is hard to see how this kind of symmetry could be realized in the known interaction types seen in the sub-atomic particles. At
first sight, all mass terms are forbidden. However, we could observe that all masses in the Standard Model are due to interactions, and it could be
that fields with positive mass squared are related to tachyonic fields by our symmetry. The one scalar field in the Standard Model is the Higgs
field. Its self interaction is described by a potential \(V_1(\F)=\half\l(\F^\dagger\F-F^2)^2\), and it is strongly suspected that the parameter
\(\l\) is unnaturally small. Our symmetry would relate it to another scalar field with opposite potential: \(V_2(\F_2)=-V_1(\F_2)\). Such a field
would have no vacuum expectation value, and, according to perturbation theory, a mass that is the Higgs mass divided by \(\sqrt 2\). Although
explicit predictions would be premature, this does suggest that a theory of this kind could make testable predictions, and it is worth-while to
search for scalar fields that do not contribute to the Higgs mechanism at LHC, having a mass somewhat smaller than the Higgs mass. We are hesitant
with this particular prediction because the negative sign in its self interaction potential could lead to unlikely instabilities, to be taken care of
by non-perturbative radiative corrections.

The symmetry we studied in this chapter would set the vacuum
energy to zero and has therefore the potential to explain a
vanishing cosmological constant. In light of the recent
discoveries that the universe appears to be accelerating
\cite{Riess1998,Perlmutter1998,Riess2004}, one could consider a
slight breaking of this symmetry. This is a non-trivial task that
we will have to postpone to further work. Note however, that our
proposal would only nullify exact vacuum energy with equation of
state \(w=-1\). Explaining the acceleration of the universe with
some dark energy component other than a cosmological constant,
quintessence for example, therefore is not ruled out within this
framework.

The considerations reported about in this chapter will only become
significant if, besides Maxwell fields, we can also handle
Yang-Mills fields, fermions, and more complicated interactions. As
stated, Yang-Mills fields appear to lead to difficulties.
Fermions, satisfying the linear Dirac equation, can be handled in
this formalism. Just as is the case for scalar fields, one finds
that mass terms are forbidden for fermions, but we postpone
further details to future work. Radiative corrections and
renormalization group effects will have to be considered. To stay
in line with our earlier paper, we still consider arguments of
this nature to explain the tiny value of the cosmological constant
unlikely to be completely successful, but every alley must be
explored, and this is one of them.

%% file: chapter4.tex
\chapter{Type II: Back-reaction}\label{Back-reaction}

It may be argued that any cosmological constant will be
automatically cancelled, or screened, to a very small value by
back-reaction effects on an expanding space. The effective
cosmological constant then is very small today, simply because the
universe is rather old. Often these effects are studied in an
inflationary background, where a primordial cosmological constant
is most dominant. In an inflationary background, the effects of
fields that are not conformally invariant will be most dominant

Two familiar massless particles are not conformally invariant,
massless minimally coupled scalars and gravitons. In this chapter
we will be concerned with back-reaction by a scalar field, discuss
the general setting of the renormalization group running of a
cosmological constant. In the next chapter we will discuss in
detail a back-reaction model in a purely quantum gravitation
setting, without any matter fields.

In chapter~1 we discussed a no-go theorem, derived by Weinberg
\cite{Weinbergreview}, that we will see at work in this chapter.
The theorem states that the vacuum energy density cannot be
cancelled dynamically, using a scalar field, without fine-tuning
in any effective four-dimensional theory with constant fields at
late times, that satisfies the following conditions:
\begin{enumerate}
\item General Covariance;
\item Conventional four-dimensional gravity is mediated by a
\textit{massless} graviton;
\item Theory contains a finite number of fields below the cutoff
scale;
\item Theory contains no negative norm states.
\end {enumerate}
Under these rather general assumptions the theorem states that the
potential for the compensator field, which should adjust the
vacuum energy to zero, has a runaway behavior. This means that
there is no stationary point for the potential of the scalar field
that should realize the adjustment, providing a severe difficulty
for such models.

\section{Scalar Field, Instabilities \textit{in} dS-Space}

The first attempts to cancel dynamically a `bare' cosmological
constant were made by referring to instabilities in the case of a
scalar field in de Sitter space \cite{Ford1985}. A massless
minimally coupled scalar field $\phi$ has no de Sitter-invariant
vacuum state, the two-point function in such a state does not
exist, because of an IR-divergence \cite{Ford:1977dj}. As a
consequence, the expectation value of $\phi^2$ is time-dependent:
\begin{equation}
\langle\phi^2\rangle = \frac{H^3}{4\pi^3}t.
\end{equation}
However, this breaking of de Sitter invariance is not reflected by
the energy-momentum tensor, since $T_{\mu\nu}$ only contains
derivatives and hence is not sensitive to long-wavelength modes.
This changes if one includes interactions. Consider for example a
$\lambda\phi^4$. Then:
\begin{equation}
\langle
T_{\mu\nu}\rangle\sim\lambda\langle\phi^2\rangle^2g_{\mu\nu}\propto
t^2.
\end{equation}
So in this case it is possible for $\langle T_{\mu\nu}\rangle$ to
grow for some time, until higher order contributions become
important. The infrared divergence results in a mass for the field
which in turn stops the growth of $\langle T_{\mu\nu}\rangle$, see
for example \cite{Ford1985, Ford}, comparable to what happens in
scalar field theory in flat spacetime, with a cubic
self-interaction term.

Another example of an instability with scalar particles in De
Sitter space was given by Myhrvold in \cite{Myhrvold:1983hx} with
an $\lambda\phi^4$ self-interaction term. In this case, spacetime
curvature makes the particle decay into three particles, which
again decay, in a runaway process. The interaction is crucial to
break conformal invariance, without this breaking there are always
stable de Sitter solutions \cite{Myhrvold:1983hu}.

One of the first illustrative, but unsuccessful attempts to use
such instabilities to screen the cosmological constant, was made
by Dolgov \cite{Dolgov82}. He used a rather simple classical model
for back-reaction:
\begin{equation}
\mathcal{L} =
\frac{1}{2}\left(\partial_{\alpha}\phi\partial^{\alpha}\phi - \xi
R\phi^2\right),\quad\quad\rightarrow\quad\quad\nabla_{\alpha}\nabla^{\alpha}\phi
+ \xi R\phi = 0,
\end{equation}
where $R$ is the scalar curvature, assumed to be positive, and
$\xi$ is taken to be a \textit{negative} constant. The
energy-momentum tensor is:
\begin{eqnarray}
T_{\mu\nu}(\phi)&=&\nabla_\mu\phi\nabla_\nu\phi
-\frac{1}{2}g_{\mu\nu}\nabla_\kappa\phi\nabla^\kappa\phi -
\xi\phi^2(R_{\mu\nu} - \frac{1}{2}g_{\mu\nu}R)\nonumber\\
&&- \xi\nabla_{\mu}\nabla_\nu\phi^2 + \xi
g_{\mu\nu}\nabla_{\kappa}\nabla^{\kappa}\phi^2,
\end{eqnarray}
and the energy density in the scalar field $\rho_\phi$:
\begin{equation}
\rho_\phi=\frac{1}{2}\dot{\phi}^2 +
3\xi\left(\frac{\dot{a}}{a}\right)^2\phi^2 +
6\xi\left(\frac{\dot{a}}{a}\right)\dot{\phi}\phi.
\end{equation}
In terms of the scale factor $a$, the equation of motion for
$\phi$ reads:
\begin{equation}
\ddot{\phi} + 3\frac{\dot{a}}{a}\dot{\phi} +
6\xi\left(\frac{\ddot{a}}{a} +
\left(\frac{\ddot{a}}{a}\right)^2\right)\phi = 0.
\end{equation}
Together with the Friedmann equation, describing the evolution of
the scale factor:
\begin{equation}
\left(\frac{\dot{a}}{a}\right)^2 = \frac{1}{3}\left(\Lambda_0 +
8\pi G\rho_\phi\right),
\end{equation}
where $\Lambda_0=3H^2$ stands for the effective value of the
cosmological constant during a de Sitter phase, this gives a pair
of non-linear coupled equations, that provide growing solutions
for $\phi$. At early times:
\begin{equation}\label{explosivephi}
\phi(t)=\phi_0 e^{\gamma t},\quad\quad\mbox{with}\quad\quad \gamma
=
\frac{3}{2}H\left[\left(1+\frac{16}{3}|\xi|\right)^{\frac{1}{2}}-1\right],
\end{equation}
with $t$ the comoving time in flat RW-coordinates. The energy
density in the scalar field is negative and increases:
\begin{eqnarray}
\rho_\phi &=& -A\phi^2\nonumber\\
A &=& \frac{1}{2}\left(6H^2|\xi| + 12\gamma H|\xi|
-\gamma^2\right)>0.
\end{eqnarray}

The scalar curvature $R$ and Hubble parameter $H$ become are:
\begin{equation}
R=\frac{8\pi[4\lambda + (6\xi-1)\dot{\phi}^2]}{M_{P}^2 +
8\pi\xi(6\xi-1)\phi^2},\quad\quad H^2=\frac{8\pi}{3}\frac{\lambda
+\frac{1}{2}\dot{\phi}^2 + 6H\xi\phi\dot{\phi}}{M_{P}^2 -
8\pi\xi\phi^2}
\end{equation}
where $\lambda$ has mass-dimension~4. The gravitational
back-reaction on $\phi$ slows down the explosive rise of
(\ref{explosivephi}). At late times the field grows linearly in
time $\phi\sim\sigma t$, where $\sigma$ is some constant, and the
scalefactor $a(t)$ grows as a power of time:
\begin{equation}
a(t)\sim t^{\beta},\quad\quad\mbox{with}\quad\quad\beta
=\frac{2|\xi|+1}{4|\xi|},
\end{equation}
and, consequently, $H\sim t^{-1}$ and $R\sim t^{-2}$.

Most importantly, the scalar field energy-momentum tensor at late
times approaches the form of a cosmological constant term:
\begin{equation}
8\pi G\langle T_{\mu\nu}\rangle\sim -\Lambda_0g_{\mu\nu} +
\mathcal{O}(t^{-2}).
\end{equation}
so the leading back-reaction term cancels the cosmological
constant originally present. The kinetic energy of the growing
$\phi$-field acts to cancel the cosmological constant. The order
$(t^{-2})$ part of the energy-density is:
\begin{equation}
\rho_\phi + \frac{\Lambda_0}{8\pi
G}\sim\frac{3(2|\xi|+1)2(1-2|\xi|)}{128\pi\xi^2t^2},
\end{equation}
which makes the effective cosmological constant nowadays extremely
small.

Unfortunately, not only the cosmological constant term is driven
to zero, since $\phi$ also couples to $R$, Newton's constant is
also screened:
\begin{equation}
G_{eff} = \frac{G_0}{1+8\pi G_0|\xi|\phi^2}\sim\frac{1}{t^2},
\end{equation}
where $G_0$ is the ``bare" value of $G$ at times where $\phi=0$.

Other models of this kind were also studied by Dolgov, see
\cite{Dolgov2,Dolgov3} but these proved to be unstable, leading
quickly to a catastrophic cosmic singularity. All these models do
evade Weinberg's no-go theorem, since the fields are not constant
at late times.

Models such as described above, where a screening is based on a
term $\xi R\phi^2$ have to be handled with extreme care, since
such a $\xi R\phi^2$ term can be obtained, or transformed away by
making a field transformation, and therefore is unphysical. The
metric $g_{\mu\nu}$ can be rescaled with a $\phi$-dependent scale
factor, yielding a new metric. In general, under a field
transformation $g_{\mu\nu}\rightarrow g_{\mu\nu} + \delta
g_{\mu\nu}$ the (gravitational) action transforms as:
\begin{equation}
\mathcal{L}\rightarrow \mathcal{L} + \left(G^{\mu\nu}-8\pi G
T^{\mu\nu}\right)\delta g_{\mu\nu} + \mathcal{O}(\delta g^2)
\end{equation}
If we now make the transformation $\delta g_{\mu\nu}$ dependent on
a field $\phi$:
\begin{equation}
\delta g_{\mu\nu} = \frac{1}{4}\lambda\phi^2g_{\mu\nu} + \ldots
\end{equation}
we see that a term linear in the curvature scalar appears:
\begin{equation}
\mathcal{L}\rightarrow \mathcal{L} - \lambda\left(R\phi^2 + 2\pi G
T\phi^2\right) + \mathcal{O}(\lambda^2)
\end{equation}
Subsequent transformation on $\phi$ can bring its kinetic term in
the canonical form. In the same way, a $\xi R\phi^2$-term can be
transformed away, at the cost of introducing higher dimensional
self-interaction terms in the scalar field potential. A different
way to see that such a term has no physical significance is to do
perturbative quantum gravity, see \cite{'tHooft:1974bx}. At tree
level, one can substitute the equations of motion, which in pure
gravity gives $R=0$, and if coupled to a scalar field $R=
T_{\mu}^{\mu}$. At higher loop orders these become the quantum
corrected equations of motion.

Moreover, a priori, it becomes unclear to which metric matter is
coupled. In other words, this brings an ambiguity in the
definition of the metric.

Models where a screening of a cosmological constant depend on such
terms therefore can never lead to a solution of the cosmological
constant problem.

\subsection{`Cosmon' Screening}

This is a different version of trying to screen the cosmological
constant by a scalar field $\chi$, called `cosmon' field
\cite{Wetterich1987,Wetterich:1994bg,Wetterich2002}. It is assumed
that all particle masses are determined by this scalar field.
Moreover, the renormalization group equation for this field is
assumed to be such, that at present it can play the role of
quintessence. The value of the $\chi$ field increases with time.

One starts with the following effective action, after integrating
out all standard model fields and fluctuations:
\begin{equation}\label{effcosmon}
S = \int d^4x\sqrt{g}\left[-\frac{1}{2}\chi^2R + \frac{1}{2}
\left(\sigma\left(\frac{\chi}{m}\right)-6\right)\partial^{\mu}\chi\partial_{\mu}\chi
+ V(\chi)\right],
\end{equation}
where $\sigma$ is a parameter depending on $\chi$ and the
potential is taken to be:
\begin{equation}
V(\chi)=m^2\chi^2.
\end{equation}
Note that in the action above, no cosmological constant is
written. Since everything is integrated out, a cosmological term
could have been introduced, but would necessarily be of near zero
value, in order for this effective theory to make any sense.

The normal Einstein-Hilbert term can be retrieved by making a
field transformation:
\begin{equation}
g_{\mu\nu}\rightarrow\left(\frac{M_P}{\chi}\right)^2g_{\mu\nu},
\end{equation}
after which the effective action (\ref{effcosmon}) transforms to:
\begin{equation}\label{effcosmonafterws}
S = \int d^4x\sqrt{g}\left[M_{P}^2R +
\frac{1}{2}\frac{M_{P}^2}{\chi^2}\left(\sigma\left(\frac{\chi}{m}\right)-6\right)
\partial^{\mu}\chi\partial_{\mu}\chi + \frac{M_{P}^2}{\chi^2}V(\chi)\right].
\end{equation}
Adding a cosmological constant to this action, and performing the
inverse of the above Weyl scaling, leads to:
\begin{equation}
S_{\Lambda} =
M_{P}^2\Lambda\quad\mbox{in~(\ref{effcosmonafterws})}\quad\rightarrow\quad\quad
S_\Lambda = \Lambda\chi^2\quad\mbox{in~(\ref{effcosmon})}.
\end{equation}
This term is small for small $\chi^2$, but becomes larger and
larger as the value of $\chi$ increases. A suggested additional
transformation of the $\chi$ field:
\begin{equation}\label{transchi2}
\phi/M_P = \ln(\chi^4/V(\chi)=\ln(\chi^2/m^2)
\quad\quad\rightarrow\quad\quad \chi^2=m^2e^{\phi/m},
\end{equation}
does not help much. In this case the value of $\phi$ increases,
and the cosmological constant term again becomes larger and
larger.

In ref. \cite{Wetterich1987,Wetterich:1994bg,Wetterich2002}, on
the other hand, the transformation (\ref{transchi2}) is used on
the $\chi$-field which, together with the Weyl rescaling of the
metric, transforming $V(\chi)$ to:
\begin{equation}
V(\chi,\phi)\rightarrow M_{P}^2\chi^2 e^{-\phi/m},
\end{equation}
which suggests that $V(\chi)$ decreases for increasing $\phi$.
However, using the same transformation (\ref{transchi2}) again to
remove the $\chi^2$ term, we are left with
\begin{equation}
V = M_{P}^2m^2,
\end{equation}
a strange constant term, which again is of the form of a
cosmological constant. Just putting $\chi^2=M_{P}^2$, and
therefore arguing that $V(\phi)\rightarrow 0$ for increasing
$\phi$ does not seem to be correct.

Positing a renormalization group equation, according to which the
kinetic term, parameterized by $\sigma(\chi)$ in
(\ref{effcosmon}), runs with energy $E$ as:
\begin{equation}
\frac{\partial\sigma}{\partial\chi}=\beta_{\sigma}
=E\sigma^2,\quad\quad\sigma=\frac{1}{E\ln(\chi_c/\chi)},
\end{equation}
determines late time cosmology. It is suggested that postulating
another renormalization group equation, for a parameter $g$
specifying a potential of the form $V=g\chi^4$:
\begin{equation}
\chi\frac{dg}{d\chi}=-Ag,\quad\quad A>0,
\end{equation}
will ensure asymptotically vanishing dark energy. However, not
only is there no deeper reason for such a particular,
renormalization group equation, with a minus-sign, this also still
does not control the cosmological constant, as we saw before.

Note that the `renormalization group equation' for the
$\chi$-field is important to make any statements about late time
cosmology. The running of the effective cosmological constant at
large values of $\chi$, characterized as a $\lambda(\chi/m)\chi^4$
potential term is especially important and would generally lead to
a late time cosmological constant that is too large. A `hidden
fine-tuning' needs to be reintroduced in the assumption that the
effective potential in (\ref{effcosmon}) has a flat direction.

Clearly, more is needed for this scenario to solve the
cosmological constant problem.

\subsection{Radiative Stability in Scalar Field Feedback
Mechanism}

Another approach deserves to be mentioned here, which will also be
listed under back-reaction mechanisms. This concerns a model that
does not solve the cosmological constant problem, but is intended
to provide a way to protect a zero or small cosmological constant
against radiative corrections, without using a symmetry,
\cite{MukohyamaRandall2003,Mukohyama2003}. This is achieved using
a scalar field with a non-standard, curvature dependent kinetic
term, such that in the limit where the scalar curvature goes to
zero, the kinetic term vanishes.
\begin{eqnarray}
S &=& \int d^4x\sqrt{-g}\left(\frac{R}{2\kappa^2} + \alpha R^2 +
L_{kin} - V(\phi)\right)\nonumber\\
L_{kin} &=& \frac{\kappa^{-4}K^q}{2qf^{2q-1}},
\end{eqnarray}
where $q$ is a constant that has to be $q>1/2$ for stability
reasons, and $f$ is a function of the scalar curvature $R$,
postulated to vanish at $R=0$ and that behaves near $R=0$ as:
\begin{equation}
f(R)\sim\left(\kappa^4R^2\right)^m,
\end{equation}
with $\kappa$ the Planck length. The parameter $\alpha$ is assumed
to be $\alpha >0$ to stabilize gravity at low energies, $m$ is an
integer that satisfies $2(m-1)>q(2q-1)$ and $K\equiv
-\kappa^4\partial^{\mu}\phi\partial_{\mu}\phi$.

The lowest value of $V(\phi)$, and thus the true value of the
vacuum energy density, is assumed to be negative in this approach,
not zero, but the peculiar dynamics makes the universe settle down
to a near zero energy state. The scalar field stops rolling and
its kinetic terms diverges.

The two main problems with this scenario are: 1) This specific
kinetic term is chosen by hand, not motivated by a more
fundamental theory, 2) all other fields settle to their ground
state faster than the vacuum energy, making the universe empty,
and reheating is necessary, to thermally populate the universe
again.

Other models where some dynamical feedback mechanism is proposed
based on a non-standard kinetic term can be found in
\cite{Rubakov1999,HebeckerWetterich2000,Hebecker2001,Wetterich2002}.
An interesting conjecture is made on the existence of a conformal
fixed point, possibly related to dilatation symmetry
\cite{Wetterich1987}. However, all these models still need
fine-tuning in masses and coupling constants, and it is unclear
whether they are experimentally viable, see \cite{Mukohyama2003}.

\subsection{Dilaton}

A natural scalar field candidate to screen the cosmological
constant could be the dilaton, which appears in string theory an
compactified supergravity theories. In the presence of a dilaton,
all mass scales arise multiplied with an exponential:
\begin{equation}
V_0(\phi)\sim M^4e^{4\lambda\phi},
\end{equation}
with $\phi$ the dilaton, and $\lambda$ a coupling constant. The
minimum of this potential is obtained for the value $\phi_0 =
-\infty$, which leads to the so-called `dilaton runaway problem':
couplings depend typically on $\phi$, and these tend to go to
zero, or sometimes infinity, in this limit. Moreover, all mass
scales have this similar scaling behavior, so particle masses also
vanish. Besides, the dilaton itself is nearly massless when it
reaches the minimum of its potential, leading to long-range
interactions that are severely constrained. Note that the
difference with quintessence is that the hypothetical quintessence
scalar field is assumed to couple only very weakly to the standard
model fields, contrary to the dilaton. Both scenarios however
predict varying `constants' of nature, such as the fine structure
constant $\alpha$. The current limits on varying $\alpha$ are
quite severe:
\begin{equation}
\frac{\Delta\alpha}{\alpha}\stackrel{z=2}{\sim} 10^{-5}
\end{equation}
In summary, the dynamical cancellation of a cosmological constant
term by back-reaction effects of scalar fields is hard to realize.
Next, let's focus on the other possible back-reaction mechanism, a
purely gravitational one.

\section{Instabilities \textit{of} dS-Space}

Gravitational waves propagating in some background spacetime
affect the dynamics of this background. Imposing the
transverse-tracefree gauge condition removes all gauge freedom,
and the independent components of the perturbation become
equivalent to a pair of massless scalar fields
\cite{Lifshitz:1945du,Ford:1977dj}.

This back-reaction can be described by an effective
energy-momentum tensor $\tau_{\mu\nu}$.

\subsection{Scalar-type Perturbations}\label{brandenberg}

In \cite{Brandenberger, Brandenberger1}, Brandenberger and
coworkers study back-reaction effects of scalar gravitational
perturbations. It is suggested that this could possibly solve the
CC problem.

Gravitational perturbation theory is formulated in terms of an
expansion in $g_{\mu\nu}$, see also section 7.5 of
\cite{Wald:1984rg}:
\begin{equation}
g_{\mu\nu}(\alpha)=\sum_n\frac{1}{n!}\alpha^ng_{\mu\nu}^{(n)},
\end{equation}
with $g^{(0)}_{\mu\nu}\equiv g_{\mu\nu}(0)$ the background metric.
Perturbation equations for $g_{\mu\nu}^{(n)}$ for the vacuum
Einstein equation, temporarily ignoring matter:
\begin{equation}
G_{\mu\nu}=0,
\end{equation}
are obtained by differentiating the Einstein tensor
$G_{\mu\nu}(\alpha)$, constructed from $g_{\mu\nu}(\alpha)$, $n$
times with respect to $\alpha$ at $\alpha =0$. The zeroth and
first order Einstein equations for $g_{\mu\nu}^{(0,1)}$ cancel
(this is the so-called `background field method'):
\begin{equation}
G_{\mu\nu}[g^{(0)}] = 0 \quad\quad\mbox{and}\quad\quad
G_{\mu\nu}^{(1)}[g^{(1)}] =0,
\end{equation}
where $G_{\mu\nu}^{(1)}$ is the linearized Einstein tensor
constructed from the background metric $g_{\mu\nu}^{(1)}$. The
second order equation shows the important effect:
\begin{equation}\label{gravpertwald}
G_{\mu\nu}^{(1)}[g^{(2)}]=- G_{\mu\nu}^{(2)}[g^{(1)}],
\end{equation}
where $G_{\mu\nu}^{(2)}[g^{(1)}]$ denotes the second-order
Einstein tensor, constructed from $g_{\mu\nu}^{(1)}$.

This implies that $G_{\mu\nu}^{(2)}[g^{(1)}]$ plays the role of an
`effective energy-momentum tensor', associated with the
perturbation $g_{\mu\nu}^{(1)}$, acting as a source for the second
order metric perturbation $g_{\mu\nu}^{(2)}$.

When gravity is coupled to matter, $T_{\mu\nu}\neq 0$, but the
same steps are taken. The zeroth and first order terms are assumed
to satisfy the equations of motion. Next, the spatial average is
taken of the remaining terms (a `coarse-grain viewpoint') and the
resulting equations are regarded as correction terms for a new
homogeneous metric $g_{\mu\nu}^{(0,br)}$, where the superscript
$(0,br)$ denotes the order in perturbation theory and the fact
that back-reaction is taken into account:
\begin{equation}\label{brandeneen}
G_{\mu\nu}\left(g_{\alpha\beta}^{(0,br)}\right) = -8\pi
G\left[T_{\mu\nu}^{(0)} + \tau_{\mu\nu}\right]
\end{equation}
and $\tau_{\mu\nu}$ contains terms resulting from averaging of the
second order metric and matter perturbations:
\begin{equation}\label{brandentwee}
\tau_{\mu\nu} = \langle T_{\mu\nu}^{(2)} - \frac{1}{8\pi
G}G_{\mu\nu}^{(2)}\rangle.
\end{equation}
In other words, the first-order perturbations are regarded as
contributing an extra energy-momentum tensor to the zeroth-order
equations of motion; the effective energy-momentum tensor of the
first-order equations renormalizes the zeroth-order
energy-momentum tensor.

Now work in longitudinal gauge and take for simplicity the matter
to be described by a single scalar field $\varphi$ with potential
$V$. Then there is only one independent metric perturbation
variable denoted $\phi(x,t)$. The perturbed metric is:
\begin{equation}
ds^{2} = -(1 + 2\phi)dt^{2} + a(t)^{2}(1 -
2\phi)\delta_{ij}dx^{i}dx^{j}.
\end{equation}
and $g^{(0,br)}$ is a Robertson-Walker metric. Denoting by
$\delta\varphi$ the matter perturbation:
\begin{equation}
\varphi(x,t)=\varphi_0(t) + \delta\varphi(x,t),
\end{equation}
the $\tau_{00}$ and $\tau_{ij}$ elements of the effective energy
momentum tensor are:
\begin{eqnarray}
\tau_{00}&=&\frac{1}{8\pi G}\left[12H\langle\phi\dot{\phi}\rangle
- 3\langle(\dot{\phi})^2\rangle  +
9a^{-2}\langle(\nabla\phi)^2\rangle\right]\nonumber\\
&&\nonumber\\
&&\nonumber\\
&+& \langle(\delta\dot{\varphi})^2\rangle +
a^{-2}\langle(\nabla\delta\varphi)^2\rangle\nonumber\\
&+& \frac{1}{2}V''(\varphi_0)\langle\delta\varphi^2\rangle +
2V'(\varphi_0)\langle\phi\delta\varphi\rangle,
\end{eqnarray}
where we have deliberately made a split, to capture in the first
line only contributions from the metric perturbations
$\delta\phi$, and in the lower lines contributions from the matter
perturbations $\delta\varphi$ (note however, that there are also
cross-terms). Similarly, we write
\begin{eqnarray}
\tau_{ij}&=& a^2\delta_{ij}\Big(\frac{1}{8\pi G}\Big[(24H^2 +
16\dot{H})\langle\phi^2\rangle +
24H\langle\dot{\phi}\phi\rangle\nonumber\\
&+& \langle(\dot{\phi})^2\rangle + 4\langle\phi\ddot{\phi}\rangle
- \frac{4}{3}a^{-2}\langle(\nabla\phi)^2\rangle\Big]\nonumber\\
&&\nonumber\\
&&\nonumber\\
&+&4\dot{\varphi}_{0}^2\langle\phi^2\rangle +
\langle(\delta\dot{\varphi})^2\rangle - a^{-2}
\langle(\nabla\delta\varphi)^2\rangle - 4\dot{\varphi}_0
\langle\delta\dot{\varphi}\phi\rangle\nonumber\\
&-& \frac{1}{2}V''(\varphi_0)\langle\delta\varphi^2\rangle +
2V'(\varphi_0)\langle\phi\delta\varphi\rangle\Big)
\end{eqnarray}
with $H\equiv \dot{a}/a$ the Hubble parameter, $V'=\partial
V/\partial\varphi$ and $\langle\rangle$ denotes spatial averaging.
In the long-wavelength limit, considering modes with wavelengths
longer than the Hubble radius, and ignoring terms
$\propto\dot{\phi}^2$ on the basis that such terms are only
important during times when the equation of state changes
\cite{Martineau:2005zu}, this gives:
\begin{equation}
\tau_{00}=\frac{1}{2}V''(\varphi_0)\langle\delta\varphi^2\rangle +
2V'(\varphi_0)\langle\phi\delta\varphi\rangle
\end{equation}
and
\begin{eqnarray}
\tau_{ij}&=& a^2\delta_{ij}\Big(\frac{1}{8\pi G}\left[(24H^2 +
16\dot{H})\langle\phi^2\rangle\right] +
4\dot{\varphi}_{0}^2\langle\phi^2\rangle\nonumber\\
&-& \frac{1}{2}V''(\varphi_0)\langle\delta\varphi^2\rangle +
2V'(\varphi_0)\langle\phi\delta\varphi\rangle.
\end{eqnarray}
In case of slow-roll inflation, with $\varphi$ the inflaton, these
modes contribute:
\begin{equation}
\tau_{0}^0\approx\left(2\frac{V''V^2}{V'^2}-4V\right)\langle\phi^2\rangle\approx
\frac{1}{3}\tau_{i}^i,
\end{equation}
and:
\begin{equation}
p\equiv -\frac{1}{3}\tau_{i}^i\approx - \tau_{0}^0
\end{equation}
showing the main result, that the equation of state of the
dominant infrared contribution to the effective energy-momentum
tensor $\tau_{\mu\nu}$ which describes back-reaction, takes the
form of a negative CC:
\begin{equation}
p_{br} = -\rho_{br}, \quad\quad\quad \rho_{br} < 0.
\end{equation}
This leads to the speculation that gravitational back-reaction may
lead to a dynamical cancellation mechanism for a bare CC, since
$\tau_{0}^0 \propto\langle\phi^2\rangle$, which is proportional to
IR phase space and this diverges in a De Sitter universe. Long
wavelength modes are those with wavelength longer than $H$, and as
more and more modes cross the horizon, $\langle\phi^2\rangle$
grows. To end inflation this way, however, takes an enormous
number of e-folds, see \cite{Martineau:2005zu} for a recent
discussion.

This approach is strongly debated in the literature as it is not
obvious whether one can consistently derive the equations of
motion in this way, see for example \cite{Grishchuk1994,
MartinSchwarz1997,Grishchuk1998,MartinSchwarz1998, Unruh1998}. We
believe that the obtained result that the back-reaction of long
wavelength gravitational perturbations screens the cosmological
constant, is incorrect.

A direct reason to be sceptical is purely intuitive. `Long
wavelength' in this framework refers to energy-momentum modes that
at every instant are much larger than the Hubble radius of
interest. The Hubble radius defines the observable portion of the
universe. How can one understand what happens locally? Take a box
of a cubic meter\footnote{This is a very useful example often
employed by Gerard 't Hooft to try to understand the fundamental
mechanism of an approach.}. In this box one could measure the
cosmological constant and one would find its near zero value. Why?
How can modes with wavelengths larger than the Hubble radius, have
any effect at all on what we measure in this box?

Note that we can always choose our coordinates such that locally
at a given point $P$, $g'_{\mu\nu}(x'_P)=\eta_{\mu\nu}$ and
$\partial g'_{\mu\nu}/\partial x'_{\alpha}=0$ evaluated at
$x=x_P$, simply constructing a local inertial frame at the point
$P$. The second and higher order derivatives of the metric can of
course not be made to vanish and measure the curvature. The long
wavelength perturbations are small enough that we do not notice
any deviation from homogeneity and isotropy, not even at cosmic
distances, but are argued to be large enough to alter the dynamics
of our universe, and determine the small value of the cosmological
constant that we in principle could measure at distances of, say,
a meter\footnote{The scale of a meter is used just to make the
argument more intuitive. The only constraint of course, is that it
should be larger than roughly a millimeter.}. This sounds
contradictory.

One of the delicate issues in the above derivation is gauge
invariance. As pointed out in \cite{Unruh1998}, the spatially
averaged metric is not a local physical observable: averaging over
a fixed time slice, the averaged value of the expansion will not
be the same as the expansion rate at the averaged value of time,
because of the non-linear nature of the expansion with time. In
other words, locally this `achieved renormalization', i.e. the
effect of the perturbations, is identical to a coordinate
transformation of the background equations and not a physical
effect. A similar conclusion was obtained in
\cite{KodamaHamazaki1997}.

Brandenberger and co-workers have subsequently tried to improve
their analysis by identifying a local physical variable which
describes the expansion rate
\cite{BrandenbergerGeshnizjani2003,Brandenberger2}. This amounts
to adding another scalar field that acts as an independent
physical clock. Within this procedure they argue that
back-reaction effects are still significant in renormalizing the
cosmological constant.

However, in \cite{Ishibashi:2005sj}, arguments are given, where
the above `derivation' (\ref{brandeneen},\ref{brandentwee}) of the
effective energy-momentum tensor goes wrong. We will briefly
review these here.

The central equation is (\ref{gravpertwald}). The point now is
twofold: one is that $G_{\mu\nu}^{(2)}[g^{(1)}]$ is a highly
gauge-dependent quantity, which cannot straightforwardly be
inserted as a new source of energy-momentum, and two, if
$G_{\mu\nu}^{(2)}[g^{(1)}]$ is small, its effects can be
calculated from (\ref{gravpertwald}), but if it is large enough to
apparently alter the dynamics of the universe, the third and
higher contributions to $g_{\mu\nu}(\alpha)$ will also be large
and calculating this back-reaction to second order perturbation
theory does not give reliable results.

This mechanism to drive the cosmological constant to zero,
therefore does not work, but back-reaction effects continue to be
a hot topic in cosmology. Back-reaction of inhomogeneities has
been invoked to explain an accelerating universe without a
cosmological constant, for example by Kolb et al.
\cite{Kolb:2004jg,Kolb:2005da,Kolb:2005me}. In a decelerating,
matter dominated universe, the back-reaction induced effective
cosmological constant is positive. There is no consensus on this
issue, although many critical studies have been undertaken to show
that these back-reaction effects do not lead to an accelerated
expansion. Simply put, it is just not possible to obtain an
accelerating cosmology from a decelerating universe, just by means
of back-reaction
\cite{Flanagan:2005dk,Hirata:2005ei,Giovannini:2005sy,Alnes:2005nq}.
For a large list of references and a critical examination, see
\cite{Rasanen:2005zy}. Amusingly, if these back-reaction effects
would lead to observable physical effects, they would include a
renormalization that might be in conflict with CMB measurements
\cite{Rasanen:2005zy,Flanagan:2005dk}.

A more specific criticism is that the super-Hubble perturbations
are assumed to satisfy the second order equation:
\begin{equation}\label{sndordbr}
G_{ab}^{(1)}[g^(0)] = \langle -G_{ab}^{(2)}[g^{(1)}]\rangle,
\end{equation}
while the first order perturbations are again assumed to satisfy
the equations of motion. The brackets indicate a spacetime
averaging, such that the linear terms in the second order
perturbation are eliminated, see \cite{Burnett:1989gp} for
details. In (\ref{sndordbr}) the term on the right hand side may
be very large, yet it can only be used when the wavelength of the
perturbation is much smaller than the curvature of the background
spacetime \cite{Burnett:1989gp,Ishibashi:2005sj}.

In the next chapter we will discuss another back-reaction model in
a purely quantum gravitational setting. At first sight, this
appears to be more promising.

\section{Running $\Lambda$ from Renormalization Group}\label{RenormalizationGroup}

As discussed in chapter one, the cosmological constant in a field
theory is expected to run with renormalization scale $\mu$ as any
other dimensionful parameter. These are rather straightforward
calculations and can be found in textbooks on quantum field
theory. We follow the derivation as given in \cite{Brown:1992db}

Consider the following Lagrangean:
\begin{equation}
\mathcal{L}=-\frac{1}{2}(\partial_{\mu}\phi)^2 -
\frac{1}{2}m_{0}^2\phi^2 -\frac{g_0}{4!}\phi^4 + \lambda_0 +
\rho_0\phi
\end{equation}
where $\lambda_0$ is just a constant, that becomes a vacuum term
in the energy-momentum tensor with dimension $\mbox{GeV}^4$. All
parameters $m_0$, $\lambda_0$ and the source function $\rho_0$ are
bare constants, that will be renormalized by quantum corrections.
Wave-function renormalization appears through a scale change of
this $\rho_0$. We do not need that here, so we will set $\rho_0=0$
from now on. For the time being, we also set $g_0=0$, considering
just the free scalar field. The first step is to derive the
running of the vacuum energy density with masses $m$.

The vacuum transition amplitude, from the remote past to the
distant future in Euclidean space can then be written as the
functional integral:
\begin{equation}
\langle 0+|0-\rangle=\int[d\phi]\exp\left(-\int
d_{E}^nx\left[\frac{1}{2}
(\partial_{\mu}\phi)^2+\frac{1}{2}m_{0}^2\phi^2 -
\lambda_0\right]\right)
\end{equation}
Using this expression, the variation with mass parameter $m_0$ is:
\begin{equation}\label{greenren}
\frac{\partial}{\partial m_{0}^2}\langle 0+|0-\rangle =
-\frac{1}{2}\int d_{E}^nx\langle 0+|\phi_{0}(x)^2|0-\rangle =
-\frac{1}{2}\int d_{E}^nx\triangle_{E}(0)\langle 0+|0-\rangle,
\end{equation}
with $\triangle_{E}$ is the Green's function, for two coincident
points:
\begin{equation}\label{greenfourierren}
\triangle_{E}(0)=\int\frac{d_{E}^nk}{(2\pi)^n}\frac{1}{k^2 +
m_{0}^2},
\end{equation}
and is written as a normalized expectation value in
(\ref{greenren}), when $\langle 0+|0-\rangle$ differs from unity.
In Feynman diagram language, this gives just a vacuum bubble, a
loop diagram with no external legs.

This Green's function can be evaluated in the usual way,
exponentiating the denominator:
\begin{equation}
\frac{1}{k^2+m_{0}^2}=\int_{0}^{\infty}ds\; e^{-s(k^2+m_{o}^2)}
\end{equation}
and thus:
\begin{equation}
\triangle_{E}(0)=\frac{1}{(4\pi)^{n/2}}\int_{0}^{\infty}ds
\;s^{-n/2}e^{-sm_{0}^2} =
\frac{(m_{0}^2)^{\frac{n}{2}-1}}{(4\pi)^{n/2}}\Gamma\left(1-\frac{n}{2}\right)
\end{equation}
Rewriting (\ref{greenren}) as:
\begin{equation}
\frac{\partial}{\partial m_{0}^2}\ln\langle 0+|0-\rangle =
-\frac{1}{2}\int d_{E}^nx\langle\triangle_{E}(0),
\end{equation}
this equation can trivially be integrated and gives:
\begin{equation}
\langle 0+|0-\rangle = \exp\left(-\int
d_{E}^nx\;\mathcal{E}\right)
\end{equation}
with:
\begin{equation}\label{mathcalEgamma}
\mathcal{E}=\frac{m_{0}^n}{(4\pi)^{n/2}}\frac{1}{n}
\Gamma\left(1-\frac{n}{2}\right) - \lambda_0
\end{equation}
and by fine-tuning $\lambda_0$, the vacuum energy density
$\mathcal{E}$ can be given any value. The special value
$\mathcal{E}=0$ would furthermore set $\langle 0+|0-\rangle=1$.
Note that because of dimensional reasons, no other constant can
appear in (\ref{mathcalEgamma}).

The expression (\ref{mathcalEgamma}) is just the lowest order
vacuum energy density and of course represents the infinite sum of
zero-point energies, as can be seen by performing the $p^4$
contour integration in the Fourier representation for
$\triangle_{E}(0)$, eqn. (\ref{greenfourierren}) and integrating
with respect to $m_{0}^2$. This gives:
\begin{equation}
\mathcal{E}=\int\frac{d^{n-1}k}{(2\pi)^{n-1}}\frac{1}{2}\sqrt{k^2+m_{0}^2}
- \lambda_0,
\end{equation}
a familiar expression, which we evaluated in chapter one.

Here we can use (\ref{mathcalEgamma}) instead. In four dimensions
the leading term is again a quartic divergence, which can be
traced back to the poles of the Gamma-function. It has poles at
$n=2$ and $n=4$, reflecting the logarithmic divergence in two
dimensions and quadratic divergence in four dimensions of the
momentum integral (\ref{greenfourierren}). This pole can be made
explicit using the recursion relation:
\begin{equation}
\Gamma(z+2)=z(z+1)\Gamma(z),\quad\quad\mbox{and}\quad\quad\Gamma(1)=1.
\end{equation}
The pole in four dimensions becomes:
\begin{equation}
\Gamma\left(1-\frac{n}{2}\right)\simeq\frac{2}{n-4}.
\end{equation}
Where we used the so-called minimal subtraction scheme (MS) in
which only the pole term is removed from the Gamma-function. The
residue is evaluated at $n=4$ and $\mu^{n-4}$ varies as to
maintain the correct dimensionality. The following renormalization
is obtained for the vacuum term:
\begin{equation}\label{renlambdavaryn}
\lambda_0=\mu^{n-4}\left(\frac{1}{2}\frac{m_{0}^4}{(4\pi)^2}\frac{1}{n-4}
+ \lambda\right)
\end{equation}
This makes the four-dimensional vacuum energy finite at one loop
order.

The counter term in (\ref{renlambdavaryn}) has to have integer
powers of $m_{0}^2$, since it corresponds to a subtraction of
divergent contributions coming from high momentum, or large $k^2$
of integrals behaving like $m_{0}^2/k^2$. The renormalization
group equation then reads:
\begin{equation}\label{rengroup}
\mu\frac{d\lambda}{d\mu}=(4-n)\lambda -
\frac{1}{2}\frac{m^4}{(4\pi)^2},
\end{equation}
written in terms of renormalized mass, which in the free theory,
that we used here, is the same as the bare mass. In general of
course they are not, but renormalization group equations, as eqn.
(\ref{rengroup}), should always be written in terms of
renormalized quantities. Note that the dominating effect in this
equation is simply the canonical term, required by the dimension
of $\lambda$. The higher orders contribute even less.

Substituting (\ref{renlambdavaryn}) in (\ref{mathcalEgamma}),
taking the limit $n\rightarrow 4$ and using that:
\begin{equation}
\Gamma(2-\frac{d}{2}) = \Gamma(\epsilon/2)=\frac{2}{\epsilon} -
\gamma + \mathcal{O}(\epsilon),
\end{equation}
with $\gamma\simeq 0.5772$ the Euler-Mascheroni constant, the
renormalized vacuum energy density becomes:
\begin{equation}
\mathcal{E}=\frac{1}{4}\frac{m^4}{(4\pi)^2}\left[\ln\left(\frac{m^2}{4\pi\mu^2}\right)
+ \gamma -\frac{3}{2}\right] - \lambda.
\end{equation}

\subsubsection{With Interactions}

Retaining the interaction term $\frac{g}{4!}\phi^4$, there is an
additional renormalization of the vacuum energy. To first order in
$g_0$, the correction to the vacuum amplitude is:
\begin{eqnarray}
\langle 0+|0-\rangle^{(1)}&=&-\frac{g_0}{4!}\int d_{E}^nx\langle
0+|\phi_{0}^4(x)|0-\rangle^{(0)}\nonumber\\
&=&-3\frac{g_0}{4!}\int d_{E}^nx\triangle_{E}^2(0)\langle
0+|0-\rangle^{(0)}.
\end{eqnarray}
In Feynman diagram language, this corresponds to the disconnected
double vacuum bubble, with a 4-point vertex.

The higher order contributions generate the following structure:
\begin{equation}\label{selfintcor}
\langle 0+|0-\rangle\simeq\langle 0+|0-\rangle^{(0)}\exp\left(-3
\frac{g_0}{4!}\int d_{E}^nx\triangle_{E}^2(0)\right),
\end{equation}
plus other corrections. The expression for $\mathcal{E}$ based on
(\ref{selfintcor}), is modified to:
\begin{equation}
\mathcal{E}=\frac{m_{0}^n}{(4\pi)^{n/2}}\frac{1}{n}
\Gamma\left(1-\frac{n}{2}\right)
+\frac{g_0}{8}\left[\frac{m_{0}^{n-2}}{(4\pi)^{n/2}}\Gamma\left(1-\frac{n}{2}\right)
\right]^2 - \lambda_0.
\end{equation}
Besides, now we also have to include mass renormalization, since
due to the self-interaction term for the scalar field, the bare
and renormalized mass are no longer the same. To order $g$ the
mass renormalization is given by:
\begin{equation}
m_{0}^n=m^n\left(1-\frac{n}{2}\frac{g}{(4\pi)^2}\frac{1}{n-4}\right),
\end{equation}
as $g$ is dimensionless.

Substituting this expression, we obtain:
\begin{eqnarray}
\mathcal{E}&=& \frac{m^n}{(4\pi)^{n/2}}\frac{1}{n}
\Gamma\left(1-\frac{n}{2}\right) - \frac{1}{2}\mu^{n-4}
\frac{m^4}{(4\pi)^2}\frac{1}{n-4}\nonumber\\
&& + \frac{1}{2}\mu^{n-4} \frac{gm^4}{(4\pi)^4}\left[\left(
\frac{m^2}{4\pi\mu^2}\right)^{\frac{n}{2}-2}\frac{1}{2}\Gamma\left(1-\frac{n}{2}\right)
- \frac{1}{n-4}\right]^2\nonumber\\
&& +\frac{1}{2}\mu^{n-4} \frac{m^4}{(4\pi)^2}\frac{1}{n-4}\left[1
- \frac{g}{(4\pi)^2}\frac{1}{n-4}\right] - \lambda_0.
\end{eqnarray}
The first two lines are finite for $n\rightarrow 4$, so vacuum
energy can again be rendered finite by renormalizing the
cosmological constant:
\begin{equation}
\lambda_0 =
\mu^{n-4}\left(\frac{1}{2}\frac{m^4}{(4\pi)^2}\frac{1}{n-4}\left[1
- \frac{g}{(4\pi)^2}\frac{1}{n-4}\right] + \lambda\right).
\end{equation}
Note that although there is a double pole at $n=4$, the residue of
this pole does not depend on $n$, except for the overall factor
$\mu^{n-4}$. This ensures that the counterterm only depends on
integer powers of the mass $m$.

The modified renormalization group equation for $\lambda$ becomes:
\begin{eqnarray}
\mu\frac{d\lambda}{d\mu}&=&(4-n)\lambda -
\frac{1}{2}\frac{m_{0}^4}{(4\pi)^2}\left[1+2\frac{g}{(4\pi)^2}
\frac{1}{n-4}\right]\nonumber\\
&=& (4-n)\lambda - \frac{1}{2}\frac{m^4}{(4\pi)^2},
\end{eqnarray}
showing that in fact, there is no correction to the
renormalization group equation for $\lambda$ to first order in
$g$.

The generalization of this result to arbitrary order in $g$ reads:
\begin{equation}
\mu\frac{d\lambda}{d\mu}= -(n-4)\lambda + m^4\beta_\lambda(g),
\end{equation}
with $\beta_\lambda(g)$ independent of $n$.

\subsubsection{Can Renormalization Group Running Nevertheless be
of Use to the Cosmological Constant Problem?}

In
\cite{Shapiro1994,Elizalde1994,ShapiroSola2000,ShapiroSola2004,Babic2004,ReuterWeyer2004}
an approach is studied, viewing $\Lambda$ as a parameter subject
to renormalization group running. The cosmological constant
becomes a scaling parameter $\Lambda(\mu)$, where $\mu$ is often
identified with the Hubble parameter at the corresponding epoch,
in order to make the running of $\Lambda$ smooth enough to agree
with all existing data, \cite{Espana-BonetLapuenteShapiroSola}.
The question is whether this running can screen the cosmological
constant.

Renormalization group equations typically give logarithmic
corrections, which makes it hard to see how this can ever account
for the suppression of a factor of $10^{120}$ needed for the
cosmological constant. In the above Refs., On dimensional grounds
the renormalization group equations are written:
\begin{eqnarray}
\frac{d\lambda}{d\ln\mu}=\sum_nA_n\mu^{2n},\quad\quad\rightarrow\quad\quad
\lambda(\mu)=\sum_nC_n\mu^{2n}\nonumber\\
\frac{d}{d\ln\mu}\left(\frac{1}{G}\right)=\sum_nB_n\mu^{2n},\quad\quad\rightarrow\quad\quad
\frac{1}{G(\mu)}=\sum_nD_n\mu^{2n}
\end{eqnarray}
with $A_n,B_n,C_n$ and $D_n$ coefficients, that depend on the
particle content of the standard model. The right-hand-sides of
the equations on the left define the $\beta$-functions for
$\lambda$ and $G^{-1}$, and all terms in this $\beta$-function for
$\lambda$ are of the form $\mu^{2n}m_{i}^{4-2n}$, with $m_{i}$ the
masses of different particles of the standard model. For the
cosmological constant, the common $A_0$-term, proportional to
$m_{i}^4$, has to be ignored in order to describe a
phenomenologically successful cosmology
\cite{Shapiro:2003ui,Espana-BonetLapuenteShapiroSola}. The
dominant term is then the second one, which makes the screening
still very slow, but stronger than logarithmic. We write:
\begin{equation}\label{betalam}
\frac{d\lambda}{d\ln\mu}=\frac{\sigma}{(4\pi)^2}\mu^2m_{i}^2 +
\ldots,
\end{equation}
where $\sigma = \pm 1$ depending on whether bosonic or fermionic
fields dominate below the Planck mass and $\sigma =0$ if $\mu
<m_i$. Integrating this, we find:
\begin{equation}\label{cosmoren}
\lambda(\mu) = C_0 + C_1\mu^2;\quad\quad
C_1=\frac{\sigma}{2(4\pi)^2}m_{i}^2.
\end{equation}
We can compare $m_{i}^2$ with $M_{P}^2$ by introducing a parameter
$\nu$:
\begin{equation}
\nu=\frac{\sigma}{12\pi}\frac{m_{i}^2}{M_{P}^2}
\end{equation}
where the pre-factor $1/(12\pi)$ is a convention used in refs.
\cite{Shapiro1994,Elizalde1994,ShapiroSola2000,ShapiroSola2004,Babic2004,ReuterWeyer2004}.
Note that for a standard FRW model, where $\lambda$ is spacetime
independent, we have $\nu=0$. The renormalization group parameter
$\mu$ is identified with the Hubble parameter $H(t)$, such that
the renormalization condition can be set at $\mu_0 = H_0$, i.e.
$\lambda(\mu_0)=\lambda_0$, with $\lambda_0$ the present day
observed value of the cosmological constant. This identification
is inspired by the numerical agreement that $H_{0}^2M_{P}^2\approx
10^{-47}~\mbox{GeV}^4=\lambda_0$, since $H_0\approx 10^{-33}$~eV.
With this renormalization condition, the integration constant
$C_0$ becomes:
\begin{equation}
C_0 = \lambda_{0} - \frac{3\nu}{8\pi}M_{P}^2\mu_{0}^2,\quad\quad
C_1=\frac{3\nu}{8\pi}M_{P}^2
\end{equation}

For all standard model particles the mass $m_i > \mu$, so they
decouple, and the beta-function in (\ref{betalam}) effectively
becomes:
\begin{equation}
\frac{d\lambda}{d\ln\mu}=\frac{\sigma}{(4\pi)^2}\mu^4
\end{equation}
resulting in coefficients:
\begin{equation}
C_0=\lambda_0 -C_{1}^{SM}\mu_{0}^4.
\end{equation}
Since $\mu_0=10^{-33}$~eV, this implies that there is basically no
running at all, for masses well below the Planck scale.

However, since $G$ and $\Lambda$ now effectively become
time-dependent, the energy conservation equation in a FRW universe
(see (\ref{consenergy})):
\begin{equation}
\dot\rho + 3H(\rho + p)=0,
\end{equation}
now becomes:
\begin{equation}\label{enconren}
(\rho + \Lambda)\dot{G} + G\dot\Lambda = 0.
\end{equation}
A cosmological model can be set up, since we have, in addition to
(\ref{cosmoren},~\ref{enconren}):
\begin{equation}
\rho + \Lambda = \frac{3H^2}{\pi G},
\end{equation}
which can be solved for $G(H,\nu)$:
\begin{equation}
G(H,\nu)=\frac{G_0}{1 + \nu\ln(H^2/H_{0}^2)},
\end{equation}
and $G_0=G(H_0)=1/M_{P}^2$.

The crucial term in the running of $G$ is the canonical one:
\begin{equation}
\mu\frac{d}{d\mu}\frac{1}{G} = \sum_ia_iM_{i}^2 +\cdot
\end{equation}
had this term not been taken into account, the running of $G$ and
$\Lambda$ with $H$ would become \cite{Sola:2005et}:
\begin{eqnarray}
G(H)&=&\frac{G_0}{1+\alpha G_o(H^2 -H_{0}^2)},\nonumber\\
\lambda(H)&=& \lambda_0 + \frac{3\alpha}{16\pi}(H^4-H_{0}^4)
\end{eqnarray}
and the running would be so slow, that it results in essentially
no phenomenology at all.

Keeping this term, the running is still very slow, but it could
possibly be measured as a quintessence or phantom dark energy and
be consistent with all data, as long as $0\leq |\nu| \ll 1$
\cite{Sola:2005et}. As a solution to the cosmological constant
problem, it obviously cannot help.

Besides, based on RG-group analysis, it is argued in
\cite{ReuterWeyer2003,ReuterWeyer2004} that there may be a UV
fixed point at which gravity becomes asymptotically free. If there
would be an IR fixed point at which $\Lambda_{eff}=0$ this could
shed some new light on the cosmological constant problem. This
scaling also effects $G$, making it larger at larger distances.

\subsection{Triviality as in $\lambda\phi^4$ Theory}

The Einstein Hilbert action with a cosmological constant can be
rewritten as \cite{JackiwNunezPi}:
\begin{equation}
S=-\frac{3}{4\pi}\int
d^4x\sqrt{-\hat{g}}\left(\frac{1}{12}R(\hat{g})\phi^2 +
\frac{1}{2}\hat{g}^{\mu\nu}\partial_{\mu}\phi\partial_{\nu}\phi
-\frac{\lambda}{4!}\phi^4\right)
\end{equation}
after rescaling the metric tensor as:
\begin{equation}
g_{\mu\nu}=\varphi^2\hat{g}_{\mu\nu}, \quad\quad
ds^2=\varphi^2\hat{d}s^2
\end{equation}
and defining:
\begin{equation}
\phi = \frac{\varphi}{\sqrt{G}}, \quad\quad
\Lambda=\frac{\lambda}{4G}.
\end{equation}
Now it is suggested that the same arguments first given by Wilson
\cite{Wilson1971}, that are valid in ordinary
$\lambda\phi^4$-theory, might also hold here and that this term is
suppressed quantum mechanically.

It is noted that perturbative running as in normal
$\lambda\phi^4$-theory is by far not sufficient, one has:
\begin{equation}
\mu\frac{d\Lambda}{d\mu}=\beta_1\Lambda^2,\quad\quad\Lambda=\frac{\Lambda_0}{1-\Lambda\beta_1
\log\mu},
\end{equation}
the purely gravitational renormalization group equation for a
cosmological constant. However, the hope is that there might be
some non-perturbative suppression. Similar ideas have been
contemplated by Polyakov, \cite{Polyakov2000}.

\section{Screening as a Consequence of the Trace
Anomaly}\label{traceanomaly}

As we discussed, Weinberg's no-go theorem is widely applicable to
screening mechanism. However, it was noted in e.g.
\cite{Peccei:1987mm}, that the symmetry (\ref{symnogo}) may be
broken by conformal anomalies, after which the Lagrangian obtains
an additional term proportional to
$\sqrt{g}\phi\Theta^{\mu}_{\mu}$, where $\Theta^{\mu}_{\mu}$ is
the effect of the conformal anomaly.

However, as already noted by Weinberg \cite{Weinbergreview}, this
does not provide a loophole to get around the no-go theorem. The
reason is that, although the field equation for $\phi$ now looks
like:
\begin{equation}
\frac{\partial\mathcal{L}}{\partial\phi}=\sqrt{-g}\left(T^{\mu}_{\mu}
+ \Theta^{\mu}_{\mu}\right),
\end{equation}
which may suggest an equilibrium value for $\phi$ with zero trace,
this is not sufficient for a flat space solution. The Einstein
equation for a constant metric now becomes:
\begin{equation}
0=\frac{\partial\mathcal{L}_{eff}}{\partial g_{\mu\nu}}\propto
e^{2\phi}\mathcal{L}_0 + \phi\Theta^{\mu}_{\mu},
\end{equation}
and the extra factor of $\phi$ shows that these two conditions are
not the same. The reason is that the term $\Theta^{\mu}_{\mu}$
does not simply end up in $T^{\mu}_{\mu}$.

More sophisticated are proposals in which it is argued that the
quantum trace anomaly of massless conformal fields in 4 dimensions
leads to a screening of the cosmological constant
\cite{Antoniadis:1991fa,Antoniadis:1995dy,Antoniadis:1996dj,AntoniadisMottolaMazur98,Mazur:2001aa}.
The idea is similar in spirit to the one in the previous section,
(\ref{TsamisWoodardbr}). One tries to find a renormalization group
screening of the cosmological constant in the IR, but instead of
taking full quantum gravity effects, only quantum effects of the
conformal factor are considered. See also \cite{Odintsov:1990rs}
for a related earlier study.

Recall that if an action is invariant under conformal
transformations:
\begin{equation}
g_{\mu\nu}(x)\rightarrow\Omega^2(x)g_{\mu\nu}(x) =
\bar{g}_{\mu\nu}(x),
\end{equation}
one directly finds, after varying with respect to $g_{\mu\nu}$ for
the trace of the energy momentum tensor:
\begin{equation}
T^{\mu}_{\mu}[g_{\mu\nu}(x)]=-\frac{\Omega(x)}{\sqrt{-g}}\frac{\delta
S[\bar{g}_{\mu\nu}]}{\delta \Omega(x)}\Big |_{\Omega =1} =0.
\end{equation}
This shows that if the classical action is invariant under
conformal transformations, then the classical energy momentum
tensor is traceless.

Masses and dimensionful couplings explicitly violate conformal
invariance and even if one starts out with a classical theory that
is invariant under conformal transformations, quantum corrections
generally introduce mass scales. This is closely related with the
scaling behavior of the action, and hence with the renormalization
group. The vacuum fluctuations of massless fields, for instance,
generate such a conformal anomaly.

Such a non-vanishing trace of the energy-momentum tensor couples
to a new spin-0 degree of freedom, reflected in the conformal
sector, or trace of the metric. It is argued, that fluctuations of
this conformal sector of the metric grow logarithmically at
distance scales of the order of the horizon. Moreover, their
effect on the renormalization group flow is argued to lead to an
IR stable, conformally invariant fixed point of gravity, where
scale invariance is restored. The anomalous trace leads to an
effective action that is non-local in the full metric for
gravitational interactions, although the trace anomaly itself is
given by a sum of local terms that are fourth order in curvature
invariants.

The authors conclude that the effective cosmological constant in
units of Planck mass decreases at large distances and that
$G_N\Lambda\rightarrow 0$ at the IR fixed point in the infinite
volume limit.

However, the cosmological constant problem manifests itself
already at much smaller distances and moreover, it is unclear
whether this scenario is compatible with standard cosmological
observations. Moreover, like the other approaches in this chapter,
it relies heavily on quantum effects having a large impact at
enormous distance scales. As argued in the previous section, it is
debatable whether these effects can be sufficiently significant.

\section{Summary}

Finding a viable mechanism that screens the original possibly
large cosmological constant to its small value today, is a very
difficult task. Weinberg's no-go theorem puts severe limits on
this approach. Back-reaction effects moreover, are generally
either very weak, or lead to other troublesome features like a
screened Newton's constant. These models are often studied on an
inflationary background to power the gravitational back-reaction
effects, but still typically require an enormous number of
e-folds, to see any effect at all.

Another drawback is that it is hard to understand the impact of
these back-reaction effects on local physics.

The underlying idea however that the effective cosmological
constant is small simply because the universe is old, is
attractive and deserves full attention.

%% file: chapter41.tex
\chapter{Do Infrared Gravitons Screen the Cosmological
Constant?}\label{TsamisWoodardbr}

Knowing that the proposals of the previous chapter do not work,
one may want to study a purely quantum gravitational backreaction
mechanism. It indeed had been argued that during inflation this
process is sufficient to screen the cosmological constant to zero.
This could therefore be of considerable importance to the
cosmological constant problem. Unfortunately however, we conclude
that this screening has negligible effects on the cosmological
constant.

\section{Introduction}

In a series of papers, Tsamis and Woodard \cite{TsamisWoodard0,
TsamisWoodard1, TsamisWoodard2, TsamisWoodard3, TsamisWoodard4,
TsamisWoodard5, TsamisWoodard6}, arrived at the remarkable result
that the back-reaction of inflationary produced gravitons is
sufficient to cancel a `bare' cosmological constant from roughly
$10^{16}$~GeV down to zero. In a Newtonian sense, it is argued
that the interaction energy density between these gravitons will
screen such a huge cosmological constant. Their actual
calculations involve two loop quantum gravitational processes and
are therefore very complicated.

In the remainder, we will be concerned with two questions. One is
whether indeed this interaction energy generates negative vacuum
energy and screens the cosmological constant, and the other, if
so, whether this effect can become as large as claimed by Tsamis
and Woodard in the above cited papers.

Our motivation for a close study of this proposal is not only the
cosmological constant problem, but also to understand whether the
standard paradigm that quantum gravitational effects cannot have a
major influence on large distances, may be incorrect. Note also
the previous results from Taylor and Veneziano
\cite{Taylor:1989ua}, who argued that at large distances quantum
gravitational effects act to slightly \textit{increase} the value
of a positive cosmological constant.

In general, the dominant infrared effects come from the lightest
particles self-interacting with lowest canonical field dimension.
Gravitons are massless for any value of $\Lambda$, but for
$\Lambda = 0$ their lowest self-interaction term consists of two
derivatives distributed between three graviton fields
($\sqrt{-g}R$-term) and this is why conventional quantum gravity
is indeed very weak in the infrared. However, when $\Lambda\neq
0$, the lowest dimensional self-interaction term is of dimension
three, a three-point vertex with no derivatives (corresponding to
the $\Lambda\sqrt{-g}$-term). The IR behavior of the theory with
cosmological constant could therefore perhaps be very different
from that without. Tsamis and Woodard christen it Quantum
Cosmological Gravity, or QCG for short \cite{TsamisWoodard0}.

They argue that on an inflationary background, the infrared
divergences are enhanced. The spatial coordinates expand
exponentially with increasing time, and so their Fourier
conjugates, the spatial momenta, are redshifted to zero. The IR
effects originate from the low end of the momentum spectrum, so
they are strengthened when this sector is more densely populated.

Since other particles are either massive, in which case they
decouple from the infrared, or conformally invariant, and
therefore do not feel the de Sitter redshift, gravitons must
completely dominate the far IR.

Of course, quantum gravitational effects are very weak. The
typical strength of quantum gravitational effects during inflation
at scale $M$ is:
\begin{equation}
G\Lambda=8\pi\left(\frac{M}{M_{P}}\right)^4,
\end{equation}
which for GUT-scale inflation becomes $G\Lambda=10^{-11}$ and for
electroweak-scale inflation $G\Lambda=10^{-67}$. The hope is that
these small numbers are overcome by a very large number of e-folds
of inflation.

\section{A Review of the Scenario}

It has been shown by Grishchuk \cite{Grishchuk:1974ny} that in an
expanding spacetime gravitons are being produced, a process most
efficient during inflation. When massless, virtual particles are
produced out of the vacuum, those with wavelengths larger than
$H^{-1}$ will not recombine and annihilate, but are able to escape
to infinity. This is similar to a black hole emitting Hawking
radiation. For a derivation one could also check
\cite{KolbTurner}. A more intuitive picture can be obtained using
the energy-time uncertainty principle \cite{Woodard2005}, which we
will now discuss.

Consider a particle with mass $m$ and co-moving wavevector
$\vec{k}$ in a spacetime with scalefactor $a(t)$:
\begin{equation}
E(\vec{k}, t)=\sqrt{m^2 + \|\vec{k}\|^2/a^2(t)}.
\end{equation}
The Heisenberg uncertainty principle restricts how long a virtual
pair of particles with $\pm \vec{k}$ can live. The lifetime,
$\Delta t$, is given by the integral:
\begin{equation}
\int_{t}^{t+\Delta t}dt'\;E(\vec{k},t')\sim 1.
\end{equation}
The smaller the mass of a particle, the longer it survives, and
for the fully massless case, in de Sitter spacetime with
$a(t)=\exp(Ht)$:
\begin{equation}
\int_{t}^{t+\Delta t}dt'E(\vec{k}, t')\Big|_{m=0}=
\left[1-e^{-H\Delta t}\right]\frac{k}{Ha(t)}\sim 1.
\end{equation}
Thus growth of $a(t)$ increases the time a `virtual' particle of
fixed $m$ and $\vec{k}$ can exist and, during inflation, particles
with zero mass and wavelength $k\lesssim Ha(t)$ can exist forever.

Conformal invariance suppresses the number of particles that is
being produced by a factor of $1/a(t)$ \cite{Woodard2005}, so
massless minimally coupled scalars and gravitons, which are not
conformally invariant, will be most abundant. The energy-momentum
carried by these gravitons is proportional to $H^4$. In TW's work
inflation is assumed to start at an energy of roughly
$10^{16}$~GeV, because of a large cosmological constant. So the
energy density in infrared gravitons is quite small, compared to
the vacuum energy density. More importantly, their energy-momentum
tensor can never be interpreted as vacuum energy-density, since
the latter is given by $Cg_{\mu\nu}$, with $C$ some constant.
Besides, the graviton energy-density is positive and could not
screen a positive cosmological constant.

A priori, it could have been a nice idea to argue that a
Bogolyubov transformation could bring one from one vacuum to
another, with a different value for the cosmological constant.
Similar to a black hole emitting Hawking radiation and thereby
decreasing its mass, one could think that the inflationary
produced gravitons would diminish the cosmological constant. This
analogy however fails. The Hawking effect is a one-loop effect,
and at one loop order, there can only be at best a small constant
renormalization of $\Lambda$. The Hawking effect can be derived
from a Bogolyubov transform of the vacuum state, whereas the
inflationary produced gravitons cannot in such a way produce a
decrease in the cosmological constant. The difference is exactly
that a black hole is formed from an object of a certain mass in
spacetime, whereas the cosmological constant is a property of
spacetime itself.

The argument of Tsamis and Woodard however, is that the
\textit{interaction energy density} between different gravitons,
takes on the form $T_{\mu\nu}=Cg_{\mu\nu}$, with $C$ a positive
constant in a $(-+++)$-metric, thus acting as a negative vacuum
energy density. Per graviton pair, this interaction energy density
is negligible as well, and, moreover, the gravitons produced this
way soon leave the Hubble volume. Pairs produced in different
Hubble volumes are not in causal contact with each other, and
hence have no gravitational interaction with each other. However,
the argument is that at the end of inflation the past lightcone is
growing larger and larger, so eventually all the inflationary
produced gravitons do come into causal contact with each other,
making the effect large enough to completely screen the huge bare
cosmological constant to zero. Their calculations indicate that
roughly $10^7$ e-folds are necessary to build up a large enough
effect, after which inflation stops rather suddenly over the
course of a few (less than ten) e-folds.

Two additional questions arise in general in these back-reaction
mechanisms. First, is it possible at all to generate a mechanism
that makes the value of the cosmological constant time-dependent
and therefore, observer-dependent? And secondly, how can such a
mechanism account for a small counterterm $\delta\Lambda$ of order
$1/\mbox{cm}^2$? We will return to both of these questions in
section (\ref{renormpqg}). But, as is well-known, and was
discussed in section (\ref{nogowein}), in order to make the
cosmological constant time-dependent, one has to introduce a new
field. This is typically a scalar field to maintain Lorentz
invariance. Let us stress here that Tsamis and Woodard do not
introduce a scalar field. However, it appears that their mechanism
could be modelled using a growing scalar field to act as the
growing interaction energy density of the produced gravitons.

\subsection{A Newtonian Picture}

To get a better feeling for the suggested proposal, consider first
the rough Newtonian derivation, as given by Woodard in
\cite{Woodard:2001xn}. We will present our comments in the next
section.

The energy-density in inflationary produced infrared gravitons is:
\begin{equation}
\rho_{IR}\sim H^4,
\end{equation}
with $H$ the Hubble parameter during inflation: $H^2=\Lambda/3$.
Starting with initial radius $H^{-1}$, the physical radius of the
universe, is exponentially growing with co-moving time $t$:
\begin{equation}
r(t)\sim H^{-1}e^{Ht}.
\end{equation}
The total energy per volume, denoted by $M$ is taken to be:
\begin{equation}\label{totalenvoltw}
M(t)\sim r^3(t)\rho_{IR}\sim He^{3Ht}.
\end{equation}
Note that this assumes a continuous production of gravitons to
balance the growing volume. We will return to this relation in the
next section, for now let's continue the discussion as given in
\cite{Woodard:2001xn}.

If this mass would self-gravitate, the Newtonian interaction
energy would be:
\begin{equation}\label{inactdentw}
E_N = -\frac{GM^2(t)}{r(t)}\sim - GH^3e^{5Ht},
\end{equation}
assuming that the gravitons are on average a distance $r(t)$
apart. This gives an interaction energy density,
$E_N/r^3=-GH^6e^{2Ht}$. However, most of the inflationary produced
gravitons will not be in causal contact with each other, as they
soon leave their Hubble volume. If one assumes, as Tsamis and
Woodard do, that their potentials $V=-GM/r$ remain, the rate at
which they accumulate is estimated to be:
\begin{equation}
\frac{dV}{dt}\sim -GH^3e^{2Ht}.
\end{equation}

The growth of the Newtonian interaction energy density during a
short time interval is:
\begin{eqnarray}
\rho(t)&\sim& \rho_{IR}V(t)\sim -GH^6Ht\nonumber\\
&=& - \frac{\Lambda}{8\pi G}(G\Lambda)^2Ht,
\end{eqnarray}
using $H^2\sim\Lambda$. After a long period of inflation, the
potential's accumulation rate is assumed to be very small, such
that:
\begin{equation}
|\dot{\rho}(t)|\ll H|\rho(t)|,
\end{equation}
and, using energy conservation:
\begin{equation}\label{encons}
\dot{\rho}(t)=-3H\left(\rho(t)+p(t)\right),
\end{equation}
this implies that the induced interaction pressure $p(t)$ must be
nearly opposite to the interaction energy density. This would
imply, that the interaction energy approaches the form of negative
vacuum energy and hence, according to Tsamis and Woodard, would
screen the positive cosmological constant.

\section{Evaluation}

The back-reaction of the inflationary produced gravitons is a
purely non-local effect. For a local observer, inside a Hubble
volume, there are too few gravitons to have any observable effect
on the cosmological constant. About one infrared pair emerges per
Hubble time in each Hubble volume \cite{Abramo:2001dd}. An
important assumption made in the above section was that even after
the gravitons have left their Hubble volume their potentials
remain. It is the accumulation of these potentials that, in a
Newtonian sense, causes the build-up of negative vacuum energy
density, and hence the screening.

The result that has led to a large effect on the cosmological
constant, appears to hinge on certain assumptions. In terms of the
Newtonian argument - which was admitted to be somewhat intuitive -
one could suspect that the graviton energy density in an expanding
universe decreases by a factor of $a^{-4}$, like any other
massless relativistic species. If this were the correct
prescription, one would have to replace the total energy per
volume, denoted $M$, as given in (\ref{totalenvoltw}) by:
\begin{equation}\label{totalenvolsn}
M(t)\sim r^3(t)\rho_{IR}e^{-4Ht}\sim He^{-Ht},
\end{equation}
The total energy density would decrease! The number of Hubble
volumes grows like $e^{3Ht}$ but the energy density per graviton
in that case decreases like $e^{-4Ht}$ and thus can never grow
larger. The interaction energy density is proportional to $M^2$
and we have to correct (\ref{inactdentw}), unless one assumes an
ever increasing production rate of gravitons, to:
\begin{equation}\label{inactdensn}
V = -\frac{GM^2(t)}{r(t)}\sim - GH^3e^{-3Ht},
\end{equation}
which obviously cannot have any effect in screening the
cosmological constant.

Another issue is that in order to make the statement that the
interaction energy density acts like negative vacuum energy, the
standard equation expressing conservation of energy has been used.
This equation is derived from the Einstein equations, more
specifically, from demanding that the energy-momentum tensor is
covariantly conserved: $\nabla_{\mu} T^{\mu\nu}=0$. However, in
this setup the energy density is not a local energy density and it
is therefore not a priori clear that its energy momentum tensor is
covariantly conserved, nor that it satisfies the \textit{local}
Einstein equations.

Furthermore, where the Newtonian argument concludes that after a
long time the interaction energy density does not change much
anymore, $\dot{\rho}=0$, the question must be raised at which
point this equation sets in. This would be the point where, using
(\ref{encons}), the potential energy due to the gravitons
satisfies the equation of state for vacuum energy density:
$p=-\rho$.

This seems remarkable: if the interaction energy density generates
an energy-momentum tensor $Cg_{\mu\nu}$ characteristic of vacuum
energy, than one might argue that it would do so from the
beginning. If the interaction energy density between a single
graviton pair cannot be interpreted as vacuum energy density, than
the interaction energy density of $N$ graviton pairs also cannot
be interpreted as vacuum energy density.

The real answers to these questions must, of course, come from the
more explicit calculations, which will be scrutinized in the
following sections.

Another issue is that gravitational potentials are of course not
gauge invariant, which implies that at this point the physical
reality of the interaction energy density is not obvious. In the
full quantum gravitational framework, this amounts to imposing the
correct renormalization condition, and we will therefore pay
special care to this issue in the following sections.

\section{The Full Quantum Gravitational Calculation}

To be more concrete, the actual computation to calculate the IR
effects, involves the expectation value of the invariant element
in the presence of a homogeneous and isotropic, initially free de
Sitter vacuum:
\begin{equation}
\langle 0|g_{\mu\nu}(t,\vec{x})dx^{\mu}dx^{\nu}|0\rangle.
\end{equation}
The easiest way to do this is first to calculate the amputated
expectation value, and then add the external leg.

\begin{figure}[ht]
\begin{center}
\includegraphics[width=12cm]{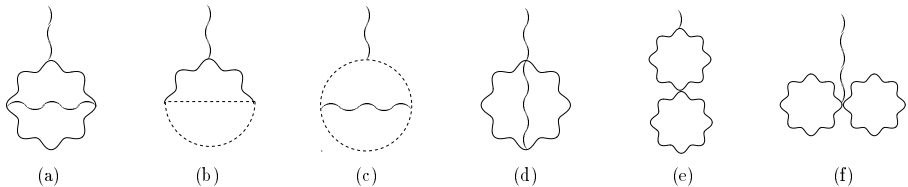}
\caption{\small Two-loop contributions to the background geometry.
Gravitons reside on wavy lines and ghosts on segmented lines, from
\cite{TsamisWoodard1}.}\label{twolooptwfig}
\end{center}
\end{figure}

The production of gravitons is a one-loop effect, so their
back-reaction at the metric starts at two-loop. The calculation
therefore focusses on the two-loop 1-point function. The effect of
the one-loop 1-point function is absorbed into a local counterterm
plus a time-dependent redefinition of the coordinate system
\cite{TsamisWoodard3}. Note that this is the general strategy for
removing the tadpole diagrams. They are removed by a substitution:
\begin{equation}
h_{\mu\nu}\rightarrow h_{\mu\nu} + a_{\mu\nu}
\end{equation}
with $a_{\mu\nu}$ necessarily time dependent, see e.g.
\cite{Veltman:1975vx}\footnote{Note that in flat spacetime, in for
example the Higgs mechanism, this can be accomplished by a purely
constant shift.}. So a time dependent cosmological counterterm
absorbs their effect. This is a very convenient bookkeeping
device, since otherwise the graviton propagators would look very
ugly. We will return to this later, since in their work, Tsamis
and Woodard actually calculate the two-loop tadpoles, without
subtracting them with a counterterm, see figure
(\ref{twolooptwfig}).

The classical background in conformal coordinates is:
\begin{eqnarray}\label{ormetricds}
ds^2 = -dt^2+e^{2Ht}d\vec{x}\cdot d\vec{x}=\Omega^2\left(-du^2+
d\vec{x}\cdot d\vec{x}\right)\\
\Omega\equiv\frac{1}{Hu}=\exp(Ht)=R(t)
\end{eqnarray}
with $H^2\equiv\frac{1}{3}\Lambda$ and $R(t)$ the scalefactor. For
convenience, perturbation theory is formulated in terms of a
`pseudo-graviton' field $\psi_{\mu\nu}$:
\begin{equation}
g_{\mu\nu}\equiv\Omega^2\tilde{g}_{\mu\nu}\equiv\Omega^2(\eta_{\mu\nu}
+ \kappa\psi_{\mu\nu})
\end{equation}
where $\kappa^2\equiv 16\pi G$.

Because of homogeneity and isotropy of the dynamics and the
initial state, the amputated 1-point function, can be written in
terms of two functions of conformal time $u$:
\begin{equation}
D_{\mu\nu}^{\rho\sigma}\langle0|\kappa\psi_{\rho\sigma}(x)|0\rangle
= a(u)\bar{\eta}_{\mu\nu} + c(u)\delta^{0}_{\mu}\delta^{0}_{\nu},
\end{equation}
where $D_{\mu\nu}^{\rho\sigma}$ is the gauge fixed kinetic
operator, and a bar on $\eta_{\mu\nu}$ indicates that temporal
components of this tensor are deleted:
\begin{equation}
\eta_{\mu\nu} = \bar{\eta}_{\mu\nu}+
\delta_{\mu}^{(0)}\delta_{\nu}^{(0)}.
\end{equation}
The explicit form for $D_{\mu\nu}^{\rho\sigma}$ is found to be:
\begin{eqnarray}\label{kinop}
D_{\mu \nu}^{\rho \sigma} & = & \left(\frac12 {\overline \delta}_{
\mu}^{~(\rho} \; {\overline \delta}_{\nu}^{~\sigma)} - \frac14
\eta_{\mu \nu} \; \eta^{\rho\sigma} - \frac12 \delta_{\mu}^{~0} \;
\delta_{\nu}^{~0} \;
\delta_0^{~\rho} \; \delta_0^{~\sigma} \right){\rm D}_A \nonumber \\
& & \mbox{} + \delta_{(\mu}^{~~0} \; {\overline \delta}_
{\nu)}^{~~(\rho} \; \delta_0^{~\sigma)} \; {\rm D}_B +
\delta_{\mu}^{~0} \; \delta_{\nu}^{~0} \; \delta_0^{~\rho} \;
\delta_0^{~\sigma} \; {\rm D}_C \; .
\end{eqnarray}

The pseudo-graviton kinetic operator $D_{\mu\nu}^{\rho\sigma}$
splits in two parts, a term proportional to
$D_A\equiv\Omega(\partial^2+\frac{2}{u^2})\Omega$, which is the
kinetic operator for a massless minimally coupled scalar, and a
part proportional to $D_B=D_C\equiv\Omega\partial^2\Omega$, the
kinetic operator for a conformally coupled scalar.

After attaching the external leg one obtains the full 1-point
function, which has the same form, but with different components:
\begin{equation}
\langle0|\kappa\psi_{\mu\nu}(x)|0\rangle = A(u)\bar{\eta}_{\mu\nu}
+ C(u)\delta_{\mu}^{0}\delta_{\nu}^{0}.
\end{equation}
The functions $A(u)$ and $C(u)$ obey the following differential
equations:
\begin{eqnarray}
-\frac{1}{4}D_A\left[A(u)-C(u)\right]&=& a(u)\nonumber\\
D_CC(u)&=& 3a(u)+c(u)
\end{eqnarray}
The functions $a(u)$ and $A(u)$ on the one hand, and $c(u)$ and
$C(u)$ on the other, are therefore related by retarded Green's
functions $G_{A,C}^{ret}$ for the massless minimally coupled and
conformally coupled scalars:
\begin{eqnarray}
A(u) &=& -4G_{A}^{ret}[a](u) + G_{C}^{ret}[3a+c](u),\nonumber\\
C(u)&=&G_{C}^{ret}[3a+c](u)
\end{eqnarray}
In terms of the functions $A(u)$ and $C(u)$ the invariant element
in comoving coordinates reads:
\begin{equation}\label{corrmettw}
\hat{g}_{\mu\nu}(t,\vec{x})dx^{\mu}dx^{\nu}=-\Omega^2\left[1-C(u)\right]du^2+\Omega^2\left[1+A(u)\right]d\vec{x}\cdot
d\vec{x}.
\end{equation}
Comparing (\ref{ormetricds}) with (\ref{corrmettw}) Tsamis and
Woodard make the following identification:
\begin{eqnarray}\label{idenTW}
R(t)&=&\Omega\sqrt{1+A(u)},\nonumber\\
dt &=&-\Omega\sqrt{1-C(u)}du,\nonumber\\
d(Ht)&=&-\sqrt{1-C(u)}d[\ln(Hu)].
\end{eqnarray}
Using this we can find the time dependence of the effective Hubble
parameter:
\begin{equation}\label{HubbleTW}
H_{eff}(t)=\frac{d}{dt}\ln\left(R(t)\right)=
\frac{H}{\sqrt{1-C(u)}}\left(1-\frac{\frac{1}{2}u\frac{d}{du}A(u)}{1+A(u)}\right).
\end{equation}
Note that a priori the signs in (\ref{idenTW}) are arbitrary.
However, since:
\begin{equation}
\frac{1}{Hu}\equiv e^{Ht}\quad\rightarrow\quad
Hdt=-\frac{1}{Hu}du,
\end{equation}
for constant $H$, so the minus sign in relating $dt$ and $du$ is
correct. A plus sign would take us back in time. Indeed, using
that, we find that the effective Hubble parameter in
(\ref{HubbleTW}) increases.

The plus-sign in the identification of the scale factor $R(t)$, is
a priori less justified. Replacing it with a minus-sign would
similarly lead to an increasing Hubble parameter. For now let us
stick with the convention of Tsamis and Woodard.

One then infers that the backreaction of the IR gravitons acts to
screen the bare cosmological constant, originally present. The
improved results\footnote{Papers before 1997 yield different
results.} in terms of:
\begin{equation}
\epsilon\equiv\left(\frac{\kappa
H}{4\pi}\right)^2=\frac{G\Lambda}{3\pi}=\frac{8}{3}\left(\frac{M}{M_P}\right)^4
\end{equation}
with $G\Lambda\sim10^{-11}$ for GUT-scale inflation and
$G\Lambda\sim10^{-67}$ for EW-scale inflation, turn out to be:
\begin{eqnarray}\label{AuCu}
A(u)&=&\epsilon^2\left\{\frac{172}{9}\ln^3(Hu) +
(\mbox{subleading})\right\} + \mathcal{O}(\epsilon^3),\\
C(u)&=&\epsilon^2\left\{\frac{57}{3}\ln^2(Hu) +
(\mbox{subleading})\right\} + \mathcal{O}(\epsilon^3)
\end{eqnarray}
Using (\ref{idenTW}) and (\ref{AuCu}) one finds:
\begin{equation}\label{HubblerelutTW}
Ht = -\left\{1-\frac{19}{2}\epsilon^2\ln^2(Hu) +
\dots\right\}\ln(Hu)
\end{equation}
using that the correction is smaller than unity. This can be
inverted to give:
\begin{equation}\label{Hubbinvert}
\ln(Hu)=-\left(1 + \frac{19}{2}(\epsilon Ht)^2 + \ldots\right)Ht,
\end{equation}
and it follows that $\ln(Hu)\sim -Ht$, for as long as perturbation
theory is valid. The equivalence is a direct result of using
conformal time $u$:
\begin{equation}
\frac{1}{Hu}\equiv e^{Ht}\quad\rightarrow\quad \ln(Hu)\equiv -Ht.
\end{equation}
This implies that for as long as perturbation theory is valid,
$H$, and therefore $\Lambda$, remain constant. Significant changes
in the value of $\Lambda$ must be non-perturbative.

However, we can now use the above equivalence to write $A(u)$ as:
\begin{equation}
A(u)=-\frac{172}{9}\epsilon^2(Ht)^3+\ldots
\end{equation}
and one arrives at:
\begin{eqnarray}
H_{eff}(t)&\approx& H + \frac{1}{2}\frac{d}{dt}\ln(1+A),\nonumber\\
&\approx&
H\left\{1-\frac{\frac{86}{3}\epsilon^2\left(Ht\right)^2}{1-\frac{172}{9}\epsilon^2\left(Ht\right)^3}\right\}
\end{eqnarray}
The induced energy density, which acts to screen the original
cosmological constant present gives:
\begin{eqnarray}
\rho(t)&\approx&\frac{\Lambda}{8\pi
G}\left\{-\frac{1}{H}\frac{\dot{A}}{1+A} +
\frac{1}{4H^2}\left(\frac{\dot{A}}{1+A}\right)^2\right\}\nonumber\\
&\approx&\frac{\Lambda}{8\pi
G}\left\{-\frac{\frac{172}{3}\epsilon^2\left(Ht\right)^2}{1-\frac{172}{9}\epsilon^2\left(Ht\right)^3}
+
\left(\frac{\frac{86}{3}\epsilon^2\left(Ht\right)^2}{1-\frac{172}{9}\epsilon^2\left(Ht\right)^3}\right)^2\right\}
\end{eqnarray}
This can be written more intuitively, to better see the magnitude
of the effect as follows:
\begin{equation}
H_{eff}(t)=H\left\{1-\epsilon^2\left[\frac{1}{6}(Ht)^2 +
(\mbox{subleading})\right] +\mathcal{O}(\kappa^6)\right\}
\end{equation}
and the induced energy density and pressure, in powers of $H$:
\begin{eqnarray}
\rho(t)&=&\frac{\Lambda}{8\pi G} + \frac{(\kappa
H)H^4}{2^6\pi^4}\left\{-\frac{1}{2}\ln^2a + \mathcal{O}(\ln
a)\right\} + \mathcal{O}(\kappa^4)\nonumber\\
p(t)&=&-\frac{\Lambda}{8\pi G} + \frac{(\kappa
H)H^4}{2^6\pi^4}\left\{\frac{1}{2}\ln^2a + \mathcal{O}(\ln
a)\right\} + \mathcal{O}(\kappa^4),
\end{eqnarray}
where in order to derive the expression for the pressure $p$,
again the stress-energy conservation equation (\ref{encons}) with
$\dot{\rho}=0$, is used. We have argued that in our opinion, it is
not a priori clear that this equation is satisfied. The energy
density and pressure generate a non-local energy-momentum tensor,
that is not covariantly conserved, since $\Lambda$ has become
time-dependent. At each instant of time, the Einstein equations
will be satisfied with a constant $\Lambda$ and a covariantly
conserved energy-momentum tensor. However, this tensor will than
also be constant.

However, let us continue the discussion. Since the effect is so
weak, the number of e-folds of inflation needed is enormous.
Recall that this quantity is defined as follows. For a pure de
Sitter phase, we have:
\begin{equation}
\frac{\ddot{a}}{a}=\left(\frac{\dot{a}}{a}\right)^2=\frac{\Lambda}{3}
\equiv H_{\Lambda}^2,\quad\rightarrow\quad
a(t)=a_1e^{H_{\Lambda}(t-t_1)}
\end{equation}
with $a(t)=a_1$ at $t=t_1$. During an inflationary period $\Delta
t$, the size of the universe increases by a factor:
\begin{equation}
Z=e^{H_{\Lambda}\Delta t}.
\end{equation}
To solve the flatness problem one must require $Z> 10^{30}$. The
number $N$ of e-folds now is defined as $Z\equiv e^N$, or,
assuming inflation lasts from time $t_0$ to time $t_1$:
\begin{equation}
N=\int_{t_0}^{t_1}Hdt.
\end{equation}
In order for $Z>10^{30}$, $N>69$. In the scenario by TW, this
number is much bigger, but that is not in conflict with any known
result.

The number of e-foldings needed to make the backreaction effect
large enough to end inflation is argued to be:
\begin{equation}
N\sim\left(\frac{9}{172}\right)^{\frac{1}{3}}\left(\frac{3\pi}{G\Lambda}\right)^{\frac{2}{3}}
=
\left(\frac{81}{11008}\right)^\frac{1}{3}\left(\frac{M_P}{M}\right)^\frac{8}{3}
\end{equation}
where $M$ is the mass scale at inflation and $M_P$ is the Planck
mass. For inflation at the GUT scale this gives $N\sim 10^7$
e-foldings. This enormously long period of inflation, much longer
than in typical inflation models, is a direct consequence of the
fact that gravity is such a weak interaction. It results in a
universe that is much bigger compared to ordinary scenarios of
inflation, but that in itself is not a problem.

It is argued that the effect might be strong enough to effectively
kill the `bare' cosmological constant, as long as such a long
period of inflation is acceptable. There do exist arguments that
the number of e-folds is limited to some~85, see
\cite{BanksFischler2003} for details, but these are far from
established. Moreover, this bound is achieved on the assumption
that at late times the acceleration is given by a pure
cosmological constant, assuming that at late times the universe
enters an asymptotic de Sitter phase which can store only a
limited amount of entropy in field theoretic degrees of freedom.
Tsamis and Woodard however, argue that eventually the cosmological
constant will be screened all the way to zero, in which case this
bound would not be applicable.

However, as we already argued, the manipulations in equations
(\ref{HubblerelutTW}) and (\ref{Hubbinvert}) are only allowed for
as long as the Hubble constant is indeed constant. Perturbative
techniques break down when the effect becomes too strong. This
makes it very difficult to conclude what happens after a large
number of e-folds. For as long as the calculations might be
reliable, nothing really happens.

Tsamis and Woodard argue that the breaking down of perturbation
theory is rather soft, since each elementary interaction remains
weak. Furthermore, a technique following Starobinski
\cite{StarobinskyYokoyama1994} is used in which non-perturbative
aspects are absorbed in a stochastic background that obeys the
classical field equations \cite{TsamisWoodard5}.

It is then argued \cite{TsamisWoodard5} that eventually the
screening must overcompensate the original bare cosmological
constant, leading to a period of deflation. This happens because
the screening at any point derives from a coherent superposition
of interactions from within the past lightcone and the invariant
volume of the past lightcone grows faster as the expansion slows
down. Now thermal gravitons are produced that act as a thermal
barrier, that grows hotter and denser as deflation proceeds.
Incoming virtual IR modes scatter off this barrier putting a halt
to the screening process. The barrier dilutes and the expansion
takes over again.

Throughout the above calculation, the `primordial' cosmological
constant $\Lambda$ was used. The mechanism, however, is argued to
screen the cosmological constant, which implies that the effective
cosmological constant should be used instead, once the Hubble
constant starts to decrease. The strength of the effect would then
be even weaker, since this is controlled by $G\Lambda$. Moreover,
although more gravitons would enter the past lightcone, once
inflation starts to end, all these gravitons are redshifted to
insignificance. It therefore appears impossible to use this
mechanism to even end inflation, let alone to argue that today's
cosmological constant is zero.

\section{Renormalization in Perturbative QG}\label{renormpqg}

As we already indicated above, there also appears to be a more
fundamental problem with this scenario. This has to do with the
very existence of growing infrared divergences and the
possibility, according to Tsamis and Woodard of quantum gravity
being capable of inducing strong effects at large distances.
Crucial in this argument is the renormalization subtraction one
should use in perturbative gravity.

The results obtained by Tsamis and Woodard are based on two-loop
1-point functions. They expand the full metric as:
\begin{equation}
g_{\mu\nu}\equiv \hat{g}_{\mu\nu} + \kappa h_{\mu\nu}
\end{equation}
where the split between $g_{\mu\nu}$ and $h_{\mu\nu}$ is
determined by requiring that the vacuum expectation value of
$h_{\mu\nu}$ vanishes. The background field $\hat{g}_{\mu\nu}$ is
assumed to satisfy the classical equations of motion
$\hat{R}_{\mu\nu}=\Lambda\hat{g}_{\mu\nu}$. It is then argued that
two-loop processes lead to growing infrared divergences which
break De Sitter invariance, such that the vacuum expectation value
of $h_{\mu\nu}$ becomes time dependent, instead of just some
number times the De Sitter metric. Any such number could have been
incorporated in the cosmological counterterm $\delta\Lambda$
\cite{TsamisWoodard3}, and no screening effect could be seen.

The growing IR divergences they get, indicates that one arrives at
a state that is filled with long wavelength particles. These are
the gravitons that continue to redshift beyond the horizon as
inflation and the graviton production mechanism continues.
However, these gravitons pass the horizon and cannot causally
influence the spacetime of the observer. Hence, they should also
not be accounted for in the setup. Note in this respect also that
the Bunch-Davies vacuum used by Tsamis and Woodard to define
perturbation theory, is the natural vacuum choice for particles of
large mass $m\gg H^{-1}$. In that case, the Compton wavelength of
the particle is small compared to the local radius of curvature of
spacetime. However, Bunch-Davies vacuum is in general not a good
choice when dealing with massless particles, as this generally
leads to unphysical infrared divergences
\cite{BirrelDavies,Vilenkin:1982wt}.

In fact, Tsamis and Woodard use the cosmological counterterm to
remove primary divergences in the amputated 1-point function only,
to make the initial Hubble constant time-independent:
$H_{eff}(0)=H$. The functions $A(u)$ and $C(u)$ in
(\ref{corrmettw}) and their first derivatives are therefore zero
at $u=H^{-1}$. They then choose \cite{TsamisWoodard3} the
condition for $\delta\Lambda$ in terms of the initial values of
$a(u)$ and $c(u)$ to be:
\begin{equation}
\left(\frac{d}{du}\right)A(u)\big|_{u=H^{-1}}=a(H^{-1})-c(H^{-1})=0.
\end{equation}
They notice that time-dependent coordinate transformations ``can
be used to impose an independent condition on $a(u)$ and $c(u)$ on
any instant $u$. A nice example would be $a(u)=-c(u)$''. Note that
this latter condition enforces exact de Sitter invariance, with
$a(u)$ and $c(u)$ constants and, because of the initial condition
$\dot{\Lambda}_{eff}(0+)$, equal to zero \cite{TsamisWoodard3}.
Insisting that the symmetries of De Sitter spacetime survive,
necessarily implies that there can be no screening.

Another disturbing point lies in the fact that the cosmological
counterterm is assumed to be time-independent. As stated before,
usually tadpoles are removed by the substitution:
\begin{equation}
h_{\mu\nu}\rightarrow h_{\mu\nu}+a_{\mu\nu},
\end{equation}
with $a_{\mu\nu}$ taken to be such that the linear terms in the
Lagrangean vanish. In gravity, this $a_{\mu\nu}$ generally has to
be time-dependent in order to achieve this. With a time-dependent
cosmological counterterm all tadpole diagrams vanish and they can
cause no physical effect.

A different setup would therefore be to impose from the start:
\begin{equation}\label{condremtdp}
\Gamma=\mathcal{L} +
\delta\mathcal{L},\quad\quad\frac{\partial\Gamma}{\partial g}=0
\end{equation}
with $\delta\mathcal{L}$ the counter Lagrangean. This condition
would immediately subtract off all tadpole diagrams to any order
in perturbation theory. Not imposing this condition would result
in very ugly looking graviton propagators, with tadpole diagrams
attached to them. However, this is not taken into account by
Tsamis and Woodard; they use the `clean' De Sitter propagator, but
without imposing the condition (\ref{condremtdp}).

Of course, when one tries to do perturbation theory on an
initially de Sitter background and forces the corrections to be
not de Sitter invariant, large IR divergences will occur. They
will be growing, since the difference from the background becomes
larger and larger. This is however not a physical effect.

We believe that in such a setup the cosmological constant does not
become a time-dependent parameter, let alone that it be screened
to zero. At a somewhat deeper level, since these gravitons all
cross the horizon as long as inflation proceeds, one might ask
whether it is possible at all that boundary effects may change the
value of the cosmological constant. We do not believe that this is
possible. One can always limit oneself to a box of dimensions
smaller than the Hubble scale of the universe at that time and
measure the cosmological constant. It is a locally determined
parameter that should not depend on boundary effects.

\section{Summary}

In this chapter we have examined an interesting proposal by Tsamis
and Woodard, that intends to use the quantum gravitational
back-reaction of inflationary produced gravitons to screen the
cosmological constant. However, we conclude that the supposed
screening is at best completely negligible in magnitude and offers
no solution to the cosmological constant problem. The infrared
divergences are non-physical and can be removed by a proper
renormalization subtraction. Besides, we believe that the induced
energy density is not of the form of vacuum energy density.

These conclusions concur with our intuition that IR effects in
gravity are virtually non-existent. Newton's constant has
dimension $[\mbox{GeV}]^2$, which implies by dimensional analysis
that it will always be accompanied by (at least) two derivatives,
cutting off any possible large distance effects.

Another remark to be made, is that one is free to add a
counterterm $\delta\Lambda$ of strength $1/\mbox{cm}^2$. When the
Hubble radius of the universe is of this length, this term will be
the dominating energy density. However, the mechanism with
inflationary produced gravitons at those scales is even weaker,
and so it seems that for the present day cosmological constant,
this scenario, even aside from our other objections, cannot work.

%% file: chapter5.tex
\chapter{Type III: Violating the Equivalence
Principle}\label{equivalenceviolatie}

An intriguing way to try to shed light on the cosmological
constant problem is to look for violations of the equivalence
principle of general relativity. The near zero cosmological
constant could be an indication that vacuum energy contrary to
ordinary matter-energy sources does not gravitate.

The approach is based not on trying to eliminate any vacuum
energy, but to make gravity numb for it. This requires a
modification of some of the building blocks of general relativity.
General covariance (and the absence of ghosts and tachyons)
requires that gravitons couple universally to all kinds of energy.
Moreover, this also fixes uniquely the low energy effective action
to be the Einstein-Hilbert action. If gravity were not mediated by
an exactly massless state, this universality would be avoided. One
might hope that vacuum energy would then decouple from gravity,
thereby eliminating its gravitational relevance and thus
eliminating the cosmological constant problem.

\section{Extra Dimensions, Braneworld Models}

Since the Casimir effect troubles our notion of a vacuum state,
the cosmological constant problem starts to appear when
considering distances smaller than a millimeter or so. This size
really is a sort of turn-over scale. Somehow all fluctuations with
sizes between a Planck length and a millimeter are cancelled or
sum up to zero. Therefore, extra dimensions with millimeter sizes
might provide a mechanism to understand almost zero 4D vacuum
energy, since in these scenarios gravity is changed at distances
smaller than a millimeter.

A lot of research in this direction in recent years has been
devoted to braneworld models in $D=4+N$ dimensions, with $N$ extra
spatial dimensions. In this setting the cosmological constant
problem is at least as severe as in any other, but new mechanisms
of cancelling a vacuum energy can be thought of. The general idea
is that the observed part of our world is confined to a
hypersurface, a 3-brane, embedded in a higher dimensional
spacetime. The standard model fields are restricted to live on
this 3-brane, while only gravitons can propagate in the full
higher dimensional space. To reproduce the correct 4-dimensional
gravity at large distances three approaches are known. Usually one
takes the extra, unseen dimensions to cover a finite volume, by
compactifying them. One of the earliest approaches was by Rubakov
and Shaposhnikov \cite{RubakovShaposhnikov} who unsuccessfully
tried to argue that the 4D cosmological constant is zero, since 4D
vacuum energy only curves the extra dimensions.

Besides, it is conceivable that the need to introduce dark matter,
and a very small cosmological constant or some other form of dark
energy to explain an accelerating universe nowadays, is in fact
just a signal of general relativity breaking down at very large
distance scales. General relativity however, works very well on
scales from $10^{-1}$~mm to at least $10^{14}$~cm, the size of the
solar system. Our challenge is to modify gravity in the IR regime
in such a way that the results of GR are not spoiled on those
intermediate distances at which it works correctly.

Extra-dimensional models, like the early Kaluza-Klein scenarios,
generically have additional degrees of freedom, often scalar
fields, that couple to the four dimensional energy-momentum tensor
and modify four-dimensional gravity. A four dimensional massless
graviton has two physical degrees of freedom, a five dimensional
one five, just like a massive 4-dimensional one\footnote{In
general, the total number of independent components of a rank 2
symmetric tensor in $D$ dimensions is $D(D+1)/2$, however, only
$D(D-3)/2$ of those correspond to physical degrees of freedom of a
$D$-dimensional massless graviton; the remaining extra components
are the redundancy of manifestly gauge and Lorentz invariant
description of the theory.}. There are however, strong
experimental constraints on such scalar-tensor theories of
gravity. One can for example calculate the slowing down of binary
pulsars due to the radiation of these gravi-scalars as they are
sometimes called. It was shown in \cite{Will:2001mx,Durrer:2003rg}
that in case of a 5-dimensional bulk, this leads to a modification
of the quadrupole formula by 20 \%, while observations agree with
the quadrupole formula to better than $\frac{1}{2}$~\%. In case of
more extra dimensions, there will be more gravi-scalars and the
problem only gets worse.

Usually, this is circumvented by giving these scalars a mass. In
infinitely large, uncompactified extra dimensions there is another
way out, since in these models the gravi-scalar represents a
non-normalizable and therefore unphysical mode.

In this chapter we will first briefly review the Randall-Sundrum
models and show why they cannot solve the cosmological constant
problem. Next we focus on the DGP-model with infinite volume extra
dimensions. This is a very interesting setup, but it also
illustrates very well the difficulties one faces in deconstructing
a higher dimensional model to a viable 4D world meeting all the GR
constraints. A rather more speculative but perhaps also more
promising approach is subsequently discussed, in which Lorentz
invariance is spontaneously broken to yield a Higgs mechanism
analog for gravity. Before concluding with a summary, we discuss
yet another option, where one considers the graviton to be a
composite particle.

\section{Randall-Sundrum Models, Warped Extra Dimensions}\label{section51}

There are in fact two different models known as Randall-Sundrum
models, dubbed RS-I and RS-II. We begin with RS-I.

This model consists of two 3-branes at some distance from each
other in the extra dimension. One brane, called the ``hidden
brane'' has positive tension, while the other one, the ``visible
brane'', on which we live, has negative tension. Both branes could
have gauge theories living on them. All of the Standard Model
fields are localized on the brane, and only gravity can propagate
through the entire higher dimensional space.

The equation of motion looks as follows:
\begin{eqnarray}
&& M_{\ast}\sqrt{G}\left(R_{AB}-\frac{1}{2}G_{AB}R\right)-M_{\ast}\Lambda\sqrt{G}G_{AB}\nonumber\\
&=&
T_{hid}\sqrt{g_{hid}}g_{\mu\nu}^{hid}\delta_{A}^{\mu}\delta_{B}^{\nu}\delta(y)
+
T_{vis}\sqrt{g_{vis}}g_{\mu\nu}^{vis}\delta_{A}^{\mu}\delta_{B}^{\nu}\delta(y-y_0),
\end{eqnarray}
where
\begin{equation}
g_{\mu\nu}^{hid}(x)=G_{\mu\nu}(x,y=0),\quad\
g_{\mu\nu}^{vis}(x)=G_{\mu\nu}(x,y=y_0),
\end{equation}
and $M_{\ast}$ is the higher dimensional Planck scale and
$\Lambda$ denotes the bulk cosmological constant.

The $y$-direction is compactified on an orbifold
$S_1/\mathbf{Z}_2$. With the above assumptions for the brane
tensions and bulk CC, it can be shown that when the bulk is taken
to eb a slice of $AdS_5$, there exists the following static
solution, with a flat 4D-metric:
\begin{equation}
ds^2=e^{-|y|/L}\eta_{\mu\nu}dx^{\mu}dx^{\nu} + dy^2
\end{equation}
with $L$ the size of the extra dimension. The minus sign in the
exponential factor occurs because of the assumption that our
visible brane has a negative tension. As a result of this
`warp-factor', all masses on the visible brane are suppressed,
compared to their natural value. For the Higgs mass for example,
one obtains:
\begin{equation}
m^2=e^{-y_0/L}m_{0}^2
\end{equation}
a small hierarchy in $y_0/L$ therefore leads to a large hierarchy
between $m$ and $m_0$, which would solve the `ordinary' hierarchy
problem, of the quadratically diverging Higgs mass.

Moreover, despite the fact that the brane tension on the visible
brane is negative, it is possible that it still has a flat space
solution. Fine-tuning is necessary to obtain this result, and
besides, this solution is not unique. Other, non-flat space
solutions also exist. Therefore, this cannot help in solving the
cosmological constant problem, but it is interesting to see that a
4D cosmological constant can be made to curve only extra
dimensions.

Alternatively, the extra dimensions can be kept large,
uncompactified, but warped, as in the Randall-Sundrum type-II
models. In this case the size of the extra dimensions can be
infinite, but their volume $\int dy\sqrt{G}\sim L$, is still
finite. Note that this cannot be obtained by a coordinate
transformation of the RS-I model, with the hidden brane at
infinity. The warp-factor causes the graviton wavefunction to be
peaked near the brane, or, in other words, gravity is localized,
such that at large 4D-distances ordinary general relativity is
recovered.

The action now reads:
\begin{equation}
S = \frac{1}{2}M_{\ast}^3\int
d^4x\int_{-\infty}^{+\infty}dy\sqrt{G}(R_5 - 2\Lambda_5)+\int
d^4x\sqrt{g}(\mathcal{E}_4 + \mathcal{L}_{SM}),
\end{equation}
where $\mathcal{E}_4$ denotes the 4D brane tension and $\Lambda_5$
the bulk cosmological constant, which is assumed to be negative.
The equation of motion derived from this action, ignoring now
$\mathcal{L}_{SM}$ is:
\begin{equation}
M_{\ast}\sqrt{G}\left(R_{AB}-\frac{1}{2}G_{AB}R\right) =
-M_{\ast}^3\Lambda_5\sqrt{G}G_{AB} +
\mathcal{E}_4\sqrt{g}g_{\mu\nu}\delta_{A}^{\mu}\delta_{B}^{\nu}\delta(y),
\end{equation}
indicating that the brane is located at $y=0$. This equation has
the same flat space solution as above, with warp-factor
$\exp(-|y|)$, where $L$ now is defined as:
\begin{equation}
L\equiv\sqrt{-\frac{3}{2\Lambda_5}},\quad\quad \mathcal{E}_4 =
\frac{3M_{\ast}^3}{L},
\end{equation}
but, again, at the expense of fine-tuning $\Lambda_5$ and
$\mathcal{E}_4$.

Gravity in the $4D$ subspace reduces to GR up to some very small
Yukawa-type corrections. Unfortunately however, with regard to the
cosmological constant problem, the model suffers from the same
drawbacks as RS-I. All fundamental energy scales are at the TeV
level, but the vacuum energy density in our 4D-world is much
smaller.

For a recent overview of brane cosmology in such scenarios, see
\cite{Durrer:2005dj,Langlois:2005nd}.

\subsection{Self-Tuning Solutions}\label{selftuning}

Transmitting any contribution to the CC to the bulk parameters, in
such a way that a 4D-observer does not realize any change in the
4D geometry seems quite suspicious. It would become more
interesting if this transmission would occur automatically,
without the necessity of re-tuning the bulk quantities by hand
every time the 4D vacuum energy changes. Models that realize this
are called \textit{self-tuning} models (see for example
\cite{Nilles}for an overview).

The literature is full of such proposals, however we have found
that all of them contain `hidden' fine-tunings. The real
difficulty is always that no mechanism can be provided to single
out flat 4D metrics from slightly curved 4D (A)dS solutions.

One of the first ideas were presented in
\cite{Arkani-Hamed:2000eg,Kachru:2000hf}, where it is argued that
the brane tension is compensated for by a change in integration
constants for fields living in the bulk. These models live in five
dimensions and contain a scalar field $\phi$ in the bulk, that is
assumed to couple to the brane vacuum energy with a potential
$\Lambda e^{-\kappa\phi}$. For any value of the brane tension a
solution with a flat brane metric can be found, with a warped bulk
geometry. However, expanding and contracting solutions are also
allowed \cite{Binetruy:2000wn} and moreover, the flat solutions
suffers from having naked singularities in the bulk. Subsequently,
efforts have been made to hide the singularities behind event
horizons in \cite{Csaki:2000dm,Csaki:2001mn}, but in this case
self-tuning of vacuum energy is lost, and the model contains
hidden fine-tunings in order to preserve a flat brane metric
\cite{Forste:2000ft,Forste:2000ps,Cline:2002fc}. These models
therefore clearly do not work.

A related approach, considering a warped higher dimensional
geometry, is studied in refs.
\cite{Verlinde1999,Verlinde:1999xm,Schmidhuber1999,Schmidhuber2000}.
It is argued that once a cosmological constant vanishes in the UV,
there exist solutions such that it will not be regenerated along
the renormalization group flow. Any vacuum energy is cancelled by
a decreasing warp factor, ensuring a flat space solution on the
brane. The renormalization group scale in 4D gauge theory is
interpreted as the compactification radius of the full gravity
theory. However, these are not the only solutions and there exists
no argument why they should be preferred. Note however, that this
is quite contrary to ordinary renormalization group behavior, as
studied in section \ref{RenormalizationGroup}.

Since five-dimensional models do not seem to work, much focus has
been put on six-dimensional approaches, with a brane located on a
conical singularity. The four-dimensional vacuum energy on the
brane creates a deficit angle in the bulk, and it is argued that
any change in brane tension will be compensated for by a change in
deficit angle, see
\cite{Carroll:2003db,Burgess1,Burgess2,Kehagias:2004fb,Burgess:2004kd,Burgess:2005wu}.

The occurrence of this deficit angle can be seen explicitly as
follows \cite{Sundrumflat98}, where for now we restrict ourselves
to two extra dimensions, and for simplicity no Einstein-Hilbert
term on the brane. The bulk Einstein equations are:
\begin{equation}
\sqrt{-G}\left(\mathcal{R}_{AB} -
\frac{1}{2}G_{AB}\mathcal{R}\right) =
\frac{\mathcal{E}_4}{4M_{\ast}^{4}}G_{A\mu}\eta^{\mu\nu}G_{\nu B}
\delta^2(X^a - Y^a),
\end{equation}
taking the 3-brane to be embedded at the point $Y^a$ and small
Latin superscripts denote coordinates running only over the two
extra dimensions. Assuming now a metric ansatz:
\begin{equation}
ds^2=\eta_{\mu\nu}dx^{\mu}dx^{\nu} +
\mathcal{G}_{ab}(X^4,X^5)dX^adX^b,
\end{equation}
Einstein's equation splits up to:
\begin{eqnarray}
\sqrt{\mathcal{G}}R^{(2)}&=&-\frac{\mathcal{E}_4}{2M_{\ast}^{4}}\delta^2(X^a
- Y^a)\nonumber\\
R_{ab} &-& \frac{1}{2}R^{(2)}\mathcal{G}_{ab}=0
\end{eqnarray}
with $R^{(2)}$ the two-dimensional scalar curvature. The first of
these equations, has as a solution $\mathcal{G}_{ab}$
corresponding to a conical geometry on the two-dimensional space,
with the tip of the cone at $Y^a$, and deficit angle given by:
\begin{equation}
\delta=\frac{\mathcal{E}_4}{4M_{\ast}^{4}},
\end{equation}
which for many values of $M_{\ast}$ is only smaller than $2\pi$ if
the brane tension is fine-tuned, which is just the re-incarnation
of the cosmological constant problem\footnote{Typically, from
stringtheory the brane tension of a 3-brane is given by:
$T=1/[(2\pi)^3(l_s)^4g_s]$ with $l_s$ the string length, typically
a Planck length, and $g_s$ the string coupling. Normally, with
$T\sim M_{W}^4$, with $M_W$ the weak scale, as often assumed in
these models, this implies that $1/g_s$ has to be extremely small,
not even in the perturbative regime. However, in this setup also
the string length $l_s$ is modified.}. So apart from exact
supersymmetry at TeV energies, one also has to require the higher
dimensional Planck mass to be of at least TeV level.

Another important issue is that in fact as far as we know, all the
models still have hidden fine-tunings. To make these explicit,
consider the bulk action of \cite{Carroll:2003db}:
\begin{equation}
S_6=\int d^6x\sqrt{G}\left(\frac{1}{2}M_{6}^4R - \Lambda_6 -
\frac{1}{4}F_{ab}F^{ab}\right)
\end{equation}
where now there is a 2-form field strength in the bulk, which is
taken to generate a magnetic flux:
$F_{ij}=\sqrt{\gamma}B_0\epsilon_{ij}$, with $\gamma_{ij}$ the
higher 2-dimensional metric, $B_0$ just a constant and all other
components of $F_{ij}$ are assumed to vanish. A flat solution with
metric:
\begin{equation}
ds^2=G_{ab}dX^adX^b =\eta_{\mu\nu}dx^\mu dx^\nu +
R_{0}^2(d\theta^2 + \alpha^2\sin^2\theta d\phi^2) ,
\end{equation}
can be obtained, with deficit angle $\delta\equiv 2\pi(1-\alpha)$,
where Greek letters run over four-dimensional coordinates $x$,
where Latin letters run also over all higher coordinates $y$. This
flat metric is obtained when:
\begin{equation}
2\Lambda_6=\frac{1}{2}B^2,\quad\quad \frac{1}{R_{0}^2} =
\frac{B^2}{2M_{4}^4}.
\end{equation}
In order to have a self-tuning solution, the deficit angle is
adjusted whenever there is a change in four-dimensional vacuum
energy. However, it can easily be seen that this is not the case,
since the magnetic flux is given by a closed form, which after
integration must be the same, before and after the change in
$\Lambda$. The deficit angle $\delta$ is related to $\Lambda_6$ as
follows:
\begin{equation}
\alpha^2 = \frac{\Phi^{2}_B\Lambda_6}{M_{4}^4},
\end{equation}
with $\Phi_B$ the magnetic flux integrated over the extra
dimensional space. There obviously is no self-tuning, the
right-hand-side has to be static if one maintains the same Planck
mass before and after a change in the 4D cosmological constant.
Similar conclusions were reached by other authors
\cite{Navarro:2003bf,Nilles,Garriga:2004tq}. In
\cite{Garriga:2004tq} on rather general terms it is shown that
many of these models fail, since the effective four-dimensional
theory has a finite number of fields below the cutoff energy, and
face Weinberg's no-go theorem again, which we discussed in section
(\ref{nogowein}).

Moreover, a severe drawback that so far all these models face is
that this scenario does not exclude `\textit{nearby curved
solutions}' \cite{Nilles}. This simply means that also solutions
for neighboring values of some bulk parameters are allowed, which
result in a curved 4D space, either expanding or contracting.
Besides, there are additional problems such as a varying effective
Planck mass, or varying masses for fields on the brane. And last
but not least, the flat space solution generically has naked
singularities in the bulk. We will return to this issue in section
(\ref{infiniteed}), which suffers from the same problem. So far no
self-tuning scenario without these drawbacks has been found. One
of the most recent papers in this direction is
\cite{Tolley:2005nu} in which it is acknowledged that this
scenario has serious flaws.

The approach of section (\ref{infiniteed}) benefits from the fact
that general relativity is modified at large distances. Among
other things, this implies that the low energy effective theory
does not have a finite number of fields below some low energy
cutoff. This might provide a way out of the drawbacks the models
in this section suffer from. But, in fact, it turns out that also
this more sophisticated model generalized to $N$ extra dimensions,
suffers from the same serious drawbacks.

\subsection{Extra Time-like Dimensions}\label{xtimelike}

For completeness let us here also briefly mention approaches using
extra time-like dimensions in $D=11$ supergravity, e.g.
\cite{ArefevaDragovicVolovich}. It is argued that classical vacuum
solutions can be obtained with zero cosmological constant and
without massless ghosts or tachyons in the low energy limit. There
are however many different solutions with different
characteristics so the predictive power seems to be minimal.
Moreover, the usual problems arise after supersymmetry breaking,
and after taking higher orders into account.

\section{Infinite Volume Extra
Dimensions}\label{infiniteed}

In
\cite{Arkani-HamedDimopoulosDvaliGabadadze,DvaliGabadadzePorrati2000,DvaliGabadadze2000,DvaliGabadadzeShifman20021,DvaliGabadadzeShifman20022,Gabadadzereview},
a model based on infinite volume extra dimensions is presented. We
will first give a qualitative idea of what this model looks like,
before going into more details. In the literature it has become
known as the DGP-model, after the founding fathers, Dvali,
Gabadadze and Porrati.

In this setup there is just one 3-brane to which all standard
model fields are confined. Only gravity can propagate in the full
higher dimensional spacetime. It is assumed that the
higher-dimensional theory is supersymmetric and that SUSY is
spontaneously broken on the brane. These breaking effects can be
localized on the brane only, without affecting the bulk, because
the infinite volume gives a large enough suppression factor. Apart
from that, an unbroken R-parity might be assumed to forbid any
negative vacuum energy density in the bulk. The bulk cosmological
constant therefore vanishes.

Gravity in the bulk is taken to be very strong; the higher
dimensional Planck mass $M_{\ast}$, is assumed to be of order
$10^{-3}$~eV. As a result, four dimensional gravity is modified in
the ultraviolet at distances $r\lesssim M_{\ast}^{-1}\sim 0.1$~mm,
where the effective field theory description of gravity breaks
down. Since gravity is so strong at these small distances, its
short-distance nature depends on the UV-completion of the theory.
This UV-completion is not known at present, but could possibly be
some string theory embedding. As argued before, this is
interesting, since this distance scale could be a defining scale
for the cosmological constant problem. In fact, graviton loops are
cutoff at $M_{\ast}$ and hence, do not contribute to the
cosmological constant problem, see also section
\ref{fatgravitons}.

Moreover, gravity is also modified in the infrared. At distances
smaller than a cross-over scale $r_{c}$, gravity looks four
dimensional, whereas at distances larger than $r_c$ it becomes
higher dimensional. The physical reason for this is that with
infinite volume extra dimensions, there is no localizable
zero-mode graviton. Instead, from a four-dimensional point of
view, the 4D-graviton is a massive metastable state that can
escape into the extra dimensions at very large distances. This can
also be understood with regard to the Einstein-Hilbert terms in
the action. Because the higher dimensional Planck mass is so
small, the bulk Einstein-Hilbert term, becomes of comparable
magnitude at very large distances, given by $r_c\sim
M_{P}^2/M_{\ast}^3$, in case of one extra dimension. Gravity thus
becomes weaker at large distances, which could possibly explain
the observed acceleration.

The cosmological constant problem is addressed by noting that the
full Einstein equations admit a flat space solution on the brane,
despite the fact that the brane tension, or 4D cosmological
constant, is non-zero. In this solution, vacuum energy will mainly
curve the higher dimensions.

Embedding our spacetime in infinite volume extra dimensions thus
has several advantages. If they are compactified, one would get a
theory approaching GR in the IR, facing Weinberg's no-go theorem
again. With infinite volume extra dimensions on the other hand, GR
is not only modified in the UV, but also in the IR. This changes
not only early cosmology, but also late-time cosmology, with
perhaps a possibility of explaining the accelerated expansion
without a cosmological constant. The fundamental, higher
dimensional Planck scale, $M_\ast$, can be much smaller in this
scenario than in ordinary models with extra dimensions. For finite
volume extra dimensions, like standard Kaluza-Klein
compactifications, or Randall-Sundrum warping (see section
(\ref{section51})): $M_{P}^2 = M_{\ast}^{2+N}V_N$, with $V_N$ the
volume of the extra dimensions, and $N$ the number of extra
dimensions. However, this relation no longer holds when
$V_{N}\rightarrow\infty$. Therefore, $M_{\ast}$ can be much
smaller than a TeV, making gravity in the bulk much stronger.

Details on how these infinite volume models circumvent Weinberg's
no-go theorem can be found in \cite{DvaliGabadadzeShifman20021}.

\subsubsection{Description of the Model}

The low-energy effective action is written:
\begin{equation}\label{infvol}
S = M_{\ast}^{2+N}\int d^4xd^Ny\sqrt{-G}\mathcal{R} + \int
d^4x\sqrt{-g}\left(\mathcal{E}_4 + M_{P}^{2}R +
\mathcal{L}_{SM}\right) + S_{GH},
\end{equation}
where $M_{\ast}^{2+N}$ is the $(4+N)$-dimensional Planck mass, the
scale of the higher dimensional theory, $G_{AB}$ the
$(4+N)$-dimensional metric, $y$ are the `perpendicular'
coordinates, and $\mathcal{E}_4 = M_{Pl}^2\Lambda$, the brane
tension, or 4D cosmological constant. Thus the first term is the
bulk Einstein-Hilbert action for $(4+N)$-dimensional gravity and
the $M_{P}^{2}R$ term is the induced 4D-Einstein-Hilbert action.
So there are two free parameters: $M_{\ast}$ and $\mathcal{E}$.
$M_{\ast}$ is assumed to be very small, making gravity in the
extra dimensions much stronger than in our 4D world.

Furthermore, $S_{GH}$ is a Gibbons-Hawking surface term. The
Einstein equations follow from varying the Einstein-Hilbert action
with a cosmological constant. Since the Ricci scalar involves
second derivatives, if one considers a compact manifold with
boundary $\partial M$, and allows variations of the metric that
vanish on $\partial M$, but whose normal derivatives do not, one
must add a surface term. This term is called the Gibbons-Hawking
term and can be written as:
\begin{equation}
S_{GH}=\frac{1}{8\pi G}\int_{\partial M}d^3x\sqrt{h}K,
\end{equation}
where $K$ is the trace of the extrinsic curvature $K_{ij}$ of the
boundary three-surface, and $h$ is the determinant of the metric
induced on the three-surface.

The 4D-Planck mass in this setup is a derived quantity, which can
be seen as follows \cite{Sakharov:1967pk}. Consider the following
low energy effective action in 4 dimensions:
\begin{equation}\label{sumdereh}
S_G=\int d^4x\sqrt{-g}\left(M_{\ast}^2R +
\sum_{n=1}^{\infty}C_n\frac{R^{n+1}}{M_{\ast}^{2(n-1)}}\right),
\end{equation}
with a derivative expansion assumed in the second term and assume
$M_{\ast}$ is much smaller than $M_P$. At distances $r\gg
M_{\ast}^{-1}$, the Einstein-Hilbert term dominates and
$G_{\ast}\sim M_{\ast}^{-2}
>> G_N$, with $G_N$ normal Newton's constant. At shorter
distances, we have to take into account the infinite series of
terms in the sum in (\ref{sumdereh}), and effective field theory
breaks down.

However, the Einstein-Hilbert term is unstable under quantum
corrections; it gets renormalized by matter loops, similar to the
Higgs mass, when this gravity action is coupled to matter. These
are loops with external gravitons with momentum $p<M_{\ast}$,
since only in this region this theory of gravity makes sense. This
generates an extra term in the action. With a cutoff on the
standard model $M_{SM}$ at GUT-scale for example, one obtains:
\begin{equation}
\Delta S_{induced}= M_{ind}^{2}\int d^4x\sqrt{-g}\left(R +
\mathcal{O}\left[R^2/M_{SM}^2\right]\right)
\end{equation}
where the value of $M_{ind}$ is given by
\cite{Zee:1981mk,Adler:1980bx,Adler:1980pg}:
\begin{equation}
M_{ind}^{2}= \frac{i}{96}\int d^4x x^2\langle
T_{SM}(x)T_{SM}(0)\rangle,
\end{equation}
with on the right-hand-side the vacuum value of the time-ordered
product of the trace of the energy-momentum tensor $T(x)$. The
total action $S_G+ \Delta S_{induced} + S_{SM}$ becomes:
\begin{equation}
S=\int d^4x\sqrt{-g}\left\{ \left(M_{\ast}^2 + M_{ind}^2\right)R +
\sum_{n=1}^{\infty}C_n\frac{R^{n+1}}{M_{\ast}^{2(n-1)}} +
\mathcal{L}_{SM}\right\}
\end{equation}
where $M_{ind}^2+M_{\ast}^2 = M_{P}^2$. Note that the higher
derivative terms are still there, but suppressed by $M_{\ast}$.
The four dimensional $M_P$ thus is a derived quantity, its value
is determined by the UV cutoff of the Standard Model, and by the
content and dynamics of the Standard Model. Scalars and fermions
of the Standard Model contribute to $M_{ind}$ with a positive
sign, and gauge fields with negative sign.

The same mechanism is used in this higher dimensional model. Since
Standard Model particles can only propagate on the brane, and not
in the bulk, any bulk loop gets cut-off at the scale $M_{\ast}$.
In this way, through the induced Einstein-Hilbert term on the
brane, gravity on the brane is \textit{shielded} from the very
strong gravity in the bulk. Submillimeter test of the $1/r^2$-law
of gravity indicate that $M_\ast\geq 10^{-3}$~eV.

\subsubsection{Recovery of 4D-Gravity on the Brane}

Gravity on the brane can be recovered either by making a
decomposition into Kaluza-Klein modes, or by considering the 4D
graviton as a resonance, a metastable state with a mass given by
$m_g\sim M_{\ast}^3/M_{Pl}^2$.

Einstein's equation from (\ref{infvol}) becomes (up to two
derivatives):
\begin{equation}
M_{\ast}^{2+N}\left(\mathcal{R}_{AB} -
\frac{1}{2}G_{AB}\mathcal{R}\right) + \delta^{(N)}M_{P}^2\left(R -
\frac{1}{2}g_{\mu\nu}R\right)\delta_{A}^{\mu}\delta_{B}^{\nu} =
\mathcal{E}_4\delta^{(N)}(y)g_{\mu\nu}\delta_{A}^{\mu}\delta_{B}^{\nu}.
\end{equation}

The higher dimensional graviton can be expanded in 4D Kaluza-Klein
modes as follows:
\begin{equation}\label{stateinfvol}
h_{\mu\nu}(x,y_n) = \int
d^Nm\epsilon_{\mu\nu}^{m}(x)\sigma_m(y_n),
\end{equation}
where $\epsilon_{\mu\nu}^{m}(x)$ are 4D spin-2 fields with mass
$m$ and $\sigma_m(y_n)$ are their wavefunction profiles in the
extra dimensions. Each of these modes gives rise to a Yukawa-type
gravitational potential, the coupling-strength to brane sources of
which are determined by the value of $\sigma_m$ at the position of
the brane, say $y=0$:
\begin{equation}
V(r)\propto\frac{1}{M_{\ast}^{2+N}}\int_{0}^{\infty}dmm^{N-1}|\sigma_m(0)|^2\frac{e^{-rm}}{r}.
\end{equation}
However, in this scenario there is a cut-off of this integral;
modes with $m>1/r_c$ have suppressed wavefunctions, where $r_c$ is
some cross-over scale, given by $r_c=M_{Pl}^2/M_{\ast}^3\sim
H_{0}^{-1}$. For $r\ll r_c$ the gravitational potential is $1/r$,
dominated by the induced 4D kinetic term, and for $r\gg r_c$ it
turns to $1/r^2$, in case of one extra dimension. In ordinary
extra dimensional gravity, all $|\sigma_m(0)|=1$, here however:
\begin{equation}\label{sigmainfvol}
|\sigma_m(0)|=\frac{4}{4+m^2r_{c}^{2}},
\end{equation}
which decreases for $m\gg r_{c}^{-1}$. Therefore, in case of one
extra dimension, the gravitational potential interpolates between
the 4D and 5D regimes at $r_c$. Below $r_c$, almost normal 4D
gravity is recovered, while at larger scales it is effectively
5-dimensional and thus weaker. This could cause the universe's
acceleration.

To derive (\ref{sigmainfvol}), consider first the scalar part of
the action (\ref{infvol}) for one extra dimension
\cite{Gabadadze:2003ii}. The potential can be written:
\begin{equation}\label{scalarpotinfvol}
V(r) = \int dt G_{R}(x,y=0;0,0),
\end{equation}
where $G_R$ is the retarded Green's function, which after Fourier
transformation becomes:
\begin{equation}
G_{R}(x,y,0,0) \equiv
\int\frac{d^4p}{(2\pi)^4}e^{ipx}\tilde{G}_R(p,y).
\end{equation}
The equation for this Green's function in Euclidean momentum space
is:
\begin{equation}
\left(M_{\ast}^3(p^2 - \partial_{y}^2) +
M_{P}^2p^2\delta(y)\right)\tilde{G}_R(p,y)=\delta(y)
\end{equation}
with $p^2$ the square of Euclidean 4-momentum. The solution, with
appropriate boundary conditions becomes:
\begin{equation}
\tilde{G}_R(p,y) = \frac{1}{M_{P}^2p^2 + 2M_{\ast}^3p}e^{-p|y|}.
\end{equation}
With proper normalization, the potential (\ref{scalarpotinfvol})
takes the form:
\begin{equation}
V(r) =
-\frac{1}{8\pi^2M_{P}^2}\frac{1}{r}\left\{\sin\left(\frac{r}{r_c}\right)
\mbox{Ci}\left(\frac{r}{r_c}\right) +
\frac{1}{2}\cos\left(\frac{r}{r_c}\right)\left[\pi-2\mbox{Si}\left(\frac{r}{r_c}\right)\right]\right\},
\end{equation}
where $\mbox{Ci}(z)\equiv\gamma + \ln(z) +
\int_{0}^z(\cos(t)-1)dt/t$,
$\mbox{Si}(z)\equiv\int_{0}^z\sin(t)dt/t$, $\gamma$ is the
Euler-Mascheroni constant, and $r_c$ is again the critical
distance, defined as $r_c\equiv M_{P}^2/2M_{\ast}^3$. This again
shows the cross-over behavior of $r_c$: Below $r_c$, the model is
4-dimensional, while at distances larger than $r_c$ it is
5-dimensional.

The form of (\ref{sigmainfvol}) can also be derived from
consideration of the Green's function for the state
(\ref{stateinfvol}). Imposing that the states
$\epsilon_{\mu\nu}^{m}(x)$ are orthogonal to each other, we can
write for the position space Green's function:
\begin{eqnarray}
G(x-x',0)_{\mu\nu,\gamma\delta}&=&\langle
h_{\mu\nu}(x,0)h_{\gamma\delta}(x',0)\rangle\nonumber\\
&=& \int d^Nm|\sigma_m(0)|^2\langle\epsilon_{\mu\nu}^{m}(x)
\epsilon_{\gamma\delta}^{m}(x'),
\end{eqnarray}
which after a Fourier transformation becomes the momentum space
Green's function $G$. For its scalar part we have:
\begin{equation}
G(p,0)=\int dm m^{N-1}\frac{|\sigma_m(0)|^2}{m^2+p^2},
\end{equation}
which is the Kall\'{e}n-Lehman spectral representation of the
Green's function:
\begin{equation}
G(p,0) = \int
ds\frac{\rho(s)}{s+p^2},\quad\quad\mbox{and}\quad\quad
\rho(s)=\frac{1}{2}s^{\frac{N-2}{2}}|\sigma_{\sqrt{s}}(0)|^2,
\end{equation}
where $s\equiv m^2$.

Using again the equations of motion in Euclidean momentum space,
the propagator becomes:
\begin{equation}
G(p,0) = \frac{1}{M_{P}^2p^2 + M_{\ast}^{2+N}D^{-1}(p,0)},
\end{equation}
where $D^{-1}(p,0)$ is the inverse Green's function of the bulk
theory with no brane. For $N>1$ and $p\gg r_{c}^{-1}$, the
propagator becomes:
\begin{equation}
G(p,0) \simeq \frac{1}{M_{P}^2p^2},
\end{equation}
which is just the propagator of a massless four-dimensional
graviton. In case of one extra dimension:
\begin{equation}
G(p,0) = \frac{1}{M_{P}^2p^2 + 2M_{\ast}^3p},
\end{equation}
yielding the above result for (\ref{sigmainfvol}).

The full tensorial structure is a bit more involved. We take again
as a starting point the field equations, now with a brane
energy-momentum tensor and expand for small metric fluctuations
around flat empty space:
\begin{equation}
g_{AB}=\eta_{AB}+h_{AB}.
\end{equation}
Now we have to pick a gauge, for example harmonic gauge in the
bulk:
\begin{equation}
\partial^Ah_{AB}=\frac{1}{2}\partial_Bh_{A}^A,
\end{equation}
where $h_{\mu 5}=0$ and:
\begin{equation}
\Box^{(5)}h_{5}^5=\Box^{(5)}h_{\mu}^{\mu},
\end{equation}
with $\Box^{(5)}$ the 5D d'Alembertian. The $\mu\nu$-component of
the field equations then becomes after rearranging some terms
\cite{Gabadadzereview}:
\begin{equation}
\left[\frac{1}{r_c}(\nabla^2 - \partial_{y}^2) +
\delta(y)\nabla^2\right]h_{\mu\nu} =
-\frac{1}{M_{P}^2}\left\{T_{\mu\nu}-\frac{1}{3}\eta_{\mu\nu}T_{\alpha}^{\alpha}\right\}\delta(y)
+ \delta(y)\partial_{\mu}\partial_{\nu}h_{\alpha}^{\alpha}
\end{equation}
The second term on the r.h.s. vanishes when contracted with a
conserved energy-momentum tensor, and hence is unimportant at the
level of one graviton exchange. The potential $h_{\mu\nu}$
becomes:
\begin{equation}
\tilde{h}_{\mu\nu}(p,x^5=0)=\frac{8\pi}{M_{P}^2p^2 +
2M_{\ast}^3p}\left[\tilde{T}_{\mu\nu}(p)-\frac{1}{3}\eta_{\mu\nu}\tilde{T}_{\alpha}^{\alpha}(p)\right],
\end{equation}
where the massive graviton behavior can be seen from the
coefficient $1/3$, see (\ref{massivegravitons}).

\subsubsection{Possible Solution to the Cosmological Constant
Problem?}

The question now is whether there exist solutions such that the 4D
induced metric on the brane is forced to be flat:
$g_{\mu\nu}=\eta_{\mu\nu}$. Einstein's equation from
(\ref{infvol}) reads:
\begin{equation}
M_{\ast}^{2+N}\left(\mathcal{R}_{AB} -
\frac{1}{2}G_{AB}\mathcal{R}\right) + \delta^{(N)}M_{P}^2\left(R -
\frac{1}{2}g_{\mu\nu}R\right)\delta_{A}^{\mu}\delta_{B}^{\nu} =
\mathcal{E}_4\delta^{(N)}(y)g_{\mu\nu}\delta_{A}^{\mu}\delta_{B}^{\nu}.
\end{equation}
In case of one extra dimension it is not possible to generate a
viable dynamics with a flat 4D metric \cite{Gabadadzereview}. For
$N=2$ an analytic solution has been obtained which generates a
flat 4D Minkowski metric \cite{Sundrumflat98}. The 4D brane
tension is spent on creating a deficit angle in the bulk. The
derivation of the occurrence of this deficit angle was given in
section (\ref{selftuning}). As also argued there, one has to
fine-tune this tension in order not to generate a deficit angle
larger than $2\pi$, since in this case $M_{\ast}\ll$~TeV. So also
the $N=2$ model does not work.

For $N\geq 2$, solutions of the theory can be parameterized as:
\begin{equation}\label{solinfvolsin}
ds^2=A^2(y)g_{\mu\nu}(x)dx^{\mu}dx^{\nu} - B^2(y)dy^2 -
C^2(y)y^2d\Omega^2_{N-1},
\end{equation}
where $y\equiv \sqrt{y_1^2 + \dots +y_n^2}$ and the functions
$A,B,C$ depend on $\mathcal{E}_4$ and $M_{\ast}$:
\begin{equation}\label{singinfvol}
A,B,C=\left(1-\left(\frac{y_g}{y}\right)^{N-2}\right)^{\alpha,\beta,\gamma},
\end{equation}
where $\alpha,\beta,\gamma$ correspond to $A,B,C$ respectively,
and depend on dimensionality, and $y_g$ is the gravitational
radius of the brane:
\begin{equation}
y_g\sim
M_{\ast}^{-1}\left(\frac{\mathcal{E}_4}{M_{\ast}^4}\right)^{\frac{1}{N-2}}\quad\quad\mbox{for}\quad\quad
N\neq 2.
\end{equation}

For $N>2$ consistent solutions of the form
(\ref{solinfvolsin}),(\ref{singinfvol}), do exist with a flat 4D
metric, without the problem of having a too large deficit angle.
Their interpretation however, is rather complicated because of the
appearance of a naked singularity in the bulk at $y=y_g$.
Spacetime in $4+N$ dimensions looks like $\Re_4\times
S_{N-1}\times R_+$, where $\Re_4$ denotes flat spacetime on the
brane, and $S_{N-1}\times R_+$ are Schwarzschild solutions in the
extra dimensions. The Einstein equations cannot be satisfied at
this singularity.

The final physical result is argued to be:
\begin{equation}\label{resultinfvol}
H\sim y_{g}^{-1}\sim M_{\ast}
\left(\frac{M_{\ast}^4}{\mathcal{E}_4}\right)^{\frac{1}{N-2}}.
\end{equation}
According to the 4D result, $N=0$, the expansion rate grows as
$\mathcal{E}_4$ increases, but for $N>2$ the acceleration rate $H$
decreases as $\mathcal{E}_4$ increases. In this sense, vacuum
energy can still be very large, it just gravitates very little; 4D
vacuum energy is supposed to curve mostly the extra dimensions.

This scenario has been criticized for different reasons. One
immediate call for concern is the appearance of a naked
singularity in the bulk. The situation can be compared with the
perhaps more familiar cosmic strings. As we discussed in the
previous section, a setup with a 3-brane embedded in two extra
dimensions is rather similar to an ordinary cosmic string in the
usual four dimensions. In both cases, the tension of the
brane/string creates a deficit angle in the space around it. For
more general embeddings, i.e. more than two extra dimensions,
higher dimensional cosmic string examples provide good analogies.

Interestingly, in the case of local and global strings in four and
six dimensions \cite{Cohen:1988sg} solutions are known similar to
the $N>2$ DGP-model, which also suffer from a singularity. In that
case, it was shown that the singularity is smoothed out, if the
worldvolume is assumed to be inflating, rather than static
\cite{Gregory:1996dd,Cho:1998xy,Gregory:1999gv}; the singularity
is replaced by a horizon. The situation in the $N>2$ DGP-model is
similar. In this case, the requirement of a flat-space solution on
the brane is too strict.

The inflating solutions are indeed conjectured to be the only
non-singular solutions
\cite{Chodos:2000tf,DvaliGabadadzeShifman20021}. However, the
exact analytic expressions for them have not yet been found, which
makes it hard to make definite statements about them. In any case,
the line element becomes:
\begin{equation}\label{metricassinf}
ds^2 = -n^2(t,y)dt^2 +a_{2}^2(t,y)\gamma_{ij}(x)dx^idx^j +
b^2(t,y)dy^2.
\end{equation}
The inflating brane metric is argued to generate a horizon exactly
at $y_g$, if one would neglect the Einstein-Hilbert term on the
brane, maintaining the desirable solution (\ref{resultinfvol}).

However, including the effect of the induced Einstein-Hilbert
term, amounts to making a shift $\mathcal{E}_4\rightarrow
\mathcal{E}_4-M_{P}^2H^2$ \cite{DvaliGabadadzeShifman20021},
resulting in an inflation rate $H^2\sim \mathcal{E}_4/M_{P}^2$,
the same as in ordinary cosmology, and opposite to the result in
\ref{resultinfvol}. For this inflating solution there is no upper
limit to the four-dimensional cosmological constant. The
four-dimensional vacuum energy again curves strongly the
four-dimensional worldvolume. Therefore we conclude that this
scenario as it stands, does not solve the cosmological constant
problem. In generating a well-behaved solution, i.e. one without
naked singularities, immediately the cosmological constant problem
returns.

Besides, even if the flat space solution were a genuine physical
solution, this would not be enough. One would have to find a
dynamical reason why the flat space solution would actually be
preferred over the inflating ones. Such an argument does not (yet)
exist.

Moreover, since gravity has essentially become massive in this
scenario, the graviton has five degrees of freedom, and especially
the extra scalar degree of freedom, often leads to deviations of
GR at small scales. It indeed has been shown that the DGP-model
suffers from this strong-coupling problem. We discuss this in more
detail in the next section.

\subsection{Massive Gravitons}\label{massivegravitons}

A much studied approach to change general relativity in the
infrared, which is not simply a variation of a scalar-tensor
theory, is to allow for tiny masses for gravitons, like the
Fierz-Pauli theory of massive gravity \cite{FierzPauli1939} and
the example above\footnote{In \cite{Will:2004xi} experimental
bounds on graviton masses are discussed.}. The Fierz-Pauli (FP)
mass term is the only possibility in four dimensions that has no
ghosts in the linear theory. The mass term takes the form:
\begin{equation}
S_{PF}\propto M_{P}^2m_{g}^2\int d^4x\left(h_{\mu\nu}^2 -
\frac{1}{2}(h^{\mu}_{\;\mu})^2\right),
\end{equation}
where $h_{\mu\nu}\equiv g_{\mu\nu}-\eta_{\mu\nu}$. With this
definition, the mass term is regarded as an exact term, not as a
leading term in a small $h$ expansion. Note in passing that due to
mass terms, gravitons might become unstable and could possibly
decay into lighter particles, for example photons. If so, gravity
no longer obeys the standard inverse-square law, but becomes
weaker at large scales, possibly leading to accelerated cosmic
expansion.

In general, the extra degrees of freedom, extra polarizations of a
massive graviton, could also become noticeable at much shorter
distances, putting severe constraints on such scenarios. In the UV
the new scalar degrees of freedom may become strongly coupled,
where the effective theory breaks down and the physics becomes
sensitive to the unknown UV-completion of the theory.

Note that this is a general phenomenon. If we modify a non-Abelian
gauge theory in the IR, by providing a mass $m$ for the gauge
boson, this introduces a new degree of freedom, the longitudinal
component of the gauge field, which becomes strongly coupled in
the UV at a scale $m/g$, determined by the scale of the
IR-modification and $g$, the coupling constant. Only in a full
Higgs mechanism, the couplings will remain weak, since this
provides a kinetic term for the extra scalar degree of freedom.

A severe obstacle massive gravity theories have to overcome is
something known as the Van Dam, Veltman, Zakharov, or (vDVZ),
discontinuity \cite{vanDam:1970vg,Zakharov}, which precisely is a
manifestation of the strong coupling problem. The gravitational
potential, represented by $h_{\mu\nu}=g_{\mu\nu}-\eta_{\mu\nu}$,
generated by a static source $T_{\mu\nu}$, like the sun, is given
by:
\begin{eqnarray}
h_{\mu\nu}^{massive}(q^2)&=&-\frac{8\pi G}{q^2+m^2}\left(T_{\mu\nu}-\frac{1}{3}\eta_{\mu\nu}T^{\alpha}_{\alpha}\right)\nonumber\\
h_{\mu\nu}^{massless}(q^2)&=&-\frac{8\pi G}{q^2}\left(T_{\mu\nu}-\frac{1}{2}\eta_{\mu\nu}T^{\alpha}_{\alpha}\right)\nonumber\\
\end{eqnarray}
in the massive and massless case. vDVZ argued that in the massive
case, even with extremely small graviton mass, the bending of
light rays passing near the sun would be too far off from
experimental results, that the mass of the graviton has to be
exactly equal to zero. The physical reason indeed being, that even
in the limit where the mass of the graviton goes to zero, there is
an additional scalar attraction, which distinguishes the theory
from Einstein's GR. However, Vainshtein proposed that this
discontinuity is just an artifact of using a linear approximation
\cite{Vainshtein:1972sx} and claimed that in the full nonlinear
theory, the discontinuity no longer persists. The perturbative
expansion in $G$ breaks down. However, other problems also arise.
Minkowski space does not seem to be stable and ghosts appear at
the linear level \cite{Gabadadze:2003jq,Gruzinov:2001hp}.

Since there exists no fully non-linear, generally covariant theory
for massive gravity, Vainshtein calculated the non-linear
completion of only the first linearized term, which becomes just
the Einstein tensor, and the field equation reads:
\begin{equation}
G_{\mu\nu} + m_{g}^2(h_{\mu\nu} -
\eta_{\mu\nu}h_{\alpha}^{\alpha})=8\pi GT_{\mu\nu}.
\end{equation}
Solutions to this equation are fully continuous and reproduce
results of massless Einstein gravity, in the zero mass limit.
However, deviations of standard Einstein gravity, become important
at distance scales shorter than the Compton wavelength of the
graviton, as one might have expected. The critical distance
$r_{\ast}$ is:
\begin{equation}
r_{\ast}=\frac{(m_gr_g)^{1/5}}{m_g}
\end{equation}
where $r_g\equiv 2GM$ is the gravitational radius of the source of
mass $M$.

However, this theory is still unsatisfactory, as it is not
generally covariant. Higher dimensional models could possibly
circumvent these problems and allow for graviton masses. When the
extra dimensions are compactified, as in the conventional
Kaluza-Klein mechanisms, then from a four-dimensional perspective,
there is one massless spin-2 state and a tower of massive
gravitons. In the linearized approximation, these massive states
have the FP-form. In this case a consistent theory is obtained as
a result of an infinite number of 4-dimensional reparametrization
invariances. Truncating this tower of massive states to any finite
order leads to an explicit breakdown of these gauge invariances
and again to inconsistencies \cite{Nappi:1989ny,Duff:1989ea}.

In the DGP model, the extra dimensions are infinitely large, and
in the literature, there is an ongoing discussion whether this
model is experimentally viable and capable of avoiding the massive
gravity difficulties, see
\cite{Luty:2003vm,Rubakov:2003zb,Porrati:2004yi,Nicolis:2004qq,Deffayet:2005ys,Lue:2005ya}
for criticism. It appears that indeed also this model suffers from
strong interactions at short distances due to the scalar
polarization of the massive graviton, that can be understood in
terms of a propagating ghosts-like degree of freedom.

The IR deviations of GR take place at distances set by $r_c\equiv
M_{P}^2/M_{\ast}^3$, in case of one extra dimension. The
brane-to-brane propagator in DGP-gravity is very similar to that
of massive gravity. The one-graviton exchange approximation breaks
down at distances $R_{\ast}\sim (R_s r_{c}^2)^{1/3}$
\cite{Luty:2003vm}, called the Vainshtein scale, with $R_S$ the
Schwarzschild radius of the source. $R_{\ast}$ is very large for
astrophysical sources, and it has been shown explicitly that at
shorter distances the full non-linear solution approaches that of
GR \cite{Deffayet:2001uk,Gruzinov:2001hp,Porrati:2002cp}, which
suggests that the DGP model may describe our universe. For
distances larger than $R_{\ast}$ gravity deviates significantly
from GR, yet for smaller distances it should yield (approximately)
the same results. Yet, the large distance at which deviations from
GR are felt, indicates that the DGP-model has hidden strong
interaction scales \cite{Luty:2003vm}. Their effects become
important at much smaller distance scales, given by:
\begin{equation}
r_{crit}=\left(\frac{r_{c}^2}{M_{P}}\right)^{1/3},
\end{equation}
which can be as small as a 1000~km, for $r_c\sim H\sim
10^{28}$~cm. These strong interactions can be traced back to the
appearance of a longitudinal Goldstone mode, that only obtains a
kinetic term through mixing with the transverse polarization
states of the graviton. This is in line with our general argument
at the beginning of this section, that the absence of kinetic
terms tends to result in strong coupling regimes. This mode, as
observed by \cite{Luty:2003vm}, can be interpreted as a brane
bending mode, in the sense that it cannot be removed by a gauge
transformation, without shifting the boundary. In other words,
this mode excites the extrinsic curvature of the boundary, giving
its shape as seen by a higher-dimensional observer.

Strong coupling, resulting in noticeable deviations of GR, at
distances smaller than $10^3$~km, is of course disastrous. This
result, however, also argued for in
\cite{Rubakov:2003zb,Gorbunov:2005zk,Nicolis:2004qq}, depends on
the UV-regulator used, since it depends on the UV properties at
the boundary. Since the UV-completion of the DGP-model is unknown,
basically no-one knows what happens exactly and one might hope
that the UV-completion will miraculously cure the model.

However, to make things even worse, Luty et al \cite{Luty:2003vm}
also pointed out a classical instability in the DGP-model,
resulting from negative energy solutions. A classical solution on
the brane that satisfies the dominant energy condition
($T_{\mu\nu}$ on the brane with $p= w\rho$, satisfying $-1\leq w
\leq 1$) has negative energy for a higher dimensional observer.
The line element is:
\begin{equation}
ds^2 = -f^2(r)dt^2 + \frac{1}{f^2(r)}dr^2 + r^2d\Omega_{3}^2,
\end{equation}
for one extra dimension, with:
\begin{equation}
f^2(r)=1-\frac{R_{S}^2}{r^2}.
\end{equation}
This is of the form of a five-dimensional black hole metric, where
DGP allows the values:
\begin{equation}
0 > R_{S}^2 > -r_{crit}^2.
\end{equation}
These negative solutions thus arise exactly when the DGP model
becomes strongly interacting, which is another indication that the
model is out of control at distances smaller than this critical
distance. Interestingly, Porrati (The `P' in the DGP-model), even
wonders whether these obstacles indicate a no-go for IR
modifications of GR \cite{Porrati-talk}.

The Schwarzschild solutions in the DGP model are also heavily
debated and it is not yet clear what the correct way is to
calculate these, and whether they will eventually lead to
consistent phenomenological behavior. For a recent study and
references, see \cite{Gabadadze:2005qy}.

In \cite{Kogan:2000uy,Porrati:2000cp} it is argued that the vDVZ
discontinuity can be consistently circumvented in (A)dS
backgrounds at the classical level. Quantum corrections however,
will induce the discontinuity again
\cite{Dilkes:2001av,Duff:2001zz}. Furthermore, note that in
supergravity theories there is a similar discontinuity for a
spin-$3/2$ gravitino field coupled to a conserved fermionic source
\cite{Deser:1977ur} in flat space. Also in this case, it is argued
that the discontinuity does not arise in AdS space, in two
particular limits where the gravitino mass $m\rightarrow 0$ and
$m\rightarrow\sqrt{-\Lambda/3}$ \cite{Deser:2000de}.

To summarize, as in the case of massive gravity, the DGP-model
appears to deviate too strongly from GR at intermediate distances.
The possible appearance of the vDVZ discontinuity is a hot topic
in braneworld scenarios, and for a good reason. If indeed the
discontinuity cannot be circumvented in these models, this would
provide a powerful argument against many such approaches. However,
here and with the appearance of singularities in the bulk, the
UV-regulation is very important. Since the UV-completion, a full
string theory embedding, is not known at present, one might
therefore hope that all the drawbacks that we encountered, will be
resolved in the final formulation.

\subsection{Non-Zero Brane Width}

The DGP model of section (\ref{infiniteed}) has also been
criticized on different grounds. If one allows for a non-zero
brane width (so-called fat branes) to regularize the graviton
propagator, the flat space graviton propagator exhibits new poles,
that correspond to very light tachyons with negative residues
\cite{Ellwanger:2003fd,Dubovsky:2002jm,Middleton:2002qa,Ellwanger:2005mn}.

This is however, a UV regularization dependent phenomenon, and it
is not known at present, whether these poles persist in for
example a string theory embedding \cite{Gabadadzereview}.

\subsection{Non-local Gravity}\label{nonlocal}

From a 4D-perspective, this approach can also be viewed as to make
the effective Newton's constant frequency and wavelength
dependent, in such a way that for sources that are uniform in
space and time it is tiny \cite{ArkaniHamedDimDvaGab}:
\begin{equation}\label{Ghighpass1}
M_{Pl}^2\left(1+\mathcal{F}(L^2\nabla^2)\right)G_{\mu\nu} =
T_{\mu\nu}.
\end{equation}
Here $\mathcal{F}(L^2\nabla^2)$ is a filter function:
\begin{eqnarray}
\mathcal{F}(\alpha)\rightarrow 0\quad &\mbox{for}&\; \alpha\gg
1\nonumber\\
\mathcal{F}(\alpha)\gg 1\quad &\mbox{for}&\; \alpha\ll 1
\end{eqnarray}
$L$ is a distance scale at which deviations from general
relativity are to be expected and
$\nabla^2\equiv\nabla_{\mu}\nabla^{\mu}$ denotes the covariant
d'Alembertian. Thus (\ref{Ghighpass1}) can be viewed as Einstein's
equation with $(8\pi G_{N}^{eff})^{-1} = M_{Pl}^2(1+\mathcal{F})$.
It is argued that for vacuum energy $\mathcal{F}(0)$ is large
enough, such that it will barely gravitate:
\begin{equation}
M_{P}^2\left(1+\mathcal{F}(0)\right)G_{\mu\nu}=\left(M_{P}^2 +
\bar{M}^2\right)G_{\mu\nu},\quad\quad\mbox{and}\quad\quad R=-
\frac{4\mathcal{E}_4}{M_{P}^2 + \bar{M}^2}.
\end{equation}
To reproduce the observed acceleration a value $\bar{M}$ is needed
$\bar{M}\sim 10^{48}$~GeV for a vacuum energy density of TeV
level, and a $\bar{M}\sim 10^{80}$~GeV for $\mathcal{E}_4$ of
Planck mass value, which is about equal to the mass of the
universe.

In terms of the graviton propagator, it gets an extra factor
$(1+\mathcal{F}(k^2L^2))^{-1}$ and therefore goes to zero when
$\mathcal{F}(0)\rightarrow\infty$, instead of generating a
tadpole.

In the limit $L\rightarrow\infty$ one arrives at:
\begin{equation}
M_{Pl}^2G_{\mu\nu} - \frac{1}{4}\bar{M}^2g_{\mu\nu}\bar{R} =
T_{\mu\nu},
\end{equation}
just the zero mode part of $G_{\mu\nu}$, which is proportional to
$g_{\mu\nu}$, where
\begin{equation}
\bar{R}\equiv\frac{\int d^4x\sqrt{g}R}{\int d^4x\sqrt{g}}
\end{equation}
$\bar{R}$ thus is the spacetime averaged Ricci curvature, which
vanishes for all localized solutions, such as stars, black holes
and also for FRW models. For de Sitter space however, $\bar{R}\neq
0$., but a constant and equal to $\bar{R}=R_{\infty}$, with
$R_\infty$ the asymptotic de Sitter curvature.

At the price of losing 4D-locality and causality, the new averaged
term is both non-local and acausal, a model is constructed in
which a huge vacuum energy does not lead to an unacceptably large
curvature. That is at least the idea. The Planck scale is made
enormous for Fourier modes with a wavelength larger than a size
$L$. It is however argued that the acausality has no other
observable effect. Moreover, it has been claimed that non-locality
should be an essential element in any modification of GR in the
infrared that intends to solve the cosmological constant problem
\cite{Arkani-HamedDimopoulosDvaliGabadadze}. The argument is that
it takes local, causal physics a time $1/L$ to respond to
modifications at scale $L\sim 10^{28}$~cm, and thus in particular
to sources which have characteristic wavelength larger than
$H_{0}^{-1}$, ``such as vacuum energy"
\cite{DvaliGabadadzeShifman20022}. Note that in Feynman diagram
language, the only contributions to the renormalization of the
cosmological constant come from diagrams with infinite wavelength
gravitons, $k=0$, on the external legs.

The non-localities in this case appear in the four dimensional
truncation of the $4+N$-dimensional theory of section
(\ref{infiniteed}). There is an infinite number of degrees of
freedom below any non-zero energy scale. Therefore, in order to
rewrite the model as an effective four dimensional field theory,
and infinite number of degrees of freedom have to be integrated
out. This results in the appearance of non-local interactions,
despite the fact that the full theory is local.

Another idea based on a model of non-local quantum gravity and
field theory due to Moffat \cite{Moffat,Moffat1}, also suppresses
the coupling of gravity to the vacuum energy density and also
leads to a violation of the Weak Equivalence Principle.

\subsection{A 5D DGP Brane-World Model}

To finish our discussion of the DGP-proposal, we will briefly
discuss a 5D model for dark energy. This model thus has no
intention to solve the (old) cosmological constant problem, and,
moreover, it suffers from the strong coupling problems discussed
in section (\ref{massivegravitons}). However, it is interesting in
the sense that an accelerated expansion can be obtained, although
there is no cosmological constant or other spurious form of dark
energy. This model has been extensively discussed in the
literature. The so-called `self-accelerating phase' the model
exhibits, was first noted by Deffayet et al. in
\cite{Binetruy:1999ut,Deffayet:2000uy}, and detailed discussions
followed in for example
\cite{Lue:2005ya,Sawicki:2005cc,Dick:2001np}.

The important result is that the Friedmann equations obtain an
additional term:
\begin{equation}\label{niewfried}
H^2 \pm \frac{H}{r_c} = \frac{8\pi G}{3}\rho,
\end{equation}
with $H=\dot{a}/a$, as usual. The second term on the left is the
new DGP generated term, with an ambiguous sign. As we discussed in
section (\ref{infiniteed}), in the DGP model, gravity looks higher
dimensional past some critical cross-over scale, taken to be of
order of the current Hubble radius. This is reflected here; at
high Hubble scales, the ordinary Friedmann equation is a very good
approximation, which is however substantially altered when $H(t)$
approaches $r_c$.

The upper sign in (\ref{niewfried}) gives a transition between a
phase $H^2\sim\rho$ and a phase $H^2\sim\rho^2$. The lower sign is
more interesting, and this phase is dubbed the `self-accelerating'
phase. With gravity becoming weaker at large distances,
gravitational attraction can be overcome, leading to an asymptotic
de Sitter phase. The suggestion is that this modified
gravitational behavior could possibly explain the current
accelerated expansion of the universe, without the need of
introducing dark energy. Instead of the `old' cosmological
constant problem, one could hope to solve the new cosmological
constant problem.

In \cite{Sawicki:2005cc,Ishak:2005zs} however, details of this
cosmological model have been compared with WMAP data. It is
concluded that it gives a `significantly worse fit to the
supernova data and the distance to the last-scattering surface'
than the ordinary $\Lambda$CDM model. Most likely therefore, the
DGP-model, interesting as it is, does not solve either of the
cosmological constant problems. It especially shows how hard it is
to consistently modify GR, without destroying its well tested
results.

\section{Ghost Condensation or Gravitational Higgs Mechanism }

One could also think of infrared modifications of gravity, as a
result of interactions with a `ghost condensate'. This also leads,
among other things, to a mass for the graviton, see
\cite{Arkani-Hamed-ghost}.

Assume that a scalar field $\phi$ develops a time dependent vacuum
expectation value, such that:
\begin{equation}
\langle\phi\rangle = M^2t,\quad\rightarrow\quad\phi = M^2t + \pi,
\end{equation}
where $\pi$ is the scalar excitation on this new background. So
the $\phi$-field is changing with a constant velocity. Assume
furthermore that it obeys a shift symmetry $\phi\rightarrow\phi +
a$ so that it is derivatively coupled, and that its kinetic term
enters with the wrong sign in the Lagrangian:
\begin{equation}
\mathcal{L}_{\phi} =
-\frac{1}{2}\partial^{\mu}\phi\partial_{\mu}\phi + \dots
\end{equation}
The consequence of this wrong sign is that the usual background
with $\langle\phi\rangle = 0$ is unstable and that after vacuum
decay, the resulting background will break Lorentz invariance
spontaneously.

The low energy effective action for the $\pi$ has the form:
\begin{equation}
S\sim\int
d^4x\left[\frac{1}{2}\dot{\pi}^2-\frac{1}{2M^2}(\nabla^2\pi)^2
+\dots\right],
\end{equation}
so that the $\pi$'s have a low energy dispersion relation like:
\begin{equation}\label{ghostdispers}
\omega^2\sim\frac{k^4}{M^2}
\end{equation}
instead of the ordinary $\omega^2\sim k^2$ relation for light
excitations. Time-translational invariance is broken, because
$\langle\phi\rangle = M^2t$ and as a consequence there are two
types of energy, a ``particle physics" and a ``gravitational"
energy which are not the same. The particle physics energy takes
the form:
\begin{equation}
\mathcal{E}_{pp}\sim\frac{1}{2}\dot{\pi}^2 +
\frac{\left(\nabla^2\pi\right)^2}{2M^2} +\dots,
\end{equation}
whereas the gravitational energy is:
\begin{equation}
\mathcal{E}_{grav}= T_{00}\sim M^2\dot{\pi} +\dots
\end{equation}
Although time-translation- and shift-symmetry are broken in the
background, a diagonal combination is left unbroken and generates
new ``time" translations. The Noether charge associated with this
unbroken symmetry is the conserved particle physics energy. The
energy that couples to gravity is associated with the broken time
translation symmetry. Since this energy begins at linear order in
$\dot{\pi}$, lumps of $\pi$ can either gravitate or
anti-gravitate, depending on the sign of $\dot{\pi}$! The $\pi$
thus maximally violate the equivalence principle.

If the standard model fields would couple directly to the
condensate there would be a splitting between particle and
anti-particle dispersion relations, and a new spin-dependent
inverse-square force, mediated by $\pi$ exchange, which results
from the dispersion relation (\ref{ghostdispers}). In the
non-relativistic limit:
\begin{equation}
\triangle\mathcal{L} \sim\frac{1}{F}\mathbf{S}\cdot\nabla\pi,
\end{equation}
where $F$ is some normalization constant. Because of the $k^4$
dispersion relation, the potential between two sources with spin
$\mathbf{S}_1$ and spin $\mathbf{S}_2$, will be proportional to
$1/r$:
\begin{equation}
V\sim\frac{M^4}{\tilde{M}^2F^2}\frac{\mathbf{S}_1\cdot\mathbf{S}_2-3\left(\mathbf{S}_1\cdot\hat{r}\right)}{r},
\end{equation}
when using only static sources, ignoring retardation effects.

Moreover, not only Lorentz invariance, but also CPT is broken if
the standard model fields would couple directly to the condensate.
With $\psi$ standard model Dirac fields, the leading derivative
coupling is of the form:
\begin{equation}
\triangle\mathcal{L} =
\sum_{\psi}\frac{c_{\psi}}{F}\bar{\psi}\bar{\sigma}^{\mu}\psi\partial_{\mu}\phi.
\end{equation}
As noted in \cite{Arkani-Hamed-ghost}, field redefinitions
$\psi\rightarrow e^{ic_{\psi}\phi/F}\psi$ may remove these
couplings, but only if such a $U(1)$ symmetry is not broken by
mass terms or other couplings in the Lagrangian. If the fermion
field $\psi$ has a Dirac mass term $m_D\psi\psi^c$, then the
vector couplings, for which $c_{\psi} + c_{\psi^c} =0$, still can
be removed, but the axial couplings remain:
\begin{equation}
\triangle\mathcal{L}\sim\frac{1}{F}\bar{\Psi}\gamma^{\mu}\gamma^5\Psi\partial_{\mu}\phi.
\end{equation}
After expanding $\phi = M^2t +\pi$ this becomes:
\begin{equation}
\triangle\mathcal{L}\sim \mu\bar{\Psi}\gamma^0\gamma^5\Psi +
\frac{1}{F}\bar{\Psi}\gamma^{\mu}\gamma^5\Psi\partial_{\mu}\pi,
\end{equation}
with $\mu = M^2/F$. This first term violates both Lorentz
invariance and CPT, leading to different dispersion relations for
particles and their anti-particles. A bound on $\mu$ is obtained
by considering the earth to be moving with respect to spatially
isotropic condensate background. The induced Lorentz and CPT
violating mass term then looks like:
\begin{equation}
\mu\bar{\Psi}\vec{\gamma}\gamma^5\Psi\cdot\mathbf{v}_{earth},
\end{equation}
which in the non-relativistic limit gives rise to an interaction
Hamiltonian:
\begin{equation}
\mu\mathbf{S}\cdot\mathbf{v}_{earth}.
\end{equation}
The experimental limit on $\mu$ for coupling to electrons is
$\mu\leq 10^{-25}$~GeV \cite{Heckeletal1999} assuming
$|\mathbf{v}_{earth}|\sim 10^{-3}$. For other limits on CPT and
Lorentz invariance, see
\cite{Phillipsetal2000,Bluhm2003,Cane2003}.

If there is no direct coupling, the SM fields would still interact
with the ghost sector through gravity. Interestingly, IR
modifications of general relativity could be seen at relatively
short distances, but only after a certain (long) period of time!
Depending on the mass $M$ and the expectation value of $\phi$,
deviations of Newtonian gravity could be seen at distances
1000~km, but only after a time $t_c\sim H_{0}^{-1}$ where $H_0$ is
the Hubble constant. More general, the distance scale at which
deviations from the Newtonian potential are predicted is $r_c\sim
M_{Pl}/M^2$ and their time scale is $t_c\sim M_{Pl}^2/M^3$.

To see the IR modifications to GR explicitly, let us consider the
effective gravitational potential felt by a test mass outside a
source $\rho_m(r,t)=\delta^3(r)\theta(t)$, i.e. a source that
turns on at time $t=0$. This potential is given by:
\begin{equation}
\Phi(r,t)=-\frac{G}{r}\left[1+I(r,t)\right],
\end{equation}
where $I(r,t)$ is a spatial Fourier integral over momenta $k$,
evaluated using an expansion around flat space; a bare
cosmological constant is set to zero.
\begin{eqnarray}
I(r,t)=\frac{2}{\pi}&\Big\{&\int_{0}^{1}du\frac{\sin(uR)}{(u^3-u)}\left(1-\cosh(Tu\sqrt{1-u^2})\right)\nonumber\\
&+&\int_{1}^{\infty}du\frac{\sin(uR)}{(u^3-u)}\left(1-\cos(Tu\sqrt{u^2-1})\right)\Big\}.
\end{eqnarray}
Here $u=k/m$, $R=mr$, $T=\alpha M^3/2M_{Pl}^2$, where $m\equiv
M^2/\sqrt{2}M_{Pl}$ and $\alpha$ is a coefficient of order 1. For
late times, $t\gtrsim t_c$, or $T\gtrsim 1$, the first integrand
will dominate and $I(r,t)$ can be well approximated by:
\begin{equation}
I(r,t) \simeq \frac{2}{\sqrt{\pi
T}}\exp\left(-\frac{R^2}{8T}+\frac{T}{2}\right)\sin\left(\frac{R}{\sqrt{2}}\right).
\end{equation}
For $R\ll T$, there is indeed an oscillatory behavior for the
gravitational potential, growing exponentially as $\exp(T/2)$,
while for $R\gg T$ the modification vanishes.

More general gravitational effects have been studied in
\cite{Dubovsky2004}, where moving sources were considered, and in
\cite{Arkani-Hamed2003} where inflation was studied in this
context. Moreover, the quantum stability of the condensate was
studied in \cite{Krotovetal2004}.

This highly speculative scenario opens up a new way of looking at
the cosmological constant problem, especially because of the
distinction between particle physics energy, $\mathcal{E}_{pp}$
and gravitational energy, $\mathcal{E}_{grav}$. It has to be
developed further to obtain a better judgement.

\section{Fat Gravitons}\label{fatgravitons}

\begin{figure}[h]
\includegraphics[width=15cm]{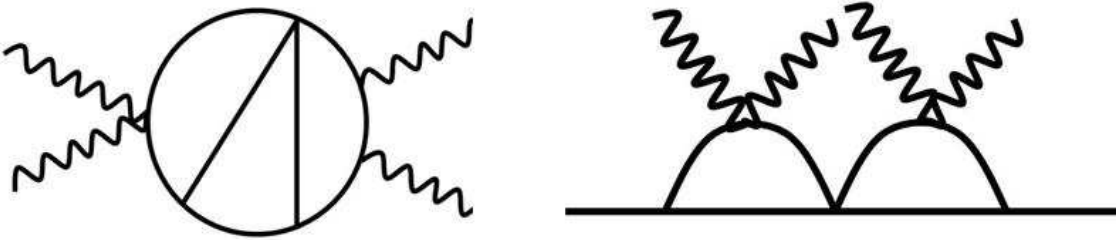}
\caption{On the left-hand-side, a typical Standard Model
contribution to $\Gamma_{eff}[g_{\mu\nu}]$. On the right, soft
gravitons coupled to loop-correction to SM self-energy. Wiggly
lines are gravitons and smooth lines are SM particles.}
\label{gravselfen}
\end{figure}

A proposal involving a sub-millimeter breakdown of the
point-particle approximation for gravitons has been put forward by
Sundrum \cite{Sundrum}. In standard perturbative gravity, diagrams
with external gravitons and SM-particles in loops (see figure
\ref{gravselfen})
give a contribution to the effective CC of which the dominant part
diverges as $\Lambda^4_{UV}$ where $\Lambda_{UV}$ is some
ultraviolet cutoff. This leads to the enormous discrepancy with
experimental results for any reasonable value of $\Lambda_{UV}$.
However, one might wonder what the risks are when throwing away
these diagrams from the effective theory
$\Gamma_{eff}[g_{\mu\nu}]$, when $|k^2|$, the momentum of the
external gravitons, is larger than some low energy cutoff.
Properties at stake are: Unitarity, General Coordinate Invariance
(GCI) and locality. In standard effective theory, diagrams where
soft gravitons give corrections to the SM self energy diagrams
(the right one in figure \ref{gravselfen}), are crucial in
maintaining the equivalence principle between inertial and
gravitational masses. However, in a theory with locality, the
value associated to the diagram on the left, should follow
unambiguously from that of the diagram on the right. Thus given
locality of the couplings of the point particles in the diagrams,
we cannot throw the first diagram away and keep the other.
Therefore, it seems progress can be made by considering a graviton
as an extended object. Note that exactly the same arguments hold
for the diagrams with just one external graviton line i.e. a
normal tadpole diagram, with matter particles running around in
the loop, and diagrams with a standard model contribution to the
self energy, with and without an external graviton leg. These
latter two diagrams are also corrections to (first) the
gravitational and (second) inertial mass. Again, the contribution
from the loop integral in the tadpole diagram can be suppressed,
if there is a loop momentum-dependent form factor to the vertex.
This would also suppress the correction to the gravitational mass
in the second diagram, but obviously not the third diagram, since
it does not contain an external graviton, thus violating the
equivalence principle. Only for an extended graviton with some
non-zero size, only the first diagram can be suppressed.

Define the graviton size:
\begin{equation}
l_{grav}\equiv \frac{1}{\Lambda_{grav}}.
\end{equation}
Such a ``fat graviton" does not have to couple with point-like
locality to SM loops, but with locality up to $l_{grav}$. Thus a
fat graviton can distinguish between the two types of diagrams,
possibly suppressing the first while retaining the second.

The value of the CC based on usual power counting would then be:
\begin{equation}
\Lambda_{eff}\sim\mathcal{O}(\Lambda^{4}_{grav}/16\pi^{2}).
\end{equation}
Comparing with the observational value this gives a bound on the
graviton size of:
\begin{equation}
l_{grav} > 20~\mbox{microns}
\end{equation}
which would indicate a short-distance modification of Newton's law
below $20$~microns. This is however not enough to suppress
standard model contributions to the cosmological constant. The
same two-point diagrams also renormalize Newton's constant,
sending it to zero; the Planck mass becomes enormous. A new model
by the same author has been proposed to take into account also
these effects, see section \ref{Energytominus}.

But there are other difficulties as well. Purely gravitational
loops will still contribute too much to the cosmological constant,
and tree level effects, such as a shift in vacuum energy during
phase transitions, are not accounted for either. Besides, such a
fat graviton would still couple normally to particles with a mass
$m < \Lambda_{grav}$, in particular massless particles. And, even
more seriously, unitarity is lost when matter couples to gravity
above this cutoff at $10^{-3}$~eV. This loss of unitarity is
Planck suppressed, but is nevertheless a severe problem for this
approach.

On a more positive note, the idea that gravity shuts off
completely below $10^{-3}$~eV is a very interesting idea. The
cosmological constant problem could be solved if one were to find
a mechanism showing that flat spacetime is a preferred frame at
distances $l < 0.1$~mm. The model of Sundrum is an approach in
this direction, and one of very few models in which gravity
becomes weaker at shorter distances. Moreover, another obvious
advantage is that it can at least be falsified by submillimeter
experiments of the gravitational $1/r^2$ law.

\section{Composite Graviton as Goldstone boson}

Another approach is to consider the possibility that the graviton
appears as a composite Goldstone boson. There exists a theorem by
Weinberg and Witten, \cite{WeinbergWitten1980}, stating that a
Lorentz invariant theory, with a Lorentz covariant energy-momentum
tensor does not admit a composite graviton. In fact, this theorem
poses restrictions on the presence of any massless spin-1 and
spin-2 particle. It states that such a massless particle cannot
exist if it is charged under the current generated by the
conserved vector or conserved symmetric 2nd rank tensor. The
photon can be massless, since it is not charged under this current
(no electric charge), the gluon can be massless because the
current is not Lorentz-covariant due to an additional gauge
freedom; the same reason for the `normal' graviton to evade this
Weinberg-Witten no-go theorem.

It is therefore natural to try a mechanism where the graviton
appears as a Goldstone boson associated with the spontaneous
breaking of Lorentz invariance. Being a Goldstone boson, the
graviton would not develop a potential, and hence the normal
cosmological constant problem is absent, see for example
\cite{Isham:1971dv,KrausTomboulis}. We will briefly review the
latter proposal.

The effective action is written:
\begin{equation}
\mathcal{L}= N\left(A^2\sqrt{-g}R - A^4V(h) +
\mbox{higher$\,$derivatives}\right) + \mathcal{L}_{matter} +
\mathcal{O}(N^0),
\end{equation}
where $N$ is some large number $N\sim 10^4$, $A$ is a cutoff, and
$h$ is defined by:
\begin{equation}
g_{\mu\nu}=\eta_{\mu\nu} + h_{\mu\nu}.
\end{equation}
The observed Newton's constant is $G_N\sim1/(NA^2)$ and general
covariance is only violated by the potential.

The point now is to consider potentials leading to a vev for
$h_{\mu\nu}$. Lorentz transformations can bring $\langle
h_{\mu\nu}\rangle$ to diagonal form and when the $h_{\mu\mu}$ are
all non-zero and distinct, the Lorentz group will be completely
broken. This leads to six Goldstone bosons, the off-diagonal
components of $h_{\mu\nu}$. Two of these are identified to form
the `ordinary' graviton.

It is clear that in such a setting, one has to rethink the
cosmological constant problem, since general covariance is
violated. Moreover, since the graviton is identified as a
Goldstone boson, it would not develop a potential.

However, so much has been ruined, that indeed the traces of broken
Lorentz invariance cannot be erased. Moreover, de Sitter and anti
de Sitter spacetime are still allowed solutions
\cite{KrausTomboulis}.

Violations of Lorentz invariance, or perhaps even violations of
general covariance, are very interesting routes to explore with
regard to the cosmological constant problem
\cite{Jenkins,Jacobson:2000xp,Graesser:2005bg,Bertolami:2005bh}.
Indeed, if flat spacetime would in some sense be a preferred frame
on short distances, say at a tenth of a millimeter, without
violating bounds on GR at larger distance scales, this could be
considered a solution. In my opinion this could very well be the
underlying reason why we observe such a small cosmological
constant. With Minkowski as a preferred frame at these rather
large submillimeter distances, one could boldly suggest that in
fact gravity does not exist at smaller distances. This would solve
the cosmological constant problem. Note that it is
phenomenologically opposite to scenarios with extra dimensions
where gravity becomes stronger at smaller distances. Such a
construction is however far from trivial, because one would have
to deal with violations of Lorentz invariance, general covariance,
and perhaps locality.

With a preferred frame, a priori Lorentz invariance and general
covariance both would be broken, but rotational invariance would
still be an exact symmetry. Mathematically, such a construction
could be built using a unit future timelike vector field $u^a$.
This generally leads to many possible observable effects, such as
modified dispersion relations for matter fields. Another important
issue is that general covariance in Einstein's equation ensures
stress-energy conservation, which is not to be violated. One
therefore often looks for a mechanism where the preferred frame
arises, while preserving general covariance. This requires the
preferred frame to be dynamical. Moreover, to prevent the theory
from instabilities, the preferred frame would have to arise from
local conditions \cite{Jacobson:2000xp}. The construction can be
setup as a vector-tensor theory of gravitation, but with an
additional constraint since the vector field is a purely unit
timelike vector field. This has for example the benefit, that
ghosts may not appear, even though a vector field is introduced
without gauge invariance. Unfortunately however, it turns out to
be very non-trivial to construct stable, viable theories of
gravity this way. The PPN-parameters give strict constraints and
often not only Lorentz invariance is broken, but also $C$, $CP$
and $CPT$ \cite{Jenkins}. Moreover, introducing just any preferred
frame leads to a renormalization of Newton's constant, through the
`aether' energy-momentum tensor, but has a priori no direct effect
on the cosmological constant. In fact, many models break Lorentz
invariance by giving a vacuum expectation value to a vector field,
which in a sense only has a negative effect on the cosmological
constant problem.

See \cite{Eling:2004dk} for a review and \cite{Will:1993ns} for a
very broad discussion of alternative theories of gravity including
their experimental consequences.

Violations of Lorentz invariance have been studied in both loop
quantum gravity \cite{Alfaro:2001rb} and string theory
\cite{Kostelecky:1988zi}. The missing GZK-cutoff could be an
experimental indication in the same direction
\cite{Coleman:1998ti}. For modern tests of violations of Lorentz
invariance, see \cite{Mattingly:2005re}.

\subsection{Summary}

Since General Relativity has only been thoroughly tested on solar
system distance scales it is a very legitimate idea to consider
corrections to GR at galactic and/or cosmological distance scales.
However, often these models are not so harmless as supposed to be:
changing the laws of gravity also at shorter scales, or leading to
violations of locality. The scenarios described in this section do
not directly solve the cosmological constant problem, but offer
new ways of looking at it.

On the more positive side, many theories that predict
modifications of GR in the IR, reproduce Einstein gravity at
smaller distances, but up to some small corrections. These
corrections are discussed in
\cite{DvaliGruzinovZaldarriagaMoon,Dvali:2003rk,Ishak:2005zs} and
could be potentially observable at solar system distance scales.
At the linearized level gravity is of the scalar-tensor type,
because the graviton has an extra polarization that also couples
to conserved energy-momentum sources. If these models are correct,
an anomalous perihelion precession of the planets is expected to
be observed in the near future.

Most likely, the DGP-model is not such a candidate. To solve the
old cosmological constant problem and arrive at a flat brane
metric solution, at least three extra dimensions are necessary.
However, with a flat brane metric, singularities in the bulk
appear unavoidable, unless there are miraculous cancellations once
the full UV completion is known. The model also suffers from
hidden strong interaction scales, which lead to modifications of
GR at distances smaller than 1000~km, for a cross-over distance to
the higher-dimensional behavior at the current Hubble length.

Besides, submillimeter experiments of Newtonian gravity set ever
more stringent bounds on both extra dimensional approaches and
composite graviton scenarios. It would be very exciting to see a
deviation of Newtonian gravity at short distances. Especially a
weakening of gravity at these distances would be very welcome with
regard to the cosmological constant problem.

On the other hand, observing no change at all, will seriously
discourage the hopes that such a mechanism might help in solving
the cosmological constant problem. The critical distance for this
problem is roughly a tenth of a millimeter and that scale is
practically within reach for experimental investigations.

%% file: chapter6.tex
\chapter{Type IV: Statistical Approaches}

In this chapter approaches to the cosmological constant problem
are studied, in which quantum effects at the beginning of the
universe play a dominant role. The first two sections deal with
so-called quantum cosmological scenarios based on the
Wheeler-DeWitt equation \cite{DeWitt1967,Wheeler1968}, and the
third section describes the anthropic principle. Common in all
three approaches is that the value of the cosmological constant is
distributed according to some governing equation and that a
probability for $\Lambda =0$ is calculated.

If the cosmological constant could a priori have any value,
appearing for example as a constant of integration as in section
(\ref{unimodular}), or would become a dynamical variable by means
of some other mechanism, then in quantum cosmology the state
vector of the universe would be a superposition of states with
different values of $\Lambda_{eff}$. The path integral would
include all, or some range of values of this effective
cosmological constant. The observable value of the CC in this
framework is not a fundamental parameter. Different universes with
different values of $\Lambda_{eff}$ contribute to the path
integral. The probability $\mathcal{P}$ of observing a given field
configuration will be proportional to $\mathcal{P}\propto\exp
(-S(\Lambda_{eff}))$ in which $\Lambda_{eff}$ is promoted to be a
quantum number.

The wave function of the universe \cite{Hartle:1983ai} would have
to satisfy an equation, analogously to the Klein-Gordon equation,
called the Wheeler-DeWitt equation. One arrives at this equation,
after canonically quantizing the gravitational Hamiltonian. This
Hamiltonian formulation is not covariant, but requires a $3+1$
split of the metric, called the Arnowitt, Deser, Misner, or (ADM),
decomposition of the spacetime metric in terms of a lapse function
$N$, a shift vector $N_i$ and a spatial metric $h_{ij}$, see
\cite{Wald:1984rg,Moss:1996ry}. In terms of these quantities, the
proper length $ds^2=g_{\mu\nu}dx^{\mu}dx^{\nu}$ becomes:
\begin{equation}\label{sixone}
ds^2 = -(Ndt)^2 + h_{ij}(N^idt + dx^i)(N^jdt + dx^j).
\end{equation}
Spacetime is decomposed into three-dimensional hypersurfaces
$\Sigma_t$; the phase space consists of 3-metrics $h_{ij}$ and
various matter fields $\phi$ on $\Sigma_t$, together with their
conjugate momenta.

The Einstein-Hilbert action is assumed to be the correct action
also at high energies and, in terms of the lapse function $N$ and
shift vector $N_i$, can be written as:
\begin{equation}\label{lagn}
\mathcal{L}(N,N_i,h_{ij})= -\sqrt{h}N\left(K^2 - K_{ij}K^{ij} -
{^3}R\right)/16\pi G,
\end{equation}
here $h_{ij}$ is the induced spatial metric, and $K_{ij}$ the
extrinsic curvature and $K$ its trace. $K_{ij}$ is defined as:
\begin{equation}
K_{ij} \equiv \frac{1}{2N}\left(N_{i|k} + N_{k|i} - \frac{\partial
h_{ij}}{\partial t}\right),
\end{equation}
where the subscript $|j$ denotes covariant differentiation with
respect to the spatial metric $h_{ij}$. Since the Lagrangian
(\ref{lagn}) contains no time derivatives of $N$ or $N_i$, their
conjugate momenta vanish:
\begin{equation}
\pi \equiv \frac{\delta\mathcal{L}}{\delta\dot{N}} = 0, \quad\quad
\pi^i \equiv \frac{\delta\mathcal{L}}{\delta\dot{N}_i} = 0,
\end{equation}
these are called `primary' constraints. Furthermore, the momentum
conjugate to $h_{ij}$ is:
\begin{equation}
\pi^{ij}\equiv\frac{\delta\mathcal{L}}{\delta\dot{h}_{ij}}=\sqrt{h}\left(
K^{ij} - h^{ij}K\right)/16\pi G.
\end{equation}
The Hamiltonian then follows in the usual way:
\begin{eqnarray}
H&=&\int d^3x\left(\pi^{ij}\dot{h}_{ij} + \pi^i\dot{N}_i +
\pi\dot{N}-\mathcal{L}\right)\nonumber\\
&=&\int d^3x\left(N\mathcal{H}_G + N_i\mathcal{H}^i\right).
\end{eqnarray}
$\mathcal{H}_G$ is defined as:
\begin{eqnarray}\label{gravham}
\mathcal{H}_G &\equiv&\sqrt{h}\left(K_{ij}K^{ij}-K^2 -
\;{^3R}\right)/16\pi G\nonumber\\
&=&\frac{16\pi G}{2\sqrt{h}}(h_{ik}h_{jl} + h_{il}h_{jk} -
h_{ij}h_{kl})\pi^{ij}\pi^{kl}-\sqrt{h}\;{^3R}/16\pi G\nonumber\\
&\equiv& 16\pi G G_{ijkl}\pi^{ij}\pi^{kl}-\sqrt{h}\;{^3R}/16\pi G
\end{eqnarray}
For a closed FRW-model, these are related as follows:
\begin{equation}
R = K^2 - K_{ij}K^{ij} - {^3R}
\end{equation}
Furthermore,
\begin{equation}
\mathcal{H}^i\equiv -2\pi^{ij}_{|j}/16\pi G.
\end{equation}
$\mathcal{H}_G$ and $\mathcal{H}^i$ have to satisfy the
`secondary' constraint $\mathcal{H}_G= \mathcal{H}^i=0$, which for
$\mathcal{H}_G$ can be written as:
\begin{equation}
\mathcal{H}_G = 16\pi G G_{ijkl}\pi^{ij}\pi^{kl}
-\sqrt{h}\;{^3R}/16\pi G=0
\end{equation}
with the definition of $G_{ijkl}$ implicitly given in
(\ref{gravham}).

So far this is just the classical treatment. Quantization is
obtained by first identifying the Hamiltonian constraint as the
zero energy Schrodinger equation:
\begin{equation}
\mathcal{H}_G(\pi_{ij},h_{ij})\Psi [h_{ij}]=0
\end{equation}
where the state vector $\Psi$ is the wave function of the
universe, and secondly by replacing the momenta $\pi$ in the
normal canonical quantization procedure by:
\begin{equation}
\pi^{ij}\rightarrow -i\left(\frac{1}{16\pi
G}\right)^{3/2}\frac{\delta}{\delta h_{ij}}.
\end{equation}
This gives the Wheeler-DeWitt equation:
\begin{equation}\label{WDW}
\left[\frac{G_{ijkl}}{(16\pi G)^2}\frac{\delta}{\delta
h_{ij}}\frac{\delta}{\delta h_{kl}} +
\frac{\sqrt{h}(^3R-2\Lambda)}{16\pi G}\right]\Psi [h_{ij}]=0
\end{equation}
One of the most difficult features of this equation is that the
wave function of the universe is independent of time. $\Psi$
depends only on $h_{ij}$ and the matter content. This makes the
interpretation of this wave function very complicated.

Probabilities are calculated, using the fact that the WKB
approximation for tunneling is proportional to $\exp(-S_E)$, with
$S_E$ the Euclidean action. In this context, it gives the
amplitude for the entire universe to tunnel from an in-state, to
an out-state. The WKB approximation is justified, since we will
consider de Sitter spaces, which are large relative to the Planck
scale. It is assumed that in this case, short distance quantum
gravity effects can be neglected.

The appropriate boundary condition is a matter of debate. Hartle
and Hawking advocate the `no-boundary' boundary condition, which
amounts to take:
\begin{equation}
\Psi_0[h_{ij}]\propto\int\mathcal{D}g\exp[-S_E]
\end{equation}
where the path integral over geometries extends over all compact
Euclidean 4-geometries, which have no boundary. $S_E(g)$ is the
Euclidean action associated with the manifold.

The `quantum cosmology' one obtains this way faces many unresolved
issues. We have, for example, used a Euclidean action throughout,
which is clearly unsuitable for gravity, considering that the
Euclidean gravitational action is unbounded from below. Besides,
when using a sum-over-histories approach, it is assumed that the
universe is finite and closed, since the relevant integrals are
undefined in an open universe. Next, the spatial topology of the
universe is assumed fixed (note that the spatial topology is
undetermined in general relativity). Moreover, there is an issue
in interpretation. What does it mean to have a wave function of
the universe? And, last but not least, the role of time, is not
exactly understood. Curiously, most of these ``difficulties'' do
not show up in the perturbative treatment of quantum gravity.

\section{Hawking Statistics}\label{Hawking}

Eleven dimensional supergravity contains a three-form gauge field,
with a four-form field strength $F_{\mu\nu\rho\sigma} =
\partial_{[\mu}A_{\nu\rho\sigma ]}$ \cite{Aurilia:1980xj}. When
reduced to four dimensions, this gives a contribution to the
cosmological constant
\cite{Duff:1980qv,Baum,GomberoffHenneauxTeitelboimWilczek,BrownTeitelboim1,BrownTeitelboim2}.
Hawking \cite{Hawking1} used such a three-form gauge field to
argue that the wave function of the universe is peaked at zero
cosmological constant. It is the first appearance of the idea that
the CC could be fixed by the shape of the wave function of the
universe.

The three-form field $A_{\mu\nu\lambda}$ is subject to gauge
transformations:
\begin{equation}
A_{\mu\nu\rho}\rightarrow A_{\mu\nu\rho}
+\nabla_{[\mu}C_{\nu\rho]},
\end{equation}
which leaves invariant the fields
$F_{\mu\nu\rho\sigma}=\nabla_{[\mu}A_{\nu\rho\sigma]}$. This field
would contribute an extra term to the action:
\begin{equation}
I = -\frac{1}{16\pi G}\int d^4x\sqrt{-g}\left( R+2\Lambda_B\right)
-\frac{1}{48}\int
d^4x\sqrt{-g}F_{\mu\nu\rho\sigma}F^{\mu\nu\rho\sigma}.
\end{equation}
The field equation for $F^{\mu\nu\rho\sigma}$ is:
\begin{equation}
D_{\mu}F^{\mu\nu\rho\sigma}=0
\end{equation}
which in four dimensions is just a constant:
\begin{equation}\label{solHawk}
\sqrt{-g}F^{\mu\nu\rho\sigma} = \omega\epsilon^{\mu\nu\rho\sigma}
\end{equation}
Such a field $F$ has no dynamics, but the $F^2$ term in the action
behaves like an effective cosmological constant term, whose value
is determined by the unknown parameter $\omega$, which takes on
some arbitrary spacetime independent value. If we substitute the
solution (\ref{solHawk}) back into the Einstein equation, we find:
\begin{equation}
T^{\mu\nu}= \frac{1}{6}\left(
F^{\mu\alpha\beta\gamma}F^{\nu}_{\alpha\beta\gamma} -
\frac{1}{8}g^{\mu\nu}F^{\alpha\beta\gamma\delta}F_{\alpha\beta\gamma\delta}\right)
= \pm \frac{1}{2}\omega^2g^{\mu\nu}
\end{equation}
using that
$\epsilon^{\mu\nu\rho\sigma}\epsilon_{\mu\nu\rho\sigma}=\pm 4!$,
where the sign depends on the metric used: in Euclidean metric
$\epsilon^{\mu\nu\rho\sigma}\epsilon_{\mu\nu\rho\sigma}$ is
positive, whereas in Lorentzian metric it is negative. In the
Euclidean action Hawking used:
\begin{equation}
R=-4\Lambda_{eff}=-4(\Lambda_B - 8\pi G\omega^2)
\end{equation}
where $\Lambda_B$ is the bare cosmological constant in Einstein's
equation. It follows that:
\begin{equation}\label{effstat}
S_{Hawking} = -\Lambda_{eff}\frac{V}{8\pi G}.
\end{equation}
with $V=\int d^4x\sqrt{g}$ is the spacetime volume. The maximum
value of this action is given when $V$ is at its maximum, which
Hawking takes to be Euclidean de Sitter space; this is just $S^4$,
with radius $r=(3\Lambda^{-1}_{eff})^{1/2}$ and proper
circumference $2\pi r$. We have:
\begin{equation}
V=\frac{24\pi^2}{\Lambda_{eff}^2},\quad\quad\rightarrow\quad\quad
S(\Lambda)=-3\pi\frac{M^{2}_{P}}{\Lambda_{eff}}
\end{equation}
and thus the probability density, which is proportional to
$\exp(-S_E)$, where $S_E$ is the Euclidean action, becomes:
\begin{equation}
\mathcal{P}\propto
\exp\left(3\pi\frac{M^{2}_{P}}{\Lambda_{eff}}\right)
\end{equation}
is peaked at $\Lambda_{eff}=0$.

Note that we have used here that the probability is evaluated as
the exponential of minus the effective action at its stationary
point. The action is stationary with respect to variations in
$A_{\mu\nu\lambda}$, when the covariant derivative of
$F_{\mu\nu\lambda\rho}$ vanishes and stationary in $g_{\mu\nu}$
when the Einstein equations are satisfied. Eqn. (\ref{effstat}) is
the effective action at the stationary point. It is a good thing
that we only need the effective action at its stationary point, so
that we do not have to worry about the Euclidean action not being
bounded from below, see for example \cite{Weinbergreview}.

However, Hawking's argument has been criticized, since one should
not plug an ansatz for a solution back into the action, but rather
vary the unconstrained action \cite{Duff:1989ah}. This differs a
minus sign in this case, the same minus sign as going from a
Lorentzian to a Euclidean metric, $\Lambda_{eff}=(\Lambda_B \pm
8\pi G\omega^2)$, but now between the coefficient of $g^{\mu\nu}$
in the Einstein equations, and the coefficient of $(8\pi
G)^{-1}\sqrt{g}$ in the action. Note that a plus sign of $\Lambda$
in the gravitational action, always leads to a minus sign in the
Einstein field equations. The correct action becomes
\cite{Duff:1989ah}:
\begin{equation}
S =(-3\Lambda_{eff} + 2\Lambda_B)\frac{-3\pi
M_{P}^2}{\Lambda^{2}_{eff}} = -3\pi M_{P}^2\frac{\Lambda_B -12\pi
G\omega^2}{(\Lambda_B - 4\pi G\omega^2)^2}
\end{equation}\
now for $\Lambda_{eff}\rightarrow 0$, the action becomes large and
positive and consequently, $\Lambda_{eff}=0$ becomes the
\textit{least} probable configuration.

Besides, in \cite{PolchinskiBousso} it is shown that this approach
has also other serious limitations. It is argued that it can only
work in the `Landscape' scenario that we discuss in section
(\ref{AP}). The reason is that the four-form flux should be
subject to Dirac quantization and the spacing in $\Lambda$ then
only becomes small enough with an enormous number of vacua.

\subsection{Wormholes}

In a related approach Coleman \cite{colemanworm} argued that one
did not need to introduce a 3-form gauge field, if one includes
the topological effects of wormholes. This also transforms the
cosmological constant into a dynamical variable. The argument
assumes that on extremely small scales our universe is in contact,
through wormholes, with other universes, otherwise disconnected,
but governed by the same physics as ours. In addition, there are
wormholes that connect our universe with itself. Both types of
wormholes are assumed be very tiny, but their end points will be
at different locations in the universe, and as such, can be
arbitrarily far apart, connecting regions that may be causally
disconnected. However, at scales larger than the wormhole size,
the only effect of wormholes is to add \textit{local}
interactions, one for each type of wormhole.

The extra term in the action has the form:
\begin{equation}\label{wormholact}
S_{wormhole} = \sum_i(a_i+a^{\dag}_{i})\int
d^4x\sqrt{g}e^{-S_i}K_i
\end{equation}
where $a_i$ and $a^{\dag}_{i}$ are the annihilation and creation
operators\footnote{We would prefer to talk about functions of
fields, since we are doing path integral calculations, and because
there is no clear definition of a Hilbert space, on which $a$ and
$a^\dag$ act. However, we will use the conventions of ref.
\cite{colemanworm}.} for a type $i$ baby universe, $S_i$ is the
action of a semi-wormhole (one that terminates on a baby
universe), and $K_i$ is an infinite series of local operators,
with operators of higher dimension suppressed by the wormhole
size. The interaction therefore, is local at distance scales
larger than the wormhole size, but non-local on the wormhole
scale. Furthermore, there is an important exponential factor that
suppresses the effects of all wormholes, except those of Planckian
size \cite{Hawking:1987mz,Strominger:1983ns,Coleman:1988cy}. This
result is obtained by treating the wormholes semi-classically and
in the dilute gas approximation. This dilute gas approximation,
provides a way of writing the functional integral over manifolds
full of baby universes and wormholes, in terms of an integral over
manifolds stripped of these.

The coefficients of these interaction terms are operators $A_i=
a_i+a_{i}^{\dag}$ which only act on the variables describing the
baby universes, and commute with everything else. The path
integral over all 4-manifolds with given boundary conditions
becomes:
\begin{equation}
\int[dg][d\Phi]e^{-S}=\int_{\mbox{No}}[dg][d\Phi]\langle
B|e^{-(S+S_{wormhole})}|B\rangle
\end{equation}
where $\mbox{No}$ means that wormholes and baby-universe are
excluded, and $|B\rangle$ is a normalized baby-universe state.
This state $|B\rangle$ can always be expanded in eigenstates of
the operators $A_i=a_i+a_{i}^\dag$:
\begin{equation}
|B\rangle=\int f_{B}(\alpha)\prod_id\alpha_i|\alpha\rangle,
\end{equation}
with $\alpha_i$ the eigenvalues of $A_i$:
\begin{equation}
(a_i+a_{i}^\dag)|\alpha\rangle=\alpha_i|\alpha\rangle,\quad\quad\mbox{and}\quad\quad
\langle\alpha'|\alpha\rangle=\prod_i\delta(\alpha_{i}'-\alpha_i),
\end{equation}
and the function $f_{B}(\alpha)$ depends on the boundary
conditions. For Hartle-Hawking boundary conditions:
\begin{equation}
a_i|B\rangle=0\quad\quad\mbox{and}\quad\quad
f_{B}(\alpha)=\prod_i\pi^{-1/4}\exp(-\alpha_{i}^2/2)
\end{equation}

Written in terms of $A$-eigenstates, the effective action becomes:
\begin{equation}\label{wormholacteigen}
S_{wormhole} = \sum_i\int d^4x\sqrt{g}\;\alpha_ie^{-S_i}K_i.
\end{equation}\
According to the Copenhagen interpretation, after performing a
measurement, the state vector of the universe collapses to an
incoherent superposition of of these $|\alpha\rangle$'s, each
appearing with probability $|f_{B}(\alpha)|^2$. The $\alpha_i$
renormalize all local operators when measured at distance scales
larger than the wormhole scale, i.e. for an observer who cannot
detect the baby universes.

This way, the effective cosmological constant also becomes a
function of the $\alpha_i$, since one of these local operators is
$\sqrt{g}$. Moreover, on scales larger than the wormhole scale,
manifolds that appear disconnected will really be connected by
wormholes, and therefore are to be integrated over. The
Hartle-Hawking wave function of the universe $\Psi^{HH}_{\alpha}$
in the presence of wormholes can be calculated, using the fact
that the no-boundary condition now also states that there are no
baby universes. In terms of the $\alpha$'s this reads:
\begin{equation}
\langle\alpha |0\rangle=e^{-\alpha^2/2},
\end{equation}
The wave function can then be written as:
\begin{equation}
\Psi^{HH}_{\alpha}(B,\alpha)=\sum e^{-S_{eff}(\alpha)} =
e^{-\alpha^2/2}\psi^{HH}_{\alpha}(B)Z(\alpha),
\end{equation}
where the sum is over all manifolds that go from no-boundary to
$B$ and $S_{eff}(\alpha)=S+S_{wormhole}(\alpha)$. This sum
factorizes, since some manifolds have components that connect to
$B$, while other components are closed, and have no boundary at
all. $\psi_{\alpha}$ is therefore given by the sum over manifolds
connected to $B$ and:
\begin{equation}
Z(\alpha)=\sum_{CM}e^{-S_{eff}(\alpha)},
\end{equation}
a sum over closed manifolds and this $Z(\alpha)$ can be
interpreted as giving the probability of finding a given value of
$\alpha$ in the Hartle-Hawking state. This expectation value can
be calculated from:
\begin{equation}
\langle\psi\rangle^{HH}=\frac{\int d\alpha\;
e^{-\alpha^2/2}\langle\psi\rangle_{\alpha}^{HH}Z(\alpha)}{\int
d\alpha\; e^{-\alpha^2/2}Z(\alpha)}
\end{equation}
Thus we find:
\begin{equation}
d\mathcal{P}=e^{-\alpha^2/2}Z(\alpha)d\alpha,
\end{equation}
with $d\mathcal{P}$ the probability distribution.

The sum of all vacuum-to-vacuum graphs is the exponential of the
sum of connected graphs, which gives:
\begin{equation}\label{sumeffact}
\mathcal{P}\propto\exp\left[\sum_{CCM}e^{-S_{eff}(\alpha)}\right],
\end{equation}
where $CCM$ stands for closed connected manifolds. The sum can be
expressed as a background gravitational field effective action,
$\Gamma$. The sum over closed connected manifolds can then be
written as a sum over topologies:
\begin{equation}
\sum_{CCM}e^{-S_{eff}(\alpha)} = \sum_{topologies}e^{-\Gamma(g)},
\end{equation}
with $g$ the background metric on each topology and each term on
the right is again to be evaluated at its stationary point. This
is progress, since the leading term in $\Gamma$ for large, smooth
universes is known, and is the cosmological constant term:
\begin{equation}\label{gammaleading}
\Gamma = \Lambda(\alpha)\int d^4x\sqrt{g} + \ldots,
\end{equation}
$\Lambda(\alpha)$ being the fully renormalized cosmological
constant. Plugging this back into (\ref{sumeffact}) gives the
final result:
\begin{equation}\label{ampworm}
\mathcal{P}\propto\exp\left[\exp\left(3\pi\frac{M_{P}^2}{\Lambda_{eff}}\right)\right],
\end{equation}
and thus is even sharper peaked at $\Lambda =0$ than in Hawking's
case. For positive CC the maximum volume is taken, like in
Hawking's case, the 4-sphere with $r=(3\Lambda^{-1}_{eff})^{1/2}$.
Furthermore, on dimensional grounds, the higher order terms in
(\ref{gammaleading}) are neglected.

An advantage of Coleman's approach is that he is able to sidestep
many technical difficulties Hawking's approach suffers from. In
particular, he uses the Euclidean path integral only to calculate
expectation values of some scalar field. These are independent of
$x$, because the theory is generally covariant. It includes an
average over the time in the history of the universe that the
expectation value for this operator was measured. This circumvents
many issues related to the notion of time in quantum gravity.

However, both Hawking's and Coleman's proposal rely strongly on
using a Euclidean path integral and since it is ill-defined, it is
unclear whether this is suitable for a theory of quantum
gravity\footnote{In \cite{Fischler:1989ka} this is made more
concrete: ``Evidently, the Euclidean path integral is so
ill-defined that it can be imaginatively used to prove
anything.''}.

There also appears to be a more direct problem with Coleman's
idea, as put forward by Fishler, Susskind and Polchinski
\cite{Fischler:1988ia,Polchinski:1989ae}, also see
\cite{Preskill:1988na,Coleman:1989ky}. The problem is that in
Coleman's scenario wormholes of every size will materialize in the
vacuum with maximum kinematically allowed density, leading to a
universe packed with wormholes of every size. The exponential
suppression factor in (\ref{wormholact}) is inconsistent with the
other assumptions that quantum gravity is described by a Euclidean
path integral, which is dominated by large scale spherical
universes connected by wormholes, where the amplitude of a large
scale universe is of order $\exp(M^{2}_{P}/\Lambda)$. In
particular, taking into account the higher order terms in
(\ref{gammaleading}), leads to a violation of the dilute gas
approximation, used by Coleman.

This can be seen as follows. The effect of wormholes is to
renormalize couplings, so the Einstein action is written:
\begin{equation}
S(a)=\int d^4x\sqrt{g}\left(\Lambda(a) + \kappa^2(a)R +
\gamma(a)R^2 + \ldots\right)
\end{equation}
where $a$ is the size of the wormholes $\kappa^2=1/(8\pi G)$ and
in the above action fluctuations up to distances $a$ are
integrated out. Since wormhole actions are $\propto a^2/\kappa^2$,
the relative amplitudes for wormholes of sizes $a$ and $a'$ is:
\begin{equation}
\frac{\exp(-a'^2/\kappa^2)}{\exp(-a^2/\kappa^2)},
\end{equation}
so large wormholes are suppressed. Keeping also the higher order
term $\propto\gamma(a)R^2$, the probability distribution
$\mathcal{P}(\alpha)$ (\ref{ampworm}) can be written:
\begin{equation}
\mathcal{P}(\alpha)\propto\exp\left[\exp\left(\frac{1}{\left[\Lambda(a)
+ \alpha_1\right]\left[\kappa^2(a) + \alpha_2\right]} +
\left[\gamma(a) + \alpha_3\right]\right)\right].
\end{equation}
This shows that the probability is enhanced not only for
$\Lambda(a) + \alpha_1\rightarrow 0$, but also for
$\alpha_3\rightarrow\infty$. However, $\gamma(a)$ cannot be too
large, or else unitarity would be lost and it has been assumed
that it would reach its maximum value, consistent with unitarity
\cite{Weinbergreview}. In \cite{Fischler:1988ia} it is shown that
this requirement is not consistent with the assumption of
dominance of only small wormholes. On the contrary, wormholes of
all scales $a$ will play a dominant role and strongly affect even
macroscopic physics.

In conclusion, wormholes should be integrated out of the
functional integral of quantum gravity. Their effect is to
renormalize the values of physical constants in our universe.
After integrating out the wormholes of all sizes, one should be
left with a local theory. If, for some reason, it is valid to only
take Planck-scale wormholes into account, this could make the
wavefunction of the universe in the Euclidean formalism, peak at
zero value of the cosmological constant. The next non-trivial
question to answer is what the physical implications of this would
be, since the formalism of a wavefunction of the universe in a
Euclidean spacetime is, to say the least, not very well defined.

\section{Anthropic Principle}\label{AP}

The anthropic principle is a way of reasoning to better understand
the circumstances of our universe. There are two different
versions. The first corresponds to the trivial or weak version,
which is just a tautology: intelligent observers will only
experience conditions which allow for the existence of intelligent
observers.

Proponents of the Strong Anthropic Principle advocate the stronger
point that the physical constants and physical laws have the
values they have exactly to make intelligent life possible. Most
physicists and cosmologists reject this latter form. In fact it
dates back to the question whether the universe has a goal or not
and leads to old philosophical discussions concerning teleology.
Basically the whole discussion is turned into a debate on cause
and consequence. Moreover, it is sometimes argued that the laws of
nature, which are otherwise incomplete, are completed by the
requirement that conditions must allow intelligent life to arise,
the reason being that science (and quantum mechanics in
particular) is meaningless without observers.

In the remainder of this section we will only discuss the weak
version. In order for the tautology to be meaningful it is
necessary that there are alternative conditions where things are
different. Therefore, it is usually assumed that there is some
process that produces an ensemble of a large number of universes,
or different, isolated pockets of the same universe, with widely
varying properties. Several inflationary scenario's
\cite{Linde3,Linde4,LindeMezhlumian,Linde:1986fc}, quantum
cosmologies, \cite{Hawking1, Coleman2, Linde5, Vilenkin2,
Garcia-BellidoLinde} and string theory
\cite{PolchinskiBousso,Susskind1,Susskind2,Kachru:2003aw,Denef:2004ze,FreivogelSusskind}
predict different domains of the universe, or even different
universes, with widely varying values for the different coupling
constants. In these considerations it is assumed that there exist
many discrete vacua with densely spaced vacuum energies.

Since the conditions for life to evolve as we know it are very
constrained, one can use a form of the anthropic principle to
select a certain state, with the right value for the cosmological
constant, the fine structure constant, etc, from a huge ensemble.

Depending on how specific the conditions for intelligent life to
form are, we can expect to find more or less bizarre looking
situations of extreme fine-tuning. For example, the only reason
that heavier elements are formed in the absence of stable elements
with atomic weights $A=5$ or $A=8$, is that the process in which
three $^4$helium nuclei build up to form $^{12}$C is resonant,
there is an excited energy level for the carbon nucleus that
matches the typical energies of three alpha-particles in a star.
Moreover, if the Higgs vev decreases by a factor of a few, the
proton becomes heavier than the neutron, and hydrogen decays. If
the vev increases by a factor of a few, nuclei heavier than
hydrogen decay, because the neutron-proton mass difference becomes
larger than the nuclear binding energy per nucleon. Insisting that
carbon should form, gives an even better determination of the
Higgs vev \cite{Agrawal:1997gf,Arkani-Hamed:2004fb}.

In this form, just setting boundary conditions to values of
physical constants, the `principle'\footnote{In Kolb \& Turner
\cite{KolbTurner} a footnote is written in which one can see a
glimpse of the conflicting opinions about this approach: ``It is
unclear to one of the authors how a concept as lame as the
``anthropic idea'' was ever elevated to the status of a
Principle''.} can be useful: a theory that predicts a too rapidly
decaying proton for example cannot be right, since otherwise we
would not survive the ionizing particles produced by proton decay
in our own bodies. Now although no one would disagree with this,
it is also not very helpful: it does not explain why the proton
lives so long and better experimental limits can be found using
different methods. See \cite{APweb} for a very general use of the
anthropic principle.

\subsection{Anthropic Prediction for the Cosmological Constant}

One of the first to use anthropic arguments related to the value
of the cosmological constant was Weinberg \cite{Weinberg:1987dv},
see also \cite{Banks:1984cw,BarrowTipler}. He even made the
prediction in 1987 that, since the anthropic bound is just a few
orders of magnitude larger than the experimental bounds, a
non-zero cosmological constant would soon be discovered, which
indeed happened.

One can rather easily set anthropic bounds on the value of the
cosmological constant. A large positive CC would lead very early
in the evolution of the universe to an exponentially expanding de
Sitter phase, which then lasts forever. If this would happen
before the time of formation of galaxies, at redshift $z\sim 4$,
clumps of matter would not become gravitationally bound, and
galaxies, and presumably intelligent life, would not form.
Therefore:
\begin{equation}
\Omega_{\Lambda}(z_{gal})\leq\Omega_M(z_{gal})\quad\quad\rightarrow\quad\quad\frac{\Omega_{\Lambda
0}}{\Omega_{M0}}\leq a_{gal}^3=(1+z_{gal})^3\sim 125.
\end{equation}
This implies that the cosmological constant could have been larger
than observed and still not be in conflict with galaxy
formation\footnote{Note that in these estimates everything is held
fixed, except for $\Omega_{\Lambda}$ which is allowed to vary,
unless stated otherwise.}. On the other hand, a large negative
cosmological constant would lead to a rapid collapse of the
universe and (perhaps) a big crunch. To set this lower anthropic
bound, one has to wonder how long it takes for the emergence of
intelligent life. If 7~billion years is sufficient, the bound for
a flat universe is $\Lambda\gtrsim -18.8~\rho_0\sim -2\times
10^{-28}~\mbox{g/cm}^3$, if 14~billion years are needed, the
constraint is $\Lambda\gtrsim -4.7~\rho_0\sim -5\times
10^{-29}~\mbox{g/cm}^3$ \cite{Kallosh:2002gg}.

It makes more sense however, to ask what the most likely value of
the cosmological constant is, the value that would be experienced
by the largest number of observers. Vilenkin's ``Principle of
Mediocrity'' \cite{Vilenkin2}, stating that we should expect to
find ourselves in a big bang that is typical of those in which
intelligent life is possible, is often used. The probability
measure for observing a value $\rho_\Lambda$, using Bayesian
statistics, can be written as:
\begin{equation}\label{BayesCC}
d\mathcal{P}(\rho_\Lambda) =
N(\rho_\Lambda)\mathcal{P}_{\ast}(\rho_{\Lambda})d\rho_\Lambda,
\end{equation}
where $\mathcal{P}_{\ast}(\rho_{\Lambda})d\rho_\Lambda$ is the a
priori probability of a particular big bang having vacuum energy
density between $\rho_\Lambda$ and $\rho_\Lambda + d\rho_\Lambda$
and is proportional to the volume of those parts of the universe
where $\rho_\Lambda$ takes values in the interval $d\rho_\Lambda$.
$N(\rho_\Lambda)$ is the average number of galaxies that form at a
specified $\rho_\Lambda$ \cite{Carroll2000}, or, the average
number of scientific civilizations in big bangs with energy
density $\rho_\Lambda$ \cite{Weinberg2000}, per unit volume. The
quantity $N(\rho_\Lambda)$ is often assumed to be proportional to
the number of baryons, that end up in galaxies.

Given a particle physics model which allows $\rho_\Lambda$ to
vary, and a model of inflation, one can in principle calculate
$\mathcal{P}_{\ast}(\rho_{\Lambda})$, see the above references for
specific models and \cite{Vilenkin:2004fj} for more general
arguments. $\mathcal{P}_{\ast}(\rho_{\Lambda})d\rho_\Lambda$ is
sometimes argued to be constant \cite{Martel:1997vi}, since
$N(\rho_\Lambda)$ is only non-zero for a narrow range of values of
$\rho_\Lambda$. Others point out that there may be a significant
departure from a constant distribution \cite{Garriga:1999bf}. Its
value is fixed by the requirement that the total probability
should be one:
\begin{equation}
d\mathcal{P}(\rho_\Lambda) =
\frac{N(\rho_\Lambda)d\rho_\Lambda}{\int
N(\rho'_\Lambda)d\rho'_\Lambda}.
\end{equation}
The number $N(\rho_\Lambda)$ is usually calculated using the
so-called `spherical infall' model of Gunn and Gott
\cite{Gunn:1972sv}. Assuming a constant
$\mathcal{P}_{\ast}(\rho_{\Lambda})$, it is argued that the
probability of a big bang with $\Omega_\Lambda\lesssim 0.7$ is
roughly 10\%, depending on some assumptions about the density of
baryons at recombination \cite{Weinberg2000,Weinberg:2005fh}.

If $\tilde{\rho}_\Lambda$ is the value for which the vacuum energy
density dominates at about the epoch of galaxy formation, then
values $\rho_\Lambda\gg\tilde{\rho}_\Lambda$ will be rarely
observed, because the density of galaxies in those universes will
be very low. Values $\rho_\Lambda\ll\tilde{\rho}_\Lambda$ are also
rather unlikely, because this range of values is rather small. A
typical observer therefore would measure
$\rho_\Lambda\sim\tilde{\rho}_\Lambda$, which is the anthropic
prediction and it peaks at $\Omega_\Lambda\sim 0.9$, in agreement
with the experimental value $\Omega_\Lambda\sim 0.7$ at the
$2\sigma$ level \cite{Garriga:2003hj}. It is argued that the
agreement can be increased to the $1\sigma$ level, by allowing for
non-zero neutrino masses \cite{Pogosian:2004hd}. Neutrino masses
would slow down the growth of density fluctuations, and hence
influence the value of $\tilde{\rho}_\Lambda$. The sum of the
neutrino masses would have to be $m_\nu\sim 1-2$~eV.

However, it has been claimed that these successful predictions
would not hold, when other parameters, such as the amplitude of
primordial density fluctuations are also allowed to vary
\cite{Banks:2003es,Graesser:2004ng}. These arguments are widely
debated and no consensus has been reached
\cite{Garriga:2005ee,Feldstein:2005bm}.

However, it has been very difficult to calculate the a priori
distribution. The dynamics, leading to a ``multiverse'' in which
there are different pocket universes with different values for the
constants of nature, is claimed to be well understood, for example
in case of eternal inflation
\cite{Vilenkin:1983xp,Linde4,Linde:1986fc}, but the problem is
that the volume of these thermalized regions with any given value
of the constants is infinite. Therefore, to compare them, one has
to introduce some cutoff and the results tend to be highly
sensitive to the choice of cutoff procedure
\cite{Linde:1993xx,Linde:1994gy,Linde:1995uf}. In a recent paper a
different method is proposed to find this distribution
\cite{Garriga:2005av}.

It should be stressed that this approach to the cosmological
constant problem is especially used within string theory, where
one has stumbled upon a wide variety of possible vacuum states,
rather than a unique one
\cite{PolchinskiBousso,Susskind1,Susskind2,Kachru:2003aw,Denef:2004ze,FreivogelSusskind,Douglas:2003um,Ashok:2003gk}.
By taking different combinations of extra-dimensional geometries,
brane configurations, and gauge field fluxes, a wide variety of
states can be constructed, with different local values of physical
constants, such as the cosmological constant. These are the 3-form
RR and NS fluxes that can be distributed over the 3-cycles of the
Calabi Yau manifold. The number of independent fluxes therefore is
related to the number of 3-cycles in the 6-dimensional  Calabi Yau
space, and can be several hundred. In addition, the moduli are
also numerous and also in the hundreds, leading to a total number
of degrees of freedom in a Calabi Yau compactification of order
1,000 or more. The number of metastable vacua for a given Calabi
Yau compactification therefore could be $10^{1000}$, and the
spacing between the energy levels $10^{-1000}M_{P}^4$, of which
some $10^{500}$ would have a vacuum energy that is anthropically
allowed. The states with (nearly) vanishing vacuum energy tend to
be those where one begins with a supersymmetric state with a
negative vacuum energy, to which supersymmetry breaking adds just
the right amount of positive vacuum energy. This picture is often
referred to as the ``Landscape''. The spectrum of $\rho_\Lambda$
could be very dense in this `discretuum' of vacua, but nearby
values of $\rho_\Lambda$ could correspond to very different values
of string parameters. The prior distribution would then no longer
be flat, and it is unclear how it should be calculated.

A review of failed attempts to apply anthropic reasoning to models
with varying cosmological constant can be found in
\cite{Garriga:2000cv}. See \cite{SmolinAnthrop} for a recent
critique. Another serious criticism was given in
\cite{Aguirre:2001zx}, where it is argued that universes very
different from our own could also lead to a small cosmological
constant, long-lived stars, planets and chemistry based life, for
example a cold big bang scenario. An analysis of how to make an
anthropic prediction is made in \cite{Aguirre:2005cj}.

Not very technical and almost foundational introductions to the
anthropic principle are for example \cite{Linde6,Weinberg:2005fh}.

\subsection{Discrete vs. Continuous Anthropic Principle}

It might be worthwhile to make a distinction between a continuous
anthropic principle and a discrete version. Imagine we have a
theory at our hands that describes an ensemble of universes
(different possible vacuum solutions) with different discrete
values for the fine structure constant:
\begin{equation}
\frac{1}{\alpha}=n+\mathcal{O}\left(\frac{1}{n}\right)
\end{equation}
such that the terms $1/n$ are calculable. An anthropic argument
could then be used to explain why we are in the universe with $n
=137$. Such a version of the anthropic principle might be easier
to accept than one where all digits are supposed to be
anthropically determined. Note that we are already very familiar
with such use of an anthropic principle: In a finite universe,
there is a finite number of planets and we live on one of the
(very few?) inhabitable ones. Unfortunately, we have no theory at
our hands to determine the fine structure constant this way, let
alone the cosmological constant.

\section{Summary}

The statistical ideas put forward by Hawking and Coleman turned
out to have serious shortcomings. Unfortunately, no consistent
model evolved from their pioneering work, that could circumvent
these problems.

As we have seen, this is the case with many (if not all) of the
proposed solutions to the cosmological constant problem. It is
therefore understandable that nowadays the majority of researchers
in this field places their bets on the anthropic principle as a
solution. These anthropic ideas however especially appear to
highlight the problem, instead of giving an explanation.

Anthropic reasoning necessarily requires an ensemble of objects or
situations in order to be meaningful. This works very well, when
applied, for example, to our planet and its distance to the sun.
There are many more planets, but we happen to live on one at the
right distance to have a temperature that makes life as we know it
possible. In general, if one starts to wonder about the size of
the earth, the sun, the solar system or the galaxy, there are two
ways to proceed. One is to look for a fundamental physical reason
why the diameter of the earth has its particular value and not
some value, say, a little bigger. The other is to realize (or
assume, in case of an Old Greek scientist) that there are more
objects like the earth and try to say something about the
properties of these objects in general. In the history of
cosmology this has turned out to be very fruitful and by now we
see that the earth, our solar system and our galaxy are by no
means special. The ultimate shift from the particular to the
general would be made by considering multiple universes. This is
the starting point of modern anthropic arguments based on some
kind of mechanism. However, as a matter of principle we can only
do experiments, and therefore statistics based on one universe.
Other pockets of our universe, or other universes where the
cosmological constant takes on a different value, will never be
accessible to experiment. Therefore, it seems very legitimate to
ask whether such an `explanation' can ever be falsified, let alone
verified.

Besides, as we have discussed, the bounds obtained on the value of
the effective cosmological constant from applying anthropic
reasoning, are not very restrictive. The probability of finding
oneself in a universe where $\Omega_\Lambda\simeq 0.7$ is only
about 10\%.

This very much debated approach offers a new line of thought, but
so far, unfortunately, predictions for different constants of
Nature, like the cosmological constant and the fine-structure
constant, are not interrelated. We continue to search for a more
satisfactory explanation.

%% file: appendixa.tex
\chapter{Conventions and Definitions}\label{AppA}

Throughout this thesis, we use a metric $(-,+,+,+)$ and write
Einstein's equations as:
\begin{equation}
R_{\mu\nu} - \frac{1}{2}Rg_{\mu\nu} - \Lambda g_{\mu\nu} = -8\pi
GT_{\mu\nu}.
\end{equation}
The minus sign with which the cosmological constant enters the
Einstein equation, means that we take the action to be
$\propto(R+2\Lambda)$. Besides, we employ natural units, in which:
\begin{equation}
\hbar=c=k_B=1.
\end{equation}
Using:
\begin{eqnarray}
\hbar&=&1.054571596\times 10^{-34}\;\mbox{Js}\nonumber\\
c &=& 2.99792458\times 10^8\;\mbox{m/s}\nonumber\\
k_B&=& 1.3806503\times 10^{-23}\;\mbox{J/K}.
\end{eqnarray}
and $E=mc^2$, only one remaining unit needs to be chosen, usually
taken to be mass. The dimensions, denoted by square brackets
become:
\begin{equation}
[\mbox{energy}]=[\mbox{temperature}]=[\mbox{mass}],\quad\quad
[\mbox{time}]=[\mbox{length}]=[\mbox{mass}]^{-1}.
\end{equation}
Masses are expressed in units of GeV, which can easily be
converted back into SI-units, using the following expressions:
\begin{table}[h!]\label{Units}
\begin{tabular}{|ll|}
\hline &\\
  energy & $1\; \mbox{GeV}=1.6022\times 10^{-10}$~J \\
 \hline &\\
  temperature & $1\; \mbox{GeV}=1.1605\times 10^{13}$~K \\
  \hline &\\
  mass & $1\; \mbox{GeV}=1.7827\times 10^{-27}$~kg \\
   \hline &\\
  length & $1\; \mbox{GeV}^{-1}=1.9733\times 10^{-16}$~m\\
   \hline &\\
  time & $1\; \mbox{GeV}^{-1}=6.6522\times 10^{-25}$~s\\
   \hline
\end{tabular}
\end{table}\\
To denote the cosmological constant, we use a capital $\Lambda$,
which has dimension $[\mbox{GeV}]^2$, and a lower case letter
$\lambda$ to denote vacuum energy density, which has dimension
$[\mbox{GeV}]^4$, and is related to $\Lambda$:
\begin{equation}
\Lambda = 8\pi G\lambda.
\end{equation}

A collection of useful quantities is listed in the following
table:
\begin{table}[h!]\label{Otherquant}
\begin{tabular}{|ll|}
\hline &\\
  Gravitational constant & $G=6.6726\pm 0.0009\times 10^{-11}\mbox{m}^3\mbox{kg}^{-1}\mbox{s}^{-2}$ \\
 \hline &\\
  parsec & $1\; \mbox{pc}=3.0856\times 10^{16}$~m \\
  \hline &\\
  Solar mass & $1\; M_{\odot}=1.989\times 10^{30}$~kg \\
   \hline &\\
  Dimensionless Hubble parameter & $h=H_0/100\;\mbox{km}\mbox{s}^{-1}\mbox{Mpc}^{-1}$\\
   \hline &\\
 Density of the Universe & $\rho_0=1.8789\times 10^{-26}\;\Omega h^2\;\mbox{kg}\mbox{m}^{-3}$\\
 & $=2.7752\times 10^{11}\;\Omega h^2\; M_{\odot}\mbox{Mpc}^{-3}$\\
 & $= 11.26\; h^2\;\mbox{protons}/m^3$\\
   \hline
\end{tabular}
\end{table}

%% file: appendixb.tex
\chapter{Measured Values of Different Cosmological Parameters}

\section{Hubble Constant}

The Hubble Space Telescope (HST) key project has determined the
value of the Hubble constant to about 400 Mpc with various
secondary indicators based on the primary cephid distance. Their
result, as well as the results of the secondary indicators are
listed in table (\ref{ValueHubbleConstant}):
\begin{table}[h!]\label{ValueHubbleConstant}
\caption{The Value of Hubble's Constant}
\begin{tabular}{|l|c|}
\hline &\\
  HST & $H_0 = 75 \pm 10 \; $km/s/Mpc \\
 \hline &\\
  Type Ia SNe & $H_0 = 71 \pm 2 \;\mbox{stat}\;\pm 6 \;\mbox{syst}$ km/s/Mpc \\
  \hline &\\
  Tully-Fisher relation &   $H_0 = 71 \pm 3 \;\mbox{stat}\;\pm 7\; \mbox{syst}\;$ km/s/Mpc \\
   \hline &\\
  Surface brightness flutuations &   $H_0 = 70 \pm 5\; \mbox{stat}\;\pm 6\; \mbox{syst}\; $ km/s/Mpc \\
   \hline &\\
  Type II SNe &   $H_0 = 72 \pm 9 \;\mbox{stat}\;\pm 7\; \mbox{syst}\;$ km/s/Mpc \\
   \hline &\\
  Fundamental plane of elliptical galaxies &   $H_0 = 82 \pm 6 \;\mbox{stat}\;\pm 9 \;\mbox{syst}\;$ km/s/Mpc \\ \hline
\hline &\\
  FINAL RESULT &   $H_0 = 72 \pm 8$ km/s/Mpc, \cite{Freedman2000}\\ \hline
\end{tabular}
\end{table}

More recent (2005) data \cite{Sanchez:2005pi} from the CMB and
2dFGRS (large scale structure) give $H_0= 74 \pm 2$.

Note that often the Hubble parameter is parameterized in terms of
a dimensionless quantity $h$ as:
\begin{equation}
H_0=100 h\mbox{km/sec/Mpc}
\end{equation}

The Sunyaev-Zel'dovich effect provides another way of determining
$H_0$, but it suffers from large systematic errors. Following this
route, the result \cite{Reeseetal2002} is: $H_0 = 60 \pm
4^{+13}_{-18} \; km/s/Mpc$, consistent with the HST-result.

\section{Total Energy Density}

From CMBR measurements
\cite{deBernardis2000,Bennett2003,Spergel2003}:
\begin{equation}
0.98 \lesssim \Omega_{tot}\lesssim 1.08,
\end{equation}
under the assumption that the Hubble parameter $h> 0.5$.

\section{Matter in the Universe}

Traditionally $\Omega_M$ is determined by ``weighing'' a cluster
of galaxies, divide by its luminosity, and extrapolate the result
to the universe as a whole. Clusters are not representative
samples of the universe, but sufficiently large that such a
procedure might work. Applying the virial theorem to cluster
dynamics typically yielded values: $\Omega_M=0.2\pm 0.1$
\cite{Dekel:1996aw,Bahcall:1997ia,Bahcall:1998ur}.

Another way to determine $\Omega_M$ is through the value of baryon
density, which would also include dark matter
\cite{White:1993wm,Fukugita:1997bi,Mohr:1999ya}. These
measurements imply: $\Omega_M=0.3\pm 0.1$

Also measurements of the power spectrum of density fluctuations
gives information on the amount of matter in the universe, but
this information is dependent on the underlying theory, and on the
specification of a number of cosmological parameters,
\cite{Peacock:1993xg,Eisenstein:1997jh}. The result is identical
to the precious method: $\Omega_M=0.3\pm 0.1$.

The total amount of baryons, again from CMBR measurements
contributes about:
\begin{equation}
\Omega_{B}=0.024\pm 0.0012\; h^{-2}\quad\quad\rightarrow \Omega_B
\approx 0.04 - 0.06,
\end{equation}
with $h=0.72 \pm 0.7$. This is a total amount of baryons, being
luminous or not. Therefore we can conclude that most of the
universe is non-baryonic.

The dark matter contribution contributes about $\Omega_{DM}\approx
0.20 - 0.35$. The need for dark matter results from a host of
observations, relating large scale structure and dynamics, for a
summary, see \cite{Peebles:2004qg}. The prime candidate consists
of weakly interacting massive particles, so-called `WIMPS'.

Combing the results for the total energy density and the matter
energy density, one is led to conclude that there must be at least
one other component to the energy density, contributing about 70\%
of the critical energy density.

\section{Dark Energy Equation of State}

An important cosmological parameter is the dark energy equation of
state $w=p/\rho$, which is exactly equal to $-1$ for a
cosmological constant. Different methods have been used to measure
this parameter and they are listed in the following table
(\ref{ValueEoS}):
\begin{table}[h!]\label{ValueEoS}
\caption{The Value of the Dark Energy Equation of State Parameter}
\begin{tabular}{|l|c|}
\hline &\\
  SNIa and CMB & $-1\leq w \leq -0.93$ \cite{CorasanitiCopeland2001} \\
 \hline &\\
  SNIa, CMB, HST, large scale structure & $w \leq -0.85$ at $1\sigma$ and $w\leq -0.72$ at $2\sigma$ \cite{Beaneal2002}\\
  \hline &\\
 X-ray clusters and SNIa & $w=0.95^{+0.30}_{-0.35}$ assuming flat universe \cite{Schueckeretal2002}  \\
   \hline &\\
  WMAP &   $w < -0.78$ at $2\sigma$ \cite{Bennett2003,Spergel2003}\\
   \hline
\end{tabular}
\end{table}

By combining results of seven CMB experiments, data on large scale
structure, Hubble parameter measurements and supernovae results,
bounds found are $-1.38 \leq w \geq -0.82$ at 95 \% confidence
level in \cite{Melchiorri:2002ux}.

\section{Summary}

This brings us to the following list of values of cosmological
parameters:
\begin{table}[h!]\label{SumValues}
\caption{Concordance Model}
\begin{tabular}{|c|}
\hline \\
$\Omega_{DE} \approx 0.7$ \\
\hline \\
$\Omega_{DM} \approx 0.26$ \\
\hline \\
$\Omega_{B} \approx 0.04$ \\
\hline \\
$\Omega_R \approx 5\times 10^{-5}$\\
\hline\\
$h \approx 0.72\pm 8$ \\
\hline\\
$-1.38 \leq w \geq -0.82$\\ \hline
\end{tabular}
\end{table}